\documentclass[11pt]{article}
\renewcommand{\baselinestretch}{1.3}

\usepackage{pdflscape}
\usepackage{multirow}
\usepackage{comment}
\usepackage{verbatim}
\usepackage{amsfonts}
\usepackage{amsmath}
\usepackage[pdftex]{graphicx}
\usepackage{longtable}
\usepackage{lscape}
\usepackage{pdflscape}
\usepackage{booktabs}
\usepackage{geometry}
\usepackage[toc,title,page]{appendix}
\usepackage{hyperref}
\usepackage[english]{babel}
\usepackage[utf8]{inputenc}
\usepackage{graphicx,kantlipsum}

\usepackage{natbib}
\usepackage{floatrow}
\DeclareFloatFont{tiny}{\tiny}  
\usepackage{rotating}
\usepackage{float}

\title{Life after (Soft) Default}

\author{Giacomo De Giorgi\footnote{GSEM-University of Geneva, BREAD, CEPR, and IPA., Uni-Mail,  40  Bd.  du  Pont  d'Arve, CH-1211 Geneva 4, Switzerland. Giacomo De Giorgi acknowledges financial support from the SNF grant $\#100018\_182243$. We thank the editors, and referees for the insightful suggestions. We also thank Yujung Hwang, Pietro Campa, Davide Pietrobon, Enrico Moretti, Harrison Wheeler, Jaromir Nosal for providing feedback and support along the way.} 
\and
Costanza  Naguib\footnote{University of Bern}.
 }

\begin{document}

\def\spacingset#1{\renewcommand{\baselinestretch}%
{#1}\small\normalsize} \spacingset{1}

\date{} 
\baselineskip18pt
\maketitle 

\begin{abstract} 
\noindent Soft default, defined as a delinquency of 90 days or more, is a relatively common event in the credit market, in 2010 such episodes affected about 3 million individuals. Yet we lack a detailed understanding of what happens afterward. We use  credit report data, on approximately 2 million individuals from 2004 to 2020, to shed light on individual trajectories after such event,  and document enduring negative impacts. These effects persist for up to ten years post-event and manifest in lower credit scores, reduced total credit limits, lower homeownership rates, lower income, and relocation to less economically active zip codes. It appears that those who are overextended in their mortgage lines, and with larger delinquent amounts, suffer the harshest consequences.

\end{abstract}



\section{Introduction}

In this paper we analyze what happens to an individual after  a soft default, defined as a delinquency of 90 days or more.\footnote{Throughout the current paper we will use the notion of soft default to distinguish that from   harsh default defined as foreclosure, chapter 7 or 13 bankruptcy.} We show that, in the short and medium run, soft defaults trigger individual relocation across ZIP codes in the US, in particular towards cheaper ones, and also to different Commuting Zones (CZ). 
A soft default has long-lasting impacts on credit score,  total credit limit, homeownership status, and income. 
Given that the occurrence of default isn't typically random,  we adopt several strategies to recover the  impact of default, such as modified event studies \cite{BorusyakJaravelSpiess2021}, and a double/debiased lasso estimation technique proposed by \cite{c2018}. 
Default and delinquencies are not uncommon, in our large   1\% sample   of the US population with  credit reports, we note that new soft defaults in 2010  affected almost 1.5\% of the population, or about 3 million people in that year. 

So what happens to these people afterward? Are their lives in jeopardy, or can they recover quickly? 
We show that the recovery is slow, painful, and in many respects only partial. In particular, after several years, up to 10, credit scores are still lower by 16 points, incomes never recover and appear to be substantially lower (by about 7,000USD or 14\% of the 2010 mean),  the defaulters live in lower ``quality'' neighborhoods (as measured by the median house value and other indicators such as proxies for average zip code income), are less likely to own a home, and are more likely to have low credit limits. We find  that the negative effects of a soft default are larger for those individuals who are overextended in their credit lines, in particular the ones of mortgage. Being indebted in a way that is unsustainable for them in the long run, such individuals have also a higher probability of a subsequent harsh default (i.e. Chapter 7, Chapter 13, foreclosure). In addition, they end up in substantially worse neighborhoods, with  lower median home values, and these moves are likely to have a substantial effect also on their labor market outlooks. 

Our interest, in this paper, is on  tracing  individuals' lives after a  default, while we do not aim to distinguish between different default motives (see for example \cite{GanongNoel2019} and the literature cited therein). More generally, our work is related to two strands of the literature, one  which focuses on the individual determinants of default, whereas the other focuses on the individual and social costs of default and on the analysis of debt relief policies.  For example, \cite{lawrence1995consumer} belongs to the first strand, she builds a theoretical model of consumer's choices with default as a possible option over the life-cycle. Similarly, \cite{GuisoSapienzaZingales2013} study the determinants of strategic default, i.e. when one does not repay the mortgage, even if she would be able to do so, because the house value has fallen  below the value of the mortgage. \cite{giliberto1989relocation} develop a theoretical model of residential mortgage default when borrowers face beneficial as well as costly relocation opportunities.  \cite{mnasri2018downpayment} finds that both income and geographical mobility are main trigger factors of default. In fact, households with a higher mobility rate (i.e. young households) are more likely to default. In general, all these studies find that default is more likely for unmarried, renters, and for those who have already moved from their birthplace.

A second strand of the literature aims at assessing what happens to individual life trajectories after a default. This literature essentially focused on the impact of a harsh default, i.e. either Chapter 7 or Chapter 13 declarations or a foreclosure. Our work sheds some light on the short and medium term consequences of a soft default, an event that is substantially more likely (e.g. 1.5 versus 1 percent in 2010). 
To this second strand of the literature belong, for example, \cite{CollinsonHumphriesMaderReedTannenbaumvanDijk2023}, who investigate the impact of eviction on low income households in terms of homelessness, health status, labor market outcomes, and long term residential instability. Similarly, \cite{currie2015there} show that foreclosure causes an increase in unscheduled and preventable hospital visits.
\cite{albanesi2018insolvency} investigate the impact of the 2005 bankruptcy reform, which made it more difficult for individuals to declare either Chapter 13 or Chapter 7. They find that the reform hindered an important channel of financial relief. \cite{diamond2020effect} analyze the negative impact of foreclosures on foreclosed-upon homeowners. They find that foreclosure causes housing instability, reduced homeownership, and financial distress. Finally, \cite{indarte2022costs} analyzes the costs and benefits of household debt relief policies. 

In the current paper, we quantify the impact of a soft default on individuals' mobility and residential choices, credit access and utilization, and income in the short and medium run. Since individuals may be hit by a soft default in different years, we are in the framework of multiple groups and multiple periods outlined in \cite{CALLAWAY2021200} (also known as a staggered design). Hence, all the results presented in Section 3 and in Section 4 of the paper have been obtained with their estimator for event studies. 

In the baseline estimates, we only control for age, age squared, to take into account standard lifecycle dynamics, and credit score, as a synthetic indicator for creditworthiness,  in the first two years of our sample. In Appendix \ref{sec:robustness} we probe the robustness by including further control variables, such as state dummies and the value of the credit score in the two years preceding the soft default, and the results do not relevantly change. 
Further, in Appendices \ref{sec:ml}, \ref{sec:DML}, and \ref{sec:long} we focus exclusively on soft defaults happening in 2010 and use the dobule/debiased Lasso estimation technique proposed by \cite{c2018}. In this second estimation exercise,  we control for up to 6 years of prior credit market behavior including past credit scores, a leading summary statistics for the probability of default, the life cycle through a quadratic function of age, and several other variables for past credit market access and usage, local labor and credit market conditions, plus location and time fixed effects. The results obtained with the two estimation methods are broadly consistent.
 
In order to avoid that our estimated impacts of a soft default are confounded by previous or contemporaneous harsh default, we exclude from the estimation sample all those who experience a harsh default either before or in the same year of a soft default.

Ultimately, we  quantify the impact of soft  defaults as hampering the ability of the individual to borrow for several years. A soft default  mechanically lowers the credit score by about 100 points (on impact), with a recovery that typically takes more than 5 years (the default flag will typically stay on a credit report for 7 years). More interestingly, such default triggers an increase in the probability of  relocation across zip codes by  3-4pp and an increase by  1-2pp in the probability of moving outside the original commuting zone. Income drops by  5,000-7,000USD in the short and medium run respectively, and the probability of opening a new mortgage falls by  10\%.
 Finally, a soft default increases, by  15-20pp, the probability of having a low revolving credit limit ($<$10,000USD), wipes away  6,000USD in the revolving credit balance, and increases by 3-10pp the probability of also experiencing a harsh default, a foreclosure or bankruptcy, in the two years following the soft default.
These effects are sizeable and long-lasting, for example in terms of income losses (about 14\%) the effects are equivalent to 1 to 2 years of schooling \citep{card2001estimating}. Ultimately, one cannot but wonder on whether basic debt relief policies, or a more accommodating treatment of soft default by credit bureaus, would be avoiding such large negative consequences. Of course, reducing the costs associated to default might have substantial consequences on credit access to start with.

Our analysis of heterogeneity of the effects, to be interpreted with   caution, offers a window into the mechanisms at play.\footnote{We thank the  editor and the referees for pushing us in this direction.} Caution is needed, as our split of the full sample into the subsamples of those who also experience a harsh default in the years after the soft default and of those who don't is not exogenously determined. 
In the same way, when we claim that those who are delinquent by a larger amount experience harsher consequences, no guarantee of causality can be given.

However, our heterogeneity analysis, even if mostly descriptive in nature, is economically interesting as it contributes to our understanding of the mechanisms underlying the harsh negative consequences of soft defaults. 
We find that the defaulters on larger amounts or with a subsequent harsh default have substantially higher penalties in terms of income and location,  they move to lower  median home values areas and to zip codes of lower economic activity. 
What seems to be happening is that there are individuals who are delinquent on smaller amounts, possibly because of uninsurable shocks, who suffer the consequences of such defaults, but substantially less than those who default on larger amounts and seek bankruptcy and other legal reliefs. The latter appear to have overextended their lines of credit, in particular on mortgages (presumably because of location choices), then gone under in their accounts and essentially diverged from their earlier life trajectories. They end up in substantially worse neighborhoods (of different CZs) with median home values that decrease about 4-times as much as those for the lower delinquent amounts/no-harsh default. 
These moves to new CZs seem also to substantially affect the labor market outlooks for this population,  their yearly income falls  by almost 10,000USD (about 5-times as much as for the low delinquent amounts). 

In the rest of the paper we present the data used in Section \ref{sec:data}, an event study  analysis in Section \ref{sec:exploratory}, heterogeneity analysis in Section \ref{sec:heterogeneity}, and a summary of the potential mechanisms at play in Section \ref{sec:taking_stock}.
Section \ref{sec:conclusions} provides some concluding remarks. A series of robustness checks and additional results are presented in the Appendices.

\section{Data}\label{sec:data}
We rely on a  proprietary dataset on credit reports from  Experian (see for example \cite{DeGiorgiHardingVasconcelos2021, bach2023born} or for  similar dataset \cite{lee2010introduction} or \cite{albanesi2019predicting}). This dataset includes information on the credit scores, and more than other 400 credit variables, plus basic socio-demographic such as date of birth, zip codes of residence, and imputed incomes, for about a 1\% sample of the total population of the US with valid credit reports in 2010. Hence, we can rely on a large sample size for our statistical analyses, as much as a panel of over 2 million individuals per year for the period 2004-2016 and then two final waves for 2019 and 2020. All the data used are annual, and they are drawn on June 30th of each year.

To be more precise, the data include a series of variables on the number of accounts  (referred to as trades) and several variables measuring credit behavior (i.e. number of bankruptcies, number of credit delinquencies, number of credit cards, average amount of credit and so on). Beyond the rich credit information, which of course includes mortgages and car loans, we have information on  age and a measure of imputed income, which has been constructed by Experian based on W2's. The exact procedure for imputation is a property of Experian, however in Figures \ref{fig:representativeness} and \ref{fig:income} in Appendix \ref{sec:reliability} we check our data reliability in terms of both  representativeness and validity of the income imputation, there we confirm that the data provided by Experian appear to be  consistent with those obtainable from the Census or from the IRS. 

In Appendix \ref{sec:reliability} we provide detailed evidence of the representativeness of the income variable imputed by Experian by checking its representativeness against Panel Study of Income Dynamics (PSID) and Census data, across ages, years, and counties.
Further, \cite{lee2010introduction} have long shown the validity and representativeness of the New York Fed Consumer Credit Panel, which is constructed on the basis of similar  data to ours  but by Equifax credit bureau. 

In Table \ref{tab:descriptives2010}, we report some basic summary  statistics for our main variables of interest, including the Median House Value. This variable is based on the House Price Index (HPI).\footnote{Data on the House Price Index are taken from the Federal Housing Finance Agency. \url{https://www.fhfa.gov/DataTools/Downloads/Pages/House-Price-Index-Datasets.aspx}}. The HPI is based on the growth rates of prices of the same housing objects (single family houses) in repeated sales and hence has no direct interpretation on the comparison between different zip codes. In order to solve this issue, we use the growth rates from FHFA and then peg them to median home values from the census (2000 or 2010) tracts, which are crosswalked to zip codes in those years.
In Appendix \ref{sec:harsh_soft} we briefly discuss how credit bureaus  treat different types of default, in particular we know that Experian keeps soft defaults for 7 years, and Chapter 7 flags for up to 10 years, while Chapter 13 (and other delinquencies) for up to 7 years, this in accordance to the Fair Credit Reporting Act.\footnote{\url{https://www.ftc.gov/system/files/documents/statutes/fair-credit-reporting-act/545a_fair-credit-reporting-act-0918.pdf}.}

To get a sense of the sample, we quickly scan through the main variables of interest for our base year (2010) in Table \ref{tab:descriptives2010} (a similar table for the entire available period (2004-2020) is  in the Appendix, Table \ref{tab:descriptives}). For example, the average credit score  is 683, with an average (nominal) income (W2 derived) of 50,770USD, and an average Median House Value of 195,809USD. The average age is 51. On average, in our sample, about 17\% of the individuals moved to a different zip code in 2010 and about 6\% moved to a different commuting zone. Both statistics are computed with respect to the residential zip code in 2009. 
The share of individuals experiencing a new 90+ days delinquency (our treatment of interest) in year 2010 is about 1.5\%.

For mortgages, the average size  is about 78,000USD (including zeros), 
and about 40\% of individuals in the sample have a mortgage. Among those with a positive mortgage amount, the average is 194,086USD.  

In the analysis, we use the following two definitions of homeownership. First, we consider an individual as a homeowner if either she ever had a mortgage or she is recorded as a homeowner according to Experian's imputation. With this comprehensive homeownership definition, about 70\% of individuals in our sample are homeowners (vs 40\%  if we consider only individuals with a positive open mortgage amount). Second, in an alternative definition of homeownership, we consider the origination of new mortgages. In this second case, we define a mortgage origination as a situation in which either the number of open mortgage trades in year $t$ is bigger than the number of open mortgage trades in year $t-1$ or the number of months since the most recent mortgage trade has been opened is lower than 12. Clearly, this definition would only capture the flow, and perhaps more importantly would miss cash purchases and wouldn't distinguish between a new mortgage and a remortgage.

\begin{table}[H]\centering \caption{Summary statistics of our main variables, 2010, balanced panel. Individuals who experienced a harsh default before or in the same year as a soft default in the sample period (i.e. from 2004 onwards) have been dropped. Special codes credit scores lower than 300 have been trimmed. Similarly, the  top 1\% of total credit limit, total balance on revolving trades and total revolving limit have been trimmed. \label{tab:descriptives2010}}
\normalsize
\begin{tabular}{l c c c c c}\hline\hline
\multicolumn{1}{c}{\textbf{Variable}} & \textbf{Mean}
 & \textbf{Std. Dev.}& \textbf{Min.} &  \textbf{Max.} & \textbf{N}\\ \hline
Credit Score & 683.3953 & 114.8561 & 300 & 839 & 2159292\\
Income & 50770.9571 & 26624.4062 & 1000 & 331000 & 2159953\\
Median House Value zip code & 195808.6002 & 145783.0996 & 0 & 1631909.75 & 1845828\\
Mortgage Bal. & 78065.0424 & 175184.4459 & 0 & 16151587 & 2162467\\
Mortgage origination &  0.0684 & 0.2524 & 0 & 1 & 2162468\\
Home-owner\footnote{This is a dummy variable that takes value 1 if either the individual has ever had a mortgage or if she is recorded as homeowner by Experian (imputed variable) and zero otherwise} & 0.7023 & 0.4572 & 0 & 1 & 2162468\\
Age & 50.706 & 15.5872 & 18 & 123 & 2162467\\
Move ZIP & 0.1677 & 0.3736 & 0 & 1 & 2162445\\
Move CZ & 0.0596 & 0.2367 & 0 & 1 & 2136613\\
Total credit limit on open trades (all)  & 151743.4705 & 221859.5747 & 0 & 17505724 & 2162467\\
Total balance on revolving trades & 11655.8136 & 41255.8252 & 0 & 5074756 & 2162467\\
Total credit limit on open rev. trades & 34211.3796 & 68332.2798 & 0 & 5566654 & 2162467\\
Prob. cred. lim $<$10k & 0.1001 & 0.2987 & 0 & 1 & 2162468\\
Soft default & 0.0143 & 0.113 & 0 & 1 & 2162467\\
Harsh default & 0.0102 & 0.1104 & 0 & 1 & 2162467\\
Number of collections & 1.8741 & 4.5258 & 0 & 90 & 2162467\\
Amount 90-180 days delinquent & 417.8958 & 5974.6461 & 0 & 3532937 & 2162467\\
Credit Card balance open & 4511.5197 & 9528.3741 & 0 & 510286 & 2162467\\
 \hline\end{tabular}
\end{table}

Average total credit limit on all open trades is about 152,000USD. This variable is defined as the total credit amount on all open trades (for example, it includes both auto and mortgage debt). 
No individual in our sample  has a credit limit equal to zero (that is almost mechanical, as to be in the credit bureau a credit line is needed). The same holds true for the total credit limit on open revolving trades (which is defined as the total credit limit on all open revolving trades with credit limit larger than zero), which has an average of about 34,000USD. Finally, the total balance on revolving trades has an average of about 12,000USD in the whole dataset, and about 90\% of the individuals in the data have a total balance on revolving trades greater than zero. Among those, the average recorded is about 16,000USD.

As far as soft and harsh defaults are concerned, as mentioned in the Introduction, in the empirical analysis (event-study) we exclude all those individuals who experienced a harsh default either in the years before or in the same year as a soft default\footnote{More precisely, in the event-study analysis we exclude all those individuals who experience at least one harsh default (defined as a new Chapter 7 or Chapter 13 declaration, or a new foreclosure) between 2004 (the first year in our sample) and the year in which they were first hit by a soft default. This means that also individuals that experience a harsh default in the same year as a soft default are excluded from the estimation sample. Eliminating individuals with a previous or contemporaneous (to a soft default) harsh default leads to dropping 16,153 individual-year observations over more than 28 mio. This corresponds to dropping less than 0.1\% of the total sample size. 
In Appendix \ref{sec:ml}, \ref{sec:DML}, and \ref{sec:long} we focus on individuals who have been hit by a soft default in 2010. Hence, in that case we drop from the sample all those who have been hit  by at least one harsh default between year 2004 and year 2010 (extremes included) and all those who have been hit by a soft default between year 2004 and 2009 (extremes included).}.


As mentioned above, we define soft default as a 90-day (or more) delinquency. In essence, as often used in the literature and by industry standard, a 90 days delinquent credit is a defaulted one.\footnote{\url{https://www.experian.com/blogs/ask-experian/what-is-the-difference-between-delinquency-and-default/} or \cite{albanesi2019predicting}.}

Finally, the average number of collections is almost 2, the average amount that is between 90 and 180 days delinquent (zeros included) is about 418USD (8,341USD if we exclude the zeros) and the average credit card balance open is about 4,500 USD (6,751USD if we exclude the zeros) in year 2010. This last variable is defined as the total balance on all open credit card trades reported in the last 6 months. In what follows, we use this variable as a proxy of consumption and hence we refer to it as "credit card consumption", this is akin \cite{gross2002liquidity} with the caveat that in credit report data we cannot distinguish pure consumption balances, the purchase of goods and services, and interest rate payment and various fees.

\section{Event study analysis}\label{sec:exploratory}

\subsection{Post soft defaults}

The aim of this Subsection is to clarify to which extent we are able to claim that each of our results is or is not causal.
We are interested in assessing the impact of a soft default on relevant outcome variables such as mobility, income, and mortgages. Of course, in many instances, default is endogenous to the outcomes of interest. For example, income may be already declining, and that may lead  to a soft default. Similarly, mobility may be the cause, and not the consequence, of a soft default. In our empirical estimations, while we recognize the fundamental descriptive nature of our exercises, we try to assuage the endogeneity concerns in the following two ways. 

First, in this Section, we present a series of event studies in the flavor of \cite{CALLAWAY2021200}. The idea is that by employing such an approach, which  reweights observations based on a set of controls to enhance comparability while aligning the event times, we are moving closer to the ideal experiment of a random default. We find this exercise, which in practice is based on the conditional parallel pre-trends between control and treated units assumption, quite instructive. Even if we were to interpret it as purely descriptive. Moreover, we find it reassuring that there is no strong evidence of pre-trends in most outcomes, as we will see later.

Second, in Appendix \ref{sec:DML} we present an analysis based on the double debiased machine learning estimation method proposed by \citep{c2018}.  We argue that with this  method, which allows for many controls, we are able to obtain soft defaults that are conditionally exogenous, and, hence, we can recover a valid causal effect of such defaults on the outcomes  of interest. In this exercise, due to the computational burden, we focus on the population of individuals who had no defaults up to 2010 and define the treatment group as those individuals who default in 2010, disregarding the other cohorts of treated. Our rich set of control variables includes  age, age squared, the amount of open mortgages and car loans, as well as the individual credit score, in each of the pre-event years, i.e. from 2004 to 2009, commuting zone fixed effects (different commuting zones are populated by different individuals), county unemployment rates (to control for local labor market conditions), number of bank closings by county in each of the pre-event years (to control for changes in local credit supply), and maximum interest rate allowed by the anti-usury laws in each state in each year of our dataset (this controls for local variation in the maximum  interest rate charged of different trades). These controls are relevant, as they are selected by the algorithm in the large majority of the cases.


The results obtained with the two estimation methods, i.e. the event studies and the DML,  are broadly consistent, hence providing  further reassurance on the estimated effects.

\subsection{Event study results}
In this Section we provide event-study like evidence on the effect of soft default on individual relocation probabilities, income, homeownership, probability of low total credit  limits (below 10,000USD), and balances on revolving credits. We limit this analysis to the years 2004-2016, as we don't have data for  2017 and 2018. 
In Figure \ref{fig:event2}, we report the result of such an approach, in which the event is a soft default.

Since different individuals may experience a soft default in different years, we are in the framework of an event study with multiple periods and multiple cohorts of treated individuals, as described by  \cite{CALLAWAY2021200} or \cite{BorusyakJaravelSpiess2021}. 

We therefore proceed in our analysis using a modified event study, allowing for heterogeneous and dynamic treatment effects as in \cite{CALLAWAY2021200}.
We define, as controls, those individuals who never experienced a soft default, in our time periods, and we apply the re-grouping by cohorts defined by the year of first default. Control and treated individuals are matched on a series of controls: age, age squared, to account for standard life cycle profiles, and credit score in 2004 and in 2005, i.e. the first two years in our sample, to control for credit worthiness.
All the reported standard errors are clustered at the individual level. In Appendix \ref{sec:control2016}, we probe the robustness of the results to a different control group, there we use as a control those who experience a default in 2016, the last year of continuous observation.

Following the original article of \cite{CALLAWAY2021200}, we define  the group-time $(g,t)$ average treatment on the treated effect as:
\begin{equation}\label{eq:callawaysantanna}
  ATT(g,t) = E[Y_t(g)-Y_t(0)|G_g=1]
\end{equation}
where $t$ is the year, that goes from 1 to $\textbf{T}$.  $G$ defines which "group" of defaulters (or calendar-year cohort, e.g. 2010) an individual belongs to. If an individual is never treated, i.e.  never  defaults in the current application, we arbitrarily set $G=\infty$. We define $G_g$ to be a binary variable that is equal to one if an individual is first treated in calendar year $g$ (i.e. $G_{i,g}=\mathbf{1}\{G_i=g\}$, see \cite{CALLAWAY2021200}, p. 203).

$Y$ is the outcome  of interest (e.g. income, credit score, probability of moving ZIP code, probability of moving commuting zone, probability of mortgage origination, probability of having a total credit limit smaller than 10,000USD, probability of experiencing a harsh default, revolving credit balance, Median Home Value of the zip code of residence).

Equation (1) defines the average treatment effect for individuals who are members of a particular group $G$ at a particular time (year) $t$, this parameter is the group-time average treatment effect. Empirically, the quantity in equation (1) is estimated nonparametrically by inverse probability weighting. More precisely, the estimand is the following object:

\begin{equation}
     ATT_{ipw}(g,t)= E \left[ \left( \frac{G_g}{E(G_g)}-\frac{\frac{p_g(X)C}{(1-p_g(X))C}}{E(\frac{p_g(X)C}{(1-p_g(X))C})} \right)(Y_t - Y_{g-1})\right],
\end{equation}
where $ipw$ stands, indeed, for inverse probability weighting, $G_g$ and $Y_t$ are defined in the same way as above, and $C$ is a binary variable that is equal to one for units that do not participate in the treatment in any time period (i.e. $C_i=\mathbf{1}\{G_i = \infty\}$). $p_g(X) = P(G_g=1 |X, G_g+C=1)$ is a propensity score which denotes the probability of being first treated in period $g$ conditional on covariates $X$ and either being a member of group $g$ or not participating in the treatment in any time period  (see \cite{CALLAWAY2021200}, p. 203). As mentioned above, in our framework $X$ includes age, age squared and credit score in 2004 and 2005. In Appendix \ref{sec:robustness} we define $X$ as including age, age squared, state fixed effects and credit score in the two years preceding the soft default.

As usual in difference-in-difference setups with multiple time periods, we are interested in estimating the treatment effect dynamics, i.e. how does the treatment effect vary over time since treatment (the event of a soft default here). Let $e$ denote the event-time, i.e., $e = t-G$ is the time elapsed since default. Recall that $G$ denotes the time period that a unit is first treated. Hence, the object in equation (1) can be aggregated with respect to $e$ as follows:
\begin{equation}
    \theta(e) = \sum_{g \in \mathcal{G}} P(G=g) ATT(g, g+e),
\end{equation}
where $\mathcal{G}$ is the set of all possible groups, and $P(G=g)$ is the probability of belonging to a specific group. The object in eq. (3) is the average treatment effect on the treated  $e$  periods after default, aggregated across all groups that  ever defaulted.  This is the object that we report in the  figures below.
The average effect of participating in the treatment "on impact" is hence  $\theta(e=0)$.

In the  figures below, we  center the event in the year before the recorded soft default. The rationale behind this lies in the annual nature of our data, drawn on June 30th of each year, while a soft credit default could occur at any time within the preceding 12 months.
First, we note that all the pre-event coefficients are  close to 0, and mostly  statistically insignificant (recall that our sample size is about 2 millions individual observed for several years, up to 13).

To facilitate the narrative, we split the analysis below by family of outcomes, i.e. mobility and income versus credit.

\subsection{Mobility and Income}

\begin{figure}[H] 
         
             	Panel (i) - Move Zip \hspace{4cm}Panel (ii) - Move CZ\\
                \includegraphics[height=4.5cm, width=7cm]{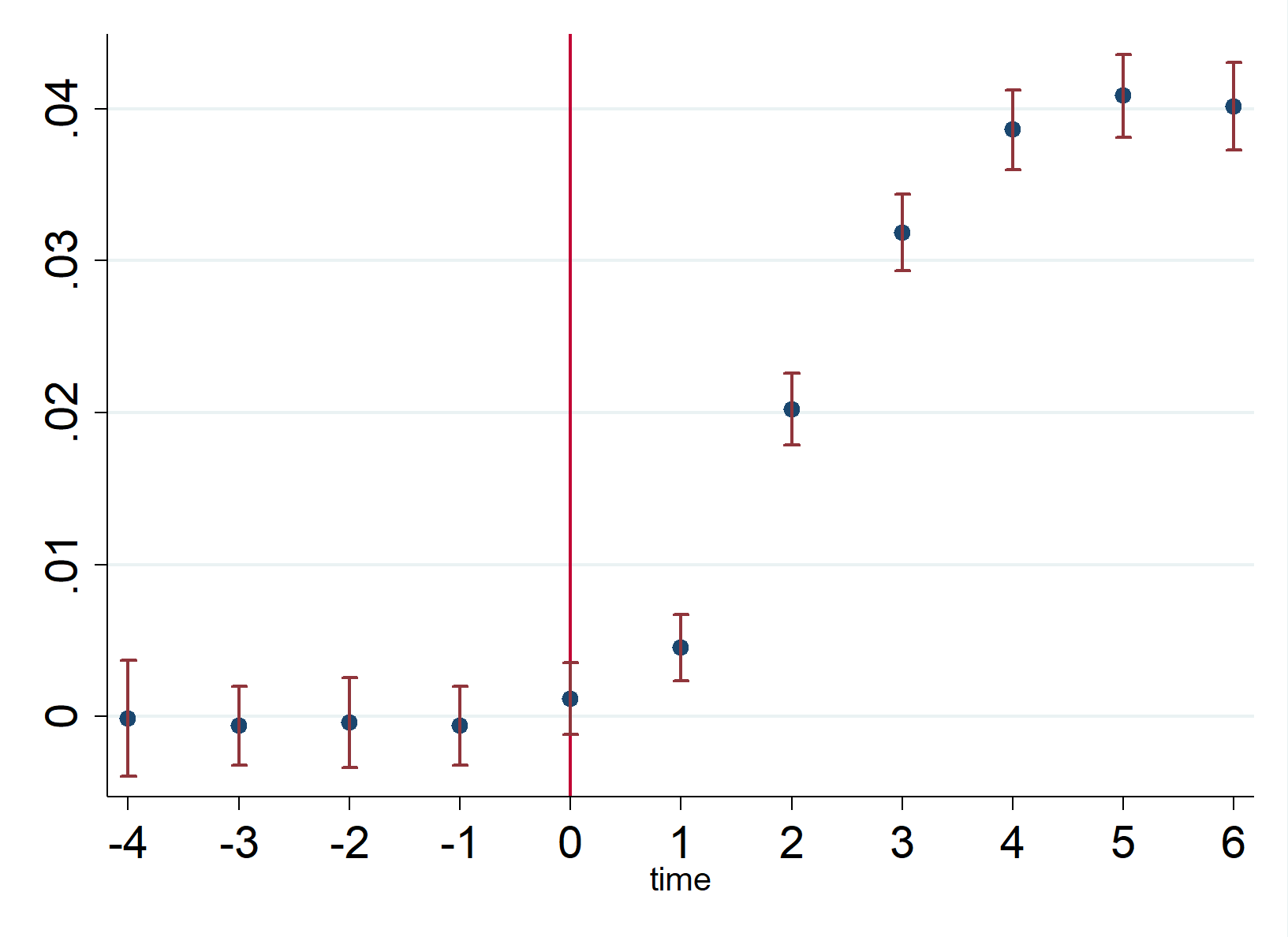}
             \includegraphics[height=4.5cm, width=7cm]{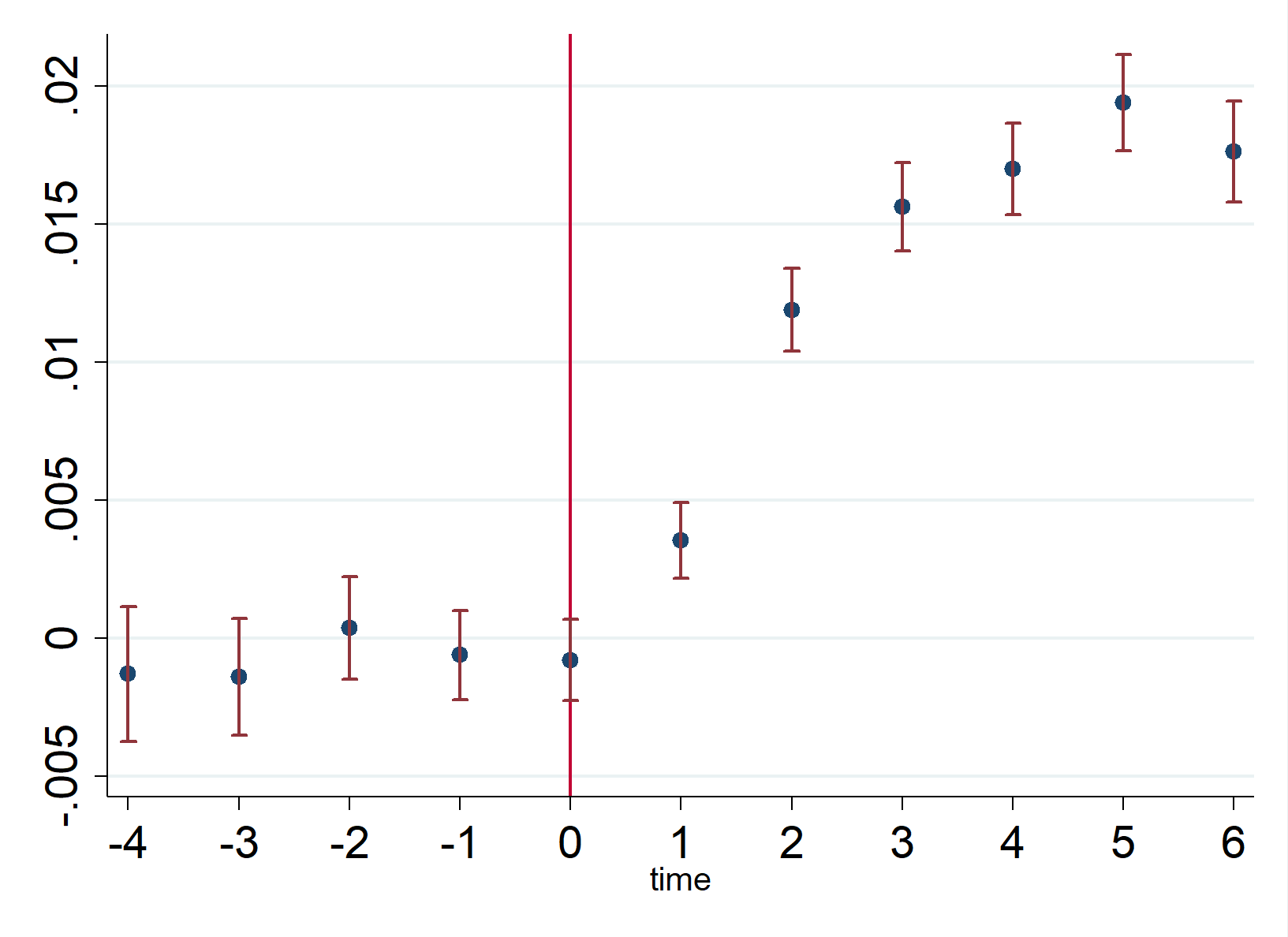}
             
                  	Panel (iii) - Median House Value \hspace{2cm} Panel (iv) - Income\\
                 \includegraphics[height=4.5cm, width=7cm]{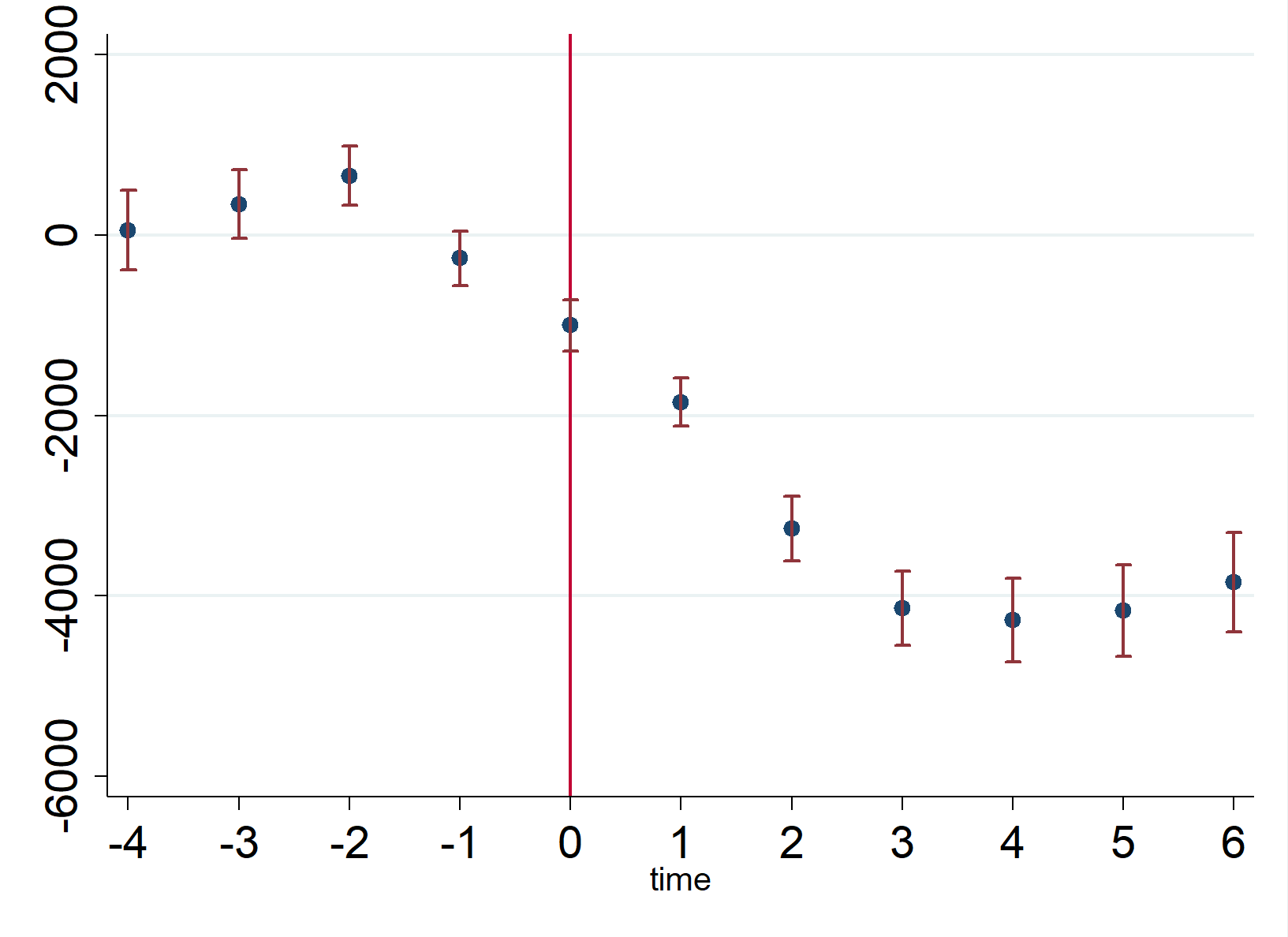}
             \includegraphics[height=4.5cm, width=7cm]{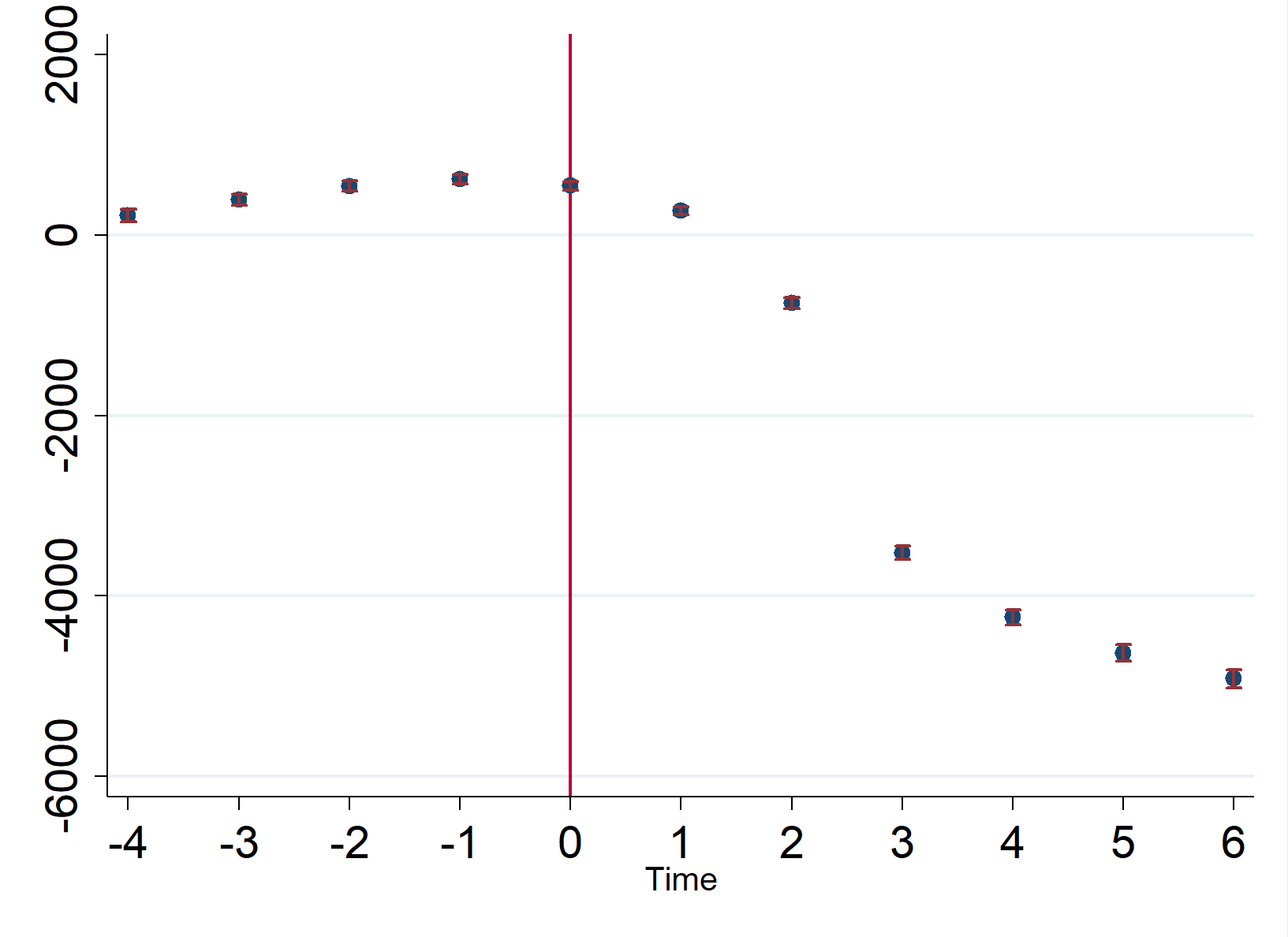}

        \caption{Event study: dependent variable is: (i) Probability of moving zip code. This variable takes value 1 if the individual is a different zip code in year $t$ than in year $t-1$, and zero otherwise, (ii) Probability of moving outside the commuting zone. This variable takes value 1 if the individual is in a different commuting zone in year $t$ than in year $t-1$, and zero otherwise, (iii) the Median House Value in the zip code of residence at year $t$ (iv) income imputed by Experian. The event considered is a soft default, i.e. a 90-day delinquency, but no Chapter 7, Chapter 13 or foreclosure taking place in the same year, neither before in the sample period. Other controls are age and age squared, credit score in 2004 and in 2005. 95\% confidence intervals around the point estimates.}   \label{fig:event2}   
     \end{figure}

Within two years, a soft default entails an increase by about 2pp in the probability of changing zip code from one year to the next (Figure \ref{fig:event2}, Panel (i)). By the third year, the effect becomes larger (i.e. +3.5-4pp) and remains essentially the same even six years after the soft default.\footnote{In all the event-study figures presented, the effect plotted $e$ years out is with respect to the event year. For example, the effect on zip code mobility is 2pp in year  two and 4pp in year four,  this signifies  an increase of 2pp two years after the event and an additional increase of 1pp four years after the event. This implies a cumulative higher mobility effect, after four years from the event, of  4pp (i.e. the coefficient plotted in the figure).} 
These are large magnitudes, considering that the year-to-year zip code mobility is about 17\%. 

A soft default is also associated with an increase by about 1pp in the probability of moving to a different commuting zone  one to two years after the event  (Figure \ref{fig:event2}, Panel (ii)). This impact increases to about 2pp three years after the event and remains approximately constant, and statistically significant, up to six years after the event (baseline CZ mobility in 2010 is 6\%). 
We will see later that on impact these moves are towards cheaper zip codes. 
The effects on mobility are large  and don't vanish over time. Soft defaults appear to trigger higher mobility both within and between commuting zones, both on impact and in the longer run. 

In Panel (iii) of Figure \ref{fig:event2}, we find evidence that a soft default entails a decline in the quality of the neighborhoods of residence for those who default. In particular, the Median House Value declines by about 4,000USD. The decline is persistent and still statistically significant six years after the event.

Let us come to the impact of a soft default on income. In Appendix \ref{sec:reliability} we assess in detail the validity of the imputed income variable provided by Experian. In particular, in the Appendix, we estimate a standard life-cycle regression of imputed income on age, age squared, credit score and credit score squared, and we verify that the residuals from that regression exhibit substantial variation and that about 70\% of the variation in imputed income cannot be explained by age and credit score alone, which would not be the case if the imputation model only relied on those variables. Further, we compare the distribution of Experian imputed income with that of income recorded in other sources (i.e. the Panel Study of Income Dynamics and the US Census) over time, age, and geography (US counties). 
We also study  the distribution of missing values and the shape of the income profile (i.e. the average income for different cohorts) obtained from the Experian imputed income with those obtained from income recorded in the PSID. All in all, we conclude that the Experian income imputation procedure is reliable.

A soft default has a negative impact of about 5,000USD on annual imputed income by year six post event. A soft default wipes away close to 10\% of annual average (imputed) income. This loss is long-lasting and increasing, as we will see also in Appendix \ref{sec:long} where we look at a longer time horizon.

As far as potential endogeneity of the soft default is concerned, we find it reassuring to see in panel (iv) of Figure \ref{fig:event2} that income was not declining before the default.

As noted, the general pre-trends of the  outcomes do not appear to be statistically and/or economically significantly different for our treatment and control groups. 

In the main text,  the treatment group is the set of individuals who defaulted between 2004 and 2016 and the control group, in the spirit of \cite{CALLAWAY2021200}, is constituted by individuals who never experience a default between 2004 and 2016. We note that the results are robust to changing the control group to those who only default in 2016, the last year of our contiguous data (see Appendix \ref{sec:control2016}). 
We interpret this as suggestive that endogeneity does not seriously bias our results.

As mentioned above, in the robustness checks, Appendix \ref{sec:robustness}, we build the propensity score ($p_g(X)$)  on a larger set of variables, i.e. age, age squared, credit score in the two years before the soft default and state fixed effects, and our main findings are confirmed.



The link between default and income is potentially complex, on the one hand income shocks can affect default, this is however not our object of interest. On the other hand, default episodes through their effects on credit availability can affect income generating opportunities in several respects. For example, it is known that non-credit actors such as potential employers, landlords, insurance companies, and mobile phone providers also make ample use of such information. Survey data show that, in the US, almost 50\% of firms check the credit information of prospective employees (\cite{bos2018labor}). As a consequence, a soft default may reduce employment opportunities, condition mobility, and the availability of essential services, for the individual and hence negatively influence her income. 

Another possibility is that a soft default triggers a relocation to a zip code that is smaller or otherwise worse in terms of job opportunities. Our results in Appendix \ref{sec:quality} seem to go in this direction, i.e. after a soft default individuals on average end up in a zip code with lower local income,  fewer employees,  fewer firms, and lower average wage. 
Further, when we analyze the potentially heterogeneous effects of different delinquent amounts in Section \ref{sec:heterogeneity}, we notice that most relocations are associated with high delinquent amounts, and the drop in the Median House Value is notably larger for those individuals with a high delinquent amount rather than for those with a low or medium delinquent amount.

Interestingly, when we compare the effects for those who will have a harsh default in the following years and for those who don't, the impacts on mobility are about the same. Indeed, we find that a soft default is followed by an increase of about 4pp in the probability of moving to a different zip code and by an increase of about 1.5pp in the probability of moving to a different commuting zone, regardless on whether this soft default is followed by a harsh default in later years.

\subsection{Credit Score}

 \begin{figure}[H] 
 
             	 Credit Score\\
                \includegraphics[ width=12cm]{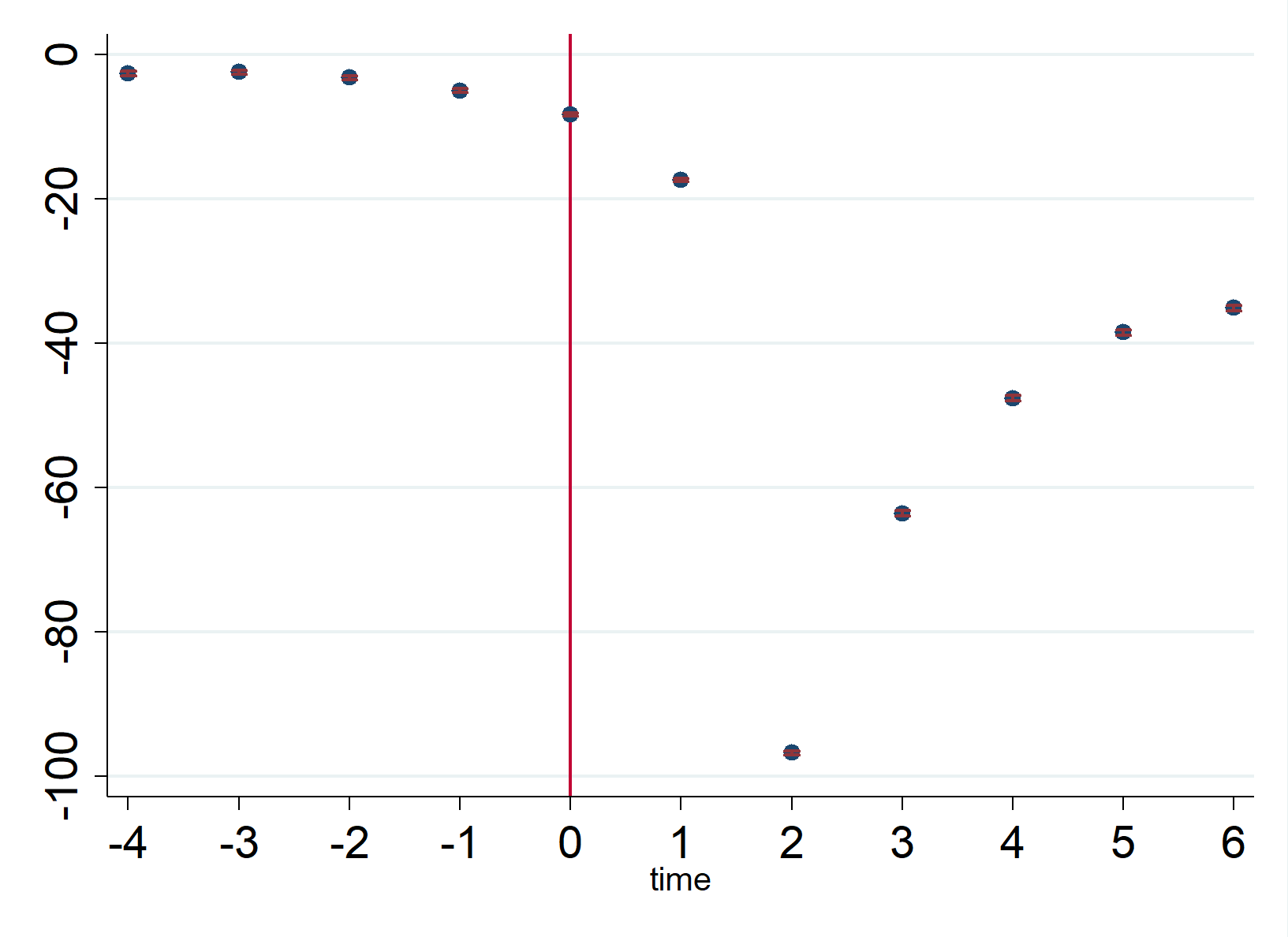}
    \caption{Event study: dependent variable is Credit Score. The event considered is a soft default, i.e. a 90-day delinquency, but no Chapter 7, Chapter 13 or foreclosure taking place in the same year, neither before in the sample period. Other controls are age and age squared, credit score in 2004 and in 2005. 95\% confidence intervals around the point estimates.}   \label{fig:cs}   
     \end{figure}
   
   From Figure \ref{fig:cs} we see that a soft default is followed by a drop in the credit score of about 100 points in the short-run. This drop is not immediate in the figure, as given the yearly frequency of the data, we do not have the exact date of default. As known, the immediate drop in the credit score is somewhat mechanical, in the sense that the default episode is used in the calculation of the credit score directly, and such episode stays in the credit report for up to 7 years. Some recovery takes place after the initial drop, and five years after the event the credit score is only about -40 points of the counterfactual value, i.e. the value for those individuals who had similar statistical and economic trajectories prior to the event. Clearly, such impact is large and would move the defaulted consumer from a prime to a subprime credit score at the sample average of 680, this large fall makes access to credit much harder for all types of credit lines and in particular for mortgages.\footnote{\url{https://www.experian.com/assets/consumer-information/product-sheets/vantagescore-3.pdf}.}

\subsection{Credit}

\begin{figure}[H] 

             	Panel (i) - Prob. new mortgage \hspace{1cm}Panel (ii) - Prob. Credit Limit $<$10k\\
                           
                            \includegraphics[height=4cm, width=7cm]{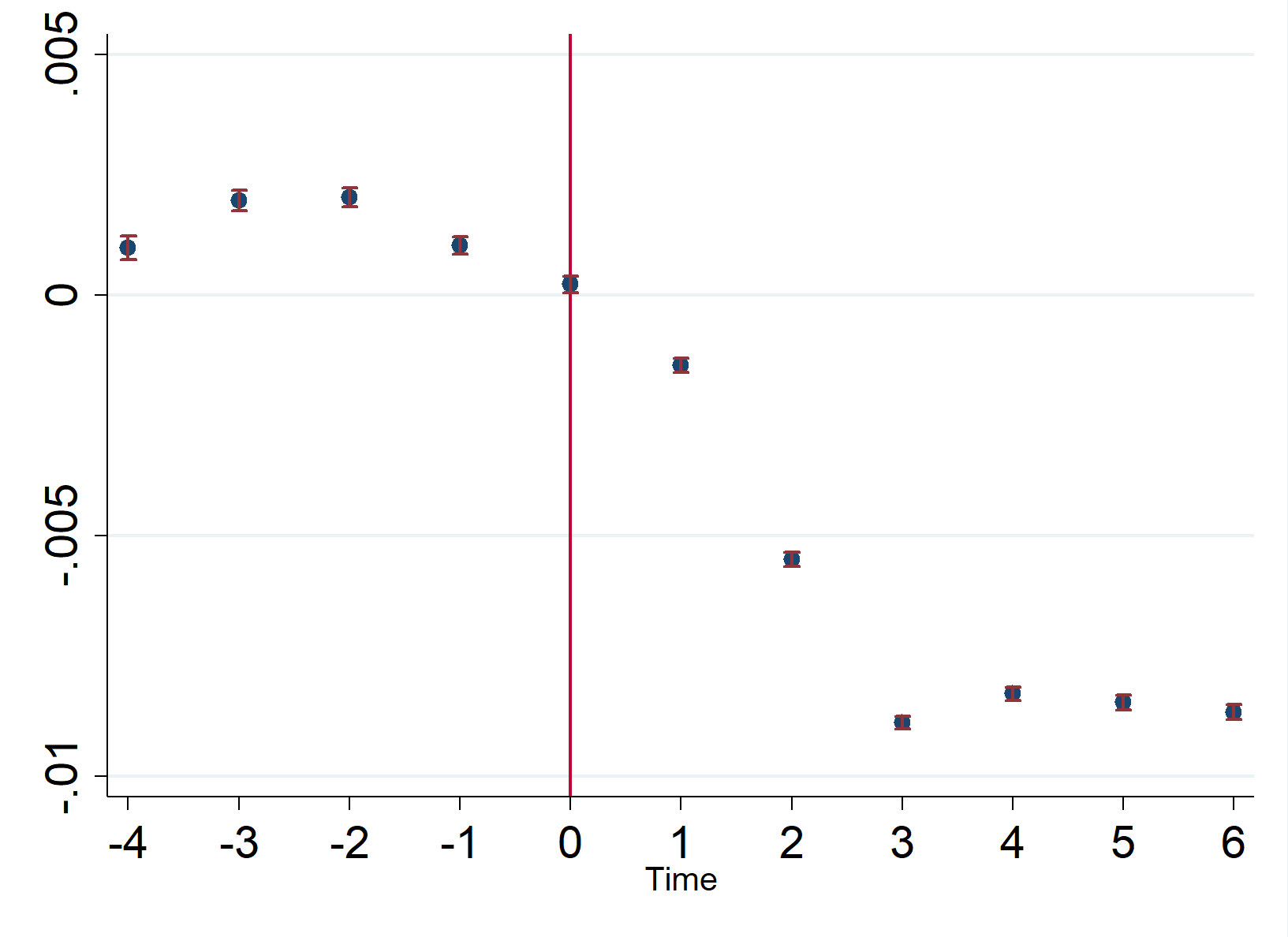}   
              \includegraphics[height=4cm, width=7cm]{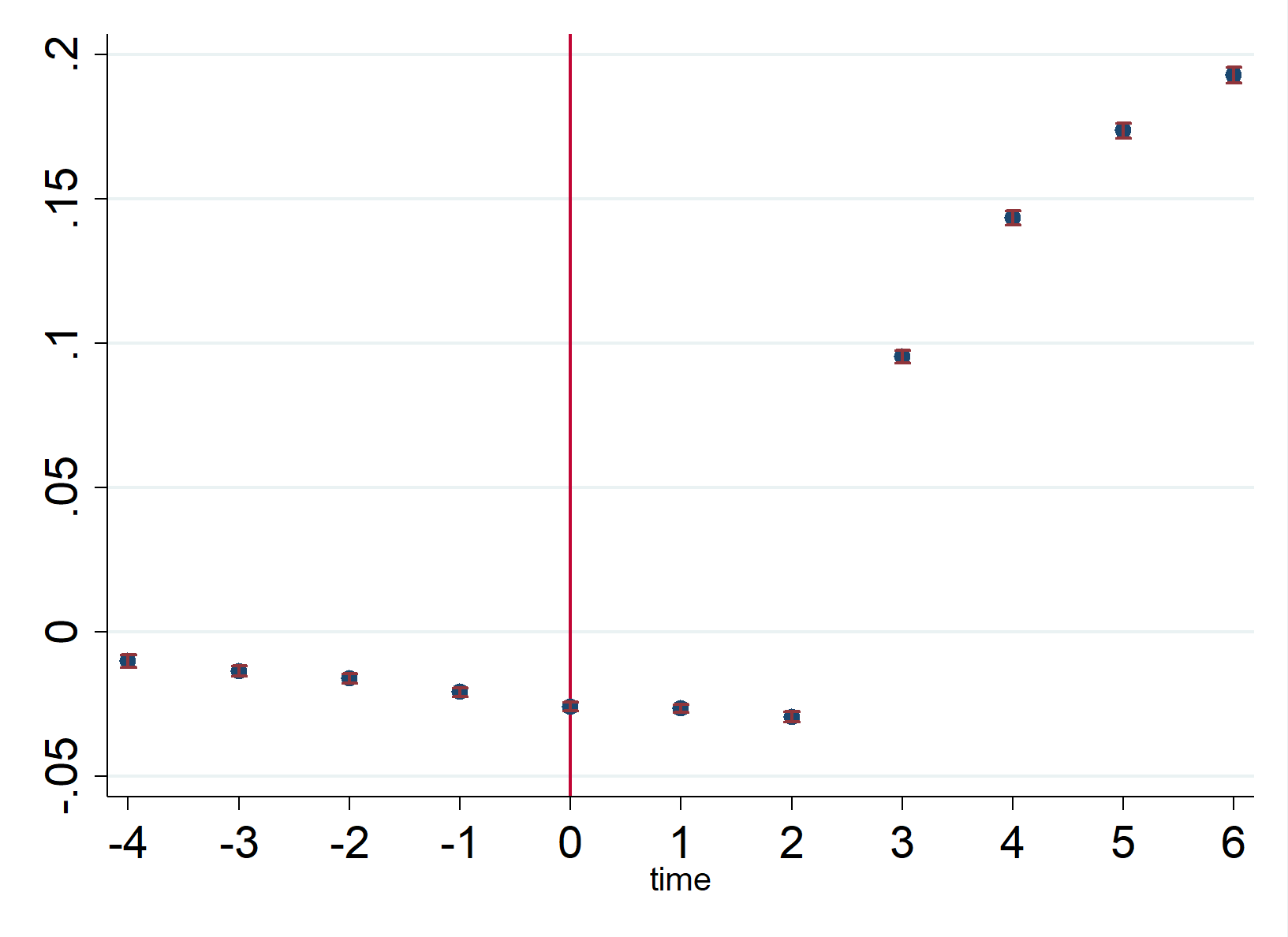}
              
                  	Panel (iii) - Revolving Balance \hspace{1cm}Panel (iv) - Harsh default\\
                     
             \includegraphics[height=4cm, width=7cm]{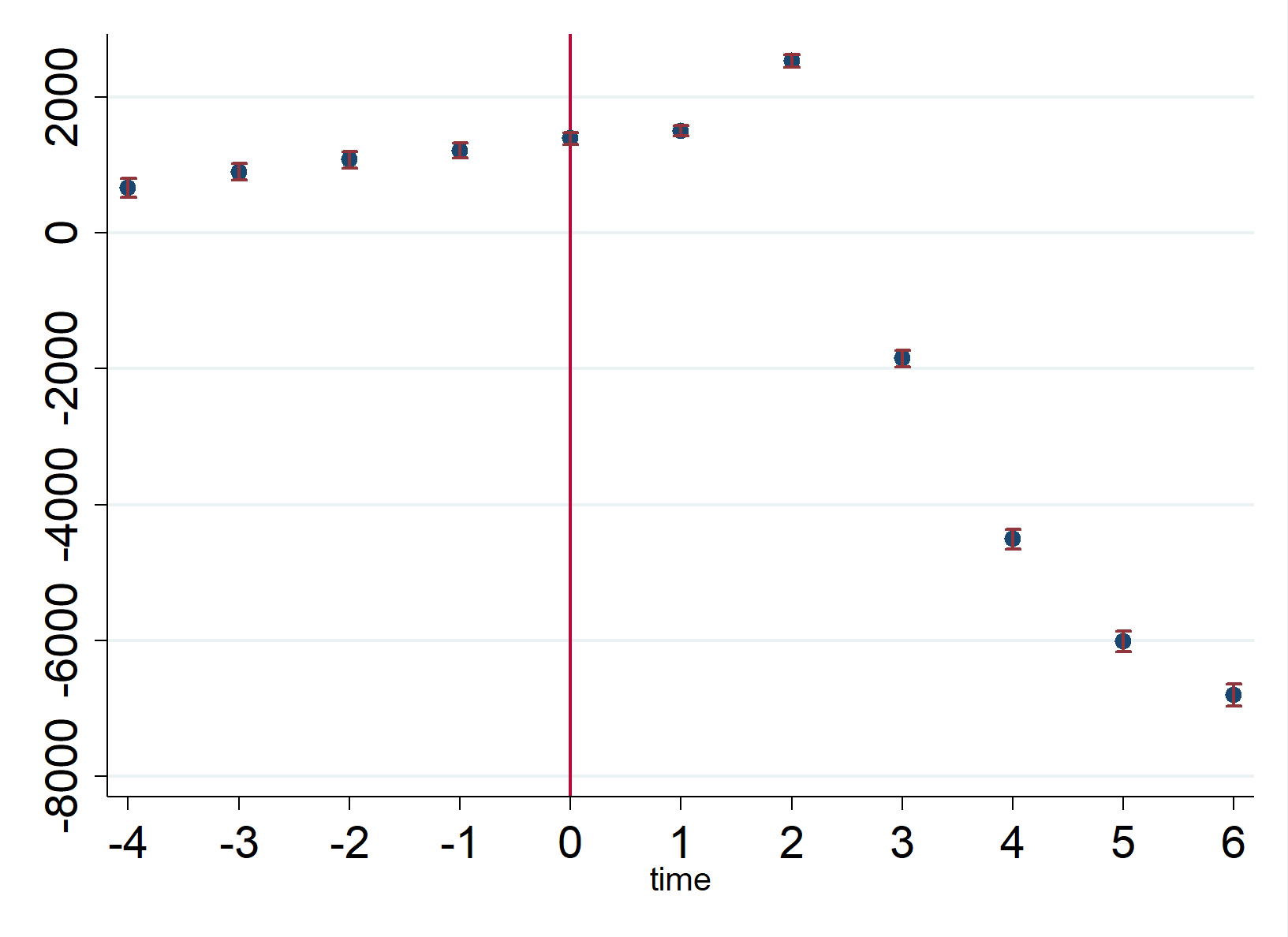}          	            	
                \includegraphics[height=4cm, width=7cm]{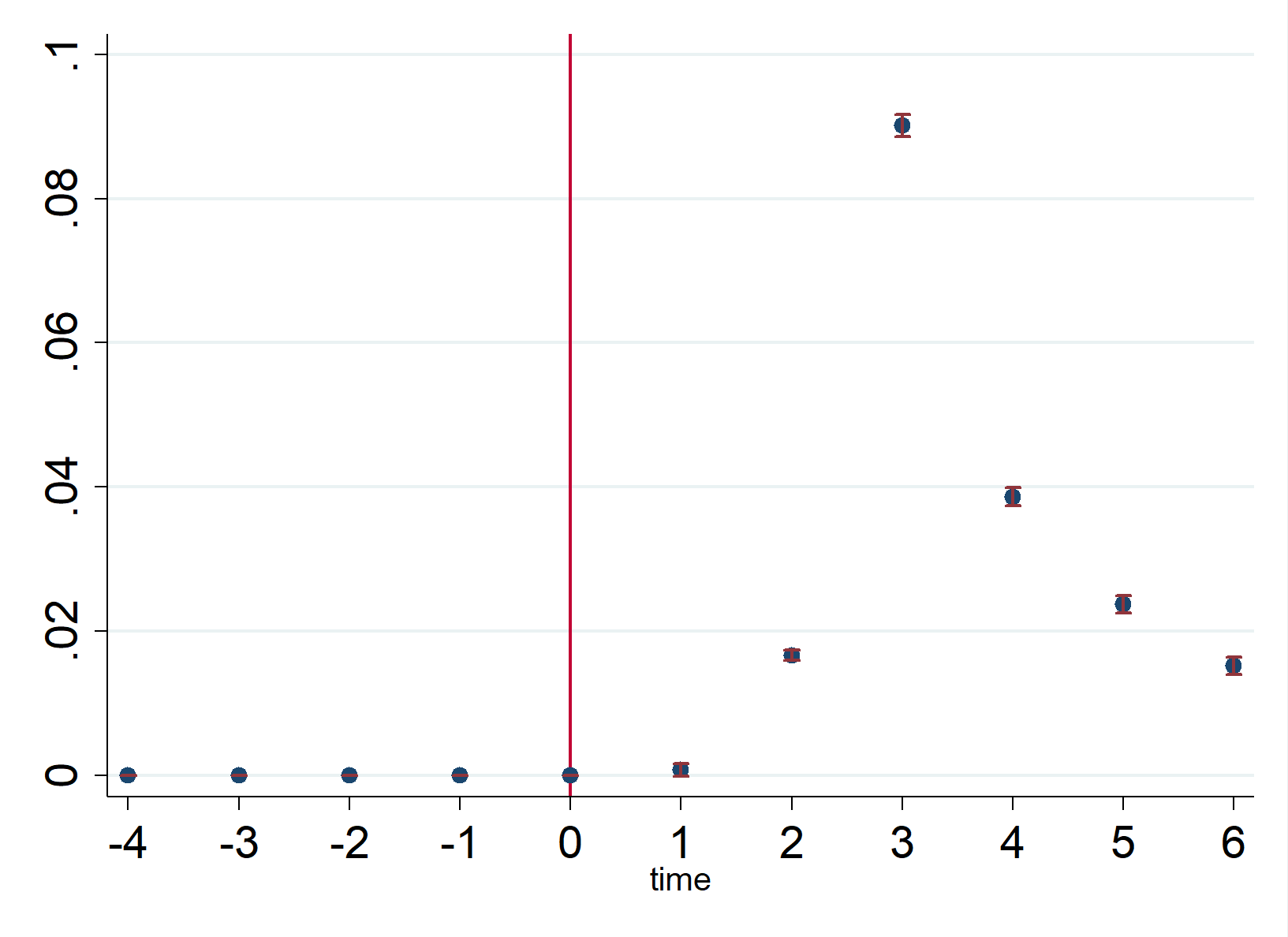}
                
                	Panel (v) - Home own \hspace{1cm}Panel (vi) - Total credit limit\\
                     
             \includegraphics[height=4cm, width=7cm]{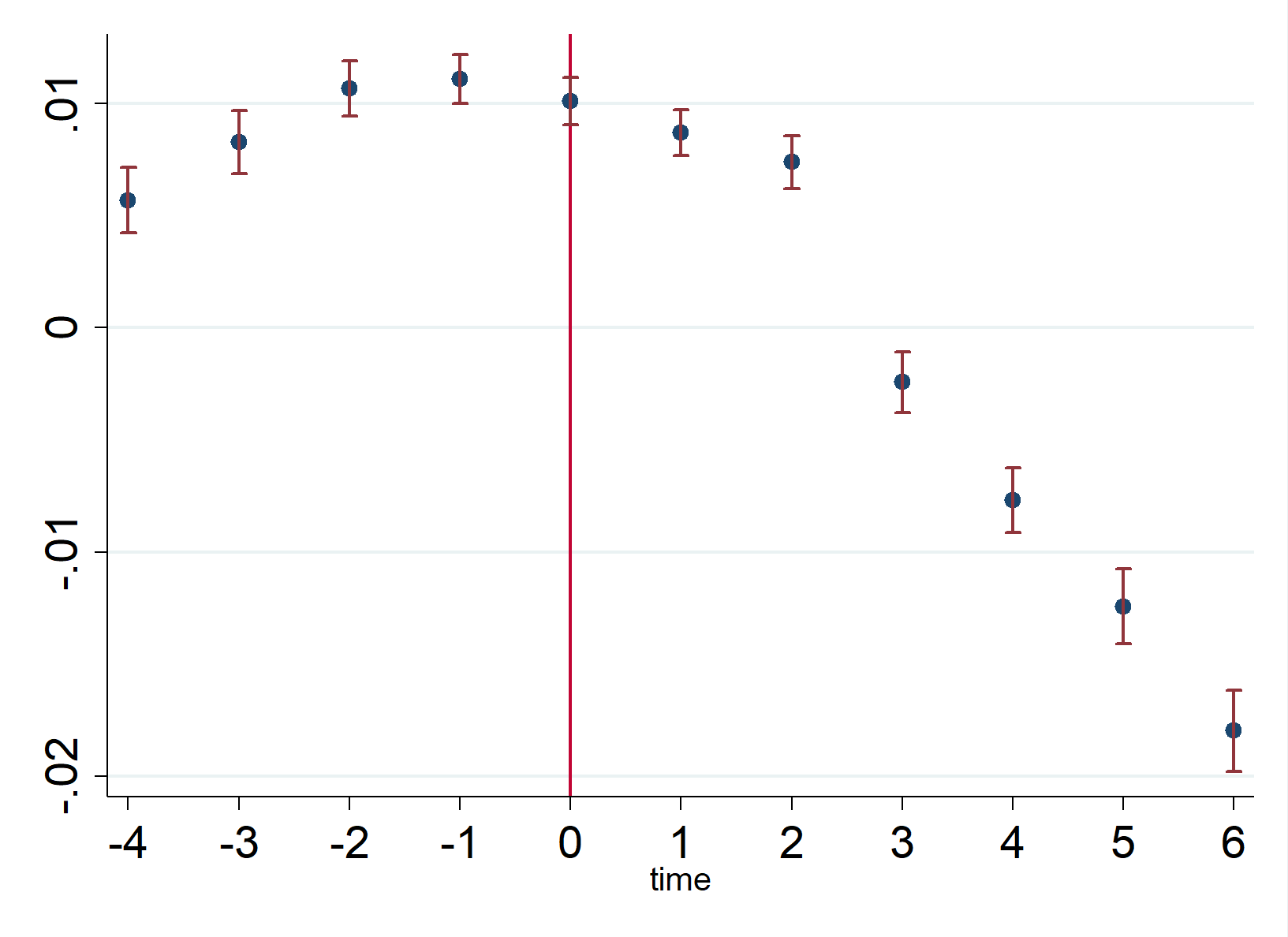}          	            	
                \includegraphics[height=4cm, width=7cm]{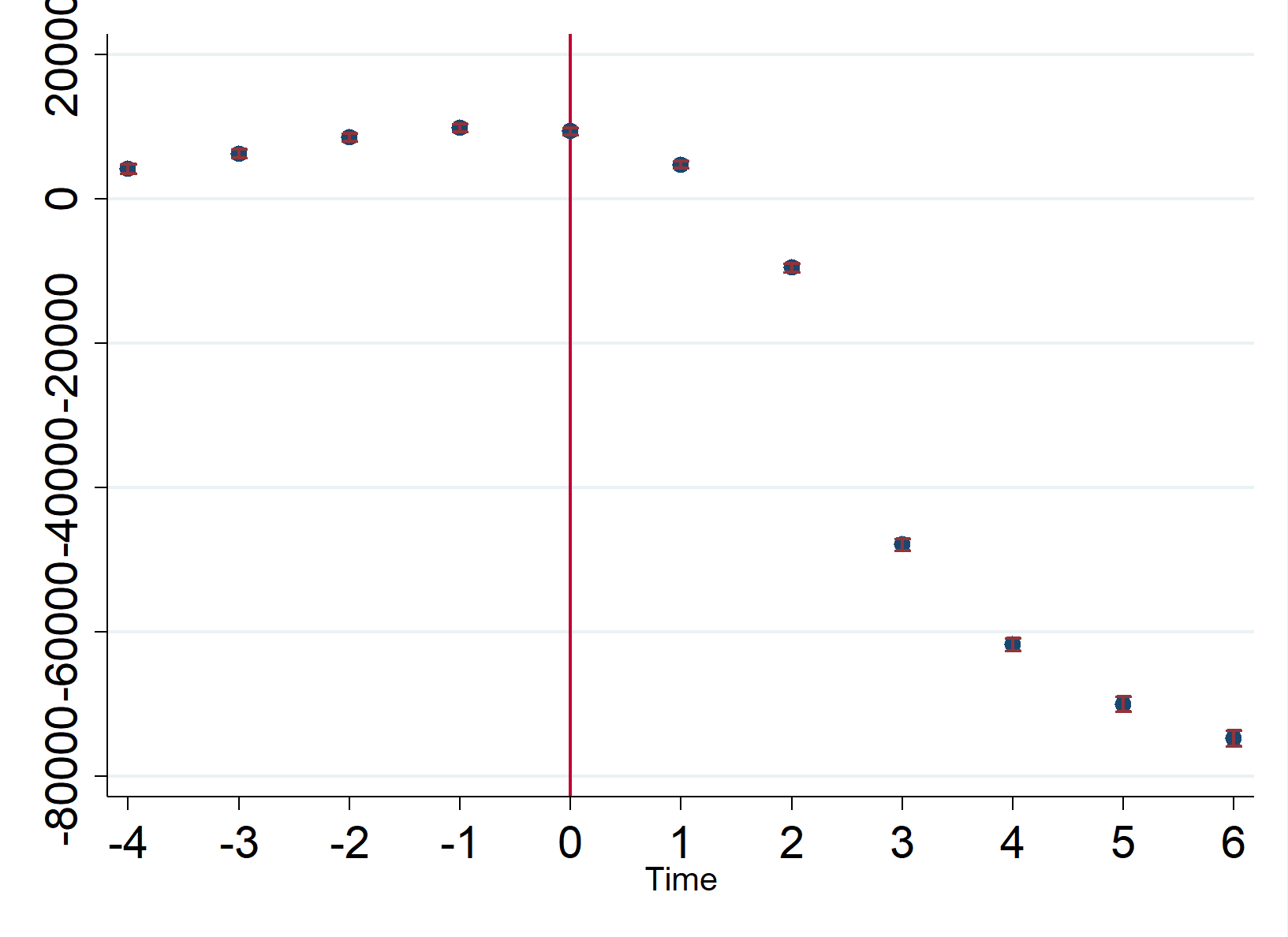}
                
                Panel (vii) - Mortgage balance open\\
                     
             \includegraphics[height=4cm, width=7cm]{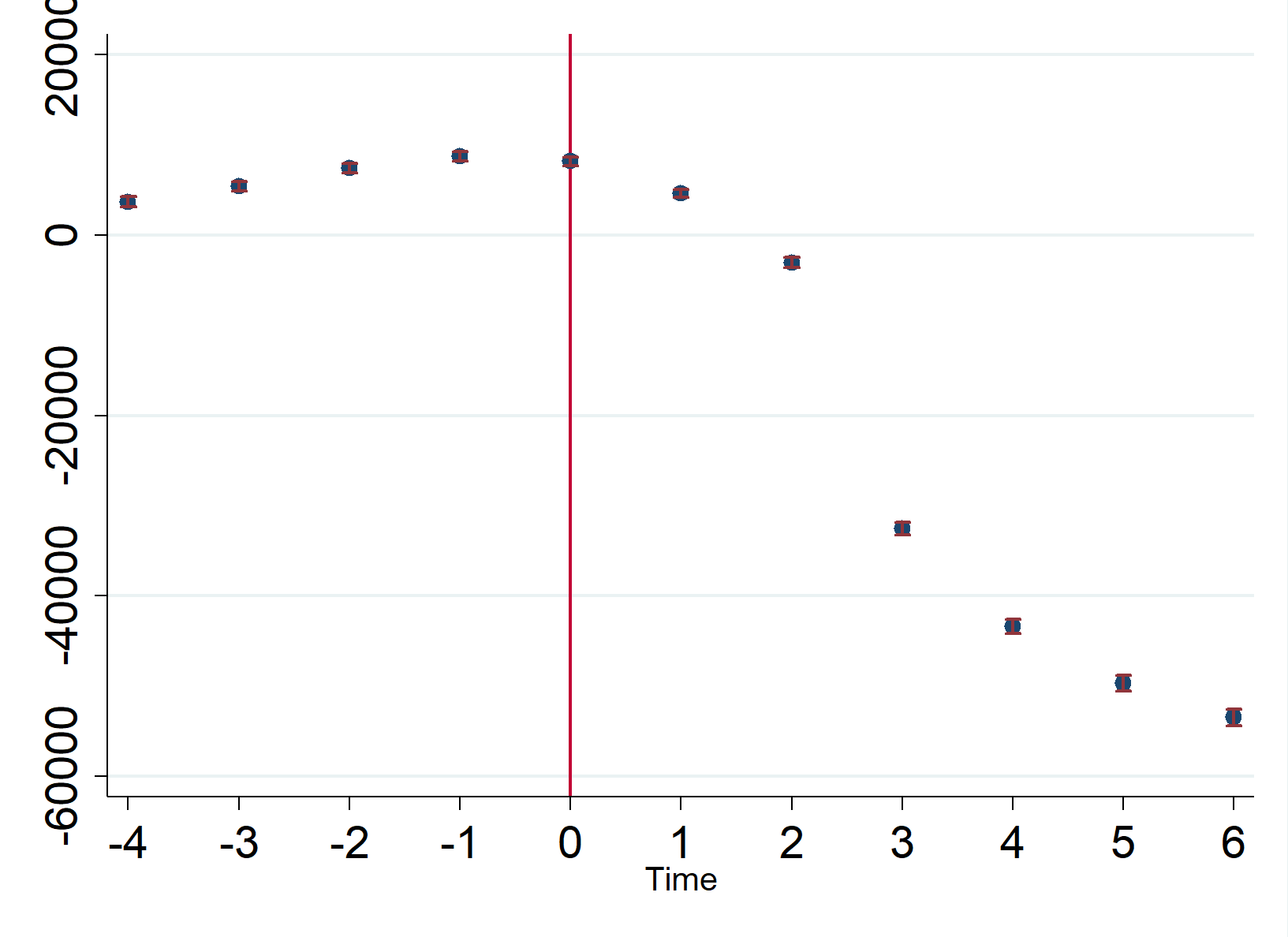}

        \caption{\small Event study: dependent variable is: (i) Mortgage origination: this variable takes value 1 if the individual has a higher number of mortgage trades in year $t$ than in year $t-1$ or if the number of months since the most recent mortgage has been opened is less than 12, and zero otherwise, (ii) probability that total credit limit is lower than 10,000USD, (iii) total amount open on all revolving credit trades, (iv) probability of experiencing a harsh default (Chapter 7, Chapter 13 or foreclosure), (v) probability of being homeowner, i.e. either being recorded as a homeowner by Experian or having ever had a mortgage open (vi) total credit limit on all trades, (vii) open amount of mortgage balance. The event considered is a soft default, i.e. a 90-day delinquency, but no Chapter 7, Chapter 13 or foreclosure taking place in the same year, neither before in the sample period. Other controls are age and age squared, credit score in 2004 and in 2005. 95\% confidence intervals around the point estimates.}   \label{fig:event3}   
     \end{figure}

In Figure \ref{fig:event3}, we study the impact of a soft default on a series of credit-derived variables.

Not surprisingly, the negative impact on the probability of having a new mortgage (Figure \ref{fig:event3}, Panel (i)) is  large, with a drop of 1pp (compared to a baseline of 7\% for 2010). In essence, a soft default makes it a lot harder to open a new mortgage, which is the instrument used by most Americans to climb the housing ladder.

Next,  we study the impact of a soft default on the probability that total credit limit is lower than 10,000USD (Figure \ref{fig:event3}, Panel (ii)), that credit limit is about the 10-th percentile of credit limits in 2010, and on revolving credit balance  (Figure \ref{fig:event3}, Panel (iii)).
For the probability of a low credit limit, a soft default is associated with an increase of 10pp to about 20pp (overtime) in the probability of having a  low total credit limit. This impact increases over time, and remains statistically significant  five years after the event. By 2016, 37\% of those who defaulted in 2010 have a total credit limit below 10,000USD while amongst the never-defaulted the same probability is about 28\%.
  For the revolving balance, such as on credit cards, typically the first source of credit for unexpected expenses (there is some debate on whether credit cards are indeed used to smooth shocks or simply to increase spending, see \cite{keys2017rainy}, \cite{NBERw26354}, and \cite{gelman2020individuals}), a soft default entails a drop by about 2,000USD shortly after the default and between 6,000 and 8,000USD, increasing over time. These effects are economically substantial given an average balance of about 12,000USD in our sample. 
   
  In Panel (iv) of Figure \ref{fig:event3} we study the impact of a soft default on the probability of experiencing a harsh default. In the first case, i.e. the impact of a soft default on the probability of a harsh one (bankruptcy, foreclosure, or declaration of chapter 7 or 13), we notice that a soft default is associated with a jump by about +10pp in the probability of experiencing a harsh default two to three years after the soft default. This is a  large magnitude  and is in line with models of opportunistic default on all loans, such as \cite{ParlourRajan2001}. However, in the following years this probability goes down, to about +2pp, five years after the soft default. 
 By 2016, 30\% of those who defaulted in 2010 incur in a harsh default while amongst the never-defaulted the same probability is about 7\%.
 
In Appendix \ref{sec:def}, we show that this increase in harsh default is mostly an increase in Chapter 7 declarations and in foreclosures (respectively +4pp each about three years after the event), whereas the increase in Chapter 13 declarations is smaller (i.e. +1.5pp). 
 Further, in Section \ref{sec:heterogeneity} we report the estimated impacts on our outcome variables of interest separately for those who will incur a harsh default afterward and for those who won't. One notable distinction that becomes apparent among individuals experiencing a harsh default and those who don't is the profile of the pre-trends for home value, where those with a subsequent harsh default appear to have a  larger pre-event home value (and mortgage, and income, for that matter).  In our context, it is plausible that individuals, who later faced a harsh default, may have overextended themselves on their mortgages, leading to a more challenging situation. A parallel pattern is also discernible for those with a substantial delinquency amount (refer to results in Section \ref{sec:heterogeneity}).
   
Panel (v) of Figure \ref{fig:event3} investigates the impact of a soft default on the probability of owning a home, while in the short-run the effect is negligible after three to six years that effect increases to almost -2pp or about -3\%. However, we note that the pre event probability of homeownership appears unbalanced, and in fact larger for those who default, so that the interpretation of the effects should be cautious. 
In the robustness checks, when we control for credit score in the two years before the soft default and state dummies in addition to age and age squared (in Appendix \ref{sec:robustness}), the pre-trend on homeownership is attenuated, but the negative impact of a soft default is still about minus 2-3pp and persistent over time.

In Panel (vi) of Figure \ref{fig:event3} we study the impact of soft default on total credit limit on all accounts, we notice a small (positive), but significant difference in the amounts prior to default of about 10,000USD, however after default the difference becomes more and more negative up to about 80,000USD five to six years after default. This is a 50\% drop in total credit limits over the 2010 average reported in Table \ref{tab:descriptives2010}.
Finally, in Panel (vii) of Figure \ref{fig:event3} we show the impact of a soft default onto intensive margin of mortgages, or the balance on open mortgages, the profile closely mimics that of total credit limit, in fact mortgages are included in that definition, with a fall overtime up to about 60,000USD, that is a lower balance by about 30\% of the sample average in 2010. Therefore, defaulted consumers have a harder time originating a mortgage, and when they do their balances are about 30\% lower, which would be consistent with all the previous findings, of lower income and lower House Value zip codes, it could also be that at origination these individuals would have to come up with a larger down-payment.

\begin{figure}
    	Panel (i) - Credit card consumption \hspace{1cm}Panel (ii) - Amount 90-180 days delinquent\\
             
             \includegraphics[height=4.5cm, width=7cm]{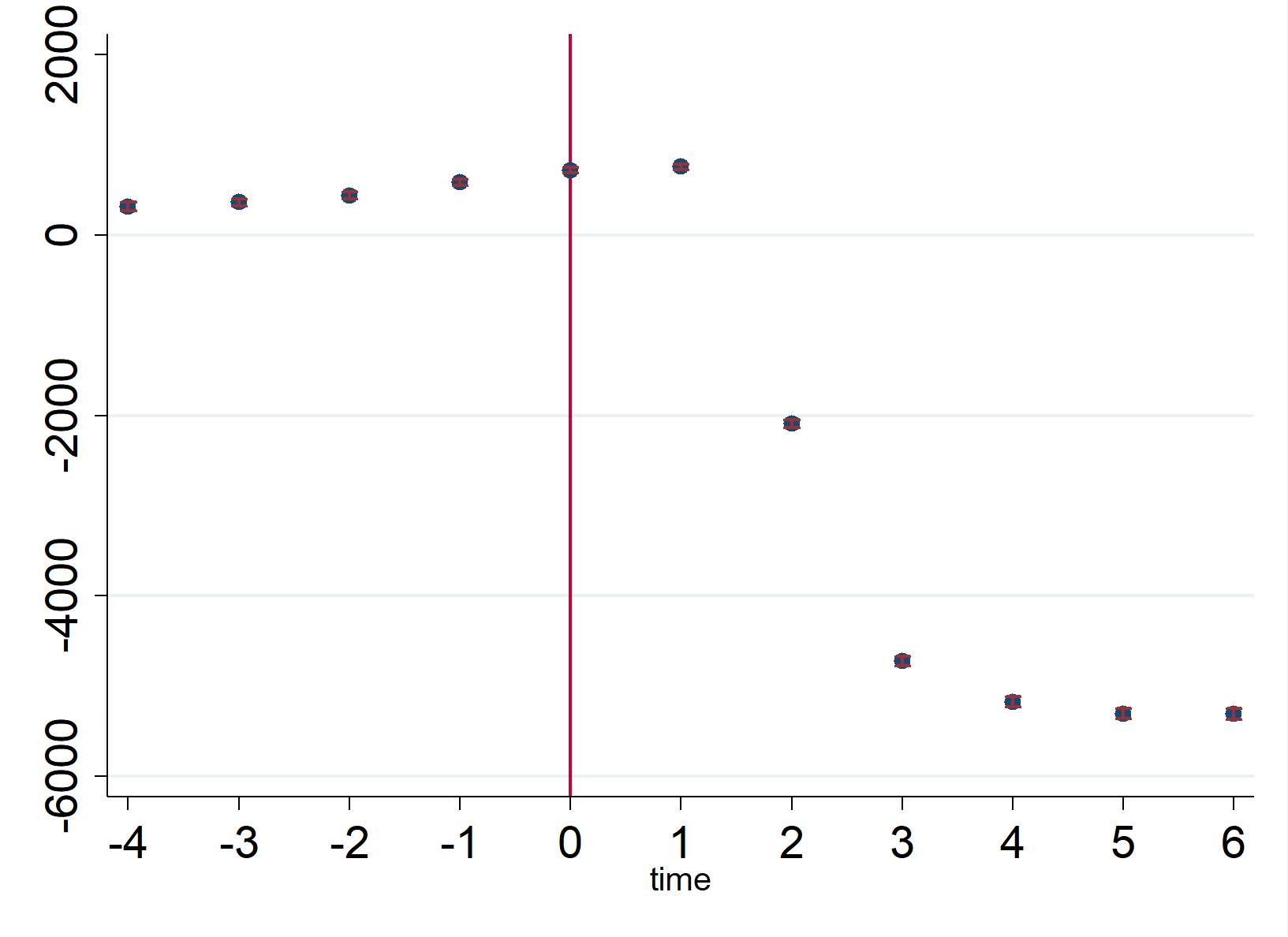}
                \includegraphics[height=4.5cm, width=7cm]{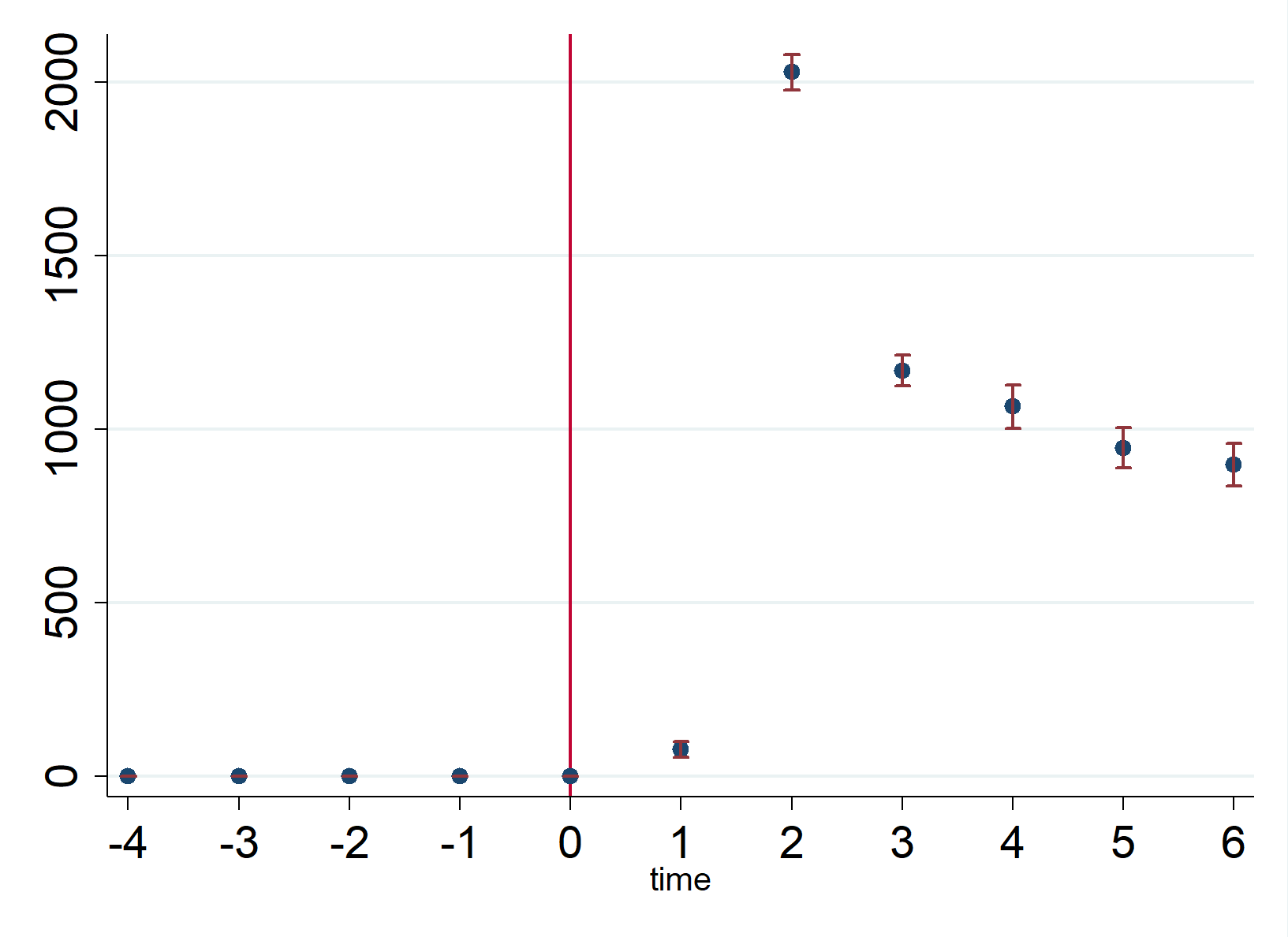}
             
                  	Panel (iii) - Number of collections\\
                              \includegraphics[height=4.5cm, width=7cm]{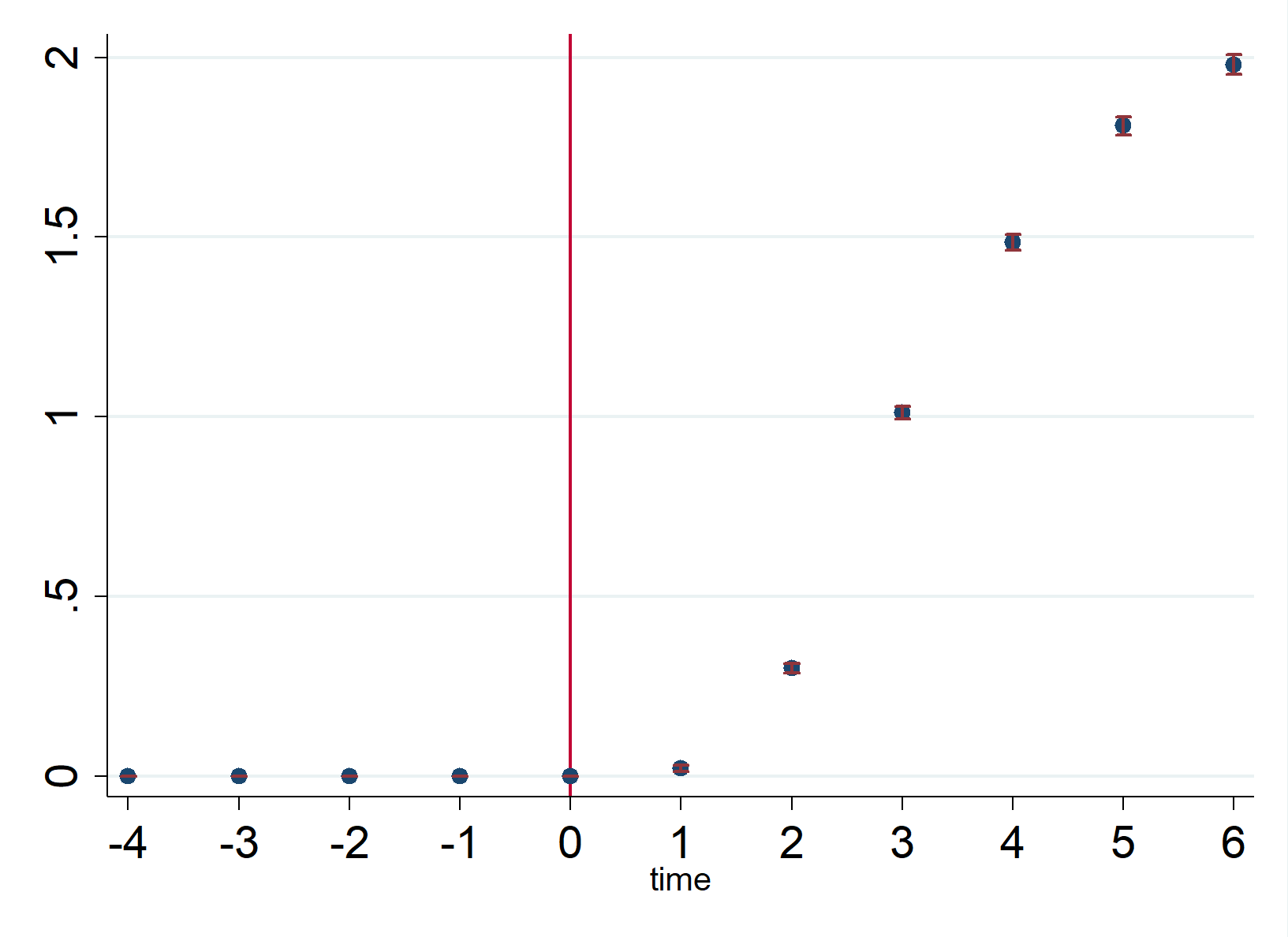}

        \caption{ Event study: dependent variable is: (i) credit card consumption: total balance on all open credit card trades reported in the last 6 months, (ii) amount 90-180 days delinquent, (iii) Number of collections. The event considered is a soft default, i.e. a 90-day delinquency, but no Chapter 7, Chapter 13 or foreclosure taking place in the same year, neither before in the sample period. Other controls are age and age squared, 2004-2005 values of credit score. 95\% confidence intervals around the point estimates.}      \label{fig:fig4}
     \end{figure}

It is important here to note that for the last three outcomes, the pre-trends do not appear well-balanced, while the post event differences are very large and negative, the pre-event differences appear much smaller and positive. When matching individuals on a larger set of controls, in Appendix \ref{sec:robustness}, the negative impact on the probability of opening a new mortgage stays statistically significant both in the short and in the medium term. Similarly, the impacts on total credit limit and on the amount of mortgage are notably close to the baseline estimation results.

From Figure \ref{fig:fig4} we find evidence that a soft default is associated with a decrease by about 4,000/5,000USD in credit card consumption (Panel (i)), with an increase by about 2,000USD in the amount delinquent two years after the event (Panel (ii)), and with an increase in the number of collections between 1 and 2 (Panel (iii)).

The overall lesson from this exercise is that soft default episodes are important determinants of lifetime trajectories in income, mobility, and on the credit  market both in the short and medium run. Episodes of default produce persistent negative effects and, as we show in Appendix \ref{sec:long}, the recovery might never happen as for most outcomes the negative impact is there up to 10 years after the original episode.

In Appendix \ref{sec:ml}, \ref{sec:DML} and \ref{sec:long} we provide further support for the results of the event studies. There we use a different identification strategy based on a double machine learning approach, and we focus on those soft defaults happening in the base year 2010 and on their long term impact in year 2020. The effects appear qualitatively  similar to those presented in the main text, and the magnitudes are also comparable. For the long term impact, we cannot run a standard event study as we do not have data for 2017 and 2018.

\section{Heterogeneity and Potential Mechanisms}
\label{sec:heterogeneity}
\subsection{Harsh vs Non-Harsh-defaulters in the Sample Period}

In Section \ref{sec:exploratory} we find evidence that a soft default leads to an increase in the probability of also incurring a harsh default, i.e. Chapter 7, Chapter 13 or foreclosure, about 3/4 years after the soft default. Hence, in this Subsection, we compare event study results obtained separately for those who experience at least one harsh default between the year of their soft default and 2016 and for those who don't. It should be clear that such split of the data is not granted by an exogenous event, and therefore should be interpreted with caution.\footnote{In all the figures presented, 95\% confidence intervals are drawn around the point estimates. However, in some cases the intervals are so narrow that they are not easy to see. This happens because the sample size is huge (i.e. millions of observations) and this allows very precise point-estimates.
It would be interesting to run formal statistical tests to check whether the differences reported among the different groups in this Section are statistically significant or not. However, since we can not implement an interacted model, that is simply not possible with the current estimator, we do not know the covariance between the estimates for the different groups. One solution would be that of a bootstrap test, i.e. running each estimate, say, at least 100 times, each time on a different generated random sample, for each subgroup, in order to get the bootstrapped
distribution of the differences of each coefficients (or an estimate of the covariance). Since the estimation of a single event study with the Callaway Sant'Anna procedure is quite computationally intensive (i.e. about 23 hours on a standard server for each subgroup and each variable), we deem that running such a bootstrap test is unfortunately infeasible, as total running time would be about 300 days.
Of course, under the hypothesis that the different subgroups are independent (which may however be a strong hypothesis), then each case in which the represented confidence intervals for different groups do not overlap could be directly interpreted as a statistically significant difference in the effects.}

The results for the harsh defaulters are generally in line with our baseline results, but some effects are larger. The increase in the probability of moving to a different zip code is still about 4pp and that of moving to a new commuting zone is about +1.5pp like in the baseline. The drops in Median House Value (minus about 4,000USD) and in income (about minus 4,000 USD per year) are rather close to the baseline impacts presented in Section \ref{sec:exploratory}. The same holds true for the drop of about 100 points in credit score presented in Figure \ref{fig:heterogeneity_harsh_cs}. The negative impact of a soft default is essentially identical (i.e. about minus 100 points in the short run) on the credit score for harsh defaults and non-harsh defaulters.

   \begin{figure}[H] 
         
             	Panel (i) - Move Zip \hspace{4cm}Panel (ii) - Move CZ\\
                \includegraphics[height=4.5cm, width=7cm]{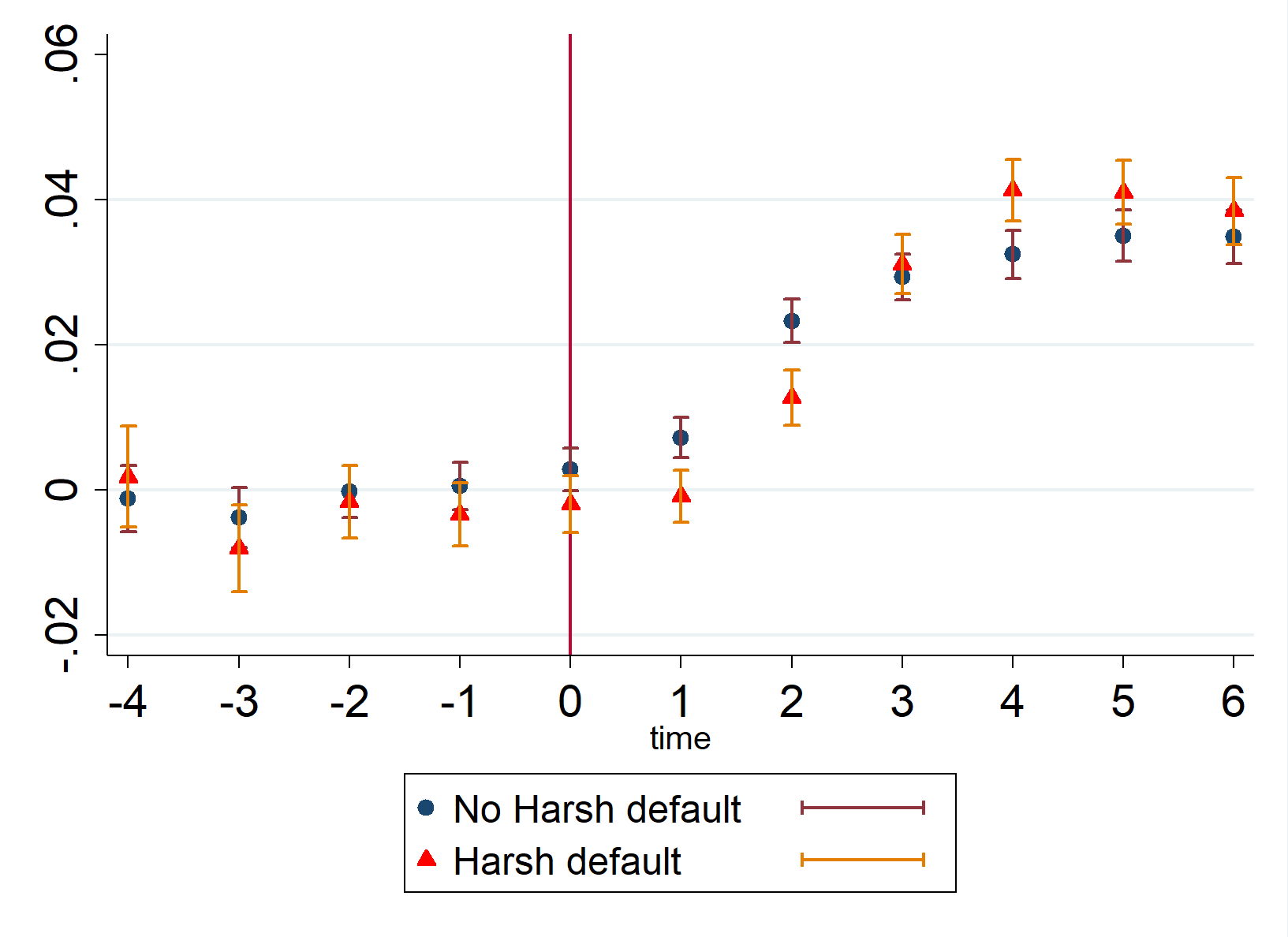}
             \includegraphics[height=4.5cm, width=7cm]{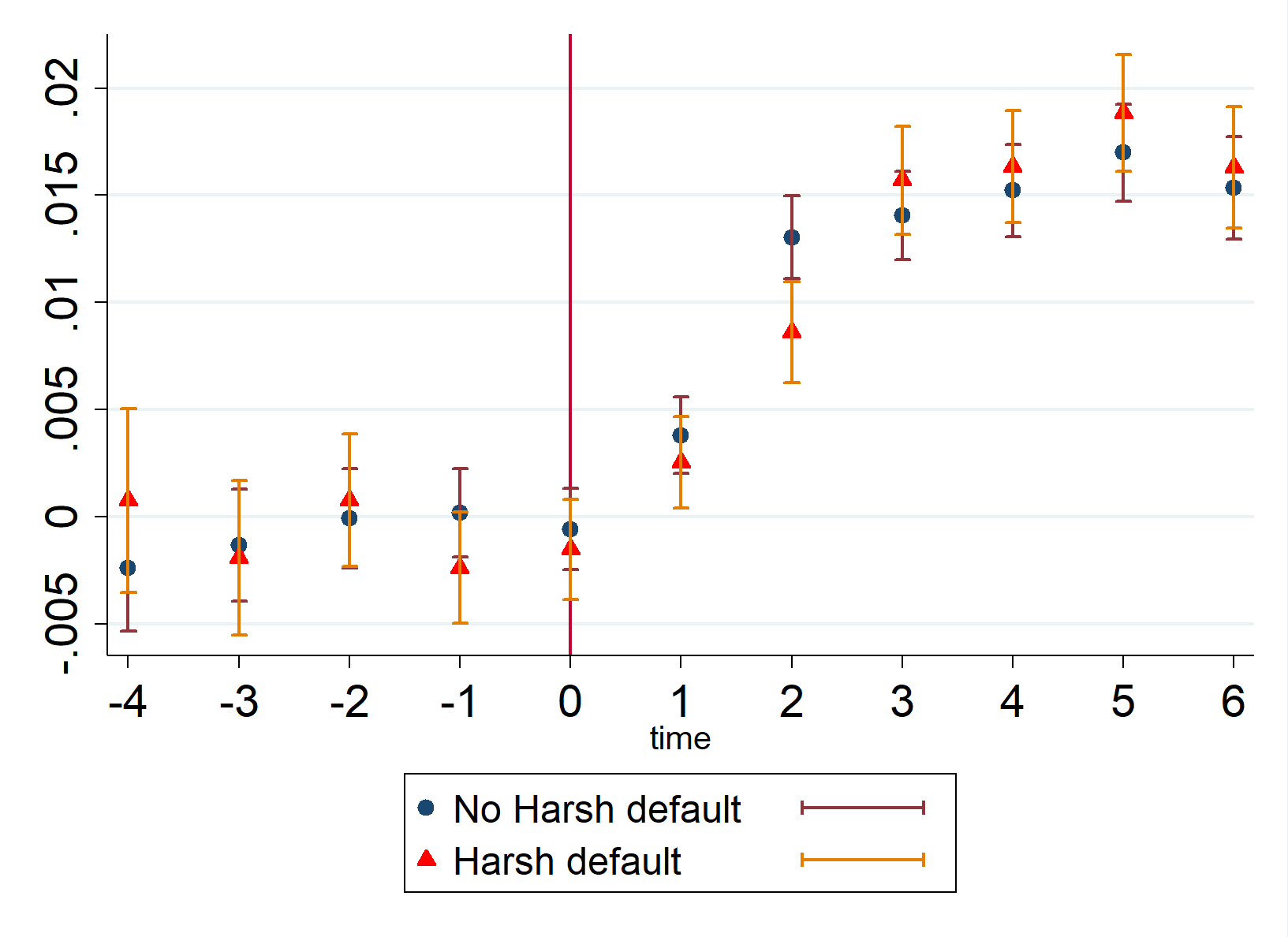}
             
                  	Panel (iii) - Median House Value \hspace{2cm} Panel (vi) - Income\\
                              \includegraphics[height=4.5cm, width=7cm]{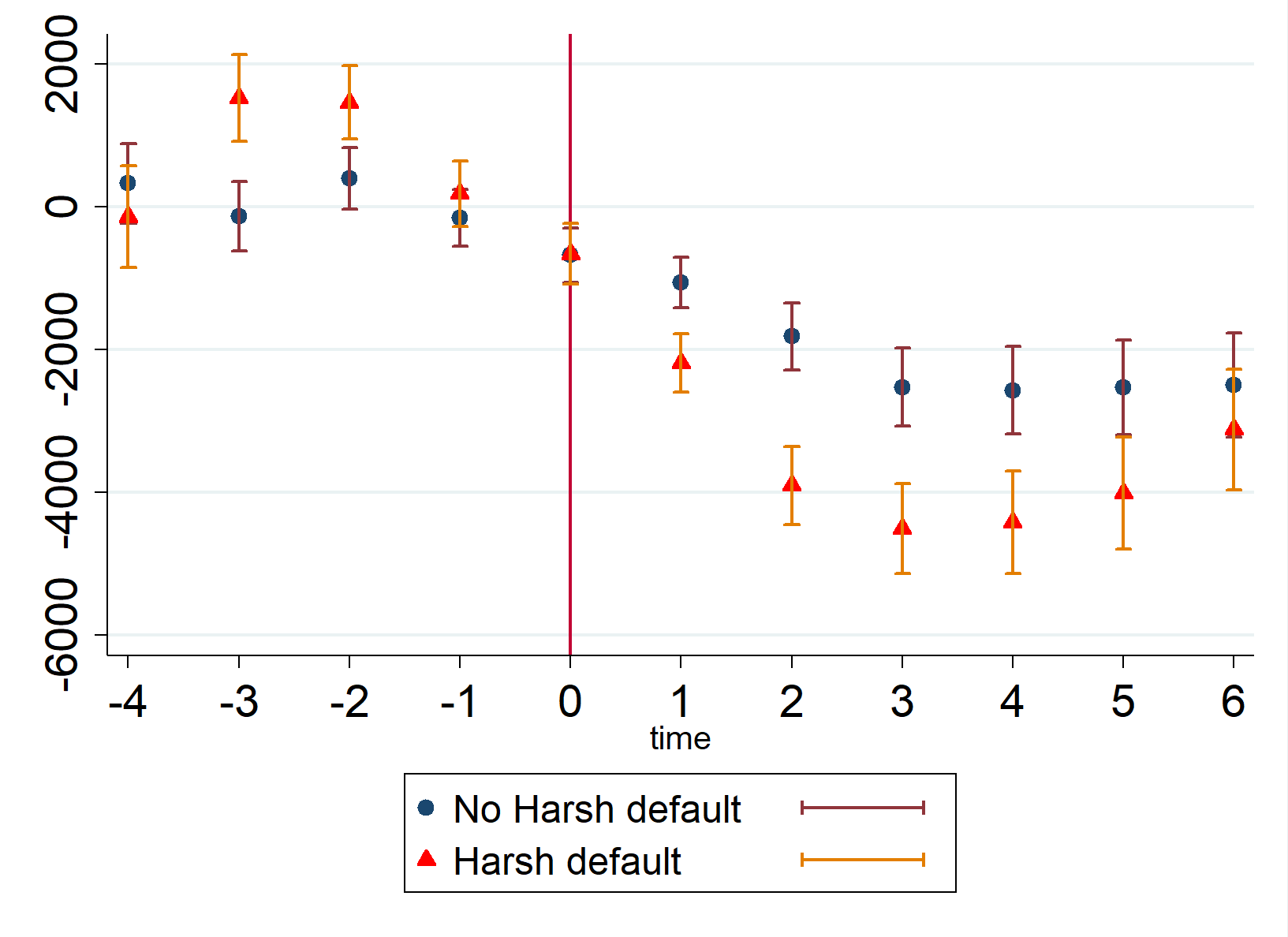}
             \includegraphics[height=4.5cm, width=7cm]{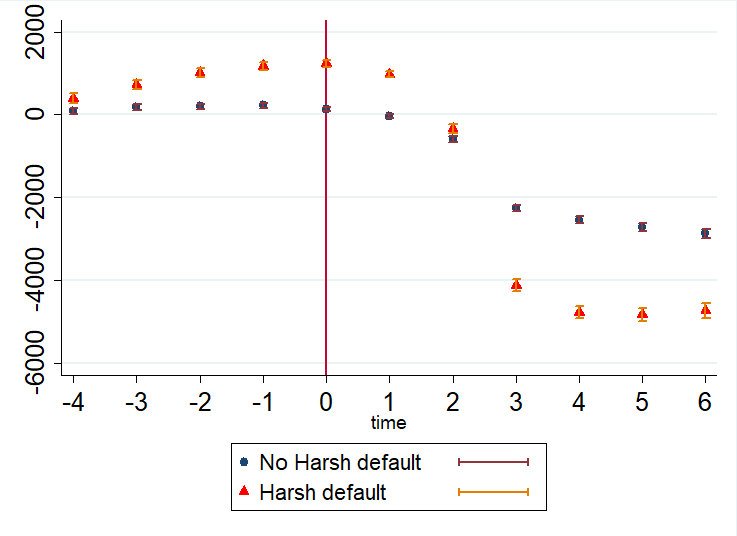}
             
              	Panel (v) - Credit card consumption\\
                              \includegraphics[height=4.5cm, width=7cm]{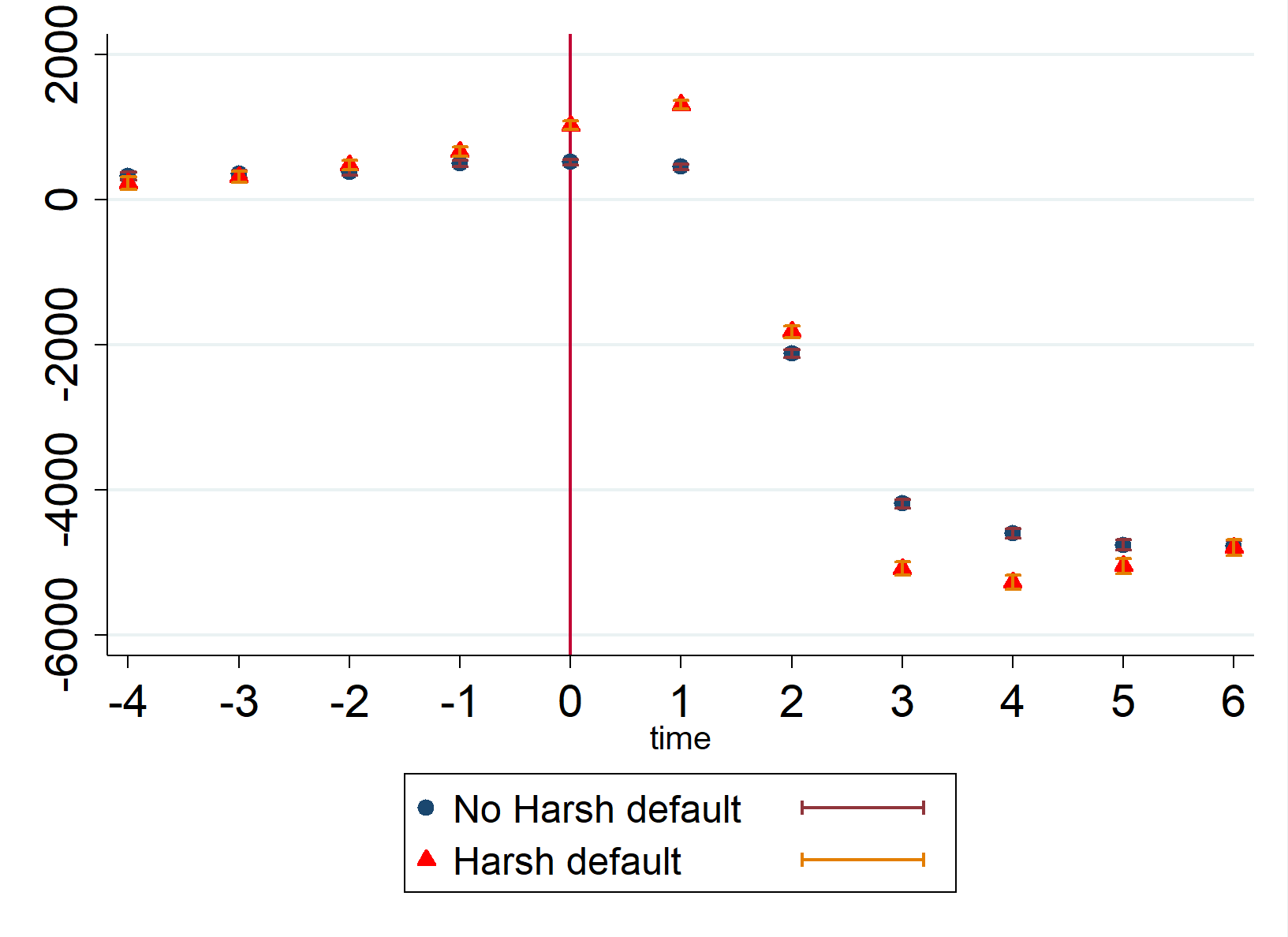}

        \caption{ \small Event study: dependent variable is: (i) Probability of moving zip code. This variable takes value 1 if the individual is a different zip code in year $t$ than in year $t-1$, and zero otherwise, (ii) Probability of moving outside the commuting zone. This variable takes value 1 if the individual is in a different commuting zone in year $t$ than in year $t-1$, and zero otherwise, (iii) the Median House Value in the zip code of residence at year $t$ (iv) income imputed by Experian, (v) credit card consumption: total balance on all open credit card trades reported in the last 6 months. The event considered is a soft default, i.e. a 90-day delinquency, but no Chapter 7, Chapter 13 or foreclosure taking place in the same year, neither before in the sample period. Other controls are age and age squared, credit score in 2004 and in 2005. 95\% confidence intervals around the point estimates.
        }     \label{fig:heterogeneity_harsh_mobility}
     \end{figure}

The negative impacts of a soft default are slightly smaller for the non-harsh-defaulters. However, such differences are moderate. Indeed, the increase in the probability of changing zip code is now about 3pp instead of 4pp, and the increase in the probability of moving to a new commuting zone is 1.5pp, as in the baseline. The drop in the Median House Value is smaller, i.e. minus 2-3,000USD vs minus 4-5,000USD for the harsh defaulters. The drop in income is also smaller, i.e. minus about 2,500 USD vs minus 4,000USD. Credit card consumption also declines less (about minus 4,000 vs minus 5,000 USD for the harsh defaulters three years after the event). However, the difference in the drop in credit card consumption vanishes over time.

    \begin{figure}[H] 
 
             	 Credit Score\\
                \includegraphics[width=12cm]{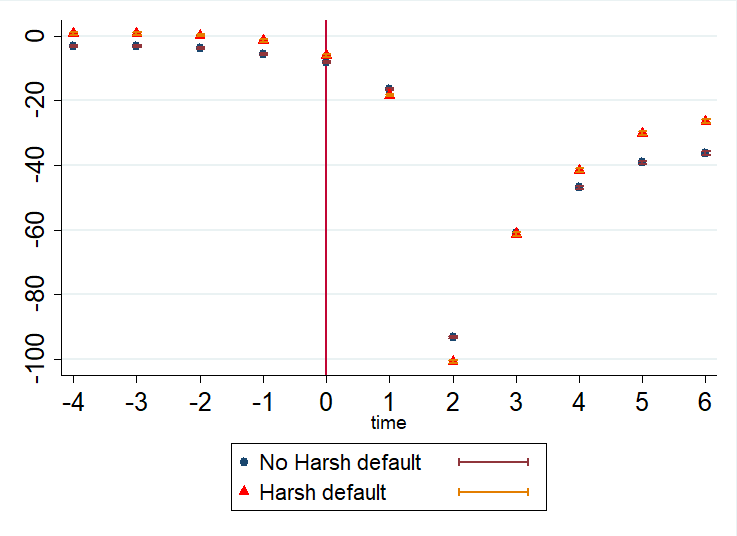}
    \caption{Event study: dependent variable is Credit Score. The event considered is a soft default, i.e. a 90-day delinquency, but no Chapter 7, Chapter 13 or foreclosure taking place in the same year, neither before in the sample period. Other controls are age and age squared, credit score in 2004 and in 2005. 95\% confidence intervals around the point estimates.}   \label{fig:heterogeneity_harsh_cs} 
     \end{figure}

\begin{figure}[H] 
        
             	Panel (i) - Prob. new mortgage \hspace{1cm}Panel (ii) - Prob. Credit Limit $<$10k\\
                          
                            \includegraphics[height=4cm, width=7cm]{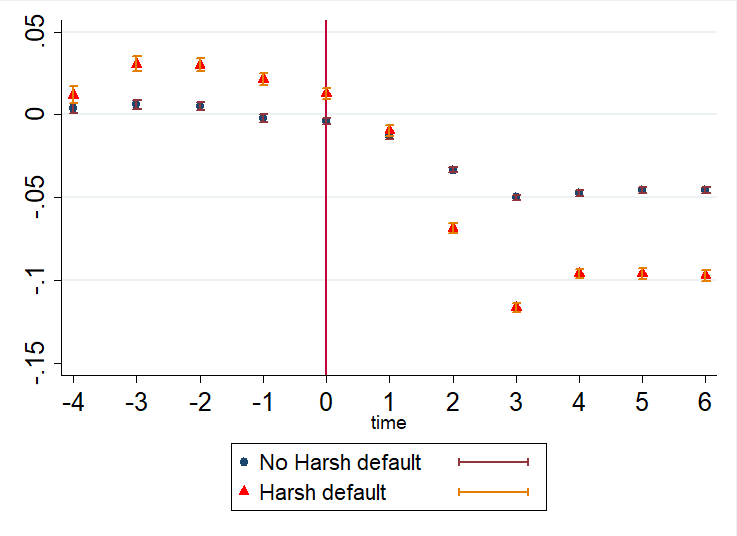}   
              \includegraphics[height=4cm, width=7cm]{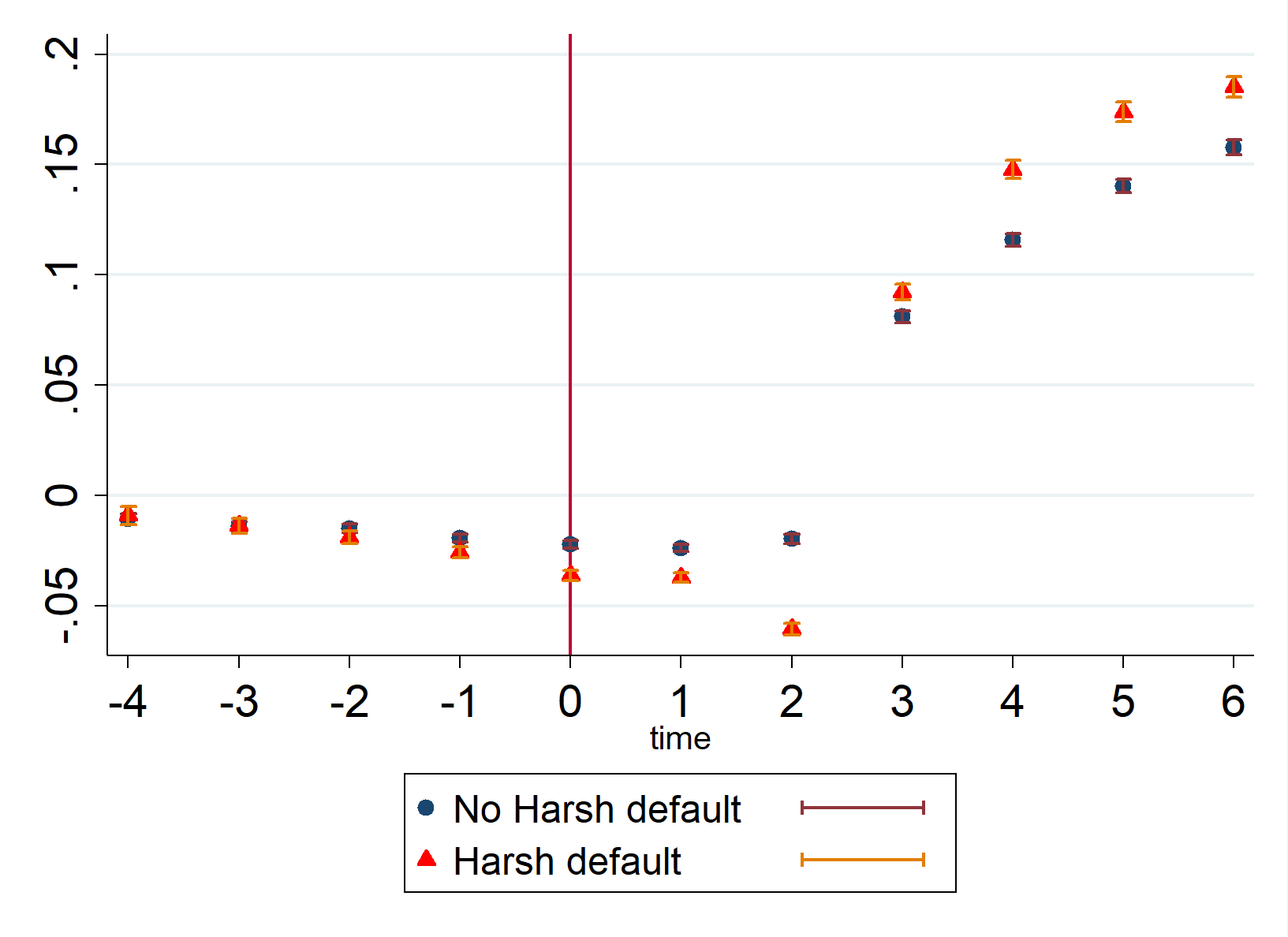}
              
                  	Panel (iii) - Revolving Balance \hspace{1cm}Panel (iv) - N. of collections\\
                     
             \includegraphics[height=4cm, width=7cm]{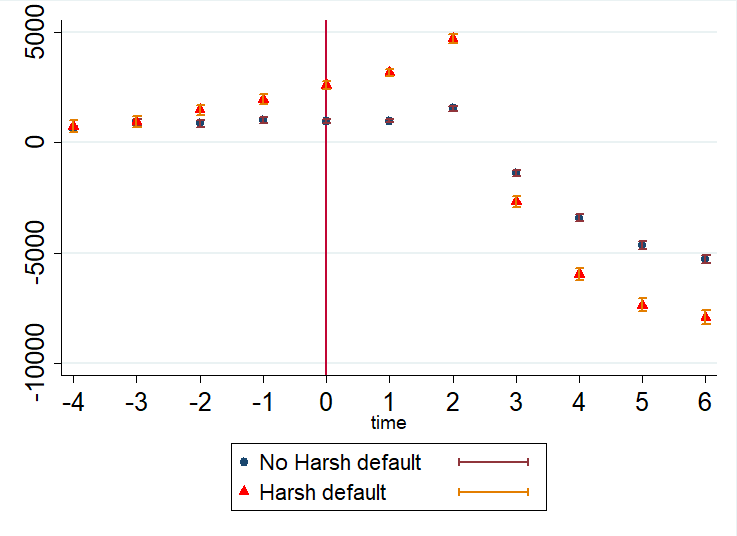}          	            	                \includegraphics[height=4cm, width=7cm]{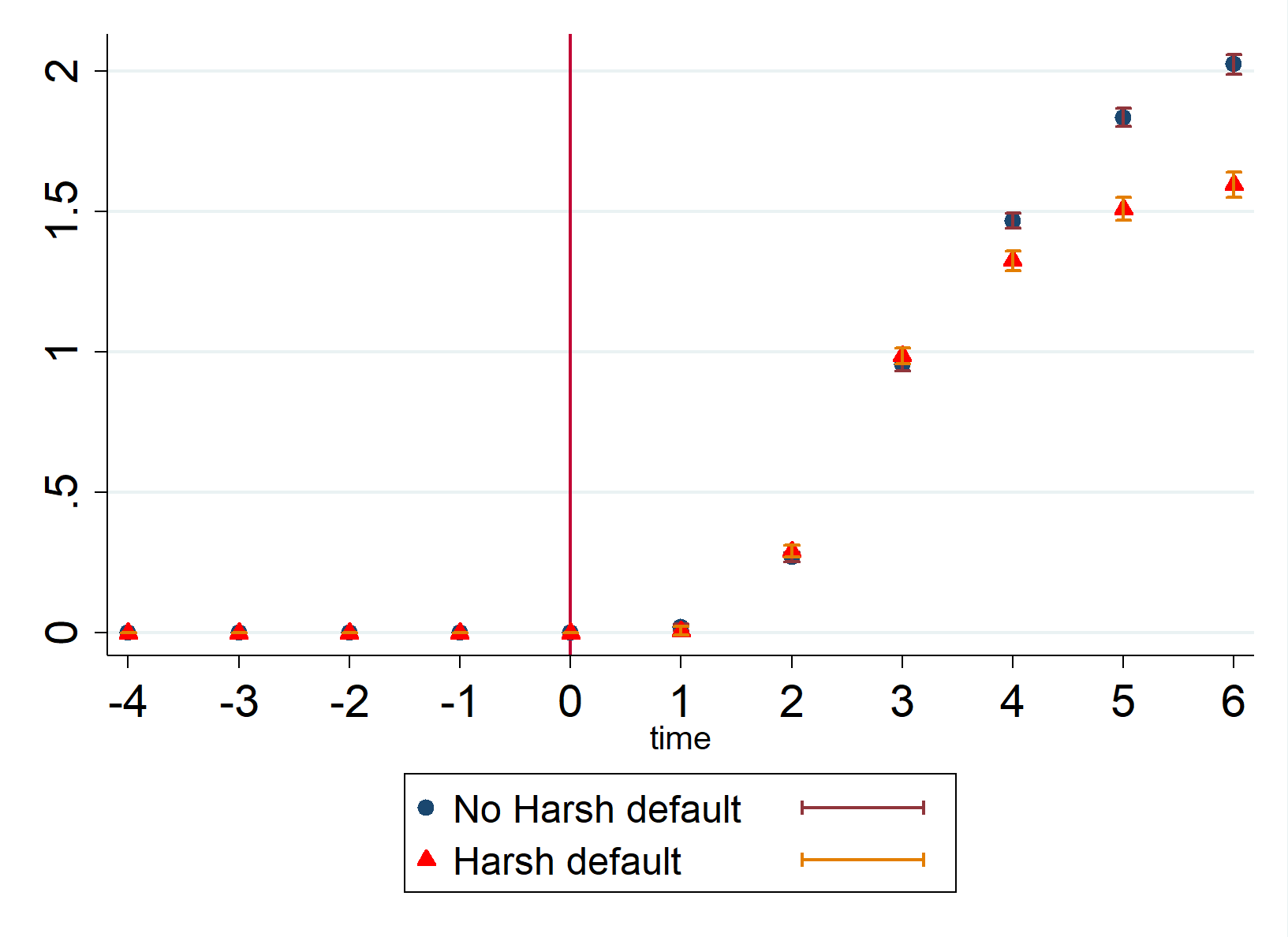}
                
                	Panel (v) - Home own \hspace{1cm}Panel (vi) - Total credit limit\\
                     
             \includegraphics[height=4cm, width=7cm]{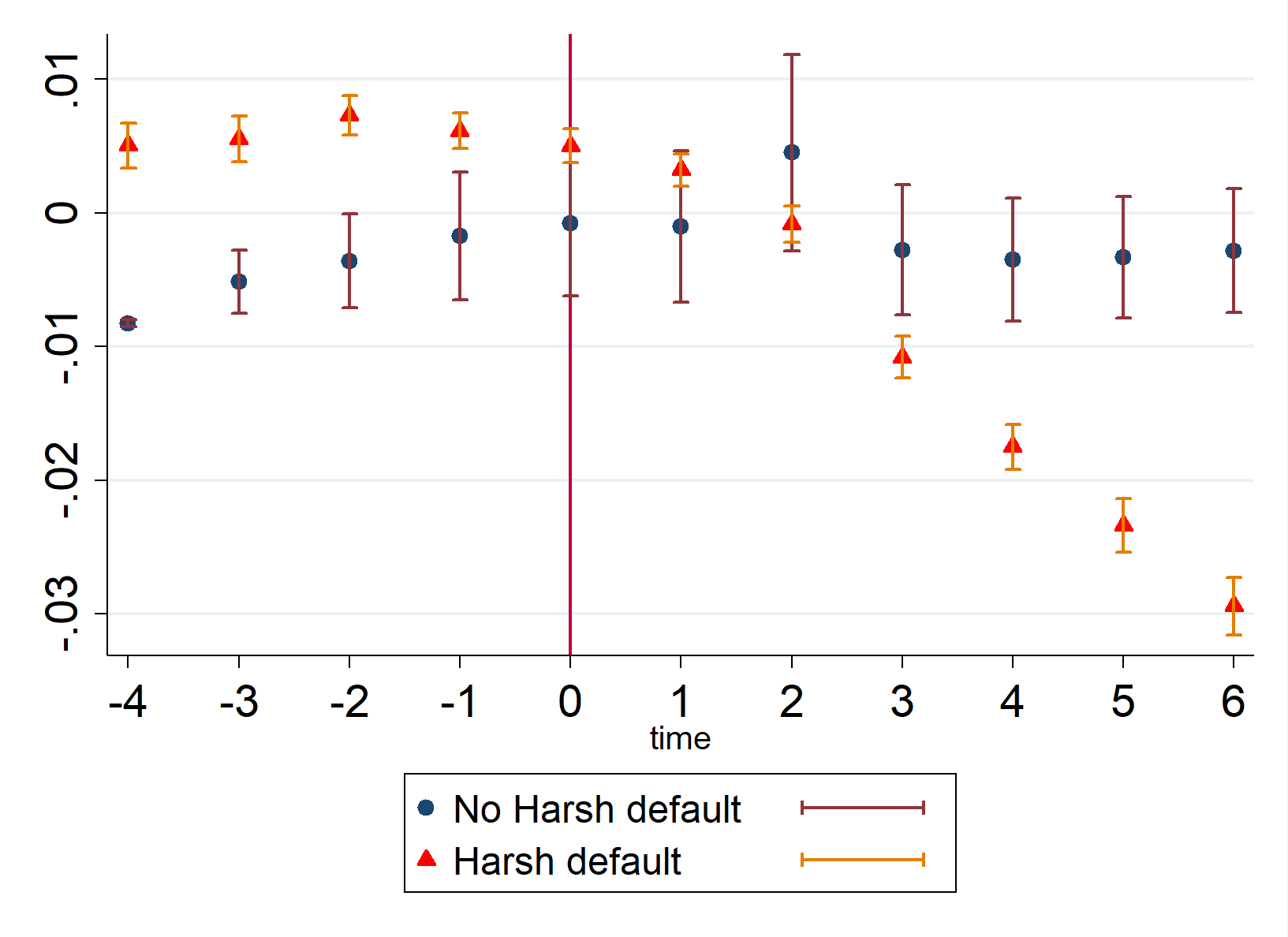}          	            	                \includegraphics[height=4cm, width=7cm]{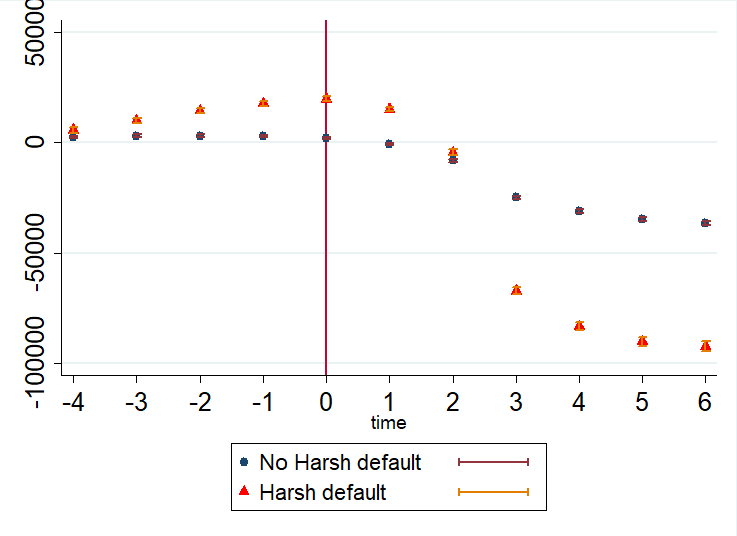}
                
                Panel (vii) - Mortgage balance open \hspace{1cm}Panel (viii) - Amount 90-180 days delinquent\\
                         \includegraphics[height=4cm, width=7cm]{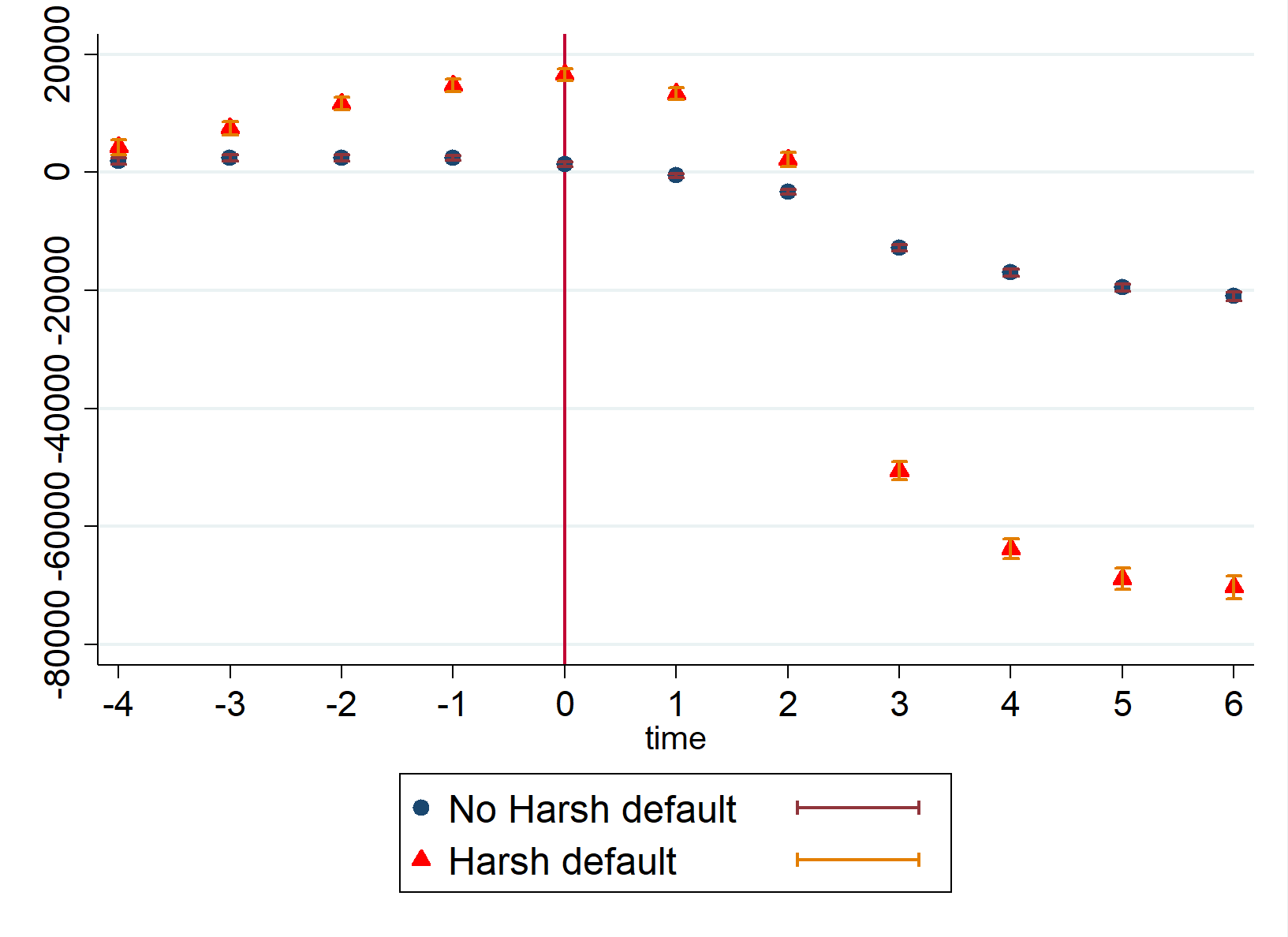}    
             \includegraphics[height=4cm, width=7cm]{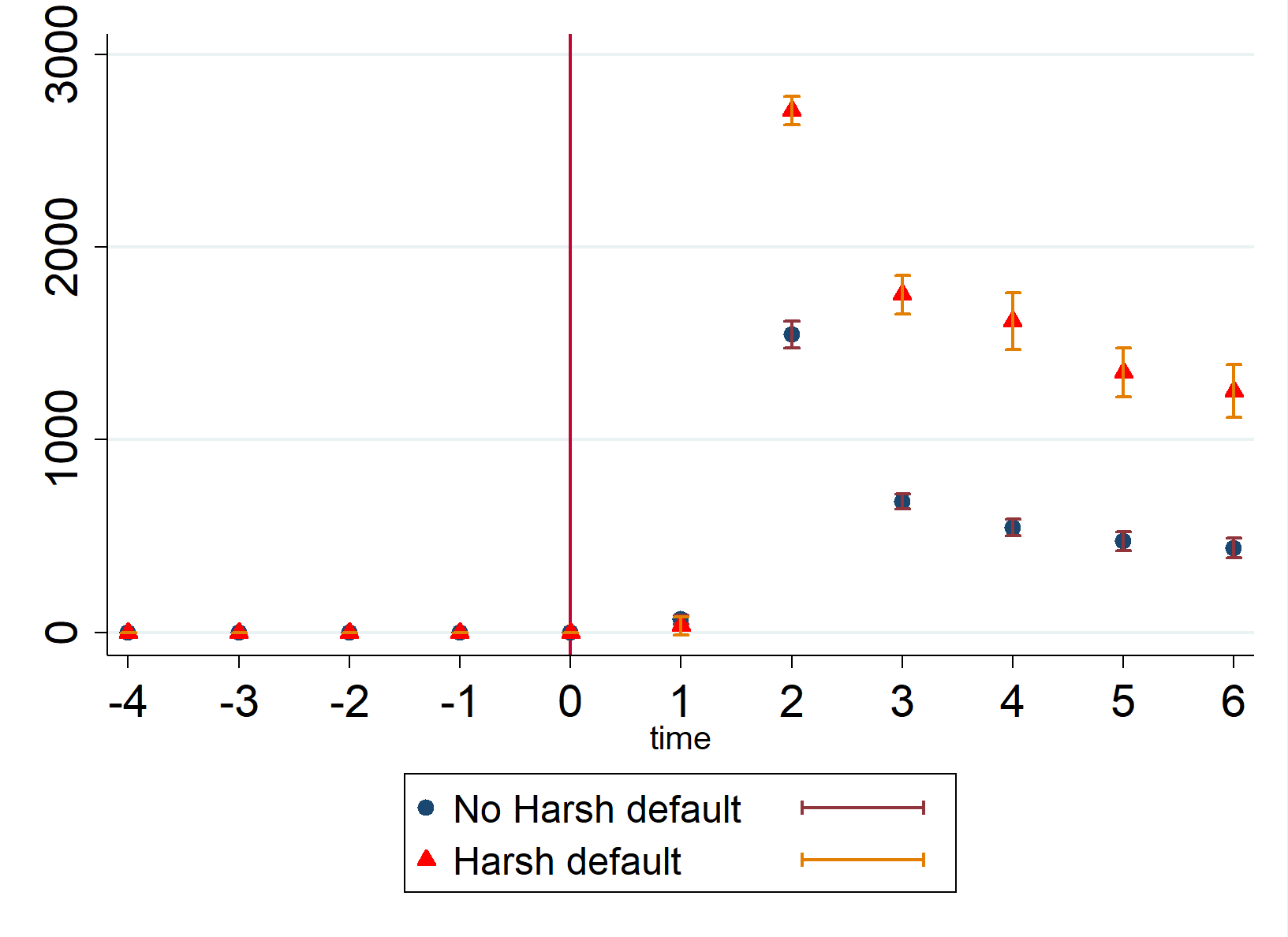}

        \caption{\small Event study: dependent variable is: (i) Mortgage origination: this variable takes value 1 if the individual has a higher number of mortgage trades in year $t$ than in year $t-1$ or if the number of months since the most recent mortgage has been opened is less than 12, and zero otherwise, (ii) probability that total credit limit is lower than 10,000USD, (iii) total amount open on all revolving credit trades, (iv) probability of experiencing a harsh default (Chapter 7, Chapter 13 or foreclosure), (v) probability of being homeowner, i.e. either being recorded as a homeowner by Experian or having ever had a mortgage open (vi) total credit limit on all trades, (vii) open amount of mortgage balance, (viii) amount 90-180 days delinquent. The event considered is a soft default, i.e. a 90-day delinquency, but no Chapter 7, Chapter 13 or foreclosure taking place in the same year, neither before in the sample period. Other controls are age and age squared, credit score in 2004 and in 2005. 95\% confidence intervals around the point estimates.
        }  \label{fig:heterogeneity_harsh_credit}  
     \end{figure}

From Figure \ref{fig:heterogeneity_harsh_credit} we find evidence that, for harsh defaulters, the probability of opening a new mortgage declines by about 10pp, i.e. about ten times more than in the baseline estimates. Further, their probability of having a low credit limit increases by more than 15pp, this last number being quite similar to that estimated for the overall sample. The number of their collections increases by about 1 or 2, whereas their revolving balance open decreases on average by more than 5,000USD (i.e. slightly less than in the baseline estimates).

The probability of being homeowners also declines by about 3pp, i.e. about twice that of our baseline estimates presented in Section \ref{sec:exploratory}. Further, the drop in total credit limit, about minus 90,000-100,000USD, is notably larger than that recorded in the baseline results, by about 20,000-30,000USD.
Finally, the amount delinquent increases by about 3,000 USD, whereas the amount of mortgage balance open declines by about 70,000USD (i.e. about 20,000 USD more than in the baseline estimates).

Also in the case of credit-derived variables, we find evidence that the impacts of a soft default are less serious for the non-harsh defaulters than for the harsh defaulters. Indeed, the probability of opening a new mortgage  declines by about 5pp, vs the minus 10pp for the harsh defaulters. However, the probability of having a low credit limit still increases by 10-15pp, i.e. a number that is close both to the baseline estimates and to the results for the harsh defaulters. The number of collections still increases by 1-2 over time and the revolving balance decreases by about 5,000 USD, i.e. for these two variables we do not record huge differences. The probability of being a homeowner essentially stays unchanged, whereas it declines by about 2-3pp in the medium run for the harsh defaulters. The drop in the total credit limit, about minus 40,000USD is about 30\% smaller than that recorded in the baseline estimates and about half of that recorded for the harsh defaulters. Further, the amount delinquent increases less than for the harsh defaulters (i.e. about 1,500 USD vs 3,000 USD for the harsh defaulters) and the amount of mortgage balance open only declines by about 20,000 USD (i.e. about half of the baseline estimates) vs minus 70,000 USD for the harsh defaulters.

\subsection{Heterogeneity across Delinquent Amounts}

\begin{figure}[H] 
         
             	Panel (i) - Move Zip \hspace{4cm}Panel (ii) - Move CZ\\
                \includegraphics[height=4.7cm, width=7cm]{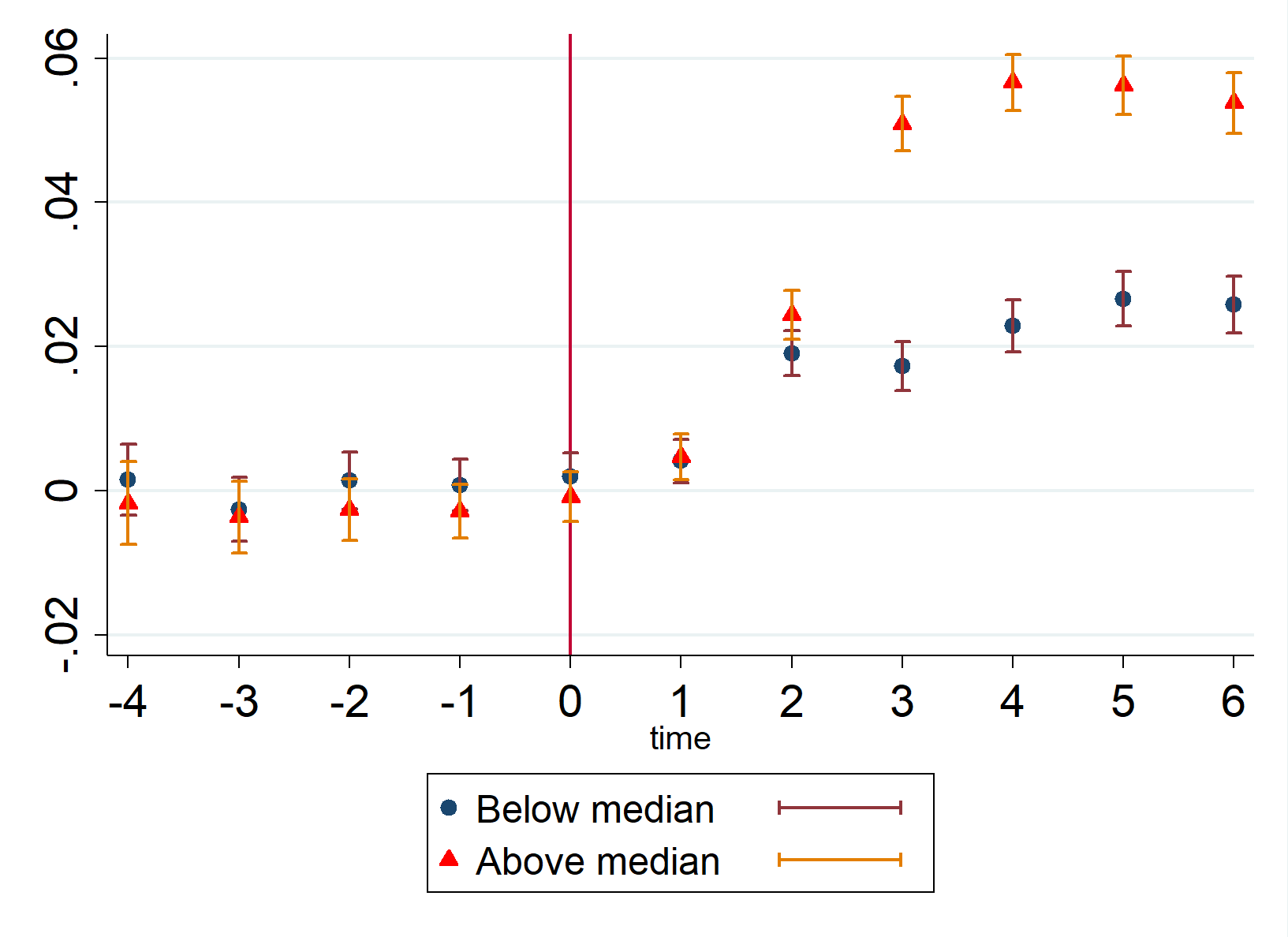}
             \includegraphics[height=4.7cm, width=7cm]{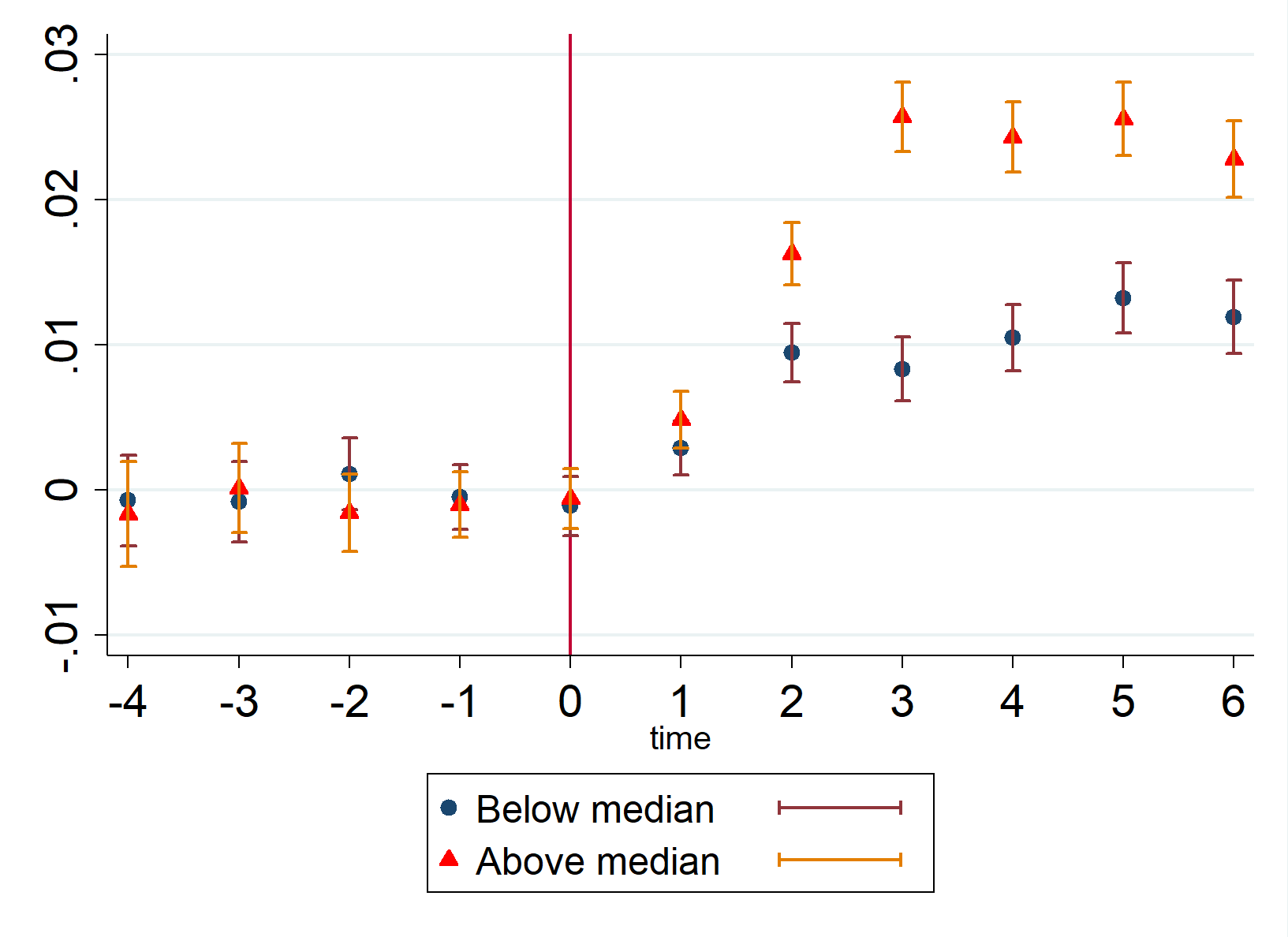}
             
                  	Panel (iii) - Median House Value \hspace{2cm} Panel (vi) - Income\\
                                \includegraphics[height=4.7cm, width=7cm]{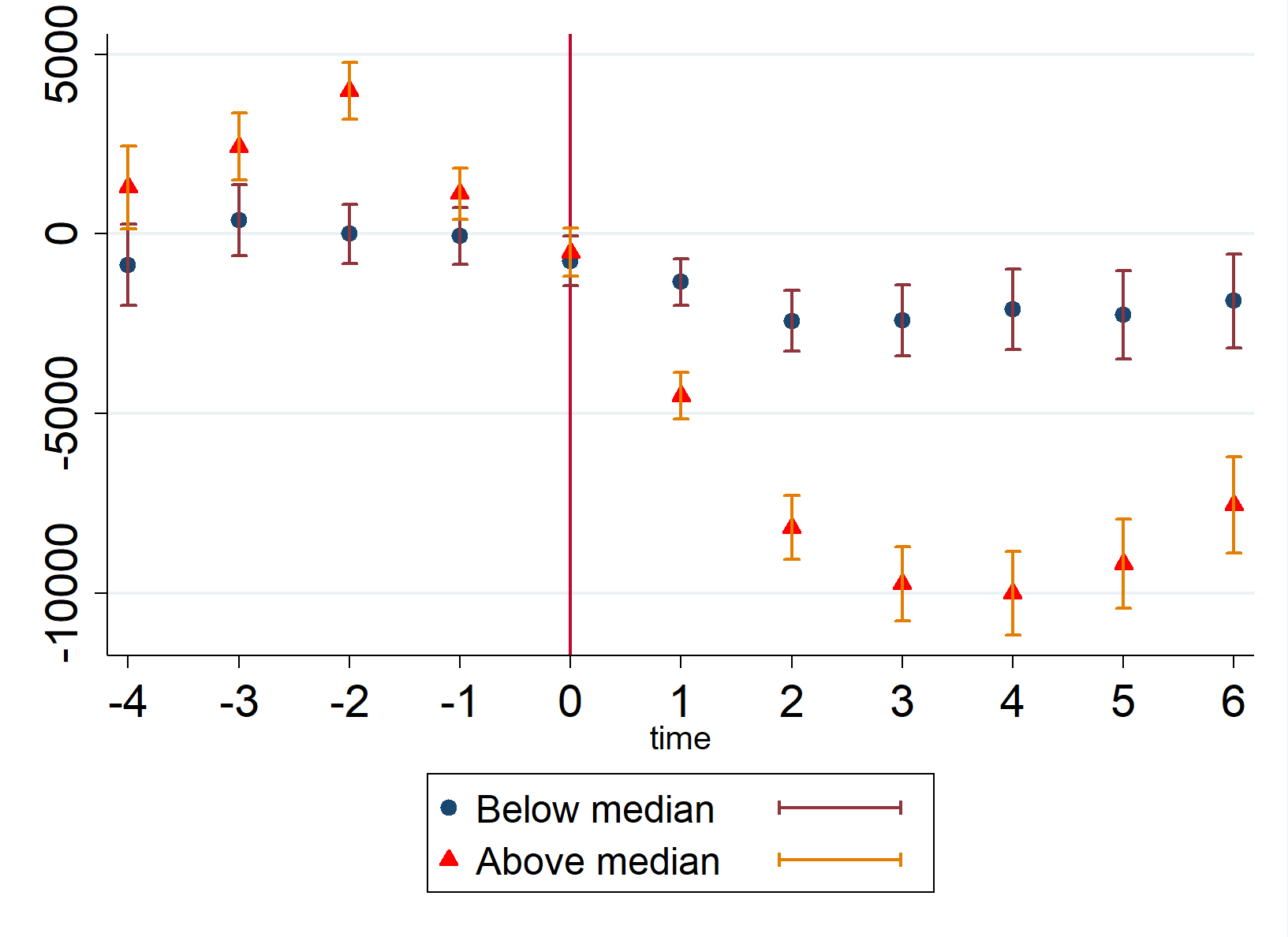}
             \includegraphics[height=4.7cm, width=7cm]{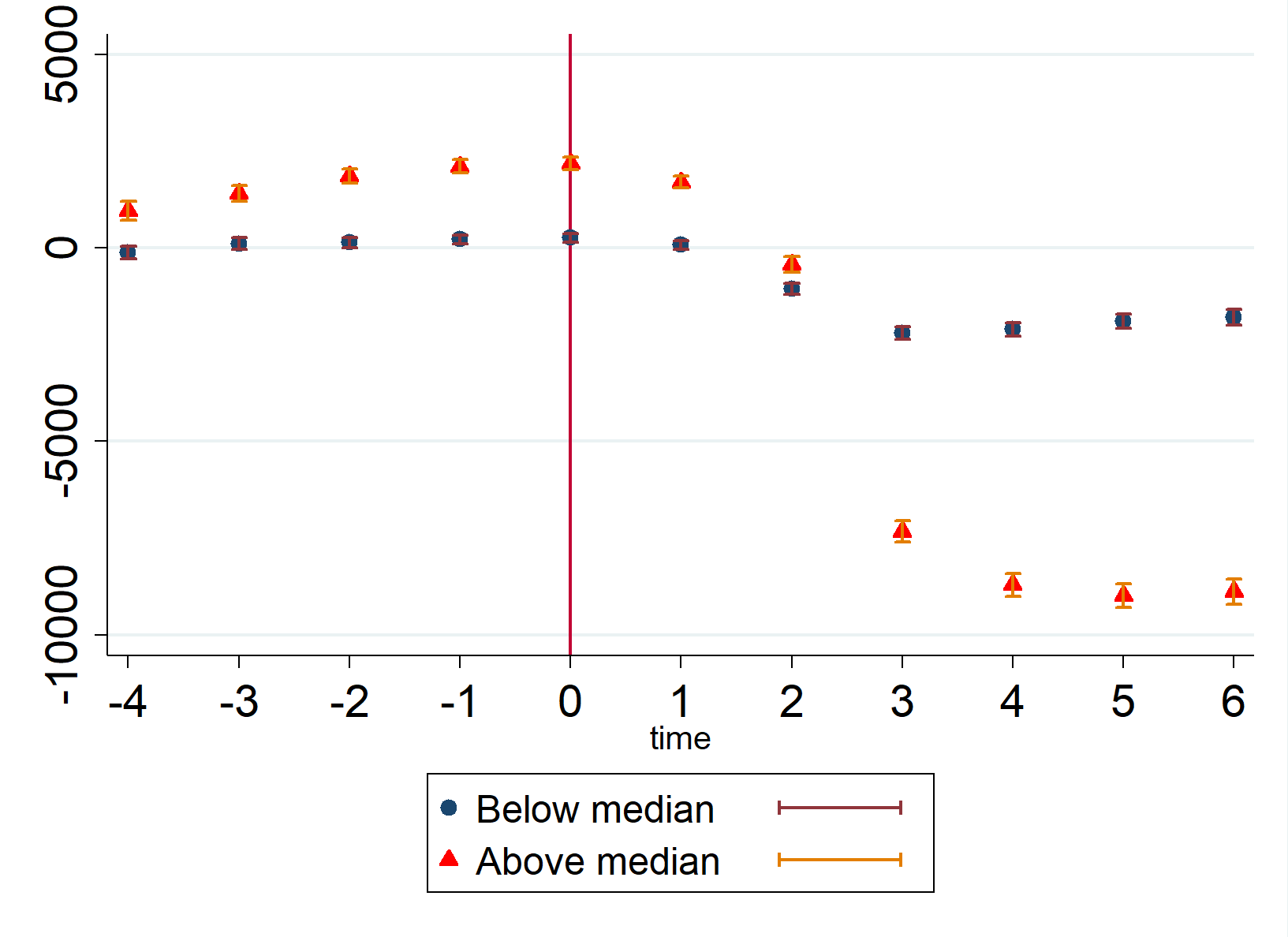}

             Panel (v) - Credit card consumption \\
                \includegraphics[height=4.7cm, width=7cm]{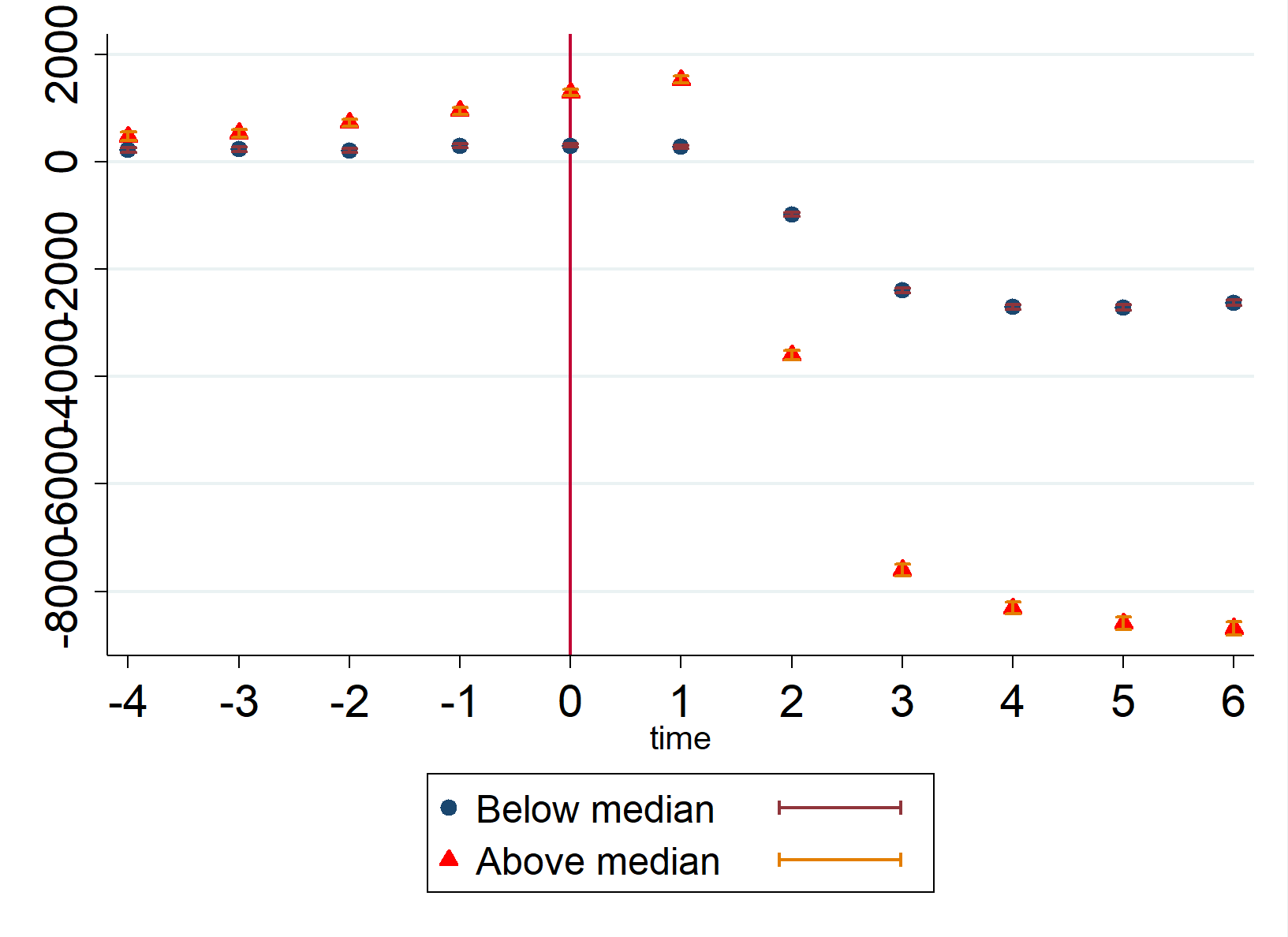}
                          
        \caption{Event study: dependent variable is: (i) Probability of moving zip code. This variable takes value 1 if the individual is a different zip code in year $t$ than in year $t-1$, and zero otherwise, (ii) Probability of moving outside the commuting zone. This variable takes value 1 if the individual is in a different commuting zone in year $t$ than in year $t-1$, and zero otherwise, (iii) the Median House Value in the zip code of residence at year $t$ (iv) income imputed by Experian, (v) credit card consumption: total balance on all open credit card trades reported in the last 6 months. The event considered is a soft default, i.e. a 90-day delinquency, but no Chapter 7, Chapter 13 or foreclosure taking place in the same year, neither before in the sample period. Other controls are age and age squared, credit score in 2004 and in 2005. 95\% confidence intervals around the point estimates.
        }  \label{fig:heterogeneity_amount_mobility}    
     \end{figure}

In this Subsection, we analyze whether the baseline results presented in the event studies in Section \ref{sec:exploratory} are notably different for individuals with a low versus high delinquent amount. We define the groups on the basis of the median of the dollar amount which is delinquent between 90 and 180 days in the year in which the first soft default takes place (923USD).
 From the results, in Figure \ref{fig:heterogeneity_amount_mobility}, it clearly emerges that the impact on our outcome variables of interest is harsher for those individuals with a higher delinquent amount. For example, the increase in the probability of moving zip code is +6pp for those above the median, vs +2-3pp for those below (Panel (i)).

 \begin{figure}[H] 
 
             	 Credit Score\\
                \includegraphics[ width=12cm]{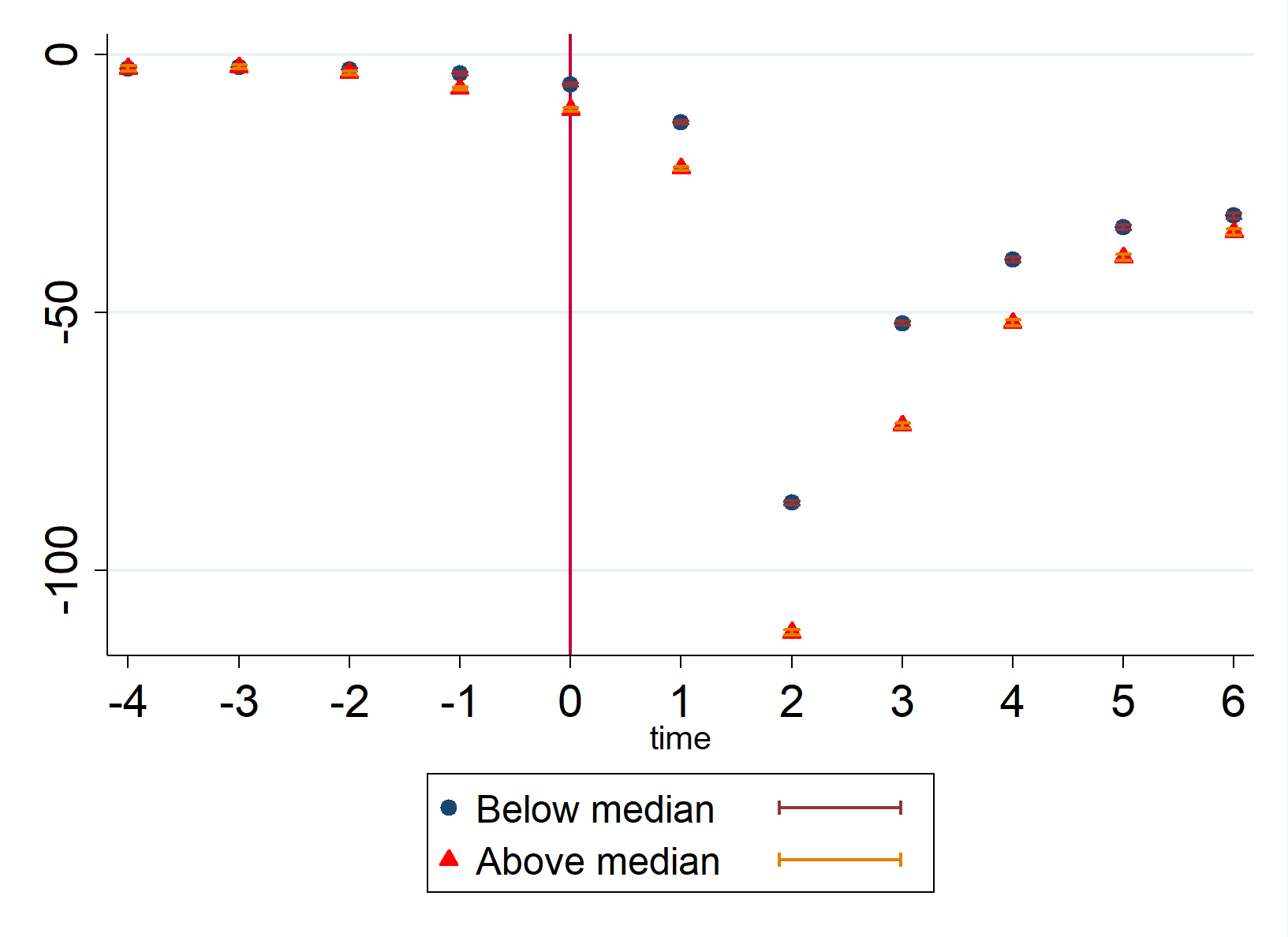}
    \caption{Event study: dependent variable is Credit Score. The event considered is a soft default, i.e. a 90-day delinquency, but no Chapter 7, Chapter 13 or foreclosure taking place in the same year, neither before in the sample period. Other controls are age and age squared, credit score in 2004 and in 2005. 95\% confidence intervals around the point estimates.}     \label{fig:heterogeneity_amount_cs}
     \end{figure}

 Similarly, the probability of moving out of the commuting zone increases by about 3pp for those above the median, but only by 1pp for those below  (Panel (ii)). In terms of median house value  (Panel (iii)), we notice that the drop for those above the median delinquent amount is larger (i.e. -10,000USD vs -2,000USD). In addition, there are some differences in the pre-trends for those above the median, that suggest that they were in more expensive areas before the soft default. Moreover,  (Panel (iv)) income drops by about 10,000USD for those with a high delinquent amount, but only by about 2,000USD for those with a low delinquent amount. Further, credit card consumption drops by only about 3,000USD for those with a delinquent amount below the median, but it declines by more than 8,000USD for those above the median  (Panel (v)).
  The differences among the two groups appear long-lasting (i.e. up to 6 years after the event). 
  For those with a high delinquent amount, getting back on track is really hard. it seems that there are effects as large as 120-130 credit score points, whereas the loss is only about 80 points for those with a low delinquent amount (Figure \ref{fig:heterogeneity_amount_cs}).

\begin{figure}[H] 
           
             	Panel (i) - Prob. new mortgage \hspace{1cm}Panel (ii) - Prob. Credit Limit $<$10k\\

                                \includegraphics[height=4cm, width=7cm]{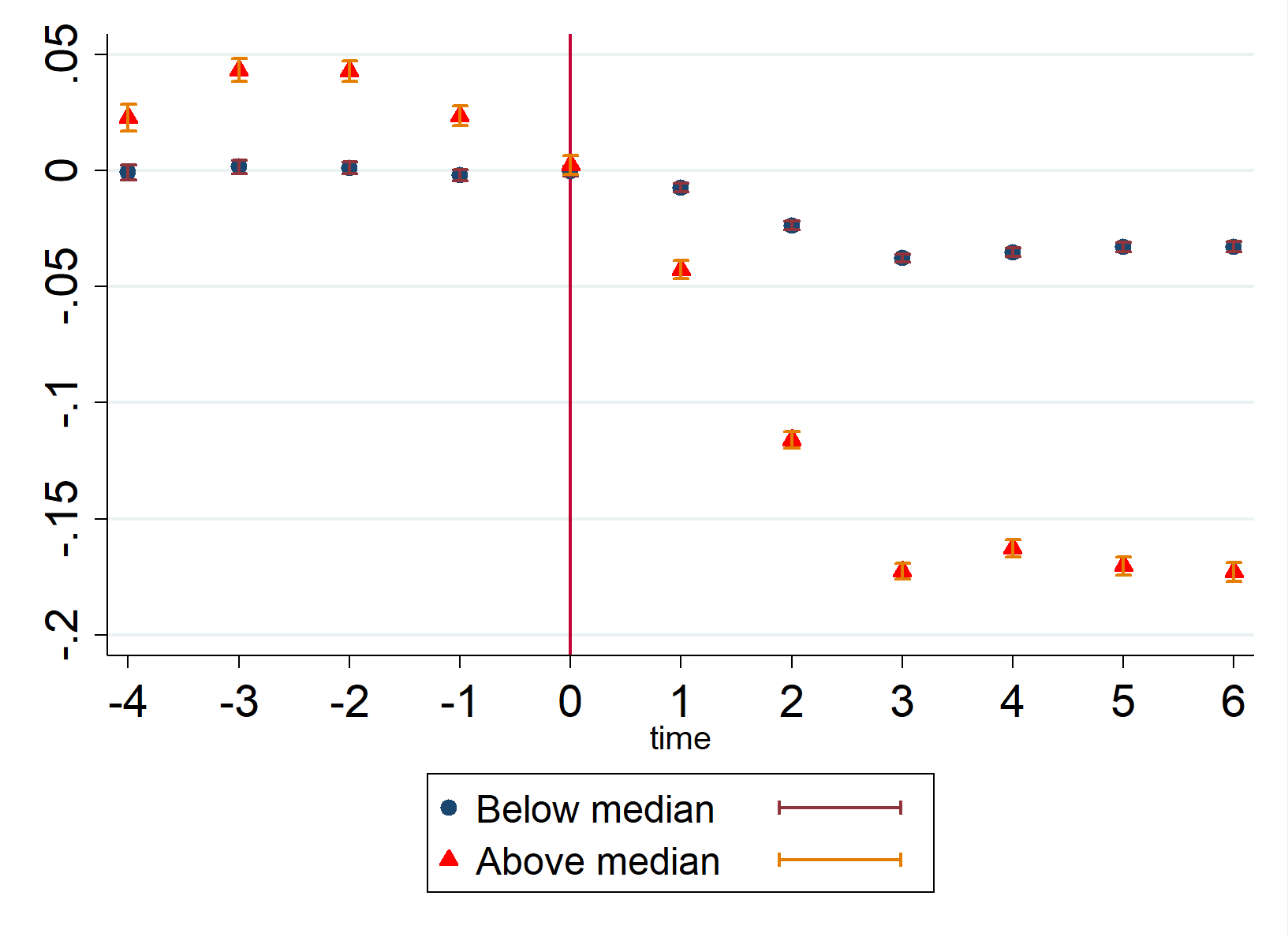}  
              \includegraphics[height=4cm, width=7cm]{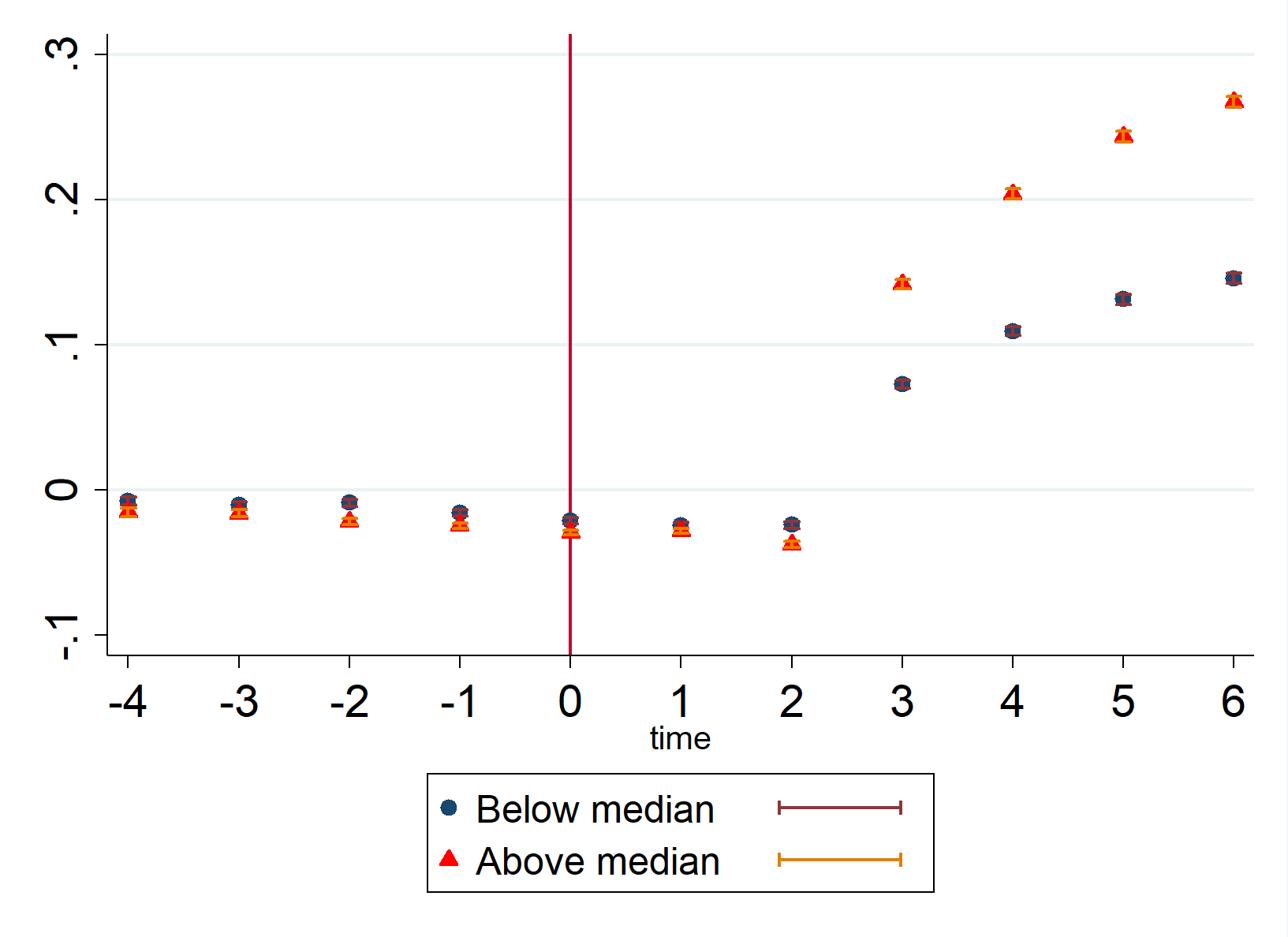}
              
                  	Panel (iii) - Revolving Balance \hspace{1cm}Panel (iv) - Harsh default\\
                     
             \includegraphics[height=4cm, width=7cm]{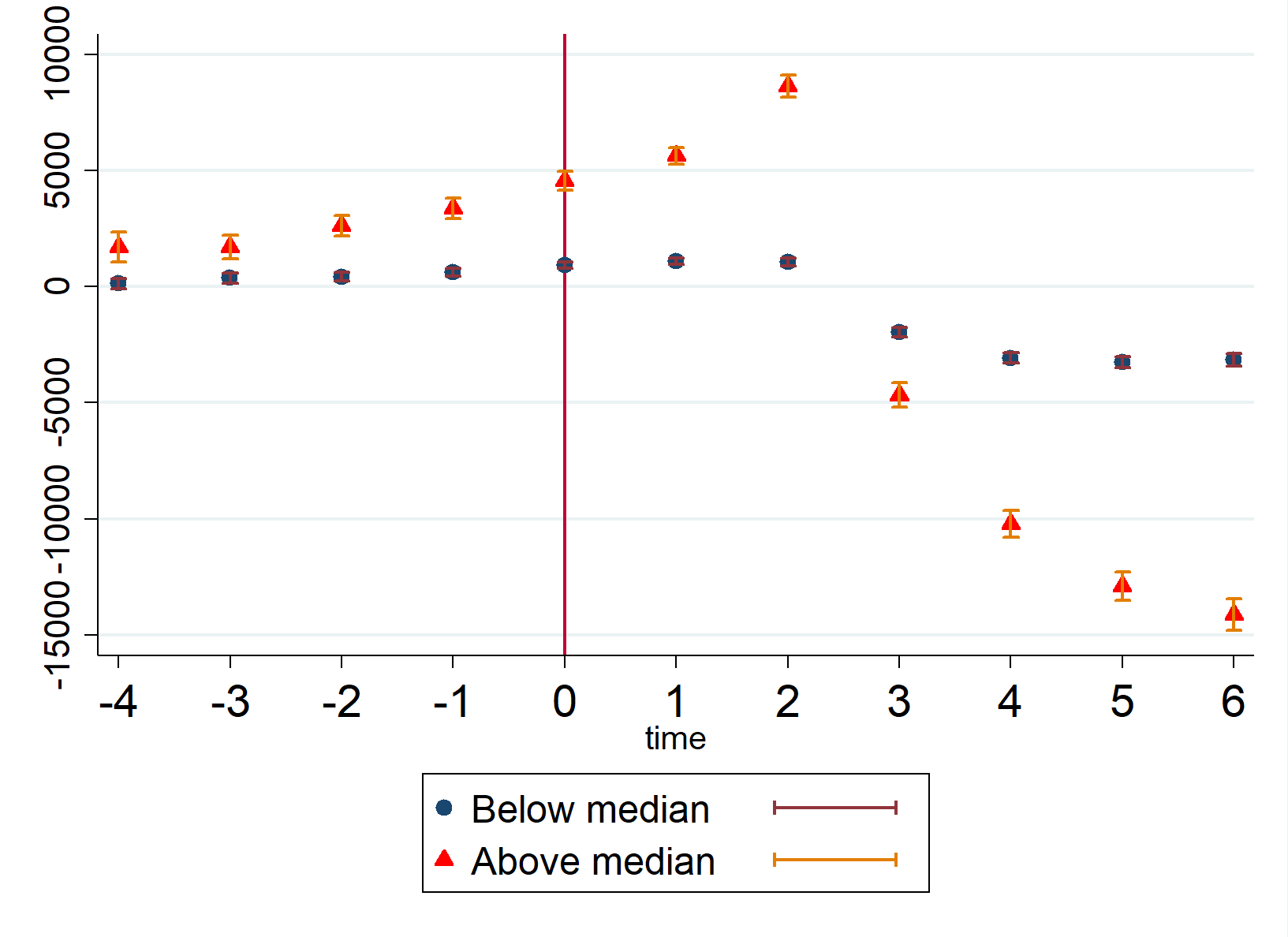}          	            	
                \includegraphics[height=4cm, width=7cm]{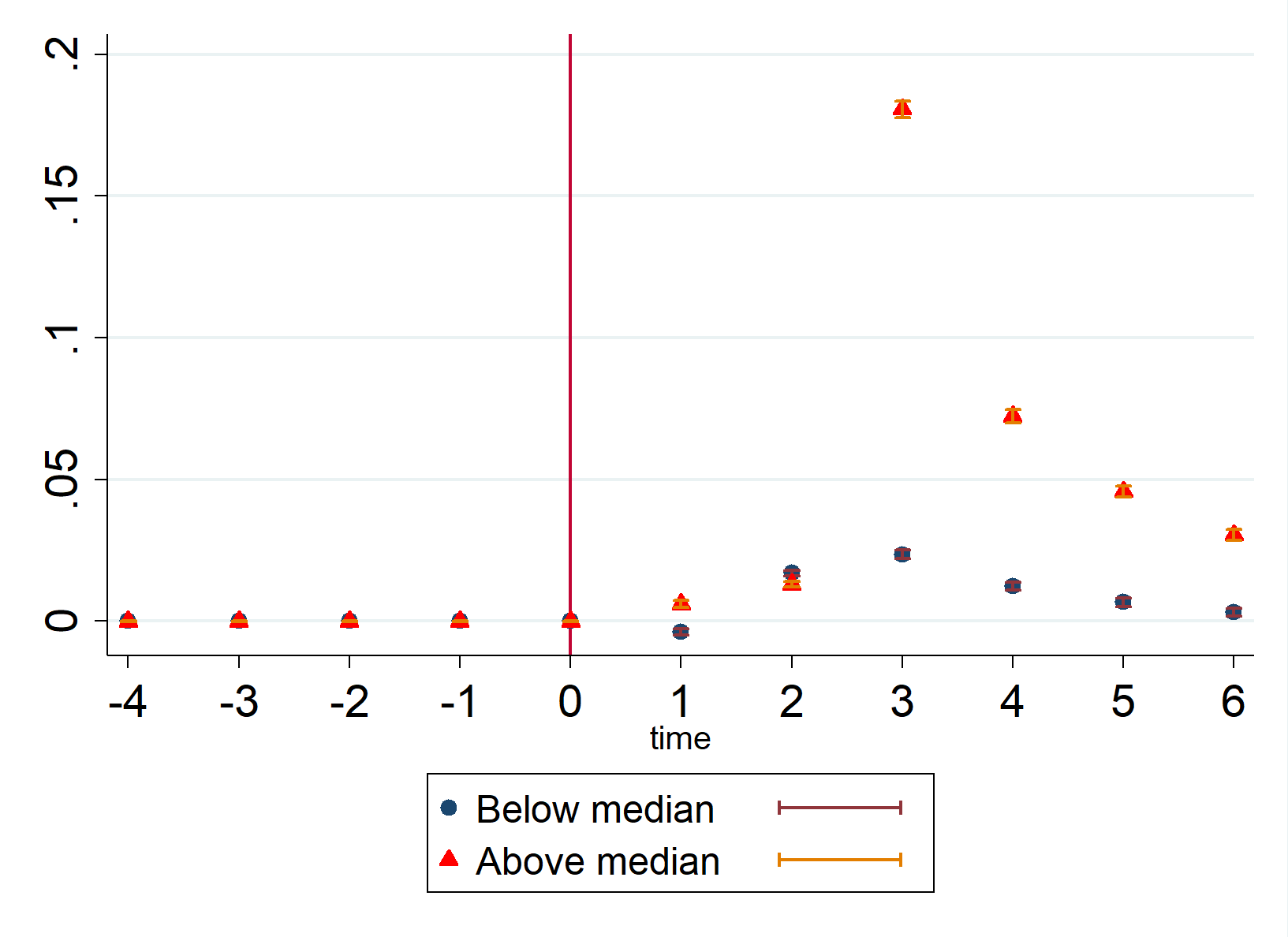}
                
                	Panel (v) - Home own \hspace{1cm}Panel (vi) - Total credit limit\\

               \includegraphics[height=4cm, width=7cm]{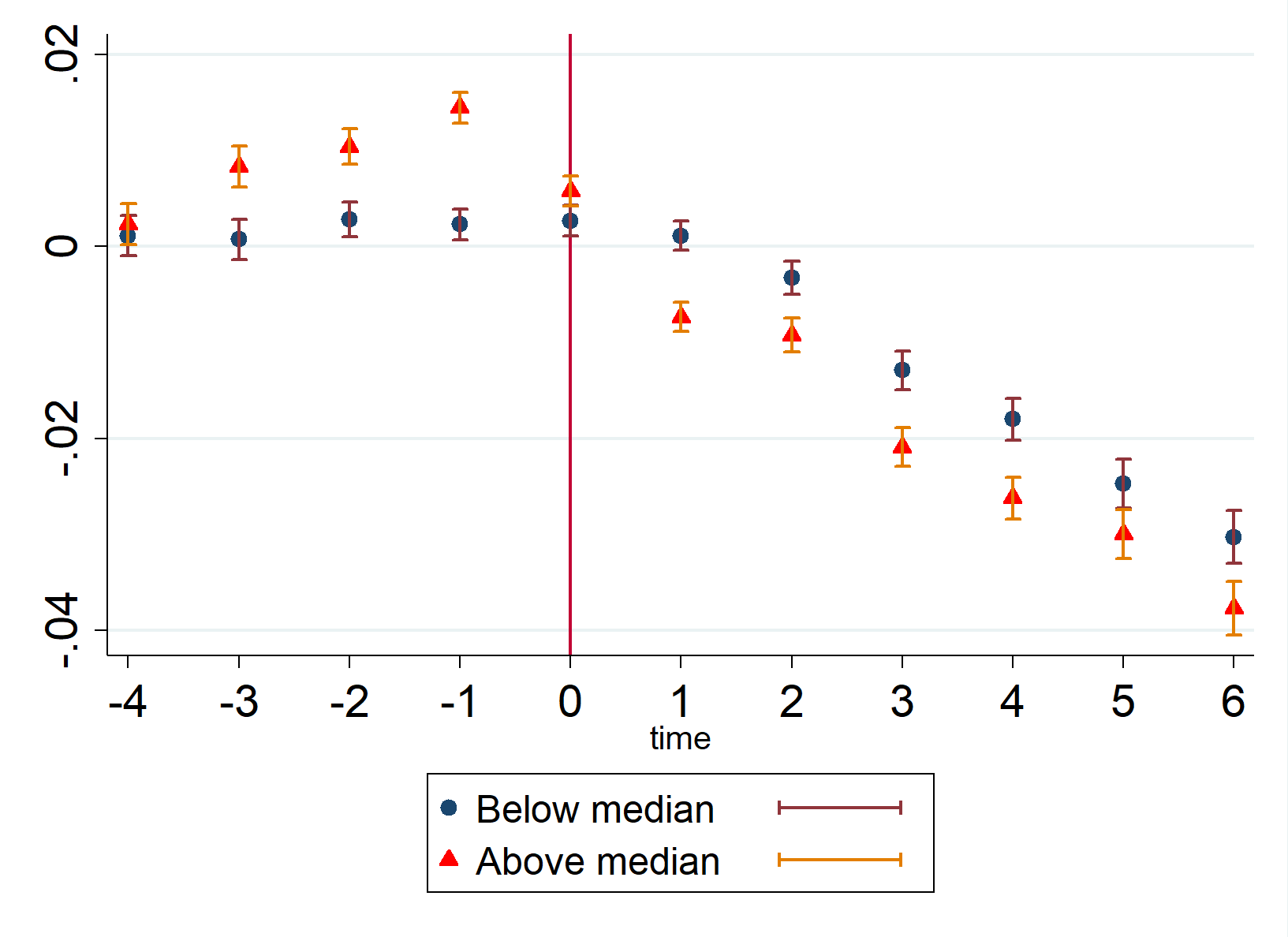}  
                \includegraphics[height=4cm, width=7cm]{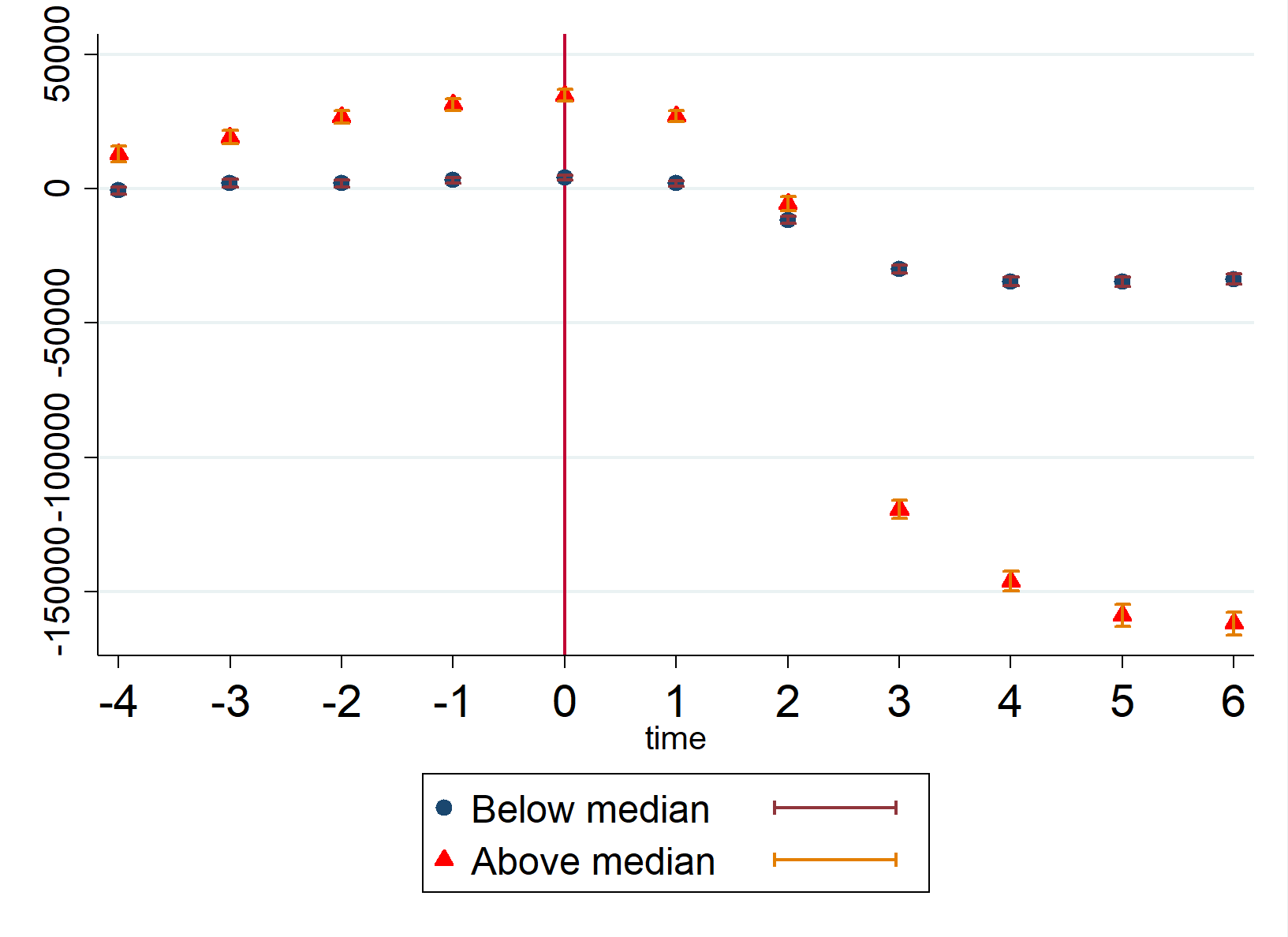}
                
                Panel (vii) - Mortgage balance open \hspace{1cm}Panel (viii) - N. of collections\\
                     
             \includegraphics[height=4cm, width=7cm]{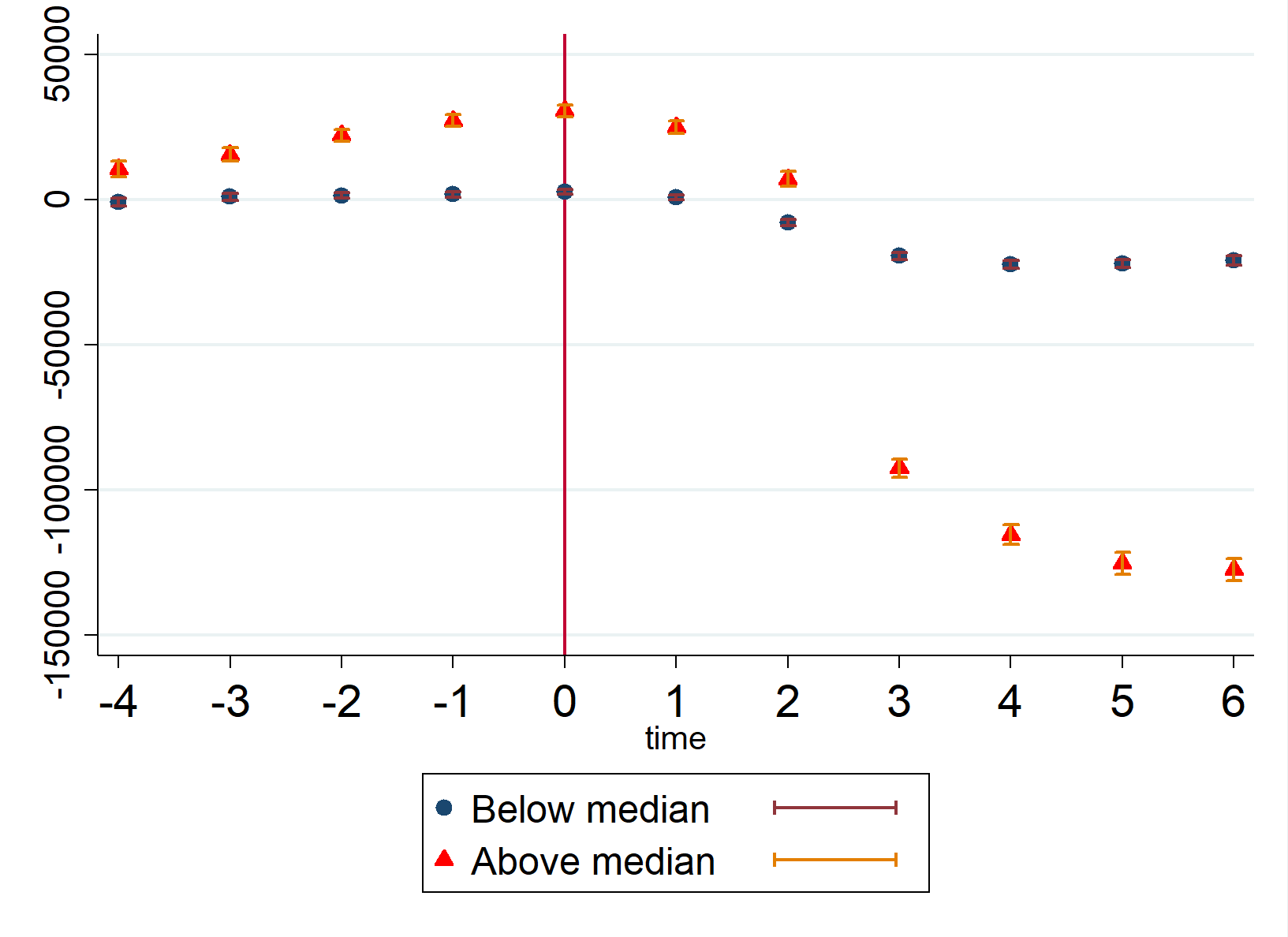}          	            	
       \includegraphics[height=4cm, width=7cm]{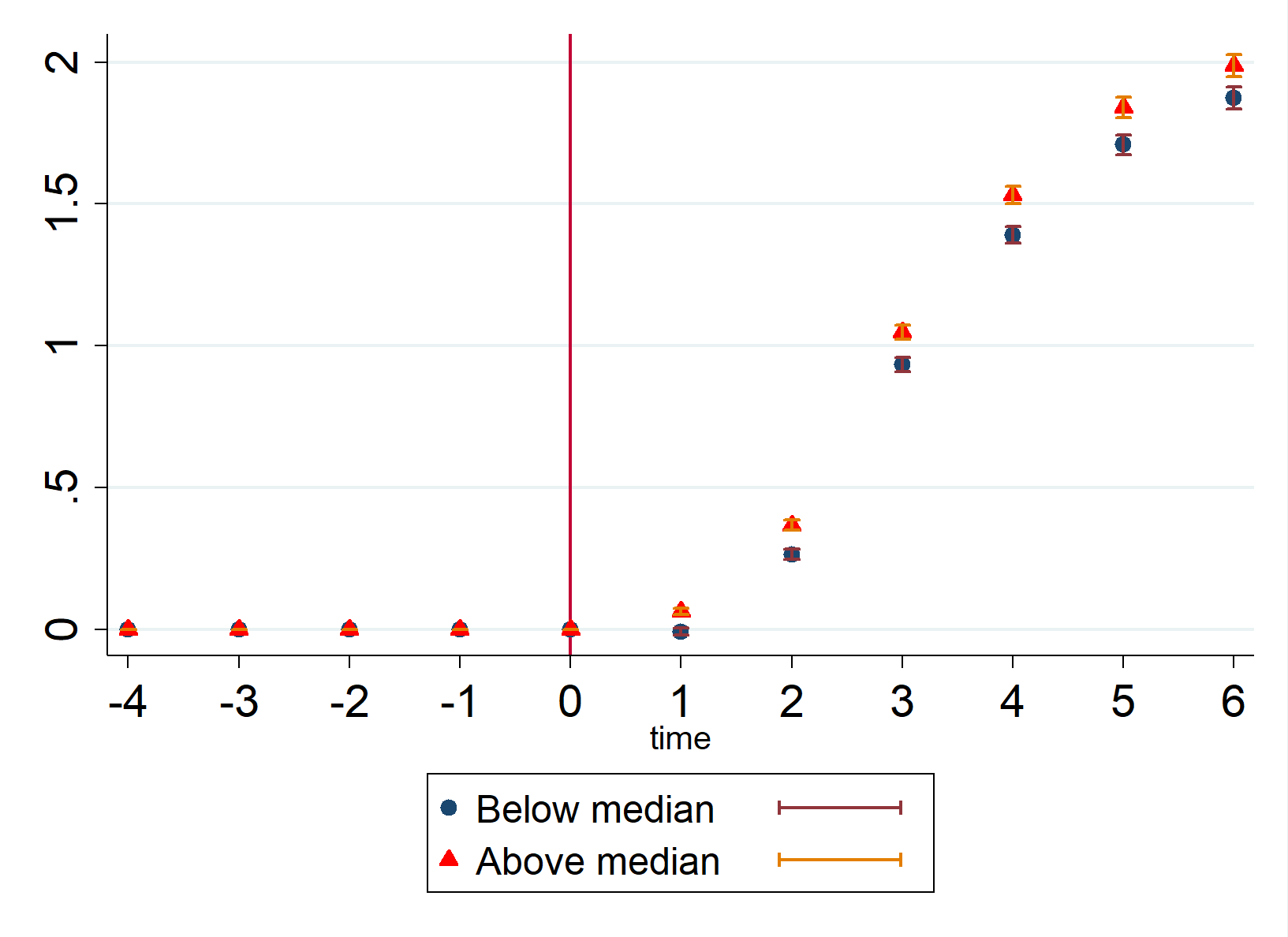}

        \caption{ \small Event study: dependent variable is: (i) Mortgage origination: this variable takes value 1 if the individual has a higher number of mortgage trades in year $t$ than in year $t-1$ or if the number of months since the most recent mortgage has been opened is less than 12, and zero otherwise, (ii) probability that total credit limit is lower than 10,000USD, (iii) total amount open on all revolving credit trades, (iv) probability of experiencing a harsh default (Chapter 7, Chapter 13 or foreclosure), (v) probability of being homeowner, i.e. either being recorded as a homeowner by Experian or having ever had a mortgage open (vi) total credit limit on all trades, (vii) open amount of mortgage balance, (viii) number of collections. The event considered is a soft default, i.e. a 90-day delinquency, but no Chapter 7, Chapter 13 or foreclosure taking place in the same year, neither before in the sample period. Other controls are age and age squared, credit score in 2004 and in 2005. 95\% confidence intervals around the point estimates.
        }     \label{fig:heterogeneity_amount_credit} 
     \end{figure}

   As far as credit-derived outcomes are concerned, depicted in Figure \ref{fig:heterogeneity_amount_credit},  we confirm the general pattern of larger effects for those defaulting on larger amounts. The probability of opening a new mortgage drops by about 15pp some years after the event for those with a high delinquent amount, whereas it only declines by about 4pp for those with a low delinquent amount (Panel (i)). Similarly, the probability of having a low credit limit increases by about 25pp for those above the median, but only about 15pp for those below  (Panel (ii)). The revolving credit balance drops by 15,000USD for those with a high delinquent amount vs minus 3,000USD for those with a low delinquent amount  (Panel (iii)). Also, we notice that those with a high delinquent amount had a higher pre-event revolving balance. This provides further evidence that those individuals were over-extending themselves, which leads to a later soft default. In addition, the probability of recording a later harsh default peaks to +18pp three years after the soft default for those with a high delinquent amount  (Panel (iv)). In contrast, the increase is about only around +2-3pp and stable over time for those with a low delinquent amount. As far as homeownership  (Panel (v)), total credit limit  (Panel (vi)), and the amount of mortgage balance open  (Panel (vii)) are concerned, in all these cases we notice the presence of differential pre-trends. Individuals with a high delinquent amount had pre-event higher probability of being home-owners, higher total credit limit and higher mortgage balance open. This suggests that those individuals engaged in buying property that was too expensive for their means and/or were hit by an increase in the interest rates that reverberates in an increase in their mortgage rates.\footnote{Unfortunately, in our dataset, we have no indication on whether a mortgage is fixed rate or variable rate, so we are not able to dig deeper into this direction of analysis.} Consequences of the soft default are larger in most cases for those above the median delinquent amount. The probability of being a homeowner drops by about 1-2pp for those with a low delinquent amount, vs minus 2-3pp for those with a delinquent amount above the median. The negative impact for those with a higher delinquent amount appears uniformly larger over time (by about 0.5pp), thus providing further indications that a potentially relevant channel for the soft default is the over-extension of individuals and families for buying (expensive) property.
   The total credit limit declines by 150,000USD vs less than 50,000 and the mortgage balance open diminishes by about 125,000USD (vs minus 25,000 for those with a low delinquent amount). Finally, the number of collection increases by around about +1-2 both for those above and for those below the median delinquent amount. This last impact is very close to our baseline estimates of the effect.

\section{Taking Stock}
\label{sec:taking_stock}

The analysis proposed thus far points towards severe and long-lasting consequences of soft-default, with effects that last well beyond the short term, and in fact they seem to be rather persistent and, for many outcomes, increasing even after five years from the event. Our analysis of heterogeneity of the effects brings some light into the mechanisms at play, of course any heterogeneity analysis might be plagued with issues and should be interpreted with caution, especially when such heterogeneity is defined post-events as in our case (in the previous Section \ref{sec:heterogeneity}).\footnote{We thank the  editor and the referees for pushing us in this direction.}
It is quite interesting, for example, that the credit score seems to recover somewhat in that time span, yet it appears on average about 40 points lower even after 6 years from the event, irrespective of the size of the delinquent amount and with no large difference for those with and without subsequent harsh defaults. 
However, the defaulters on larger amounts or with a subsequent harsh default have substantially higher penalties in terms of income and location (see Figures \ref{fig:heterogeneity_harsh_mobility} and \ref{fig:heterogeneity_amount_mobility}),  they move to lower  median home values areas and to zip codes with lower average wages and higher shares of minorities (see Appendix \ref{sec:quality}). 
What seems to be happening is that there are consumers who are delinquent on smaller amounts, possibly because of uninsurable shocks, who suffer the consequences of such defaults, but substantially less than those who default on larger amounts and seek bankruptcy and other legal reliefs. The latter appear to have overextended their lines of credit, in particular on mortgages (presumably because of location choices), then gone under in their accounts and essentially diverged from their earlier life trajectories. They end up in substantially worse neighborhoods (of different CZs) with median home values that decrease about 4-times as much as those for the lower delinquent amounts/no-harsh default. 
These moves to new CZs seem also to substantially affect the labor market outlooks for this population,  their yearly income falls  by almost 10,000USD (about 5-times as much as for the low delinquent amounts). These results are in line with the fact that the new neighborhoods appear to be of lower labor market opportunities (lower wages, fewer establishments, and jobs) as shown in Figure \ref{fig:zipcode_quality}. 

Furthermore, the large defaulters, coming from large mortgage amounts, are essentially excluded from the housing and credit market (very low credit lines, unlikely to have a mortgage, or if a mortgage is open that comes with very low amounts, and generally very low credit limits), see Figure \ref{fig:heterogeneity_harsh_credit} and \ref{fig:heterogeneity_amount_credit}.

  \section{Conclusions}\label{sec:conclusions}

In this paper we study the impact of soft default on a variety of outcomes such as credit availability, credit score, income, probability of moving to a different zip code and  commuting zone. To do this, we use credit bureau data, from Experian, covering 1\% of the US  population with valid reports in 2010 and for whom we have yearly observations  2004-2020. We adopt multiple empirical approaches to study the impact of soft default on individual trajectories. 

Our findings are that the impacts of a soft default are substantial and long-lasting, up to 10 years after the event.  
After a soft default, an individual experiences an increase in the probability of moving to a different zip code by about 4pp, and an increase in the probability of moving to a new commuting zone by about 1-2pp. 

A soft default is also linked to substantial income losses. This impact is long-lasting and statistically significant up to 10 years after the event, and equal to about -6,000USD on average, but much larger for individuals with a large delinquent amount. 
Individuals experiencing a soft default witness a surge by 15pp in the probability of having a  low credit limit (i.e. lower than 10,000USD), as well as an 8,000USD drop in their revolving balance in the medium term.

Finally, we find evidence that the impact of a default is heterogeneous at least across two dimensions, i.e. the amount delinquent and the presence or not of a harsh default in a subsequent year to the soft default. Indeed, the effects of a default on our outcome variables of interest are more marked for those who will also record a harsh default later on, as well as for those with a high (i.e. above the median) dollar delinquent amount. Our interpretation, based on the pre-trends of variables such as the Median house value of the zip code of reference or the amount of mortgage open, is that those individuals were over-extending themselves and most likely an external shock (e.g. health shock, increase in the mortgage interest rate, etc...) caused them to default. 

Our findings are policy relevant, as knowing the cost of soft defaults for the individual is essential in order to design adequate debt relief policies. 

Interestingly, in our machine learning exercise we control for local labor and credit market conditions and find that the long term effects of a soft default are rather substantial and do not seem to be driven by those conditions or other local time invariant characteristics.

\newpage

  \typeout{}
  \bibliography{Manuscript}

\newpage
\appendix

\section{Soft and Harsh Defaults in Credit Reports}
\label{sec:harsh_soft}

Positive and negative credit events are recorded by credit bureaus, i.e. Experian, Equifax, TransUnion, in individuals' credit reports. Such events stay in the report for some time, depending on the event type. 
For example, Experian keeps soft defaults for up to 7 years, chapter 7 for 10 years and Chapter 13 for 7 years.\footnote{In the case of Experian see \url{https://www.experian.com/blogs/ask-experian/how-long-does-it-take-information-to-come-off-your-report/}, for Equifax \url{https://www.Equifax.com/personal/education/credit/report/how-long-does-information-stay-on-credit-report/}.} This means that other things equal, an individual with a negative episode will have a lower credit score, of course the other things equal is not a plausible situation as the negative episode will immediately lower the score and diminish the ability of that individual to participate in credit operations. Therefore, while the flag for Chapter 7 will stay on for at most 10 years, its impact will typically diminish overtime and the individual credit score could in principle recover in a much shorter time span, e.g. opening secured credit cards (security deposit for a given credit line).

\section{Pooled Credit Data}

\label{sec:appb}
\setcounter{table}{0}
\renewcommand{\thetable}{B\arabic{table}}

\setcounter{figure}{0}
\renewcommand{\thefigure}{B\arabic{figure}}

\begin{table}[H]\centering \caption{Summary statistics of our main variables, 2004-2020. Top 1\% of total credit limit, total balance on revolving trades and total revolving credit limit have been trimmed for readability. Credit scores lower than 300 have also been trimmed. \label{tab:descriptives}}
\normalsize
\begin{tabular}{l c c c c c}\hline\hline
\multicolumn{1}{c}{\textbf{Variable}} & \textbf{Mean}
 & \textbf{Std. Dev.}& \textbf{Min.} &  \textbf{Max.} & \textbf{N}\\ \hline
 Credit score & 681.7447 & 114.4018 & 300 & 839 & 24212846\\
Income & 48860.1644 & 26281.1134 & 1000 & 337000 & 24225810\\
Median House Value & 211944.7779 & 163304.9158 & 0 & 2298873&  24421655\\
Mortgage Bal. & 52984.9853 & 144575.4774 & 0 & 23709369 & 24664952\\
Mortgage origination & 0.203 & 0.402 & 0 & 1&24664952 \\
Homeowner\footnote{This is a dummy variable that takes value 1 if either the individual has ever had a mortgage or if she is recorded as homeowner by Experian (imputed variable) and zero otherwise} & 0.5937 &    0.4912   &      0       &   1 & 24664952\\
Age & 52.4836 & 16.8479 & 18 & 130 & 24664952\\
Move ZIP & 0.1536 & 0.3605 & 0 & 1 & 23382932\\
Move CZ& 0.0592 & 0.236 & 0 & 1 & 23167015\\
New Delinq. (90+) & 0.0105 & 0.1018 & 0 & 1 & 24664952\\
Total credit limit on open trades (all)   & 101814.0926 & 191836.7158 & 0 & 83842952 & 24664952\\
Total balance on revolving trades & 8448.4735 & 31579.6963 & 0 & 13653348 & 24664952\\
Total credit limit on open rev. trades & 30190.935 & 60625.2228 & 0 & 20036200 & 24664952\\
Prob. credit limit $<$10k & 0.1072 & 0.3093 & 0 & 1 &24664952 \\
Harsh default & 0.0079 & 0.0885 & 0 & 1&24664952 \\
Number of collections & 1.4932 & 3.8632 & 0 & 90 & 24664952 \\
Amount 90-180 days delinquent & 249.7265 & 5173.9714 & 0 & 6300910 & 24664952\\
Credit Card balance open & 4737.8246 & 9884.5008 & 0 & 3682167& 24664952 \\
\hline\end{tabular}
\end{table}

\section{Data Reliability}
\label{sec:reliability}

\setcounter{table}{0}
\renewcommand{\thetable}{C\arabic{table}}

\setcounter{figure}{0}
\renewcommand{\thefigure}{C\arabic{figure}}

\begin{figure}[H]
\caption{Data representativeness}\label{fig:representativeness}
\includegraphics[width=13cm]{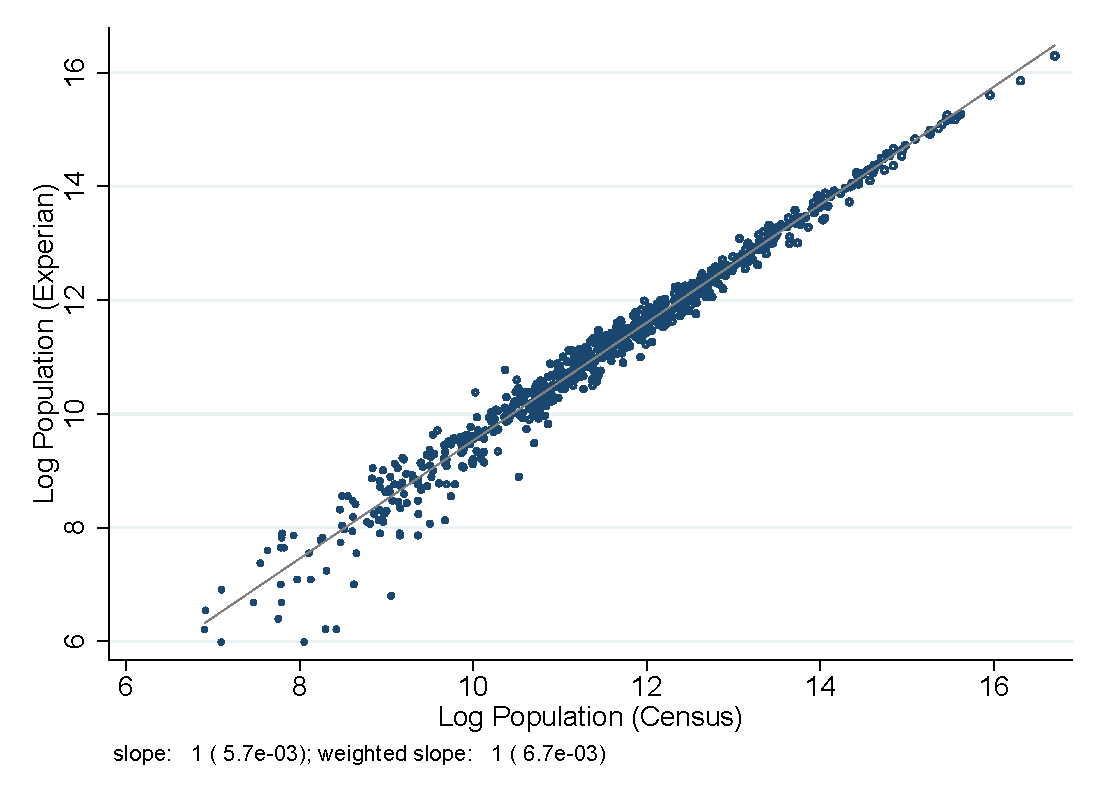} 
\end{figure}

\begin{figure}[H]
\caption{Validity of income imputation}
\includegraphics[width=13cm]{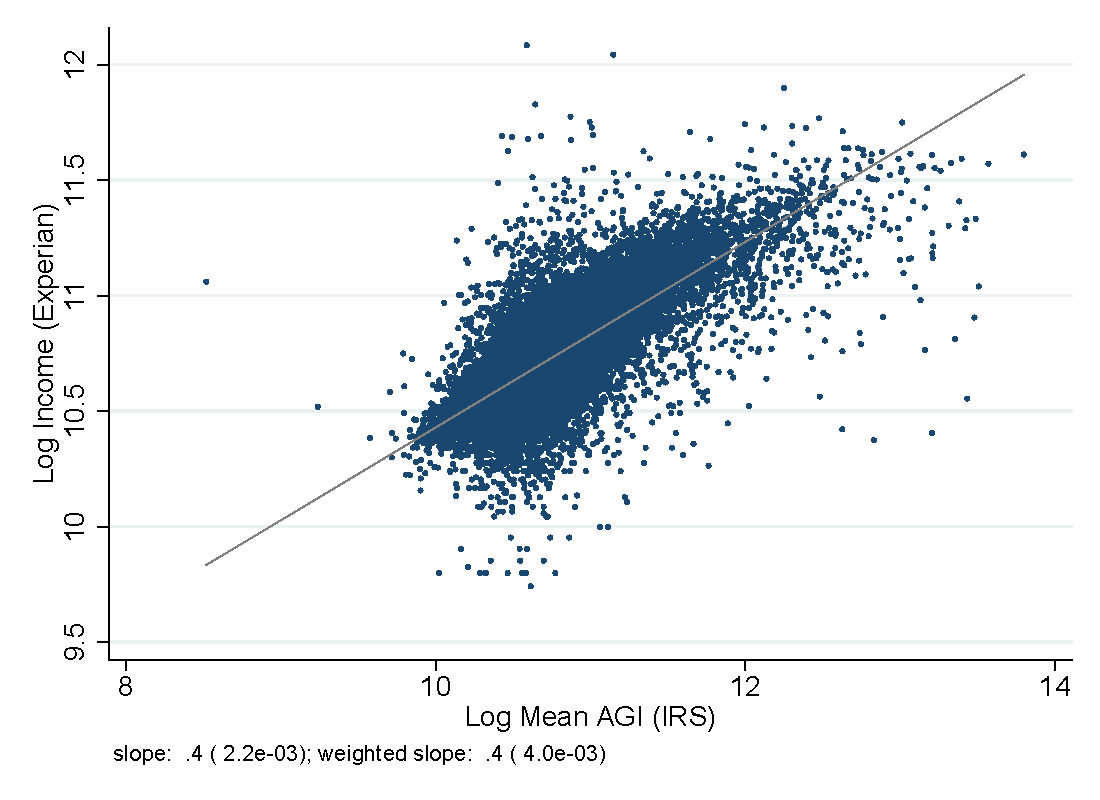} \label{fig:income}
\end{figure}


\begin{table}[htbp]\centering \caption{PSID Labor Incomes for the 2009/11 Waves. Negative or zero incomes dropped \label{tab:PSID}}
\normalsize
\begin{tabular}{l*{2}{c}}
\hline
            &\multicolumn{2}{c}{Labor Income }\\
\midrule
 Mean  &    48378.05\\
 Sd          &    82146.32\\
  Min          &          10\\
 Max           &     5210000\\ \hline
\end{tabular}
\end{table}

 
 \begin{table}[htbp]\centering \caption{(Log)Labor Income Age Profiles in PSID and Experian. We use Labor incomes in  the PSID and the W2 imputed income in Experian, we restrict the sample to the years between 2009 and 2011 and to individuals age 25 to 60.  \label{tab:PSID_age}}
\normalsize
\begin{tabular}{l*{3}{c}}
\hline
            &\multicolumn{1}{c}{Labor Income PSID}  &\multicolumn{1}{c}{Labor Income Experian}\\

 Age  &    .099 &    .067 \\
Std. error       &    (.011) & (.0003)\\
 Age Squared          &          -.001 & -.001\\
  Std. error           &          (.0001) & (.0000) \\
 N           &     10,302 & 4,390,814\\ \hline
\end{tabular}
\end{table}

    \begin{figure}[H]
   \includegraphics[height=3.5cm, width=6cm]{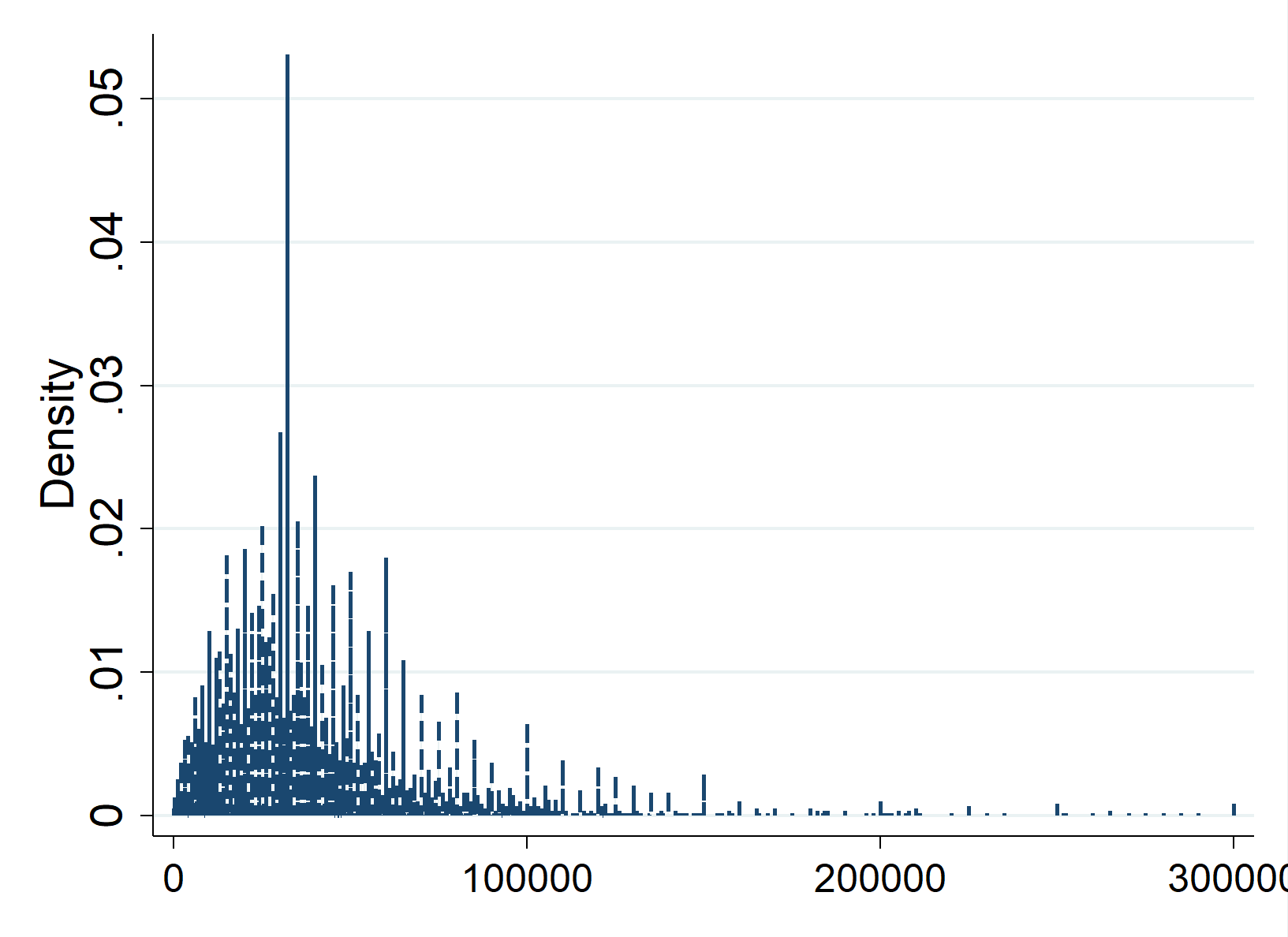}
      \includegraphics[height=3.5cm, width=6cm]{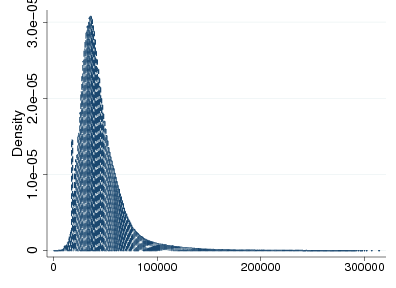} \\
      
         \includegraphics[height=3.5cm, width=6cm]{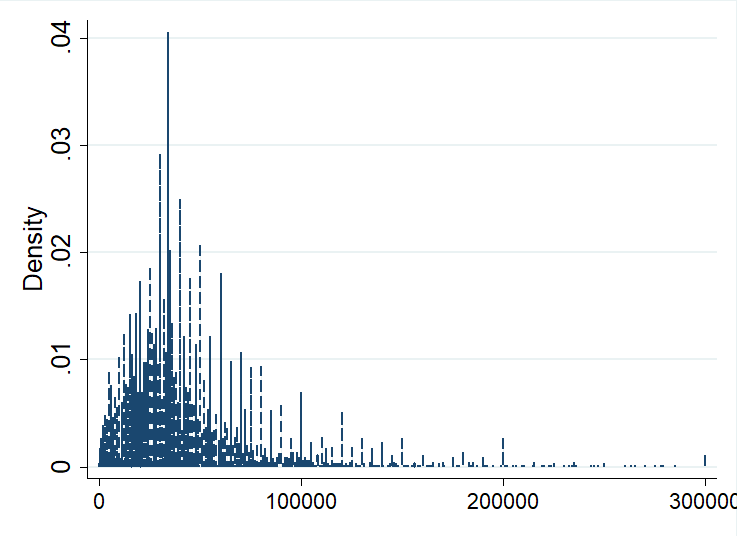}
      \includegraphics[height=3.5cm, width=6cm]{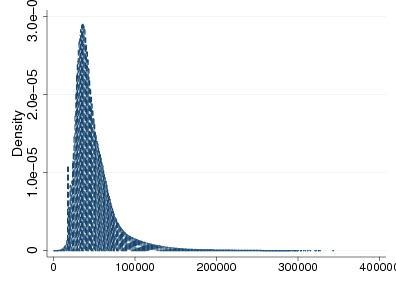} \\
      
             \includegraphics[height=3.5cm, width=6cm]{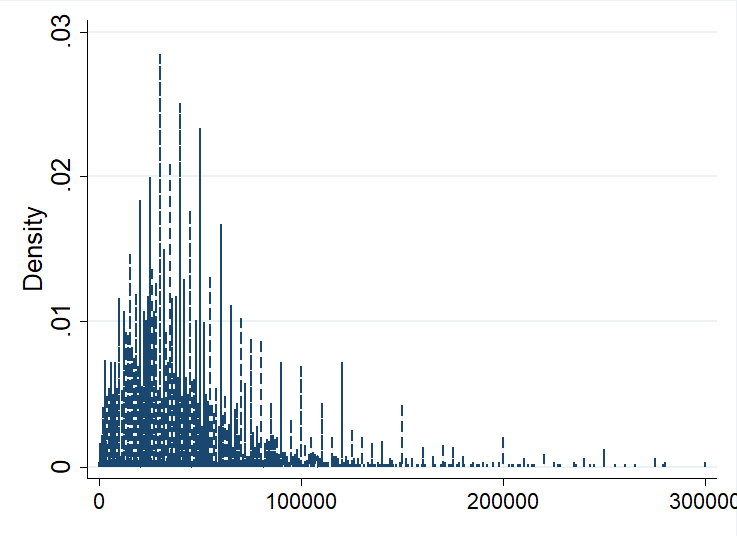}
      \includegraphics[height=3.5cm, width=6cm]{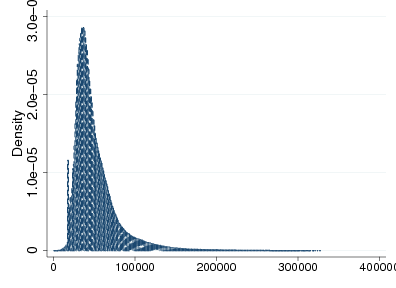} \\
      
         \includegraphics[height=3.5cm, width=6cm]{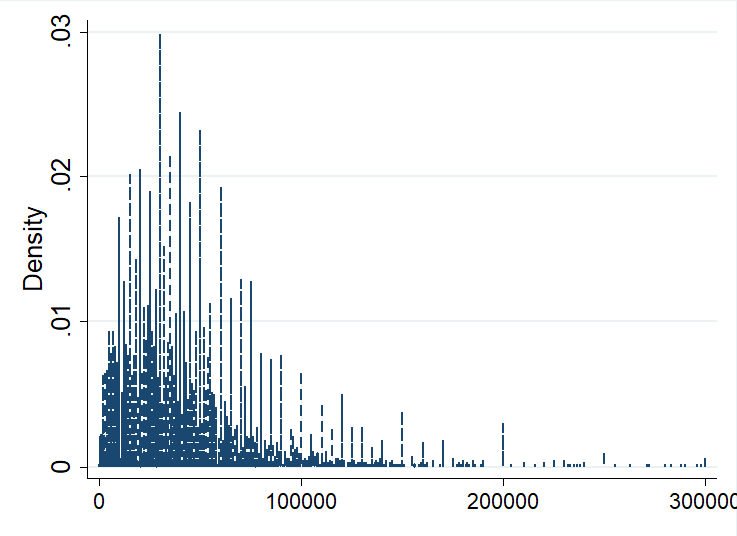}
      \includegraphics[height=3.5cm, width=6cm]{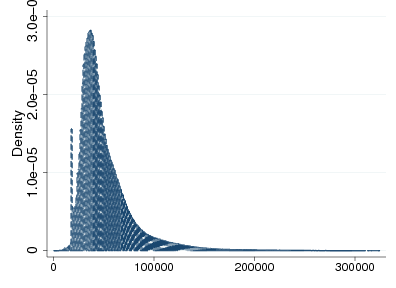} \\
      
         \includegraphics[height=3.5cm, width=6cm]{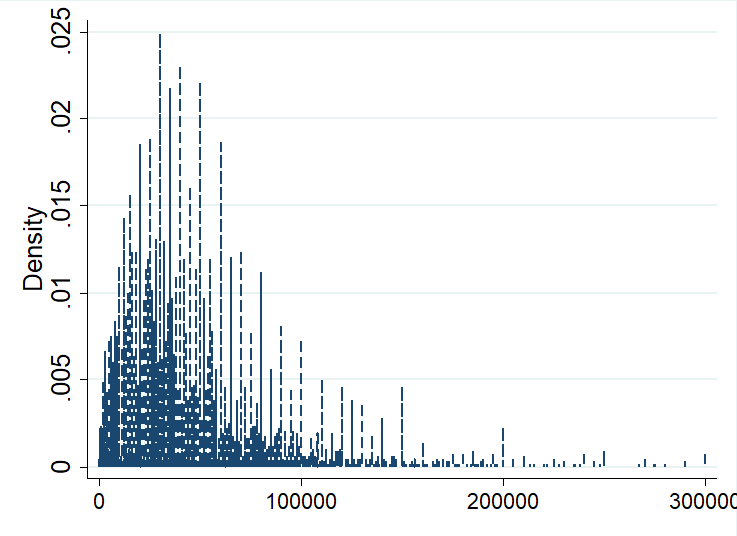}
      \includegraphics[height=3.5cm, width=6cm]{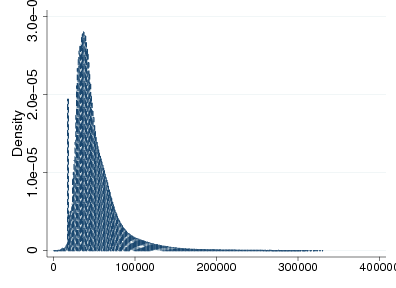} \\
      
             \includegraphics[height=3.5cm, width=6cm]{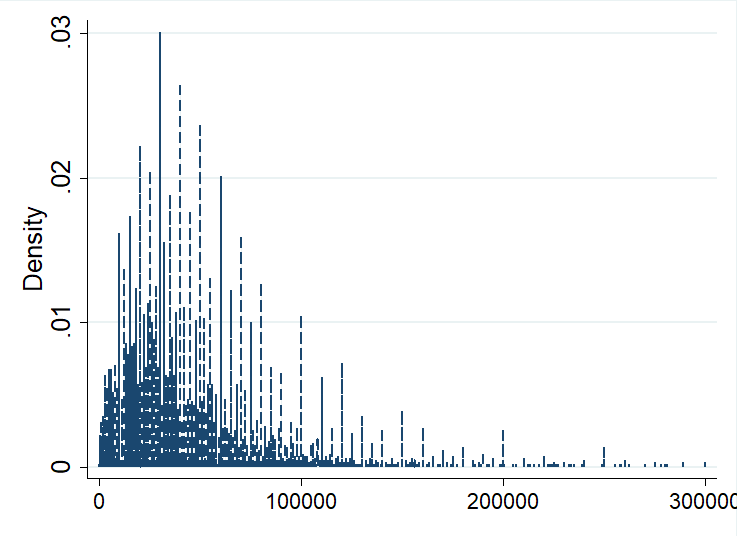}
      \includegraphics[height=3.5cm, width=6cm]{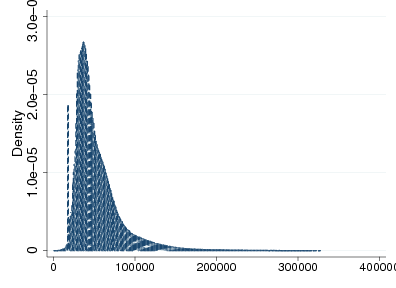}

        \caption{Reliability of Experian income imputation by year. In each left graph we report the histogram of income according to the PSID (variable: wages and salaries of reference person), in 2005, 2007, 2009, 2011, 2013 and 2015 (PSID is bi-annual), whereas in the right panel we report the histogram of income imputed by Experian in the same years. In both cases observations equal to zero have been dropped.}  
   
   \end{figure}

       \begin{figure}[H]
 
       \includegraphics[height=4cm, width=7cm]{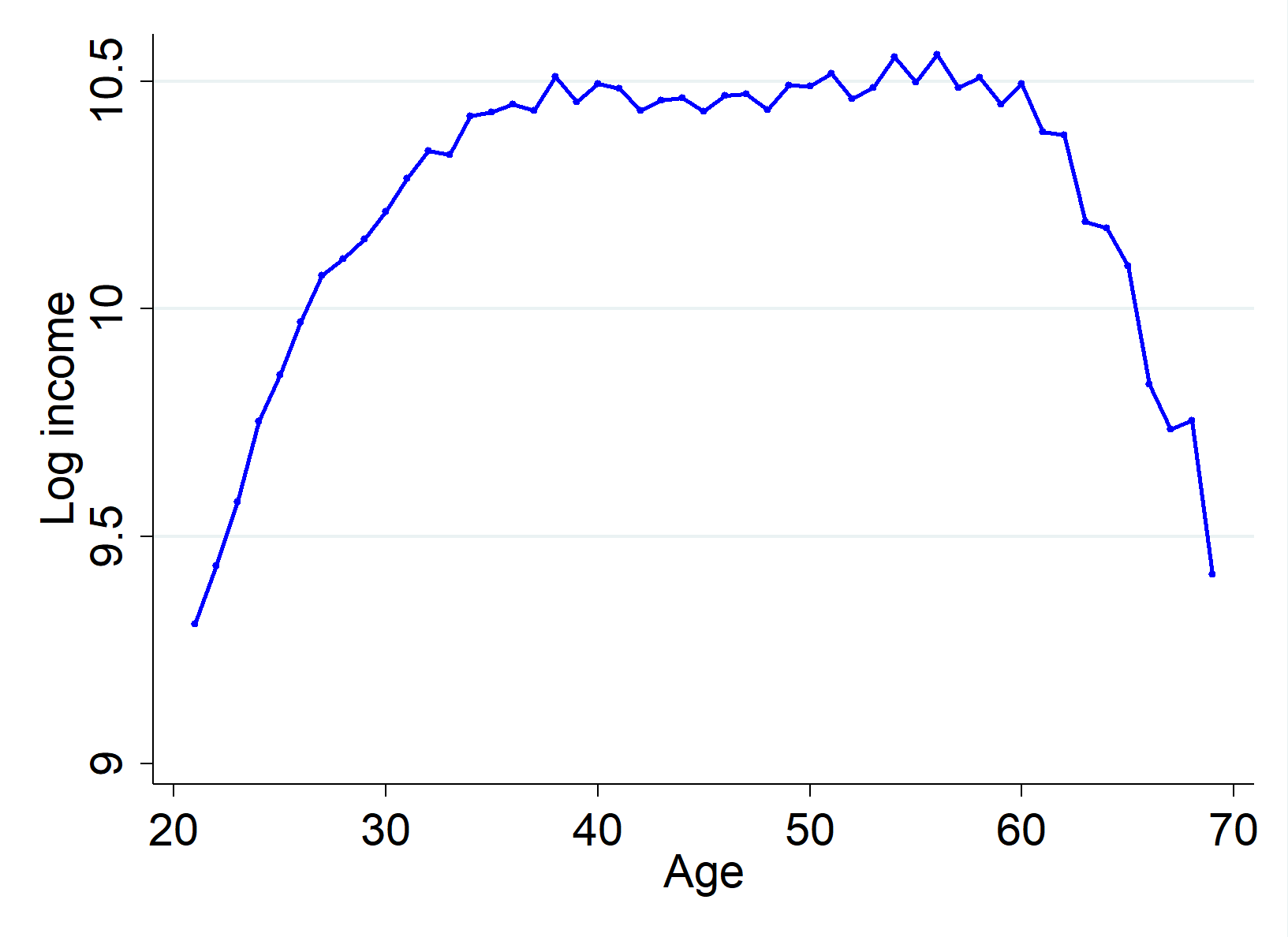}
      \includegraphics[height=4cm, width=7cm]{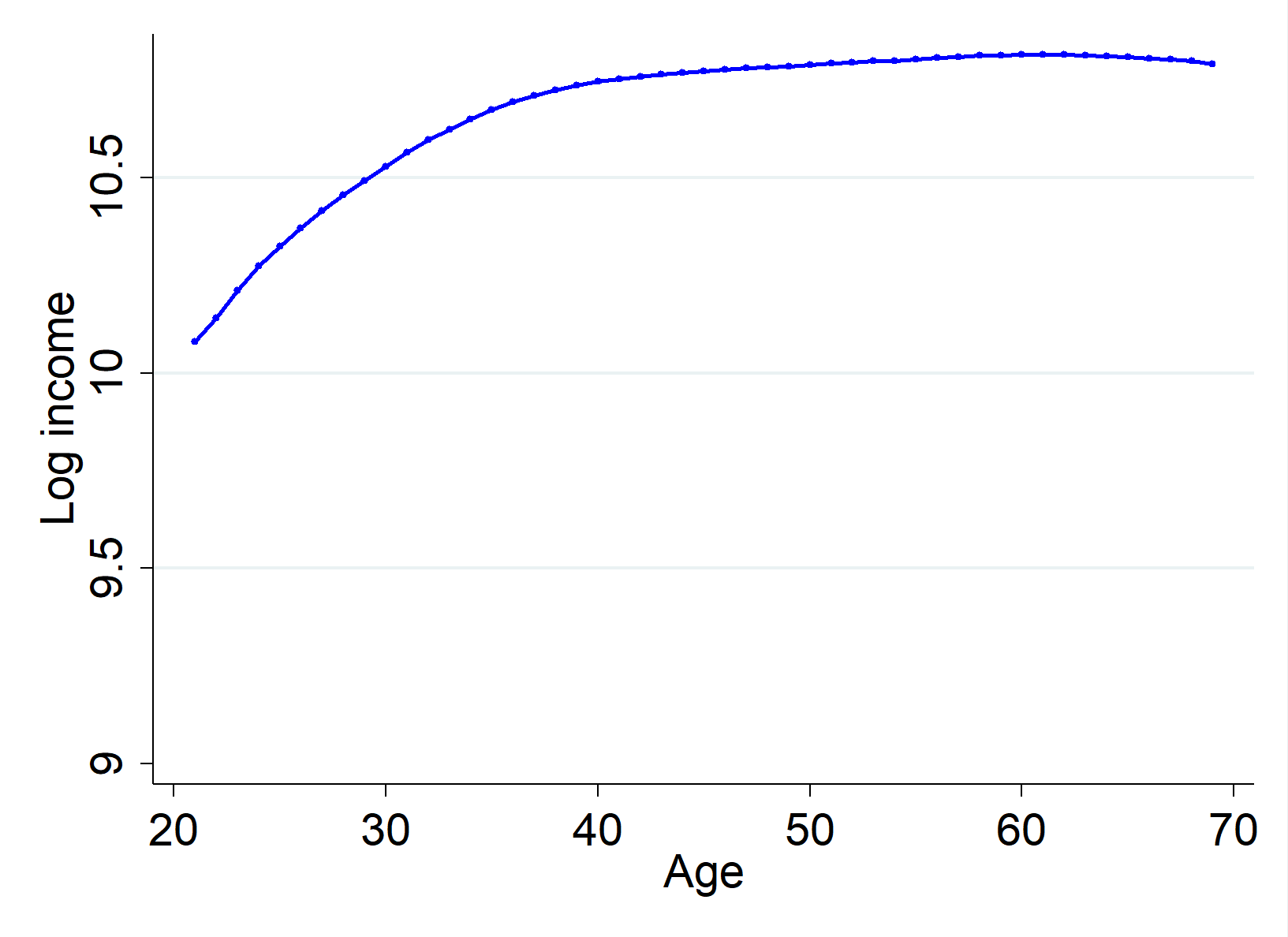} \\

        \caption{Life cycle income trajectory according to PSID data (left panel) and to Experian data (right panel). Average log labor income for individuals of each age group is plotted in both graphs.}  
   
   \end{figure}

    \begin{figure}[H]
 
       \includegraphics[height=4.5cm, width=7cm]{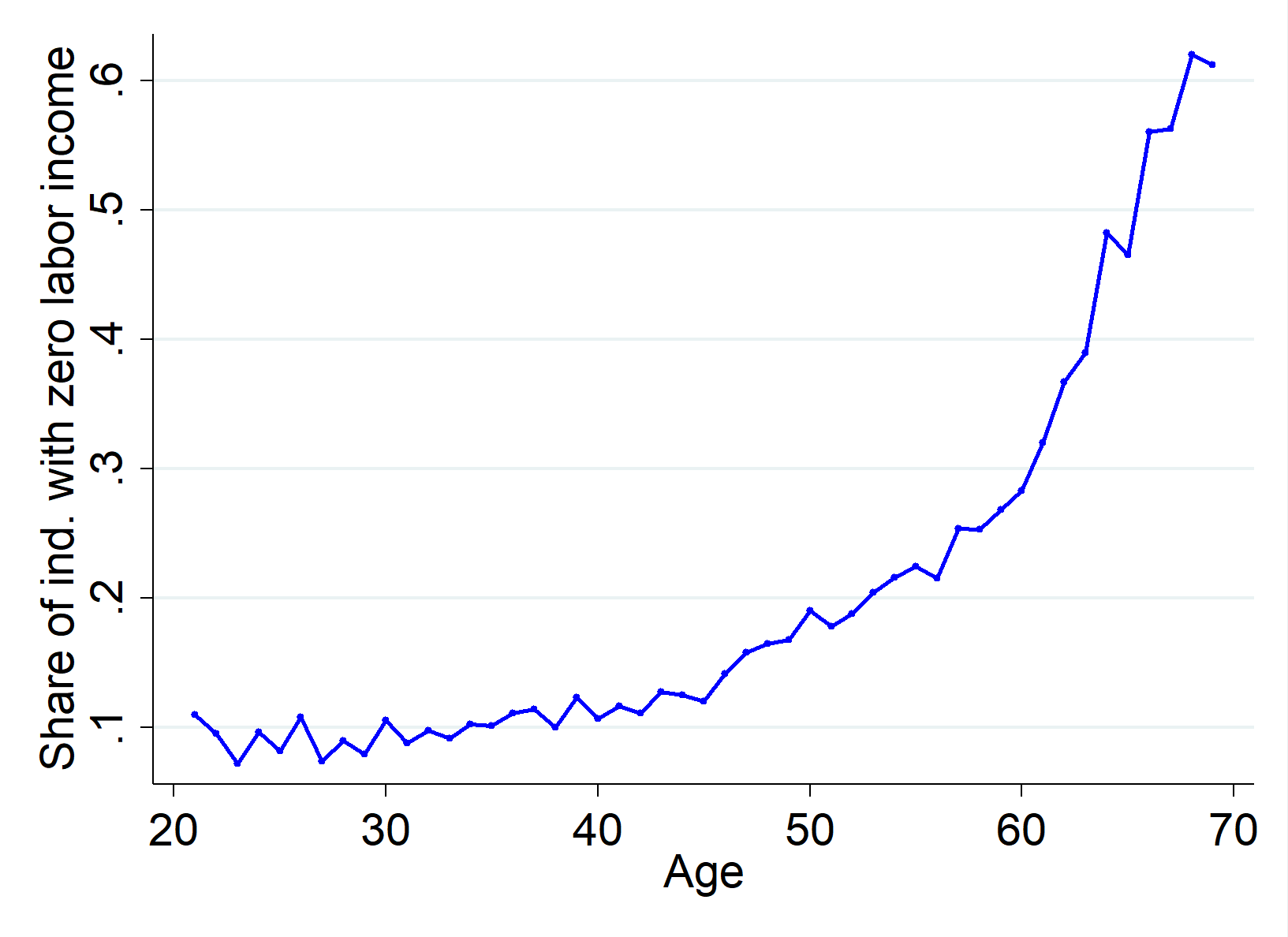}
      \includegraphics[height=4.5cm, width=7cm]{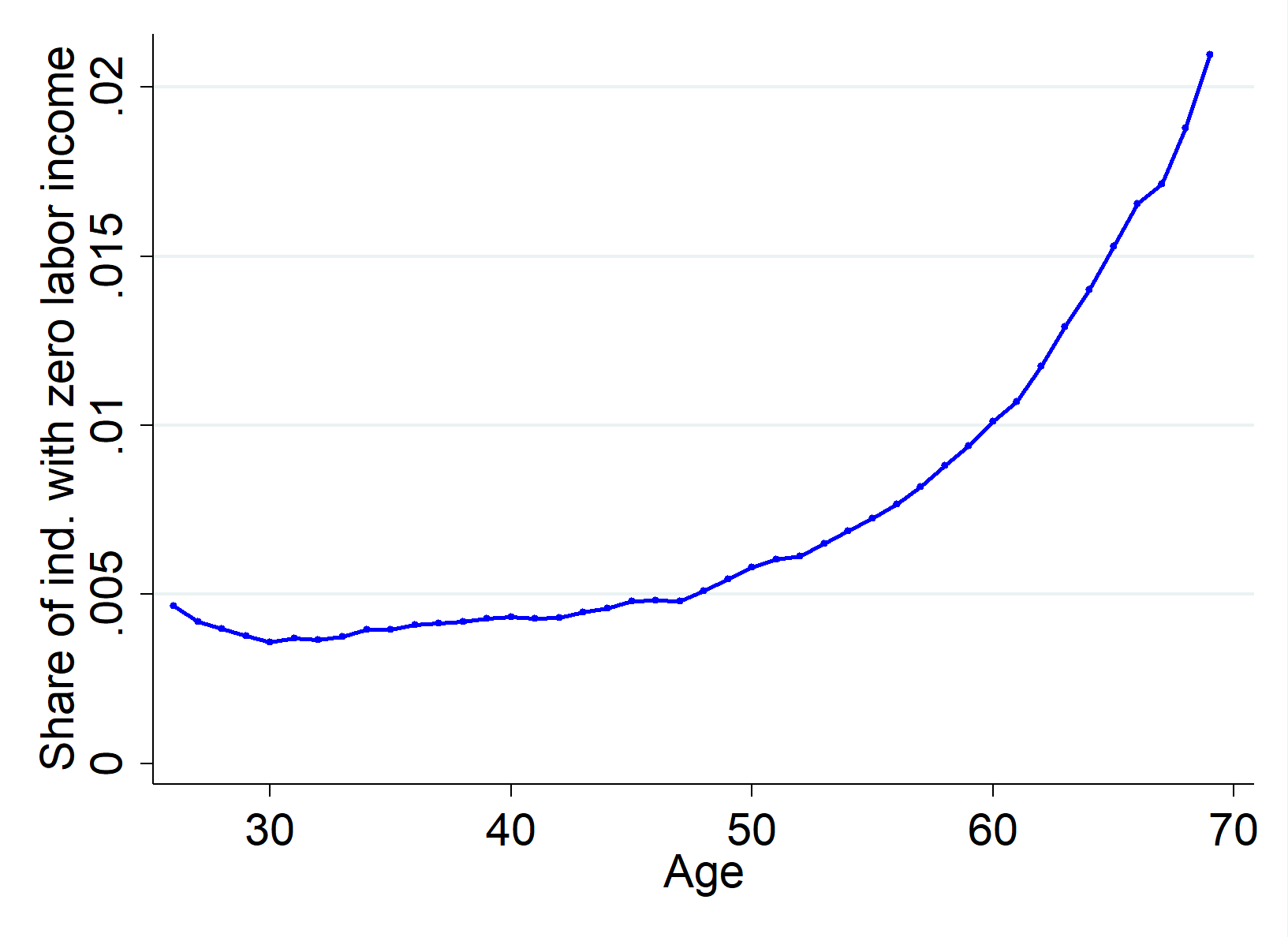} \\
      
       \includegraphics[height=4.5cm, width=7cm]{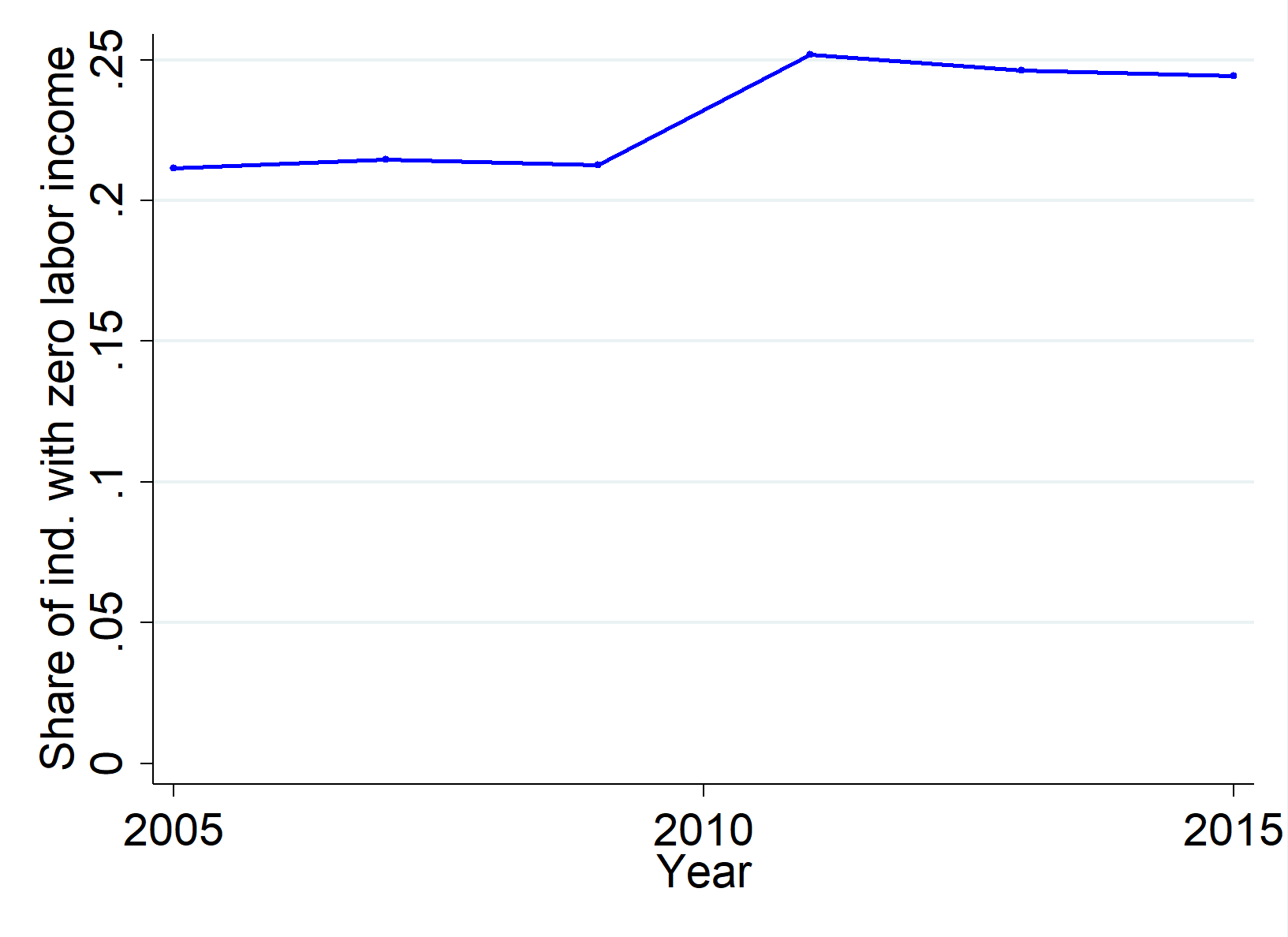}
      \includegraphics[height=4.5cm, width=7cm]{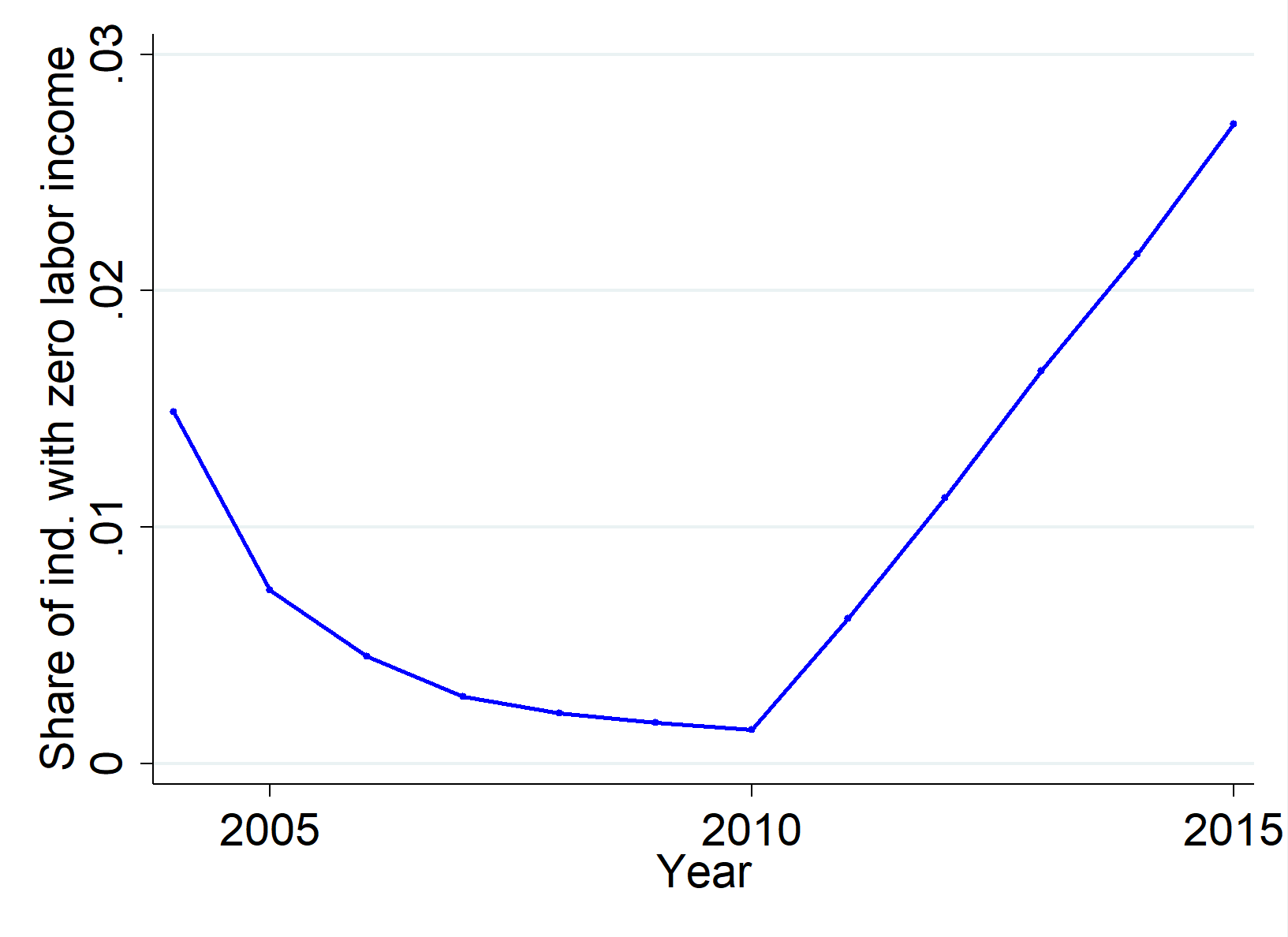} \\

        \caption{Share of individuals with zero labor income in PSID data (2005-2015) by age (left upper panel) and in Experian data by age (upper right panel). Share of individuals with zero labor income in PSID data by year (bottom left panel) and in Experian data by year (bottom right panel).}  
   
   \end{figure}

    \begin{figure}[H]

       \includegraphics[height=6cm]{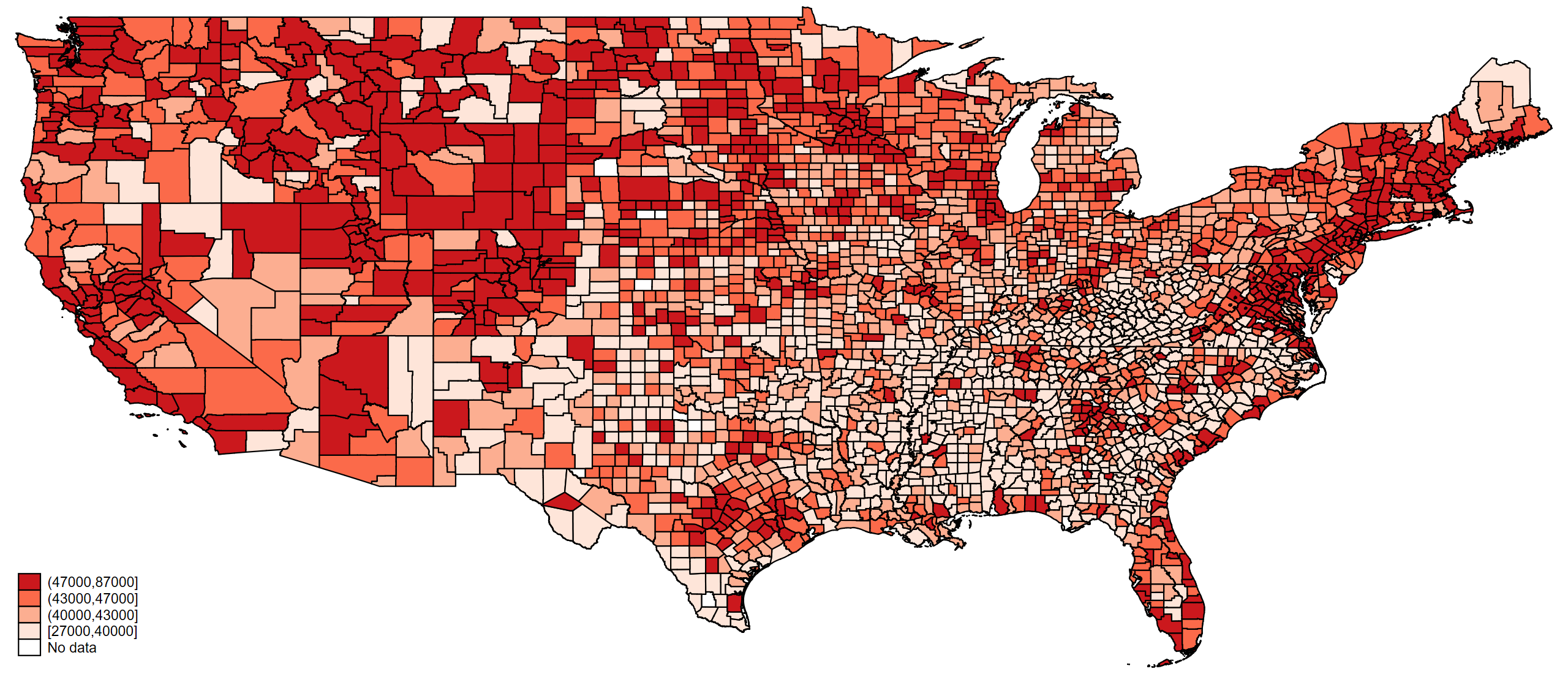}
       
     \includegraphics[height=6cm]{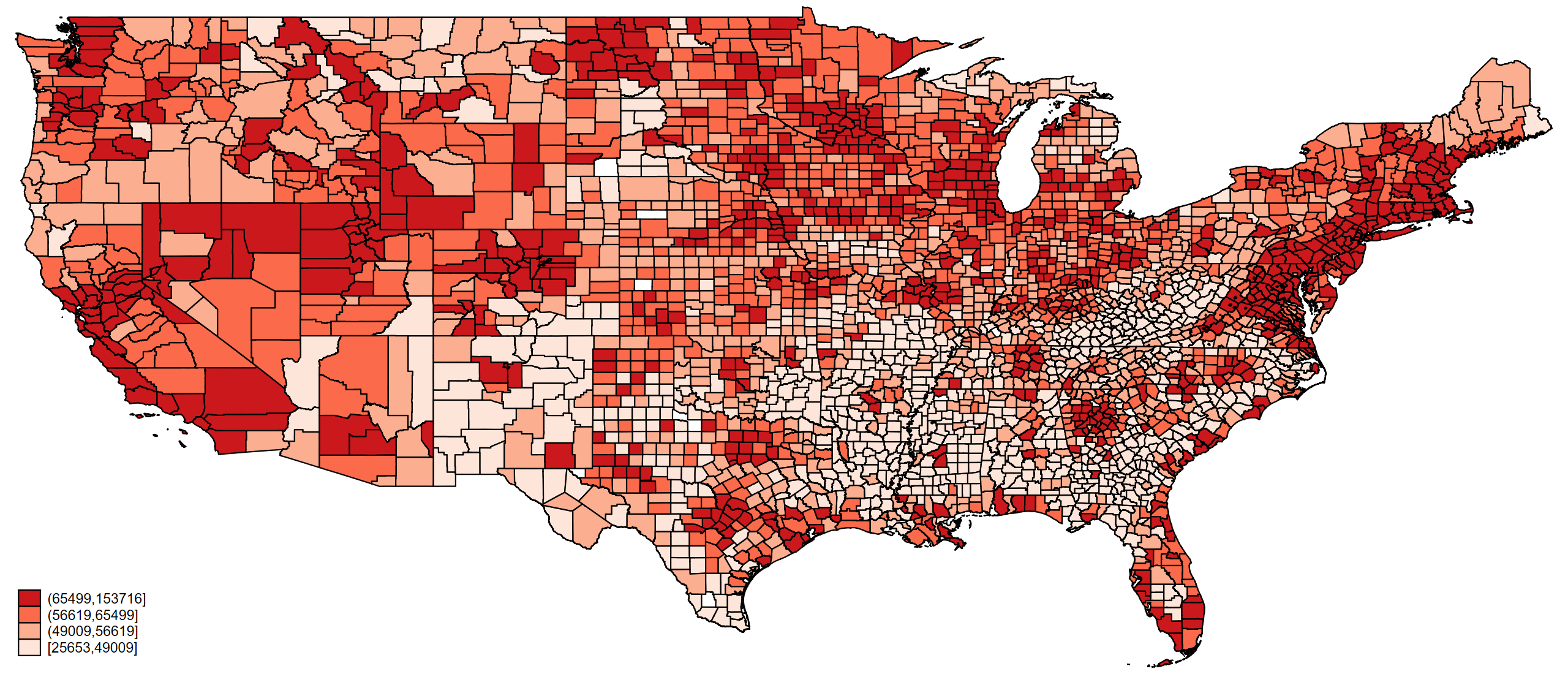} \\

        \caption{Distribution of median imputed income according to Experian data in 2019 (upper panel) and distribution of median household income according to Census data in 2021 (bottom panel), by US counties.}  
   
   \end{figure}

Further, we estimate the following regression:

\begin{equation}
    y_{it} = \alpha_0 +\alpha_1 Age_{it}+ \alpha_2 Age^2_{it} + \alpha_3 CS_{it}+ \alpha_4 CS^2_{it} + zip code_{it} +\varepsilon_{it}
\end{equation}

where $ y_{it}$ is imputed income, $Age_{it}$ and $Age^2_{it}$, respectively stand for individual age and age squared, $CS_{it}$ and $CS^2_{it}$ correspond to individual credit score and credit score squared in year $t$ and $zip code_{it}$ are zip code fixed effects. The adjusted R-squared of this regression is about 0.29. This means that the remaining about 70\% of variation in imputed income cannot simply be explained by age and credit score. This provides a further hint that imputed income exhibit enough variation to be considered as a reliable proxy for actual income. 

\begin{figure}[H]
    \centering
    \includegraphics[width=12cm]{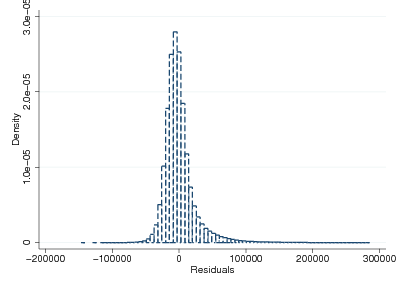}
    \caption{Histogram of the estimated residuals from equation (4)} \label{fig:resid}
  \end{figure}

Indeed, from Figure \ref{fig:resid} we notice that there is a huge dispersion of the estimated residuals from equation (4).

     \section{The Double/Debiased Lasso Estimation Method}\label{sec:ml}

 The key issue when trying to assess the impact of a soft default on financial and socio-economic variables is that defaults are not randomly assigned. Individuals with more unstable economic conditions (e.g. a low and volatile income), or of different types or with larger debts, will be more prone to have a soft default. Of course,  controlling for credit histories and life cycle would alleviate the issue, and our event study type approach is meant to capture the causal impacts of soft default on the outcomes of interest. 
 We recognize that adopting a different, and complementary, approach might be reassuring and for that we believe that using causal machine learning methods is an appropriate alternative in our context (for an overview see \cite{AtheyImbens2019}).   We therefore employ a double/debiased machine learning procedure in the spirit of \citep{c2018}, as described below. While in Appendix \ref{sec:long} we validate the results by analyzing the long-term impacts of a soft default, in Appendix \ref{sec:DML} we focus on a year-by-year approach.
Our double-debiased lasso estimates overcome the problems tackled by the Callaway and Sant'Anna's method (as we only consider soft defaults happening in year 2010 here) and allow us to control for a potentially  large number of  variables. Our argument  is that, conditional on all these controls, soft defaults can be reasonably deemed to be  exogenous. As it will become apparent, the results presented in Appendix \ref{sec:DML} are quite similar to those obtained with the event studies in Section \ref{sec:exploratory}, which we find rather reassuring.

In all the lasso estimates presented in this Section we control for  age, age squared, the amount of open mortgages and car loans, as well as the individual credit score, in each of the pre-event years, i.e. from 2004 to 2009. This should mean that we effectively control for pre-existing individual economic conditions before the (soft) default. Further, we add to the controls commuting zone fixed effects, county unemployment rates\footnote{Source: https://www.kaggle.com/datasets/jayrav13/unemployment-by-county-us}, and number of bank closings by county in each of the pre-event year (data from \cite{nguyen2019credit}), so that we also condition on local economic condition and on the phase of the local business cycle. Last, but not least,  another control that we include is the maximum interest rate allowed by the anti-usury laws in each state in each year of our dataset.\footnote{Source: https://www.findlaw.com/state/consumer-laws/interest-rates.html} Since in some states this maximum rate depends on the current rate paid by the treasury bonds, our control exhibits some variation over time, albeit limited.\footnote{An alternative, if we were interested in harsh measures of default, would be to use bankruptcy fees. This alternative control variable has more geographical detail (94 US judiciary districts vs the 50 US states), however, to the best of our knowledge, it is only available for a snapshot in time (\cite{lupica2011consumer}).}

 Since many controls need to be included in order to be able to consider soft default as exogenous, the use of the double/debiased lasso technique proposed by \citep{c2018} is  a viable approach. Indeed, this method allows to use a large number of controls, without losing power to learn about treatment effects. Flexible ML tools offer an adequate solution to the issue at hand, and in particular this method allows for robust inference in presence of many covariates and potentially many instruments (\citep{c2018}).

To be more precise, we study the effects on the probability of moving ZIP code, the probability of moving commuting zone, Median House Value, income, credit score, probability of opening a new mortgage conditional on not having one before 2010, probability of having a low credit limit, i.e. below 10,000 USD, the amount of revolving balance open, the probability of experiencing a harsh default, the probability of being a homeowner in our more comprehensive definition (i.e. Experian definition plus having ever had a mortgage), total credit limit and mortgage balance open as our outcome variable $y_i$ in the analysis.

Our treatment variable, $d_i$ is an indicator that stands for the occurrence of a soft default. The vector of raw covariates, $X_i$, consists, as mentioned above, of the 2004-2009 values of age, age squared, credit score, open mortgages, open car loans, plus commuting zone dummies, local unemployment rate, number of bank closings by county, and the maximum interest rate allowed in a given year by the state laws.
 We then effectively estimate the (second stage) model: 
 \begin{equation}
 y_i = d_i\tau + X_i'\beta+ u_i
 \end{equation}
 where $u_i$ is an error term, and  the first stage is defined as:
     \begin{equation}
     d_i = X_i'\delta + v_i
     \end{equation}
   with   $d_i$, $X_i$ defined above, and with $v_i$ an error term. 
    The double/debiased estimation procedure consists in the following steps: (i) Predict $y_i$ and $d_i$ using $X_i$ with separate Lasso regressions and obtain $\hat{\beta}$ and $\hat{\delta}$, (ii) residualize: $\hat{u_i} = y_i -X_i'\hat{\beta}$ and $\hat{v_i} = d_i - X_i'\hat{\delta}$, (iii) the  debiased estimator of the treatment effect is :
     \begin{equation}
     \hat{\tau} = \left( \frac{1}{n} \sum_{i=1}^n \hat{v}_i d_i \right)^{-1} \frac{1}{n} \sum_{i=1}^n \hat{v}_i\hat{u}_i
     \end{equation}

  \section{The Causal Effects of Soft Default }
  \label{sec:DML}

  \setcounter{table}{0}
\renewcommand{\thetable}{E\arabic{table}}

\setcounter{figure}{0}
\renewcommand{\thefigure}{E\arabic{figure}}
  
  In this Section we report our baseline double/debiased lasso results for a series of outcomes (probability of moving ZIP code, probability of moving commuting zone, Median House Value of the ZIP code of residence, credit score, income, total credit limit, revolving credit open balance, home-ownership status).

In all the estimations presented in this Section, the controls are: age, age squared, commuting zones fixed effects, credit score, the amount of mortgage balance open, the amount of car loan open, county-level unemployment rate, county-level number of bank closings, maximum interest rate allowed by the State anti-usury laws. Except for age and age squared, that are for sure exogenous, all the other controls are measured in years 2004 to 2009, i.e. pre-event.

In all the estimates presented in this Section we restrict the analysis to  those affected by a soft default in year 2010, compared to those who were not affected by any default up to that same year. This means that we drop from the sample individuals who record a soft default between between 2004 (the first year in our data) and 2009, as well as those who record a harsh default between 2004 and 2010.

\begin{figure}[H]
      	Panel (i) - Move Zip \hspace{4cm}Panel (ii) - Move CZ\\
           
                \includegraphics[height=4.5cm, width=7cm]{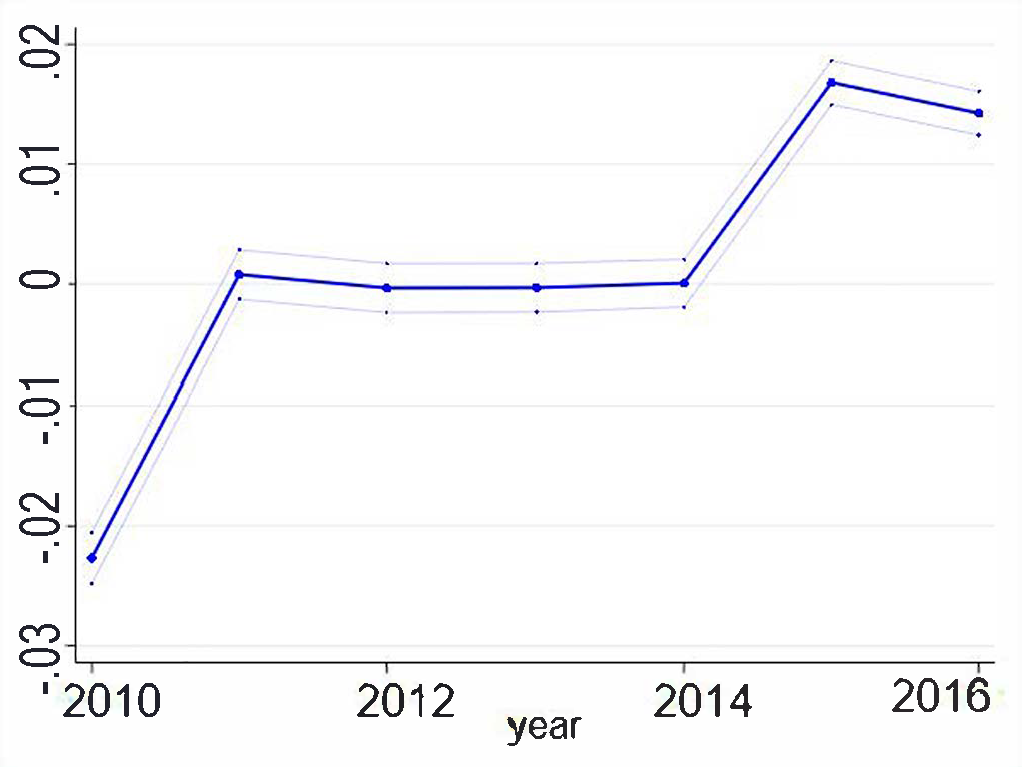} 
             \includegraphics[height=4.5cm, width=7cm]{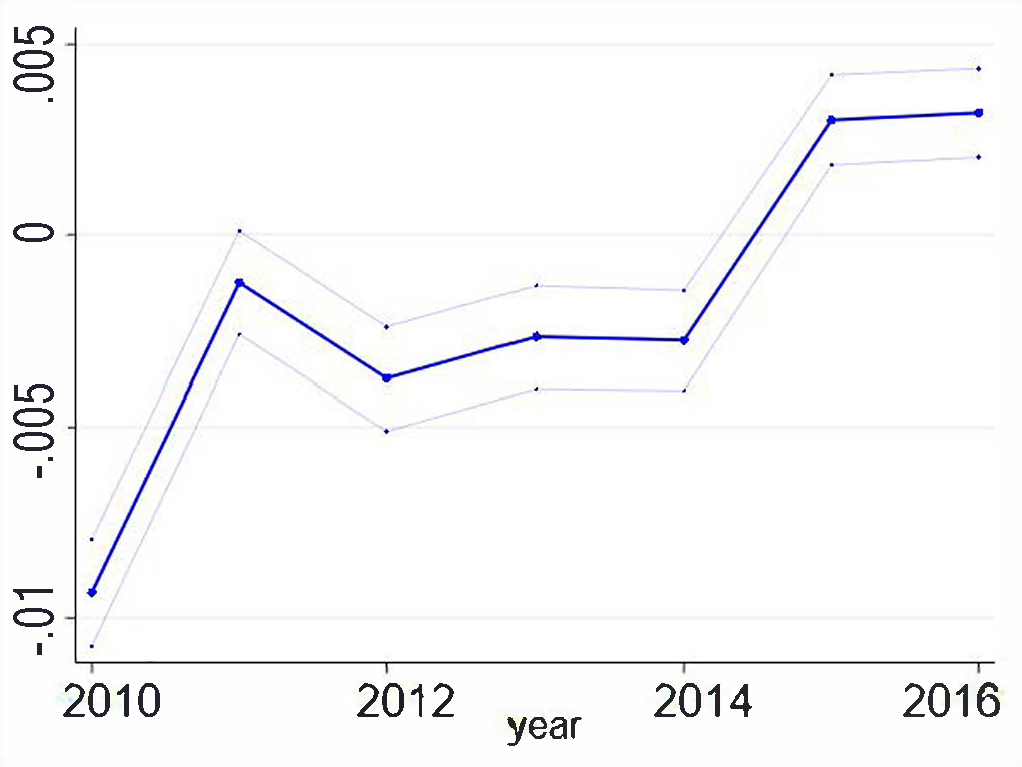} \\
          
          	Panel (iii) - Median House Value \hspace{2cm}Panel (iv) - Income\\
          	
             \includegraphics[height=4.5cm, width=7cm]{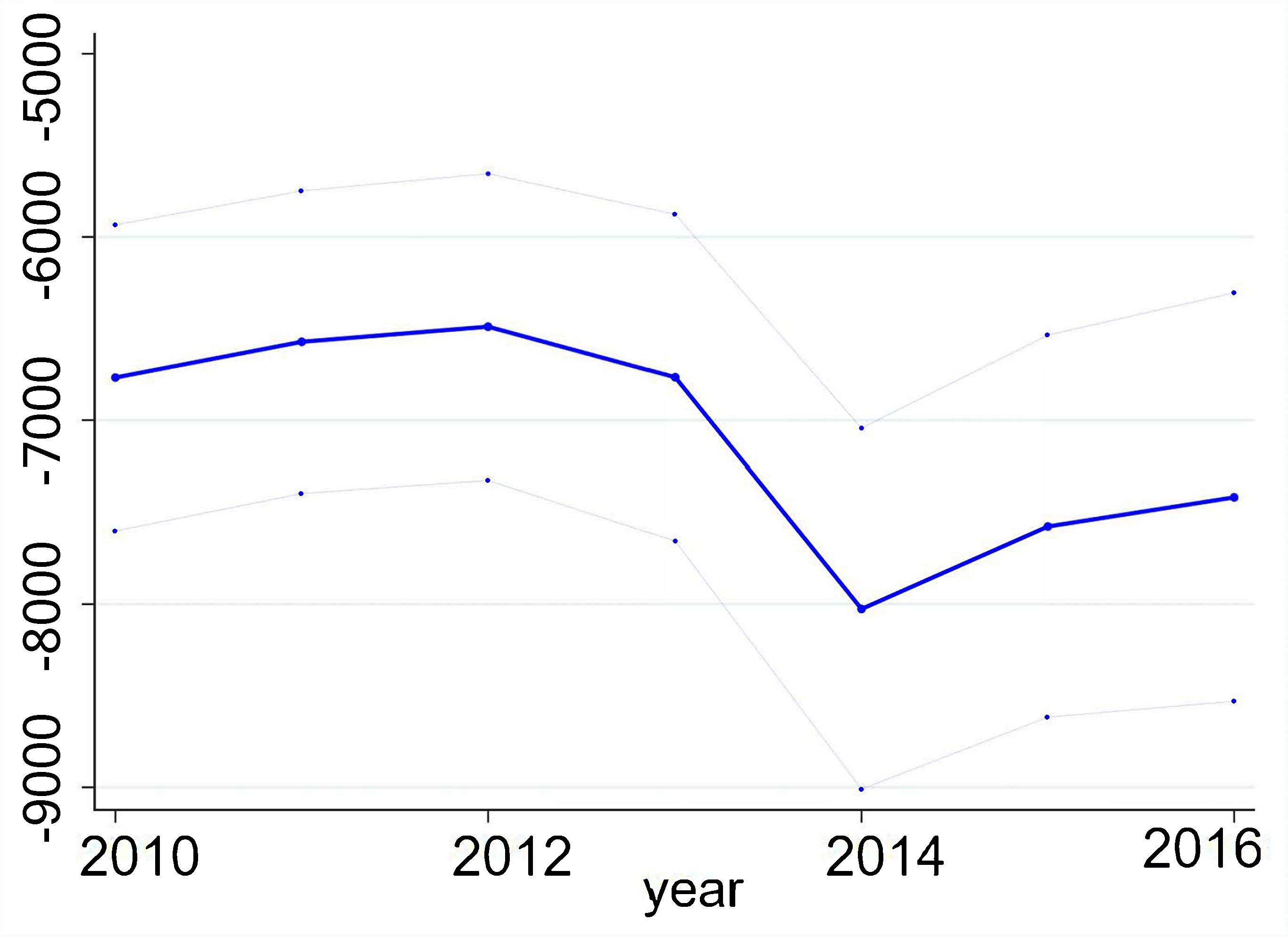} 
          \includegraphics[height=4.5cm, width=7cm]{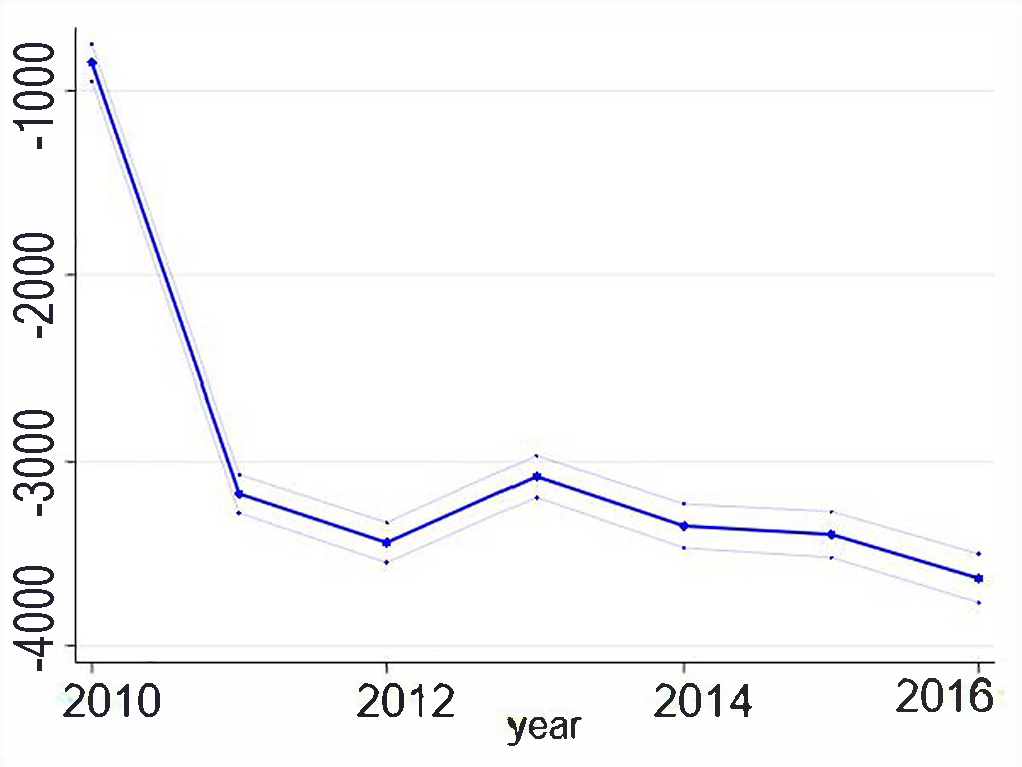}

        \caption{DML year-by-year impacts of soft (blue) default on: (i) Probability of moving zip code. This variable takes value 1 if the individual is a different zip code in year $t$ than in year $t-1$, and zero otherwise, (ii) Probability of moving outside the commuting zone. This variable takes value 1 if the individual is in a different commuting zone in year $t$ than in year $t-1$, and zero otherwise, (iii) the Median House Value in the zip code of residence at year $t$ (iv) income imputed by Experian. In all panels the event takes place in 2010 and is an absorbing state. Controls: age, age squared, commuting zones fixed effects, credit score, the amount of open mortgages and car loan, county unemployment rate, number of bank closings in the county pre-event (2004-2009), maximum interest rate allowed by law. 95\% confidence intervals around the point estimates. Individuals recording a harsh default between 2004 and 2010 (extremes included) and those recording a soft default between 2004 and 2009 (extremes included) have been dropped from the estimation sample.}
        \label{fig:DML}
   \end{figure}

In Figure \ref{fig:DML} we report, for each of our outcome variables, the DML estimated impact of a soft default in each of the post event years in our sample (i.e. from 2010 to 2016). 


From Panel (i) in Figure \ref{fig:DML}, we deduce that a soft default causes on impact a decrease by about 2pp in the probability of moving ZIP code. This impact goes to zero the year after and turns positive four years after the default, with an increase by about 2pp in the probability of moving ZIP code.
As far as the probability of moving commuting zone is concerned (Panel (ii)), a soft default causes a decrease by about 1pp in the probability of moving commuting zone on impact. This effect then goes essentially to zero afterwards. The recorded effects in both cases are not negligible if one considers that zip code and CZ mobilities are  about 17 and 6\% in 2010. 
Soft default leads to an increase by about 0.5pp in commuting zone mobility in the long run (i.e. six years after the event).

In Panel (iii) of Figure \ref{fig:DML}, we find evidence that a soft default, as expected, leads to moves towards cheaper areas. Indeed, a soft default leads to a drop by more than 8,000USD in the Median House Value of the zip code of residence.


From Panel (iv) of Figure \ref{fig:DML}, we find evidence that income also drops following a soft default. The drop is however modest, about 3,000 USD over annual income. The impact of a soft default on income becomes larger over time, reaching almost 4,000USD six years after the event. It is plausible that a default episode makes a person less employable \cite{bos2018labor}, credit score background checks are frequent, further to provide worse investment opportunities.\footnote{\url{https://www.experian.com/blogs/ask-experian/do-employers-look-at-credit-reports/}}

   \begin{figure}[H]
                      \includegraphics[ width=10cm]{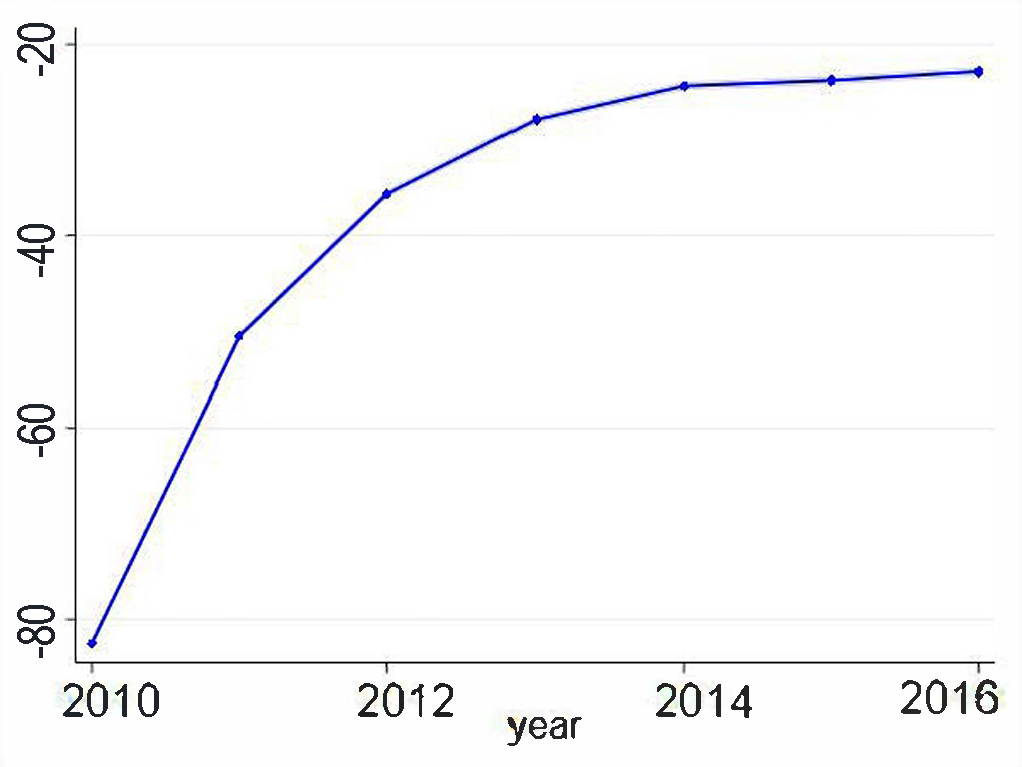} \\
              	    
        \caption{DML year-by-year impacts of soft (blue) default on credit score. In all panels the event takes place in 2010 and is an absorbing state. Controls: age, age squared, commuting zones fixed effects, credit score, the amount of open mortgages and car loan, county unemployment rate, number of bank closings in the county pre-event (2004-2009), maximum interest rate allowed by law. 95\% confidence intervals around the point estimates. Individuals recording a harsh default between 2004 and 2010 (extremes included) and those recording a soft default between 2004 and 2009 (extremes included) have been dropped from the estimation sample.}
        \label{fig:DMLb}
   \end{figure}

\begin{figure}[H]
      	Panel (i) - Mortgage origination \hspace{2cm}Panel (ii) - Prob. cred lim $<$10k\\
           
                \includegraphics[height=4cm, width=7cm]{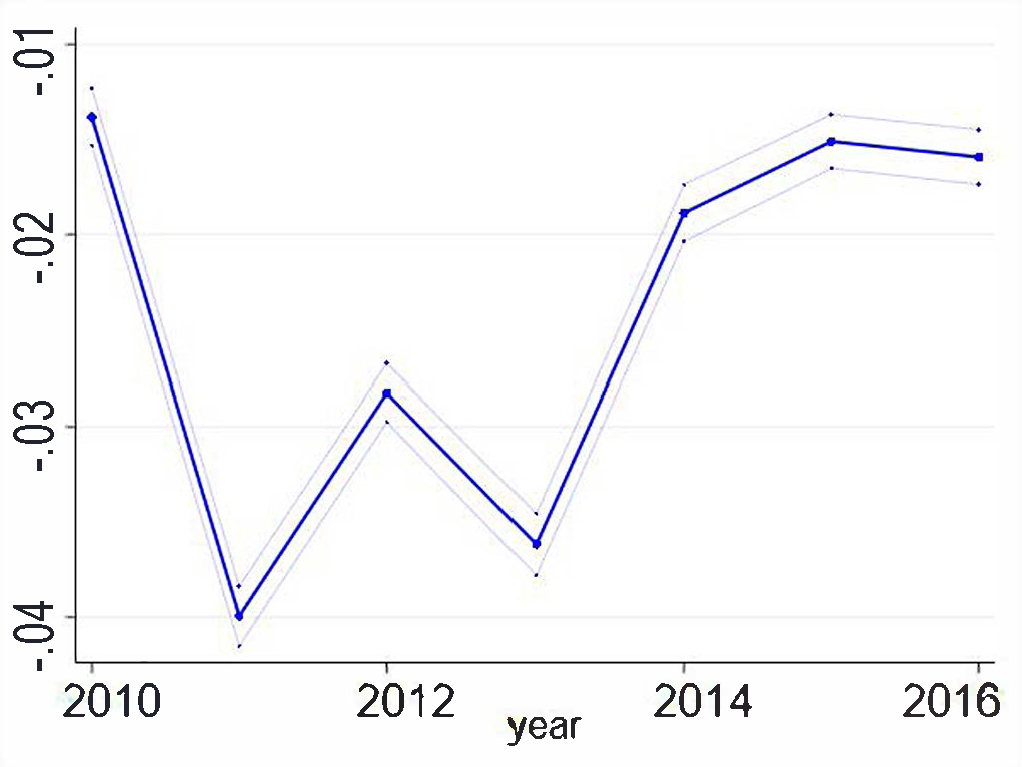} 
             \includegraphics[height=4cm, width=7cm]{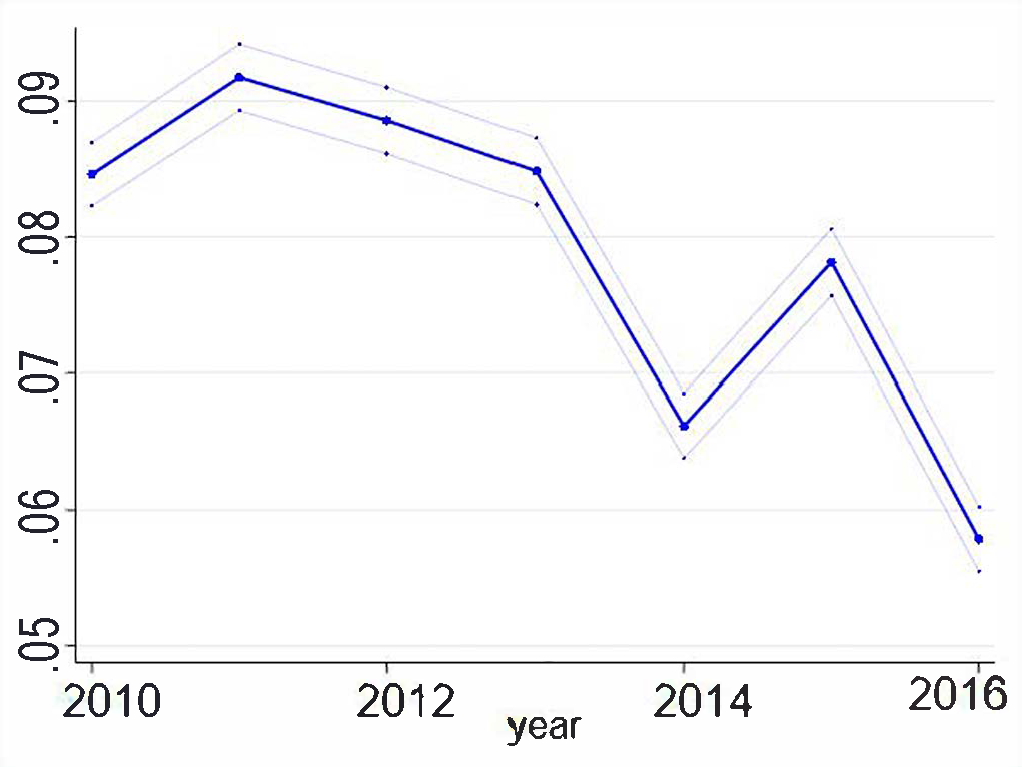} \\
          
          	Panel (iii) - Rev bal open \hspace{4cm}Panel (iv) - Harsh default\\
          	
             \includegraphics[height=4cm, width=7cm]{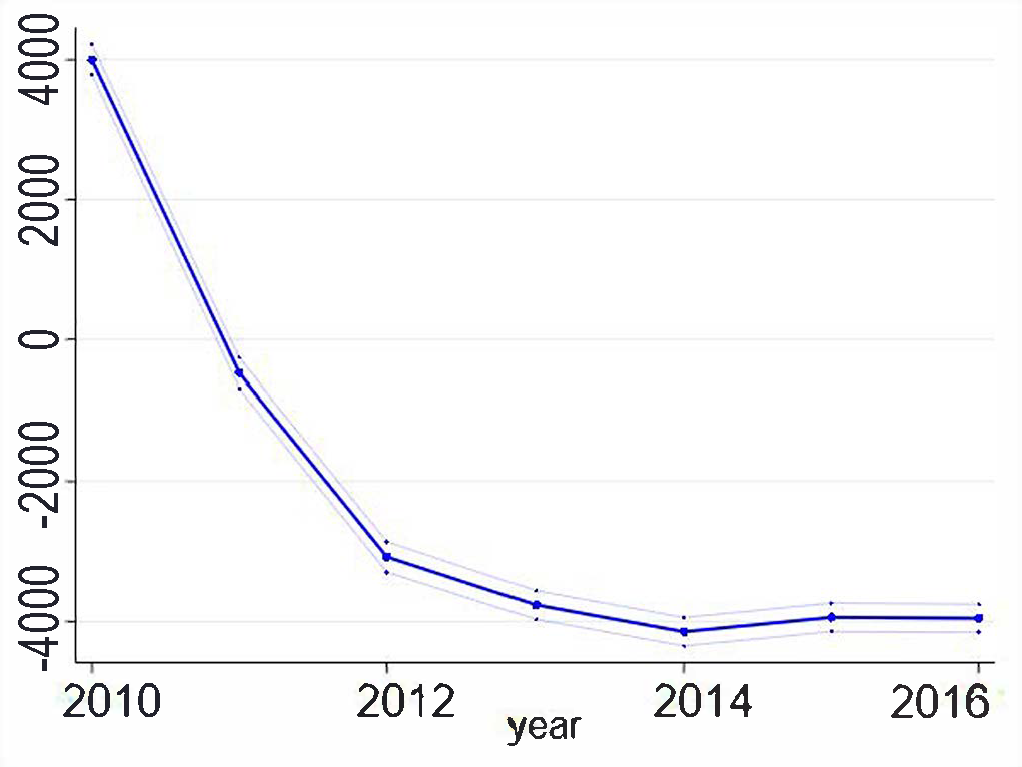} 
                \includegraphics[height=4cm, width=7cm]{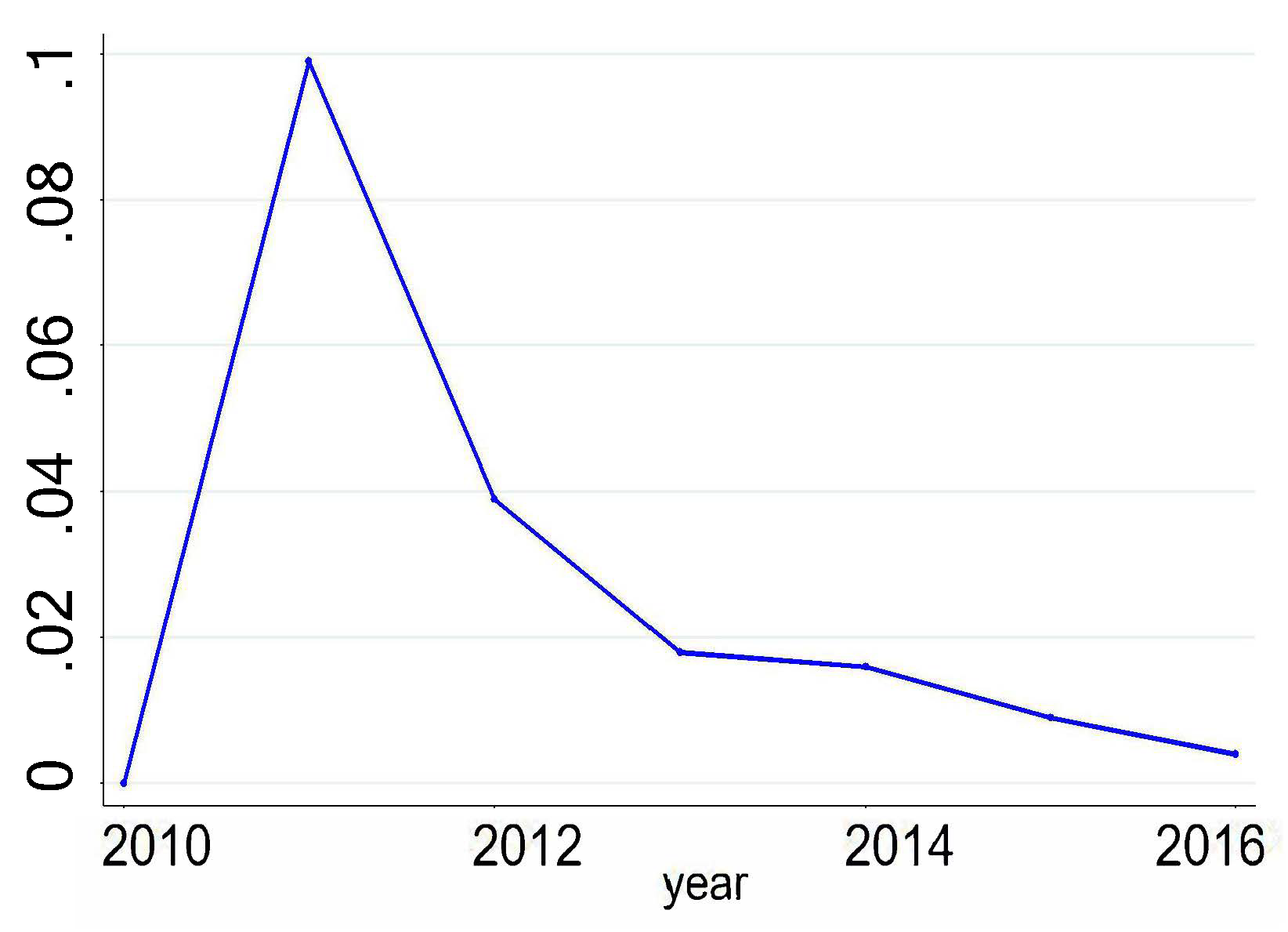} \\
             
 	Panel (v) - Homeownership \hspace{2cm}Panel (vi) - Tot cred limit (in k=1,000 USD)\\
 	
             \includegraphics[height=4cm, width=7cm]{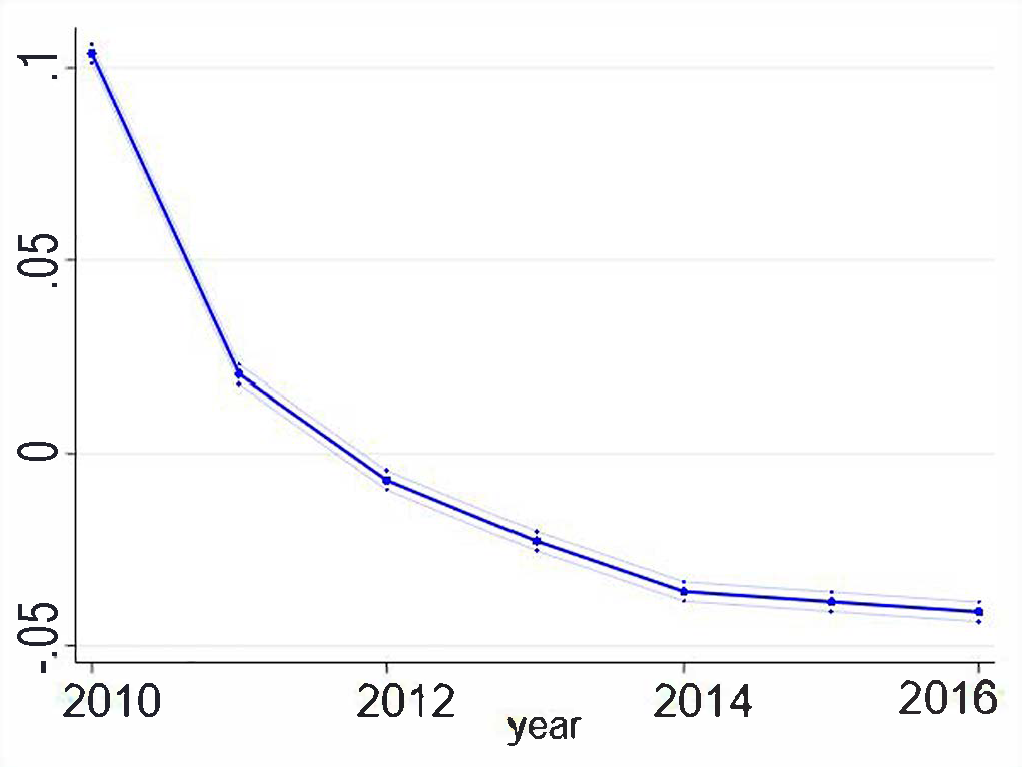} 
                \includegraphics[height=4cm, width=7cm]{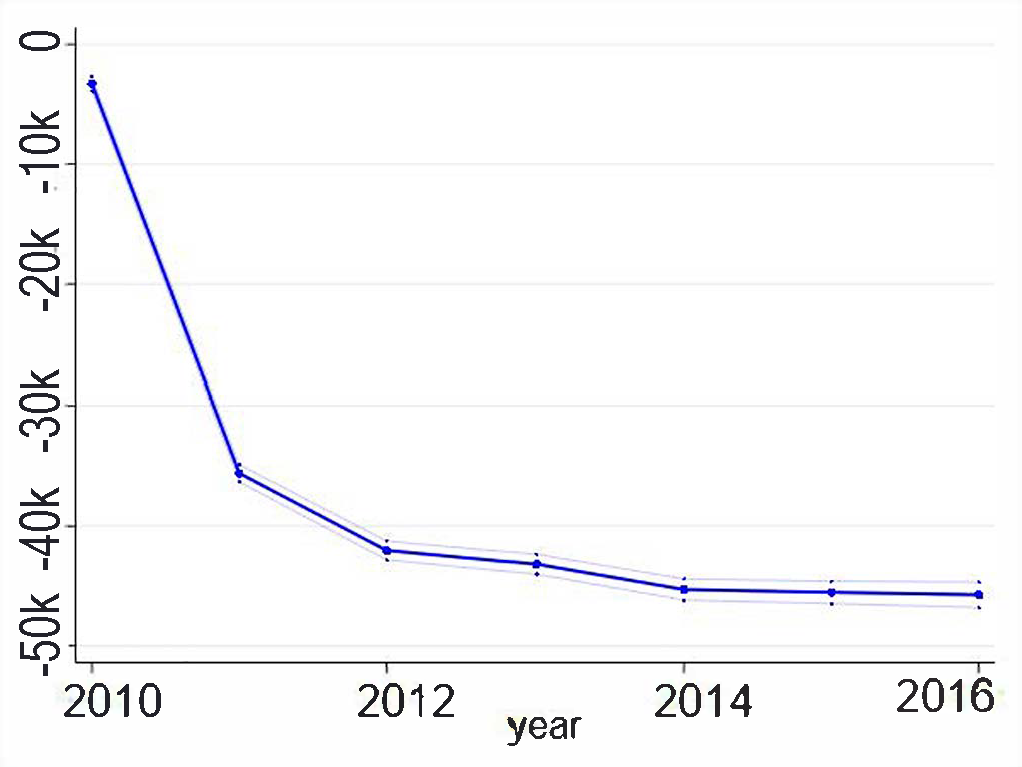} \\

                \centering                
                             Panel (vii) - Mortgage balance open
 	
             \includegraphics[height=4cm, width=7cm]{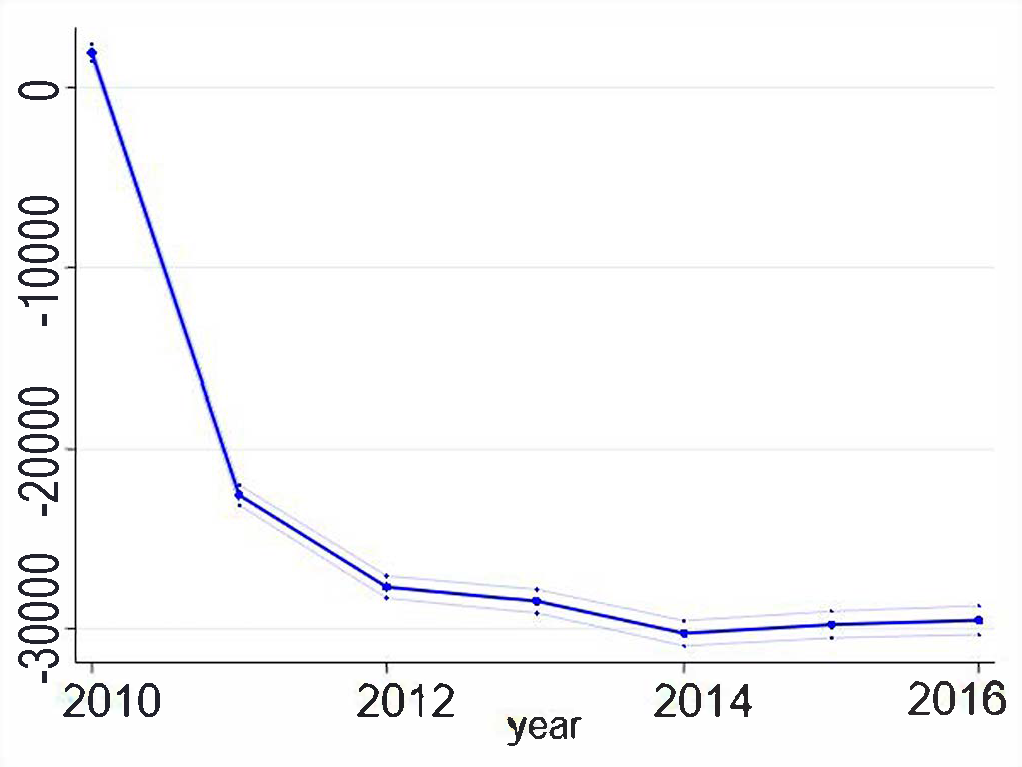} \\

        \caption{\small DML year-by-year impacts of soft (blue) default on: (i) Mortgage origination: probability of opening one or more new mortgages given that no mortgage was open before the event, (ii) probability that total credit limit is lower than 10,000USD, (iii) total amount open on all revolving credit trades, (iv) probability of experiencing a harsh default (Chapter 7, Chapter 13 or foreclosure), (v) probability of being homeowner, i.e. either being recorded as a homeowner by Experian or having ever had a mortgage open (vi) total credit limit on all trades, (vii) open amount of mortgage balance. 
        In all panels the event takes place in 2010 and is an absorbing state. Controls: age, age squared, commuting zones fixed effects, credit score, the amount of open mortgages and car loan, county unemployment rate, number of bank closings in the county pre-event (2004-2009), maximum interest rate allowed by law. 95\% confidence intervals around the point estimates. Individuals recording a harsh default between 2004 and 2010 (extremes included) and those recording a soft default between 2004 and 2009 (extremes included) have been dropped from the estimation sample.}
        \label{fig:DMLa}
   \end{figure}
 
   From Figure \ref{fig:DMLb}, we further deduce that a soft default causes a drop in the credit score by about 80 points on impact, reducing to -20 points over time (i.e. 5-6 years after the event). This is a very substantial drop, as it would take the average borrower from a 680 or a good score to a barely fair score. This drop would make some transactions unattainable, and certainly would make one's credit and daily life substantially harder (lower probability of credit approval, loans, mortgages, rentals, mobile phones contracts, worse insurance, etc.).

 Further, from Figure \ref{fig:DMLa} we deduce that a soft default leads to a decrease in the probability of opening a new mortgage (mortgage origination) by about -3/4pp, with the negative effect decreasing over time. Moreover, a soft default is associated with an increase by about 8/9pp in the probability of having a low credit limit (i.e. below 10k), a decrease over time in the revolving balance open by about 4,000USD, a surge by 4 to 10pp in the probability of also experiencing a harsh default and a decline by about 5pp in the medium run in the probability of being homeowner. Finally, a soft default also leads to a drop in the total credit limit by about 50,000USD in the long run, as well as to a decline by about 30,000USD in the amount of mortgage balance open.
 
 In Appendices \ref{sec:long}, and \ref{sec:long_termB} we analyze the long term impact of a soft default on our outcome variables. This means that we consider as the final year either 2020 or 2019, in order to assess whether the negative impacts reported here are still evident ten years after the event. We find evidence that most of the impacts are still statistically significant. 

  \section{Event study results for different types of delinquencies}
        \label{sec:types}

        \setcounter{table}{0}
\renewcommand{\thetable}{F\arabic{table}}

\setcounter{figure}{0}
\renewcommand{\thefigure}{F\arabic{figure}}
     
     In this Section we repeat the Callaway-Sant'Anna event study analysis, by distinguishing across four different types of delinquencies: (i) on auto loan, (ii) on mortgages, (iii) on bankcard trades, (iv) on revolving credit.

\begin{figure}[H] 
         
             	Panel (i) - Move Zip \hspace{4cm}Panel (ii) - Move CZ\\
               \includegraphics[height=4.5cm, width=7cm]{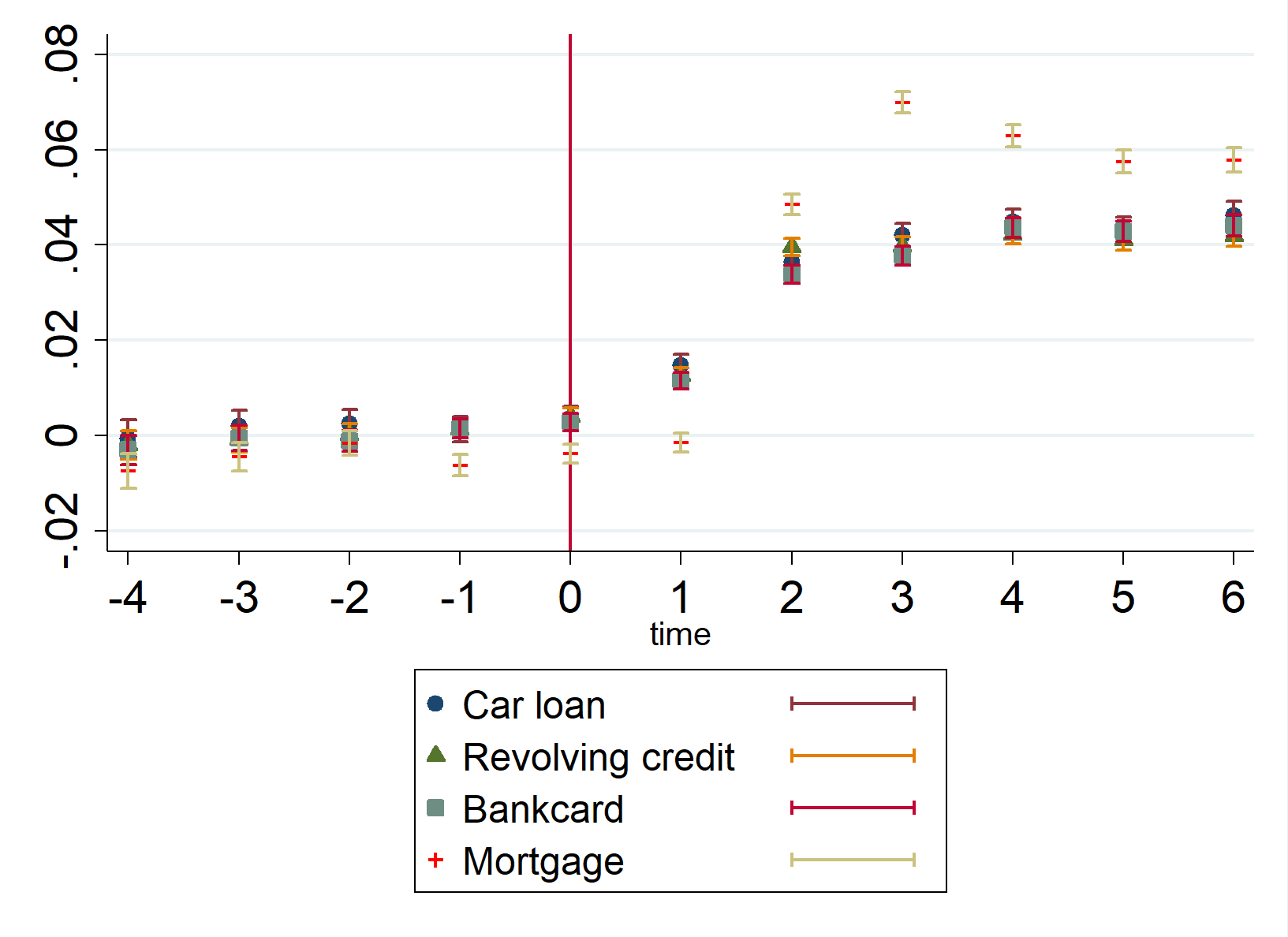}
             \includegraphics[height=4.5cm, width=7cm]{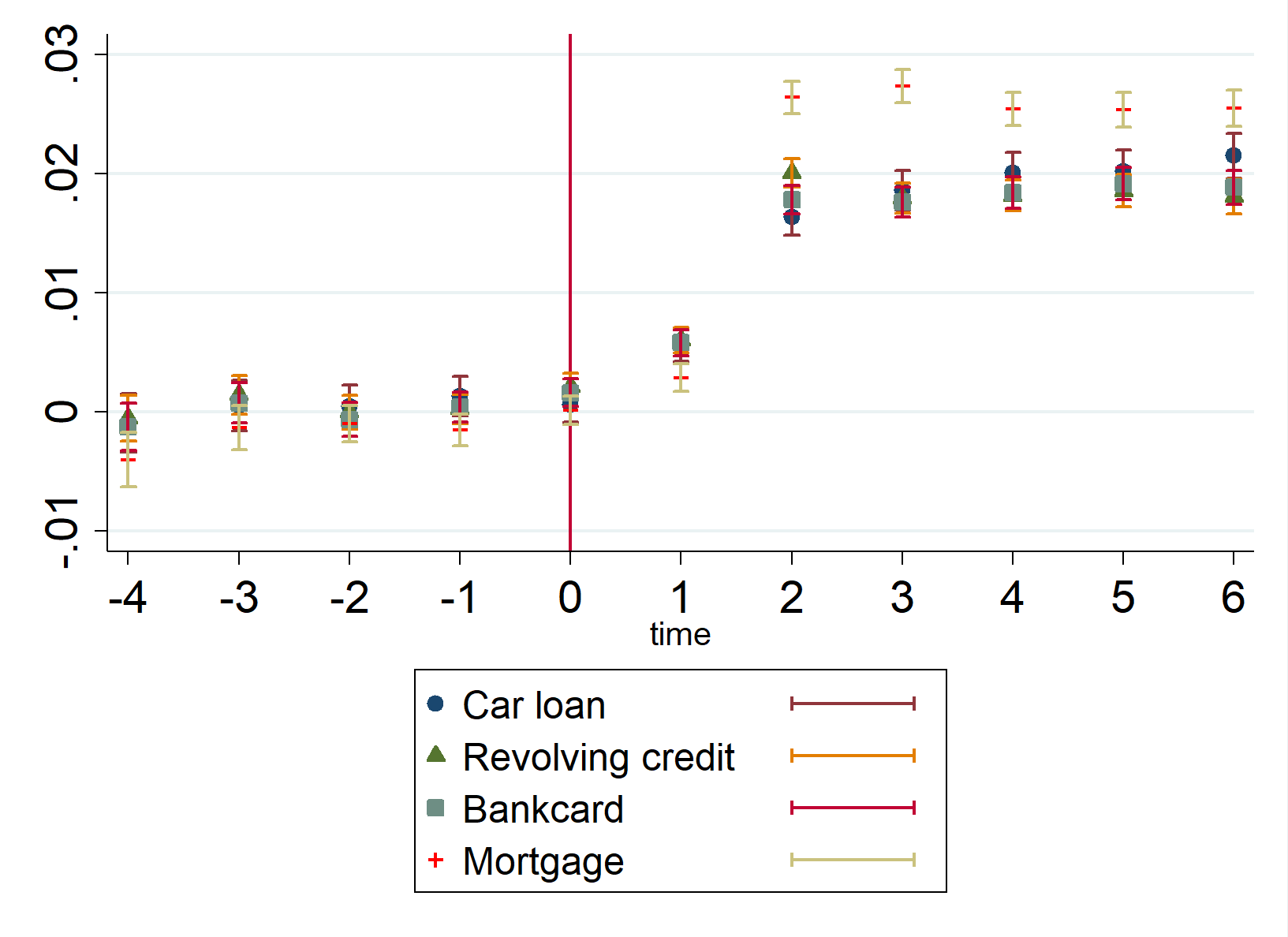}
             
                  	Panel (iii) - Median House Value \hspace{2cm} Panel (vi) - Income\\
                             \includegraphics[height=4.5cm, width=7cm]{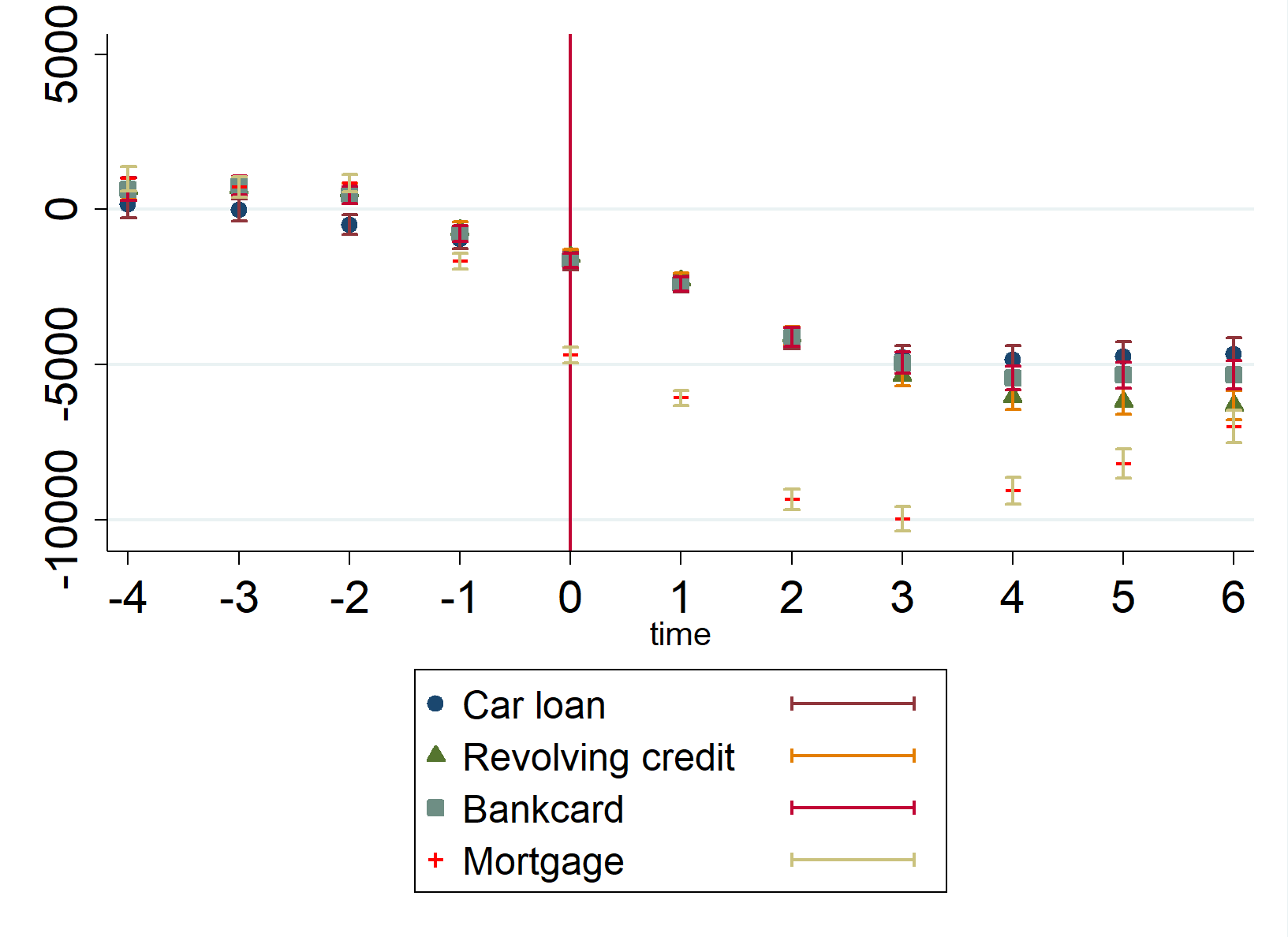}
             \includegraphics[height=4.5cm, width=7cm]{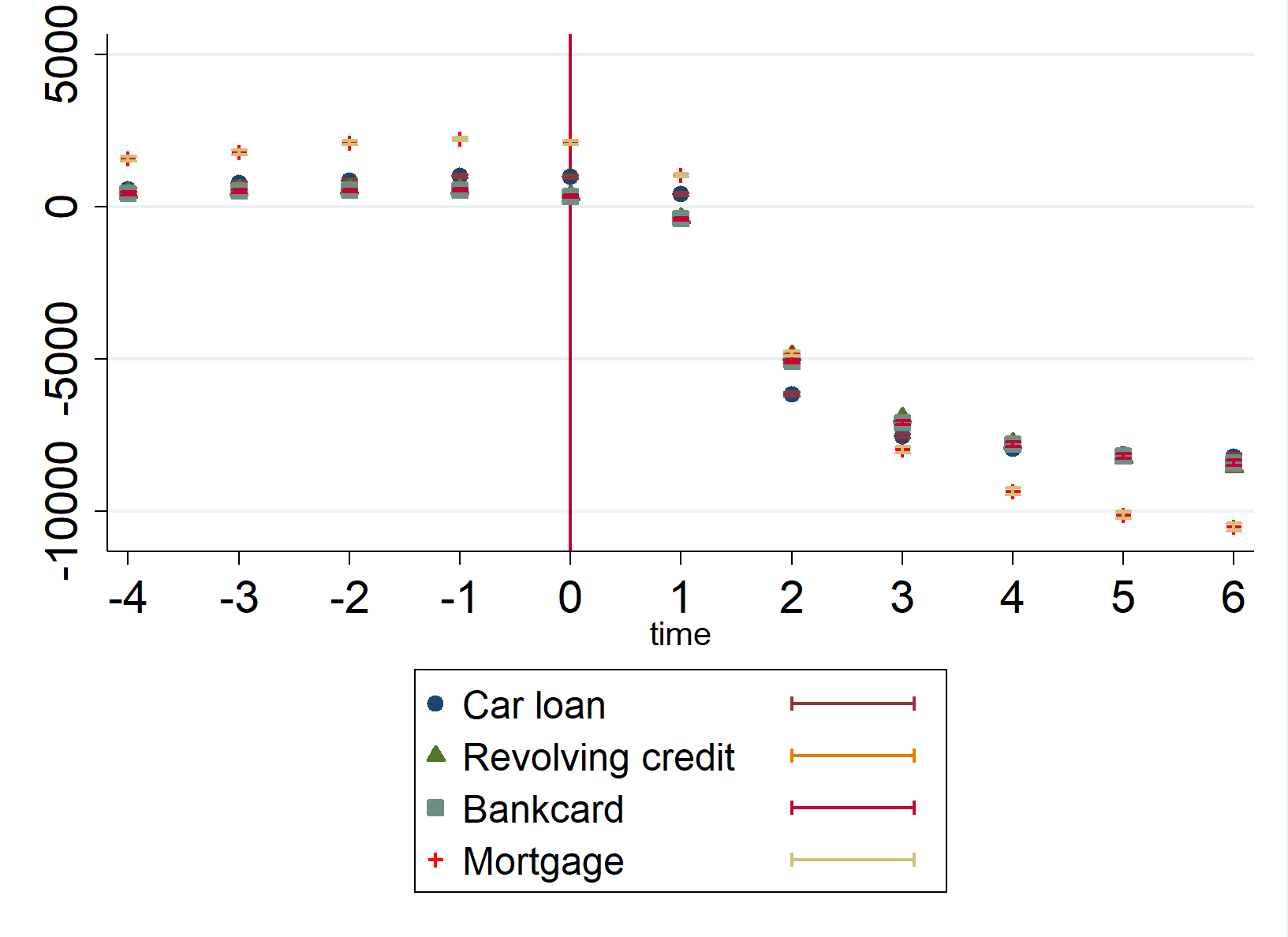}
                          
        \caption{Event study: dependent variable is: (i) Probability of moving zip code. This variable takes value 1 if the individual is a different zip code in year $t$ than in year $t-1$, and zero otherwise, (ii) Probability of moving outside the commuting zone. This variable takes value 1 if the individual is in a different commuting zone in year $t$ than in year $t-1$, and zero otherwise, (iii) the Median House Value in the zip code of residence at year $t$ (iv) income imputed by Experian. The event considered is a soft default, i.e. a 90-day delinquency, but no Chapter 7, Chapter 13 or foreclosure taking place in the same year, neither before in the sample period. Other controls are age and age squared, credit score in 2004 and in 2005. 95\% confidence intervals around the point estimates.}      \label{fig:bank}
     \end{figure}

                 From Figure \ref{fig:bank}, we find evidence that the different types of delinquencies have rather similar impacts on mobility and income. The probability of moving zip code increases by about 4pp following a car loan, revolving credit or bankcard delinquencies, whereas it increases slightly more, i.e. about 6pp, following a mortgage delinquency. Similarly, car loan, revolving credit and bankcard delinquencies are all associated with an increase by about 2pp in the probability of moving commuting zone.              
               
     As before, the impact is slightly larger, i.e. about 2.5-3pp, for a mortgage delinquency. As far as income is concerned, the four different types of delinquencies appear to have a very similar negative impact, which is equal to about minus 7,500USD in the medium term (i.e. six years after the soft default). Also in the case of the Median House Value, the impacts of car loan, revolving credit and bankcard delinquency are rather close to each other, i.e. about minus 5,000USD. On the other hand, the negative impact of mortgage delinquency is larger, i.e. about minus 7,000-10,000USD.

 \begin{figure}[H] 
 
             	 Credit Score\\
               \includegraphics[ scale=.18]{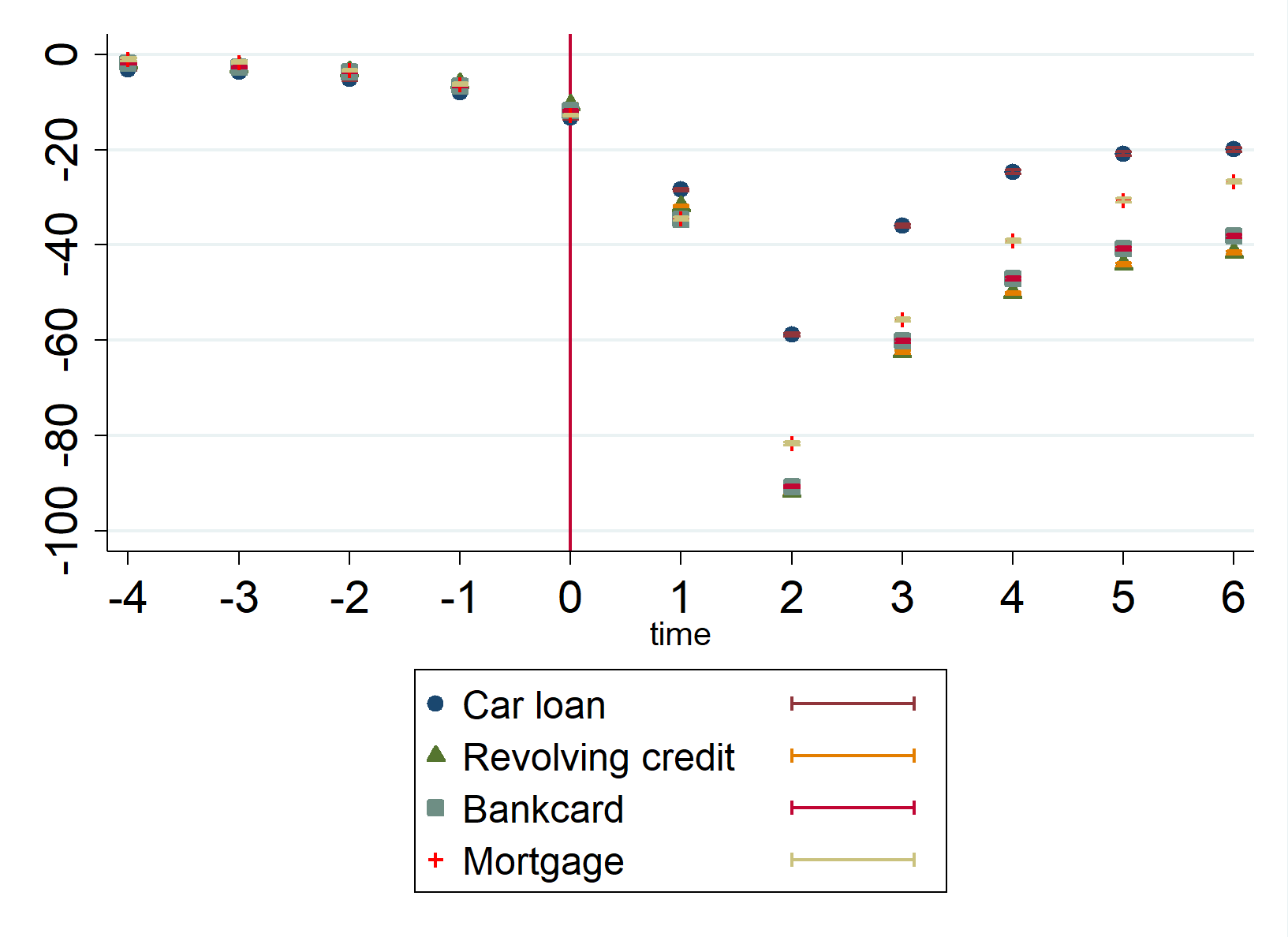}
    \caption{Event study: dependent variable is Credit Score. The event considered is a soft default, i.e. a 90-day delinquency, but no Chapter 7, Chapter 13 or foreclosure taking place in the same year, neither before in the sample period. Other controls are age and age squared, credit score in 2004 and in 2005. 95\% confidence intervals around the point estimates.}   \label{eventtypes} 
     \end{figure}

     From Figure \ref{eventtypes}, we deduce that the different types of delinquencies have a similar impact on the credit score. Importantly, a delinquency on the car loan leads to a smaller drop in the credit score (i.e. about minus 60 points), than the other three types of delinquencies (leading each to a decline of about 90 points in the credit score).

\begin{figure}[H] 
           
             	Panel (i) - Prob. new mortgage \hspace{1cm}Panel (ii) - Prob. Credit Limit $<$10k\\
                           
                           \includegraphics[height=4cm, width=7cm]{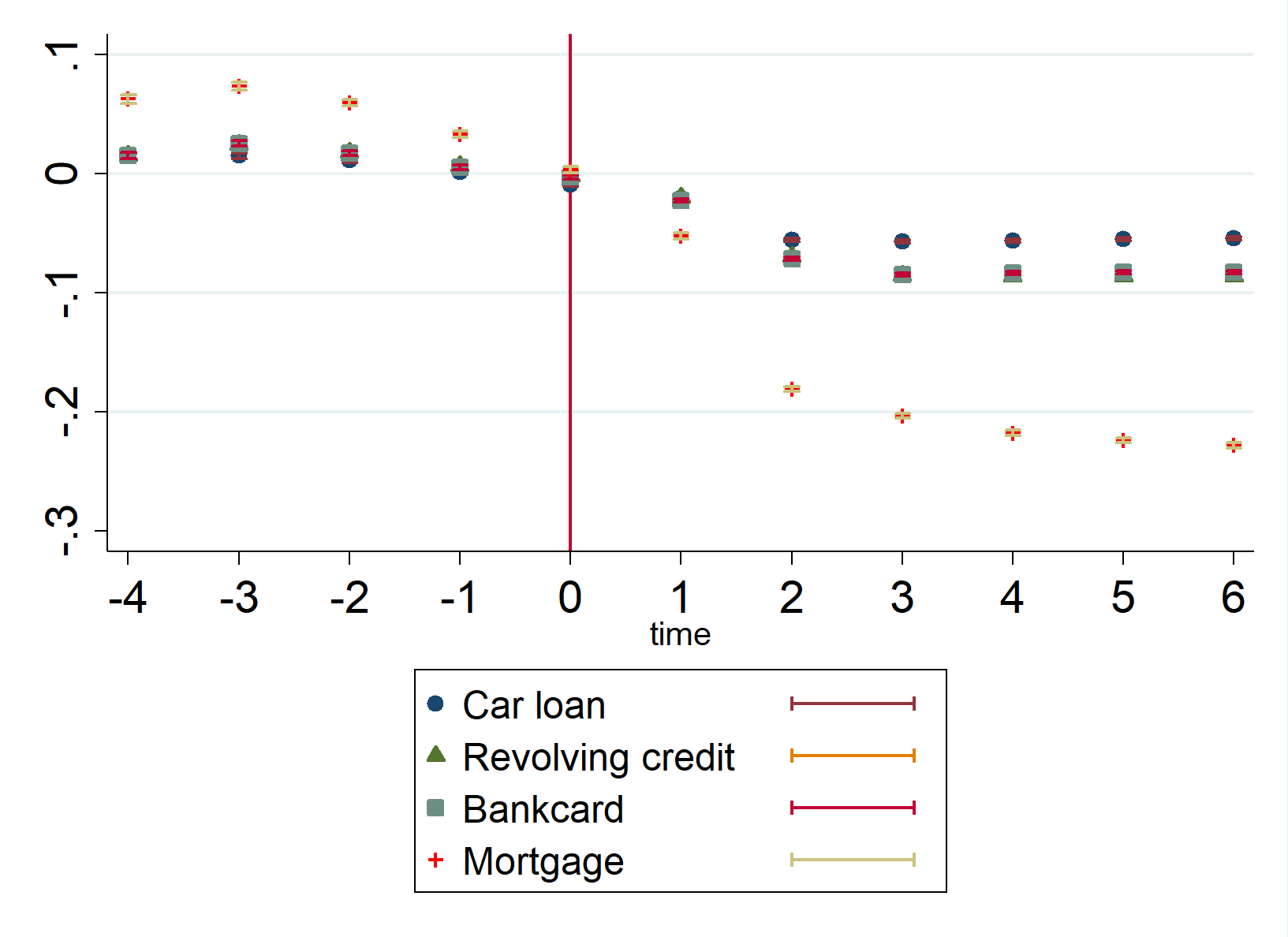}   
             \includegraphics[height=4cm, width=7cm]{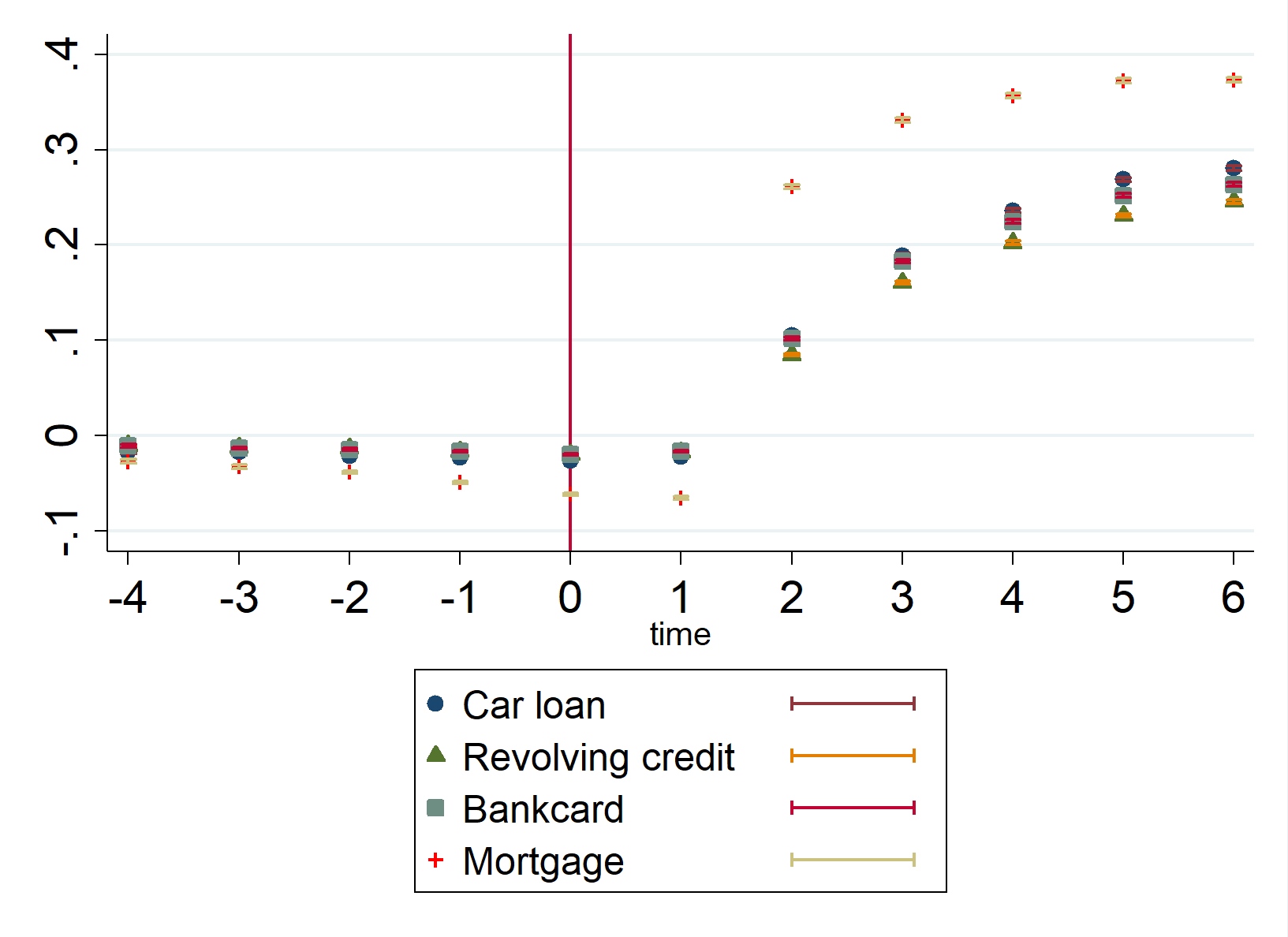}
              
                  	Panel (iii) - Revolving Balance \hspace{1cm}Panel (iv) - Harsh default\\
                     
           \includegraphics[height=4cm, width=7cm]{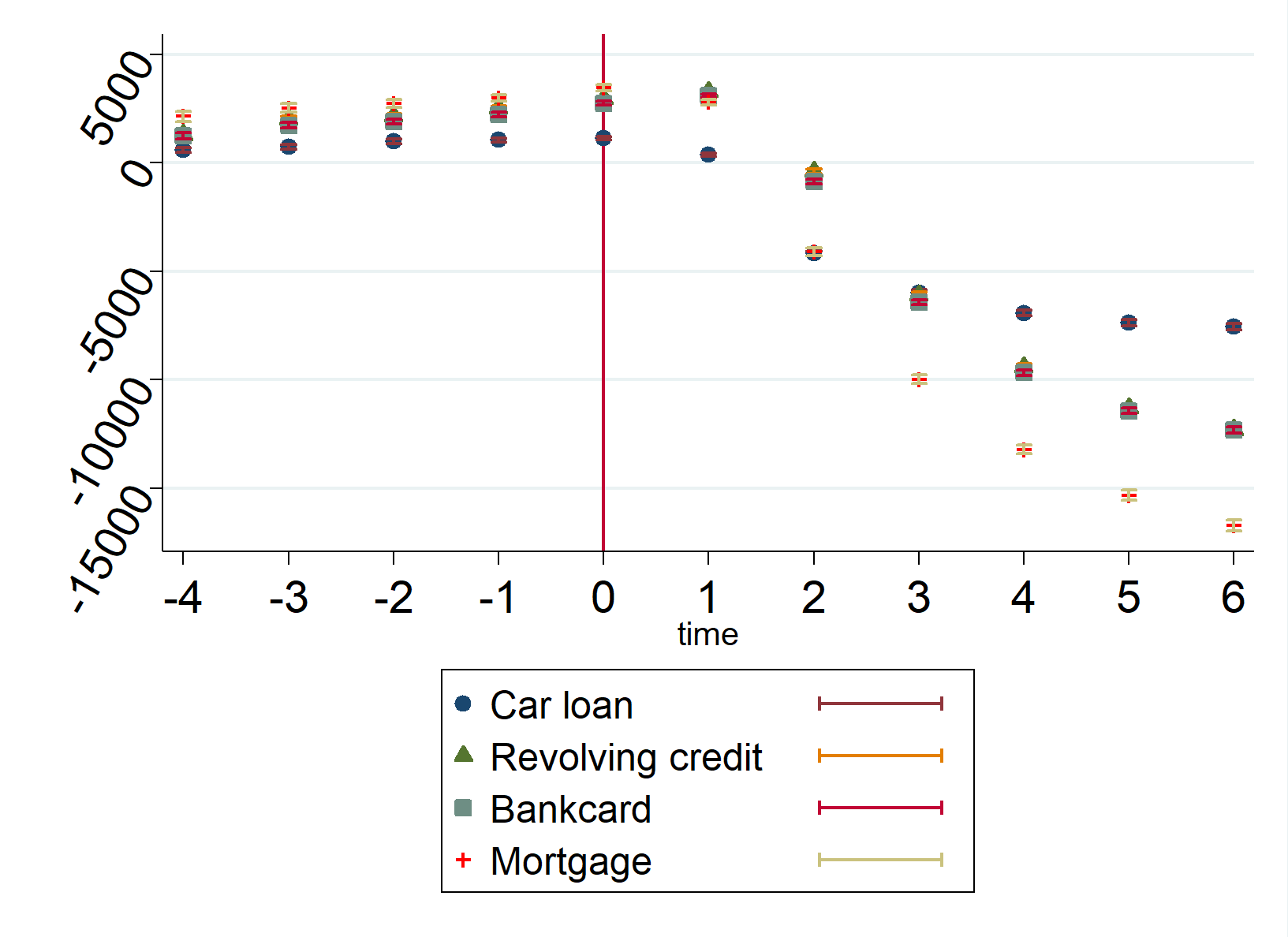}          	            	
              \includegraphics[height=4cm, width=7cm]{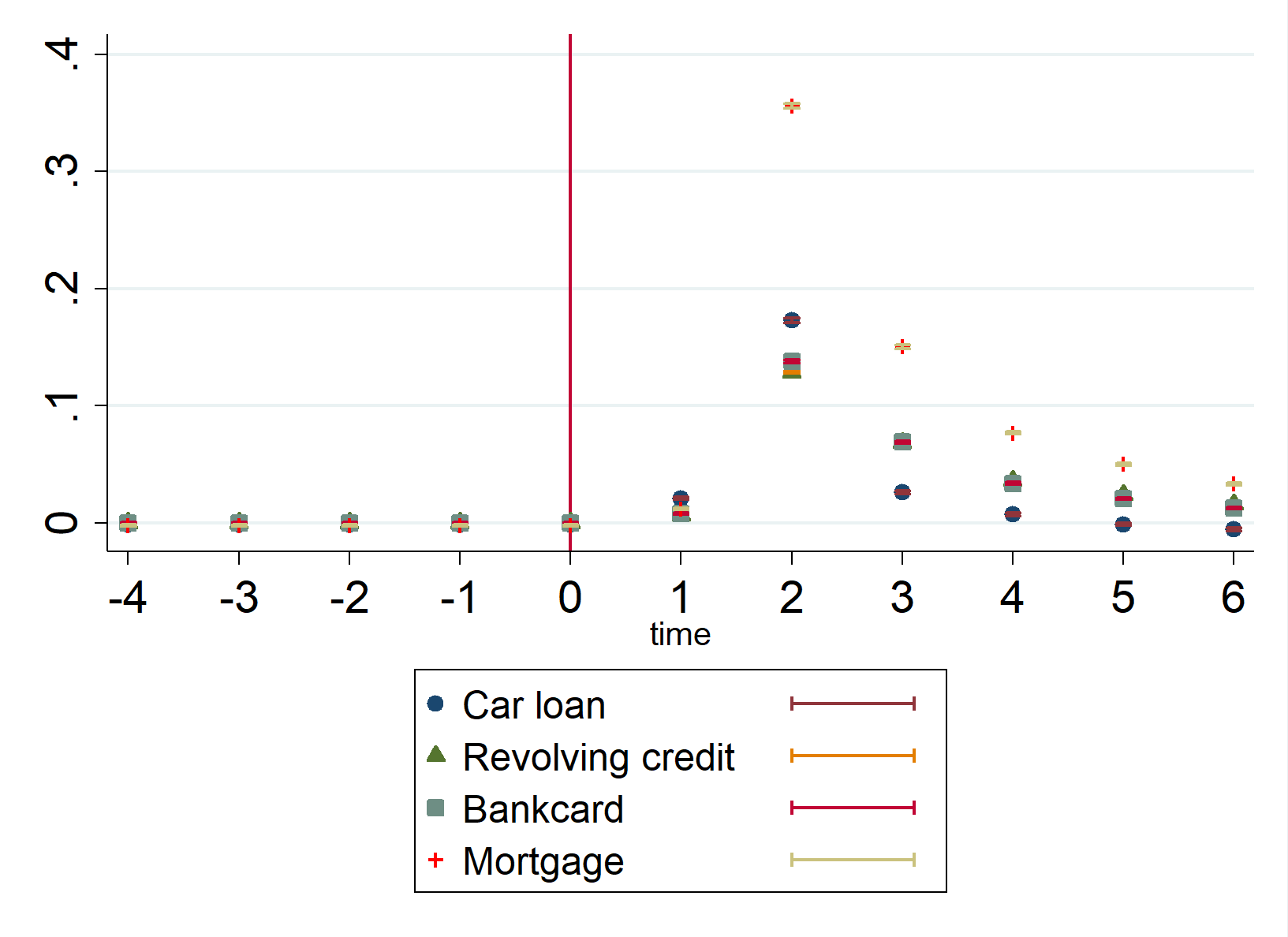}
                
                	Panel (v) - Home own \hspace{1cm}Panel (vi) - Total credit limit\\
                     
            \includegraphics[height=4cm, width=7cm]{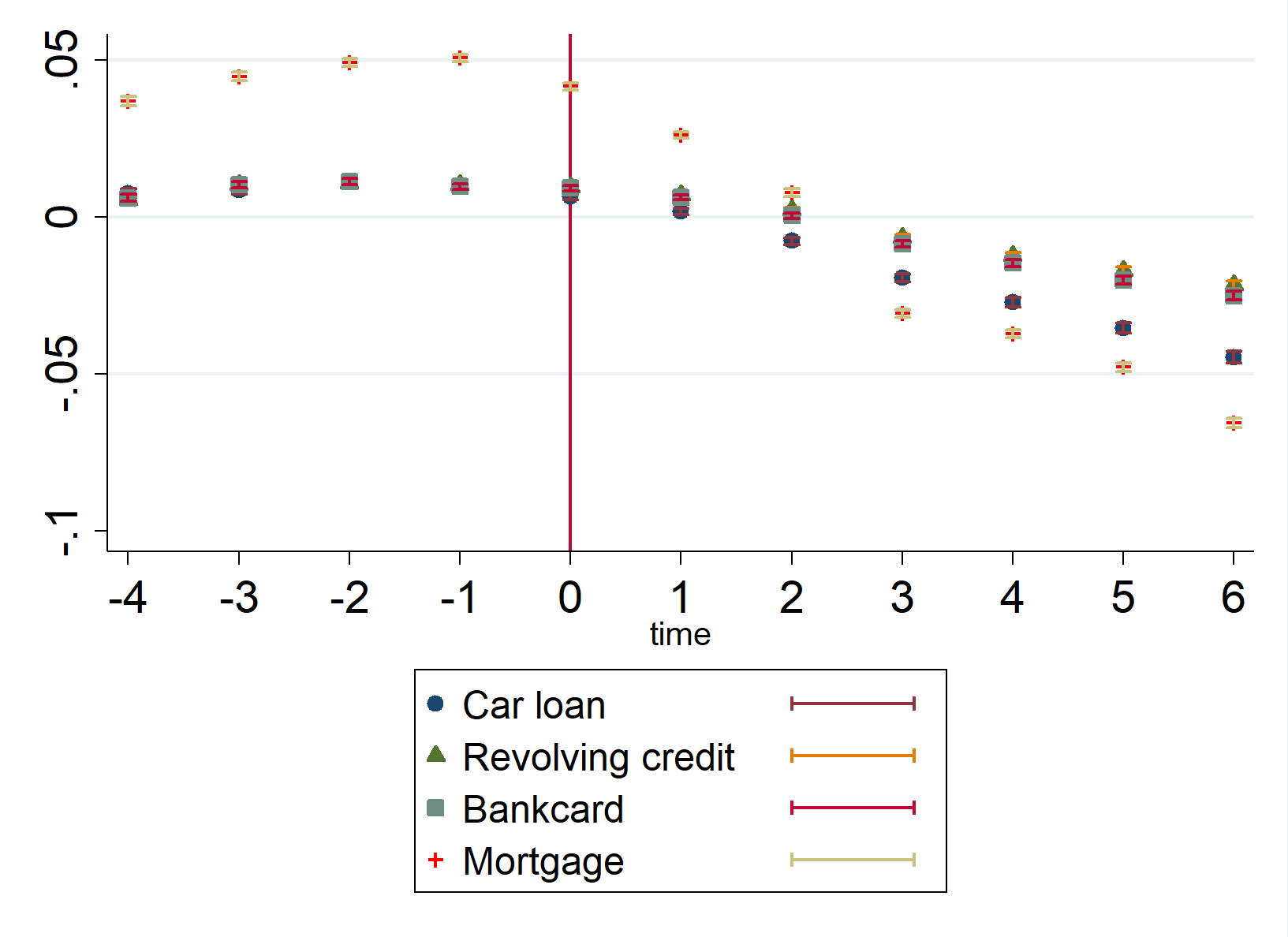}          	            	
               \includegraphics[height=4cm, width=7cm]{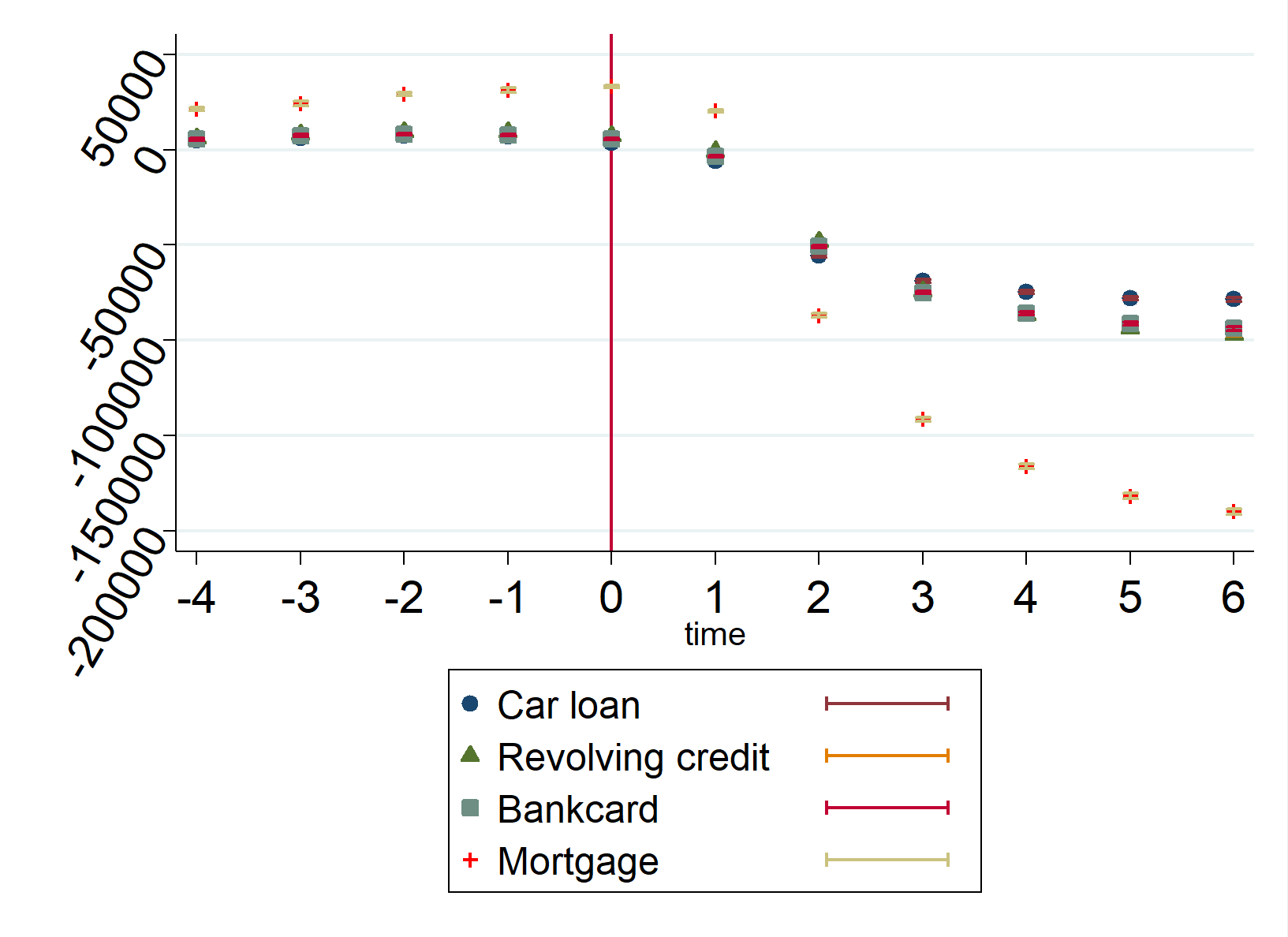}
                
                Panel (vii) - Mortgage balance open\\
                     
            \includegraphics[height=4cm, width=7cm]{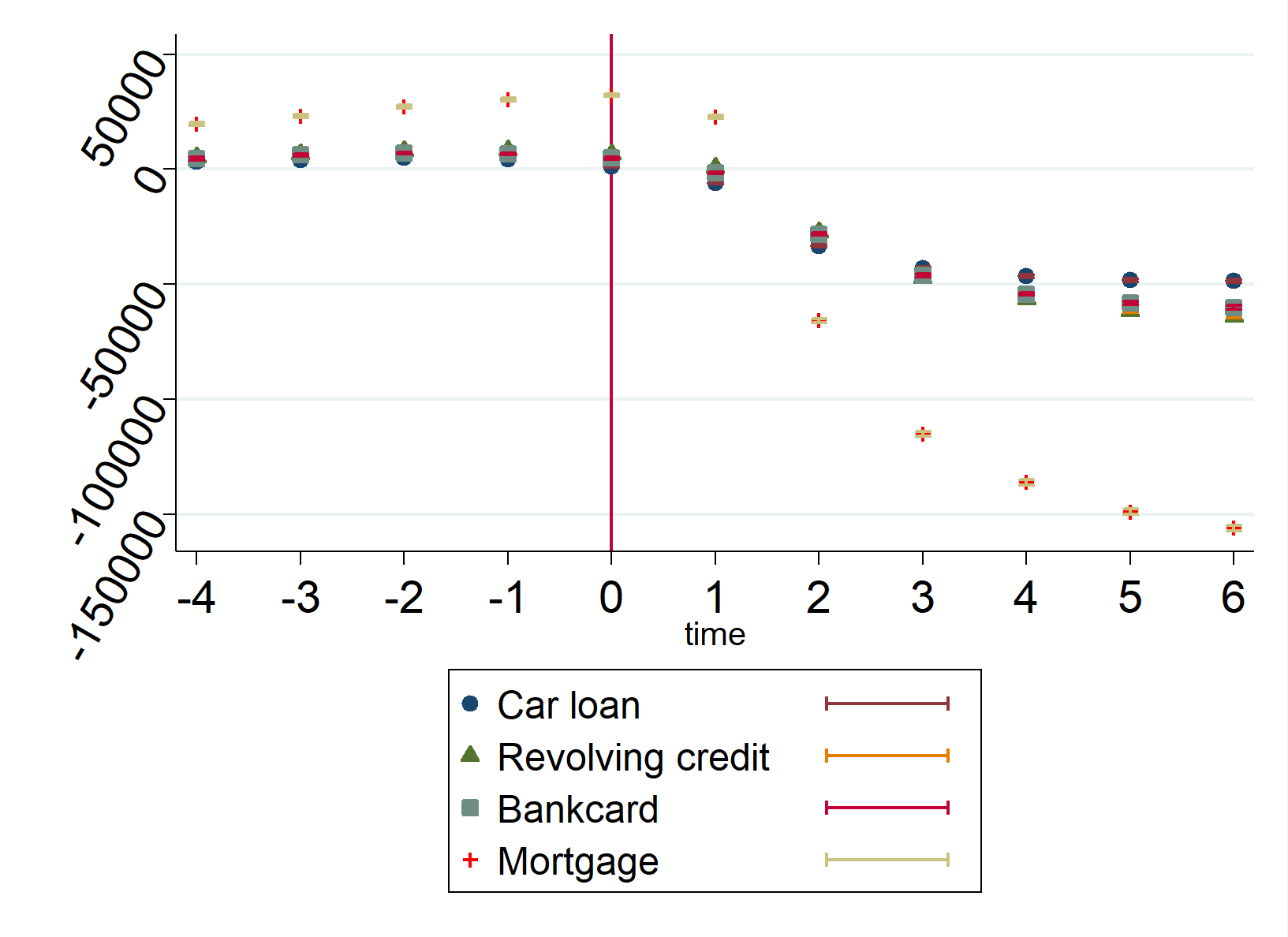}

        \caption{\small Event study: dependent variable is: (i) Mortgage origination: this variable takes value 1 if the individual has a higher number of mortgage trades in year $t$ than in year $t-1$ or if the number of months since the most recent mortgage has been opened is less than 12, and zero otherwise, (ii) probability that total credit limit is lower than 10,000USD, (iii) total amount open on all revolving credit trades, (iv) probability of experiencing a harsh default (Chapter 7, Chapter 13 or foreclosure), (v) probability of being homeowner, i.e. either being recorded as a homeowner by Experian or having ever had a mortgage open (vi) total credit limit on all trades, (vii) open amount of mortgage balance. The event considered is a soft default, i.e. a 90-day delinquency, but no Chapter 7, Chapter 13 or foreclosure taking place in the same year, neither before in the sample period. Other controls are age and age squared, credit score in 2004 and in 2005. 95\% confidence intervals around the point estimates.}     \label{credittypes}
     \end{figure}

     As far as credit-derived variables are concerned, from Figure \ref{credittypes} we find evidence that the negative effects on our outcome variables of interest of the different types of delinquencies are rather similar. However, the impact of mortgage delinquency appears to be the more serious among all the delinquency types considered. Indeed, a car, revolving credit or bank delinquency leads to a a drop by about 8-10pp in the probability of opening a new mortgage and to an increase by about 10-20pp in the probability of having a low credit limit. In the case of a mortgage delinquency these effects are, respectively, minus 20pp for the probability of mortgage origination, and plus 30-40pp for the probability of having a low credit limit.

     Further, the revolving balance open drops by about 10,000USD in the case of a revolving credit or bankcard delinquency, slightly less (about 8,000USD) in case of car loan delinquency and notably more (i.e. about 15,000USD) in case of mortgage delinquency. Similarly, a mortgage delinquency is associated with an increase by about 35pp in the probability of recording a harsh default two years after the event (i.e. the soft default), whereas this increase is only equal to less than 20pp for the other three types of delinquency. 

     Also for the probability of being homeowner, the larger negative effect (i.e. about minus 5pp) is recorded in the case of mortgage delinquencies, whereas the impact is slightly less (i.e. about 4pp) for a car loan delinquency and even smaller (about 1-2pp) for revolving credit and bankcard delinquencies. As far as the total credit limit is concerned, it drops by about 175,000USD following a mortgage delinquency. The impact on the same variable is notably smaller for the other three types of delinquency (car loan, bankcard and revolving credit), i.e. about minus 75,000USD only.

     Finally, as one could reasonably expect, a mortgage delinquency also has the largest impact on the amount of mortgage open. Indeed, following a mortgage delinquency this amount drops by about 150,000USD in the medium run (six years after the soft default), whereas the same outcome variable only declines by about 50,000USD for all the three others default types.

     To summarize, this further event study exercise shows that bankcard, revolving credit and car loan delinquencies are rather similar in their effects. Mortgage delinquencies are more serious as far as the size of the effect is concerned for most outcome variables. However, for all the four types of delinquency, the negative impacts on the outcome variables are statistically significant in the short and in the medium run, i.e. also delinquencies that are not about mortgages have a long-lasting negative impact on individual trajectories\footnote{Note that the baseline event study results presented in Section \ref{sec:exploratory} are not necessarily a weighted average of the impacts of the four different types of delinquencies presented here. This is because individuals may have multiple delinquencies (e.g. both bankcard and auto). In particular descriptive statistics on our data show that, among individuals that are delinquent, more than 75\% are delinquent in more than one of the four categories considered above.}.
        
\section{Baseline Evidence in the Long-Term}
   \label{sec:long}

   \setcounter{table}{0}
\renewcommand{\thetable}{G\arabic{table}}

\setcounter{figure}{0}
\renewcommand{\thefigure}{G\arabic{figure}}

 In this Section we analyze the long-run impact of  a soft default in 2010 by means of the double/debiased lasso estimation techniques. 
   We produce two set of estimates: one as of June 2020 (the last year we have data for) and another for June 2019.   While June 2020 is right in the midst of COVID-19, several previous studies have not found a large COVID-19 impact on credit outcomes by that time, in part this is due to  The Coronavirus Aid, Relief, and Economic Security (CARES) Act.\footnote{See  \url{https://www.consumerfinance.gov/data-research/research-reports/special-issue-brief-early-effects-covid-19-pandemic-on-consumer-credit/} for  a Consumer Financial Protection Bureau report covering our period of interest; and \url{https://www.consumerfinance.gov/about-us/blog/protecting-your-credit-during-coronavirus-pandemic/}.}  As an alternative, to probe the robustness of our 2020 analysis, we also present the same evidence for  2019 (the last pre-COVID observation we have). 
  
    It is here worth mentioning that the long-run effects estimated in this Section need not  be the same as those found in the last year of the event study approach. There are at least two reasons for this: 1. the last year in the event study is 2016, while here we look at 2020 (or 2019); 2. more importantly, the identification strategies are different and that can lead to different results. Therefore,  the long-run results lining up with our previous analysis is not a given and that offers some reassurance on the causal interpretation of our findings.
  
\subsection{Mobility, Income and Credit Score}

In the  Tables below, we present the impact of a soft default in 2010 on our outcome variables of interest in the long run. We consider here 2020 as the final year of interest. The impacts on the probability of moving zip code and on that of moving commuting zone, are not statistically significant. Given the mobility in the data, a large majority of the individuals would have changed their zip code irrespective of default in the 10 years window (see  Table \ref{tab:descriptives2010}). 
  However, the drop by about 7,000USD in income is statistically significant, and in line with the analysis in Section \ref{sec:exploratory}.
  
   In Table  \ref{tab:DML_2019} we replicate the estimates presented in Table  \ref{tab:DML_2020}, i.e. the long term impact of a soft default, but this time by considering 2019 instead of 2020 as the final year of interest. From the comparison of these two tables, we notice that the main results are qualitatively and quantitatively very similar.

  The negative impact of about -16 points in the credit score is statistically significant. This is confirmed when we use 2019 as the final year instead of 2020. These results are consistent with the previous analysis provided in Section \ref{sec:exploratory}.

  \begin{table}[H]
  \caption{DML long-run impact of a soft default 2020. Controls: age, age squared, commuting zones fixed effects, credit score, the amount of open mortgages and car loan, as well as county unemployment rate and number of bank closings in the county pre-event (years 2004-2009), plus max interest rate. Individuals recording a harsh default between 2004 and 2010 (extremes included) and those recording a soft default between 2004 and 2009 (extremes included) have been dropped from the estimation sample.\label{tab:DML_2020}}
  \footnotesize
  {
\def\sym#1{\ifmmode^{#1}\else\(^{#1}\)\fi}
\begin{tabular}{c c c c c}
\hline\hline
            &\multicolumn{1}{c}{(1)}&\multicolumn{1}{c}{(2)}&\multicolumn{1}{c}{(3)}&\multicolumn{1}{c}{(4)}\\
            &\multicolumn{1}{c}{move}&\multicolumn{1}{c}{move\_cz}&\multicolumn{1}{c}{cs}&\multicolumn{1}{c}{incomeW2}\\
\hline
Soft def &     0.00534         &    -0.00427         &     -16.33\sym{***}         &          -7090.2\sym{***}\\
            &   (0.00430)         &   (0.00343)         &     (0.510)         &        (263.8)              \\
\hline
\(N\)       &     1043406         &     1043406         &       1027786          &      1027822         \\
\hline\hline
\multicolumn{5}{l}{\footnotesize Standard errors in parentheses}\\
\multicolumn{5}{l}{\footnotesize \sym{*} \(p<0.05\), \sym{**} \(p<0.01\), \sym{***} \(p<0.001\)}\\
\end{tabular}
}
  \end{table}

     \begin{table}[H]
  \caption{DML long-run impact of a soft default  2019. Controls: age, age squared,  commuting zones fixed effects, credit score, the amount of open mortgages and car loan, as well as county unemployment rate and number of bank closings in the county pre-event (years 2004-2009), plus max interest rate. Individuals recording a harsh default between 2004 and 2010 (extremes included) and those recording a soft default between 2004 and 2009 (extremes included) have been dropped from the estimation sample.\label{tab:DML_2019}}
  \footnotesize
  {
\def\sym#1{\ifmmode^{#1}\else\(^{#1}\)\fi}
\begin{tabular}{c c c c c}
\hline\hline
            &\multicolumn{1}{c}{(1)}&\multicolumn{1}{c}{(2)}&\multicolumn{1}{c}{(3)}&\multicolumn{1}{c}{(4)}\\
            &\multicolumn{1}{c}{move}&\multicolumn{1}{c}{move\_cz}&\multicolumn{1}{c}{cs}&\multicolumn{1}{c}{incomeW2}\\
\hline
Soft def &     0.00677         &    -0.00576         &      -17.66\sym{***} &         -7342.0\sym{***}\\
            &   (0.00433)         &   (0.00339)         &      (0.523)         &       (259.9)             \\
\hline
\(N\)       &     1044486         &     1044486         &      1028231          &       1028270              \\
\hline\hline
\multicolumn{5}{l}{\footnotesize Standard errors in parentheses}\\
\multicolumn{5}{l}{\footnotesize \sym{*} \(p<0.05\), \sym{**} \(p<0.01\), \sym{***} \(p<0.001\)}\\
\end{tabular}
}
  \end{table}

\subsection{Credit}

From Table \ref{tab:DML_2020b} and Table  \ref{tab:DML_2019b}, we find evidence that a soft default entails a decrease in the probability of opening a new mortgage by about 1pp. This decrease is substantial and in line with our findings in the event studies (Section \ref{sec:exploratory}). Further, a soft default causes an increase by about 1pp in the probability of having a low total credit limit (i.e. below 10,000USD) and about a 4,000USD decrease in the amount of revolving credit open. Finally, a soft default is linked to a notable increase (i.e. +18/19pp) in the probability of experiencing at least once a harsh default in the period 2010-2020, respectively 2010-2019 when we take 2019 as the last year of the sample. All these impacts are non-negligible and statistically significant in the long run.

From Table \ref{tab:DML_2020b} we also deduce that the negative impact of a soft default on the probability of being a home owner (in our more comprehensive definition) is about -2pp, whereas the negative effect on the total credit limit is about -52,000USD, i.e. substantial. The drop in the mortgage amount open is also noticeable, i.e. about 35,000USD. These impacts are all confirmed both in sign and in size when we consider 2019 instead of 2020 as the final year of the sample (Table  \ref{tab:DML_2019b}).

  \begin{table}[H]
  \caption{DML long-run impact of a soft default 2020. Controls: age, age squared, commuting zones fixed effects, credit score, the amount of open mortgages and car loan, as well as county unemployment rate and number of bank closings in the county pre-event (years 2004-2009), plus max interest rate. Individuals recording a harsh default between 2004 and 2010 (extremes included) and those recording a soft default between 2004 and 2009 (extremes included) have been dropped from the estimation sample.\label{tab:DML_2020b}}
  \footnotesize
  {
\def\sym#1{\ifmmode^{#1}\else\(^{#1}\)\fi}
\begin{tabular}{l*{7}{c}}
\hline\hline
              &\multicolumn{1}{c}{(1)}&\multicolumn{1}{c}{(2)}&\multicolumn{1}{c}{(3)}&\multicolumn{1}{c}{(4)}&\multicolumn{1}{c}{(5)}&\multicolumn{1}{c}{(6)}&\multicolumn{1}{c}{(7)}\\
            &\multicolumn{1}{c}{Prob. mort}&\multicolumn{1}{c}{Prob. low cred lim}&\multicolumn{1}{c}{rev\_bal\_open}&\multicolumn{1}{c}{harsh def}  &\multicolumn{1}{c}{home\_own}&\multicolumn{1}{c}{tot cred lim}&\multicolumn{1}{c}{m\_bal\_open}\\
\hline
Soft def &      -0.00620\sym{*}&     0.00328         &     -3957.7\sym{***}&        0.180\sym{***}  &   -0.0187\sym{***}&    -48493.5\sym{***}&    -30409.0\sym{***}\\
            &   (0.00267)        &   (0.00268)         &     (336.1)         &   (0.00230)      &   (0.00336)         &    (2170.4)         &    (1711.0)       \\
\hline
\(N\)        &     1043406         &     1043406         &     1043406         &     1043406      &     1043406         &     1043406         &     1043406      \\
\hline\hline
\multicolumn{8}{l}{\footnotesize Standard errors in parentheses}\\
\multicolumn{8}{l}{\footnotesize \sym{*} \(p<0.05\), \sym{**} \(p<0.01\), \sym{***} \(p<0.001\)}\\
\end{tabular}
}
  \end{table}

    \begin{table}[H]
  \caption{DML long-run impact of a soft default 2019. Controls: age, age squared, commuting zones fixed effects, credit score, the amount of open mortgages and car loan, as well as county unemployment rate and number of bank closings in the county pre-event (years 2004-2009), plus max interest rate. Individuals recording a harsh default between 2004 and 2010 (extremes included) and those recording a soft default between 2004 and 2009 (extremes included) have been dropped from the estimation sample.\label{tab:DML_2019b}}
  \footnotesize
  {
\def\sym#1{\ifmmode^{#1}\else\(^{#1}\)\fi}
\begin{tabular}{l*{7}{c}}
\hline\hline
              &\multicolumn{1}{c}{(1)}&\multicolumn{1}{c}{(2)}&\multicolumn{1}{c}{(3)}&\multicolumn{1}{c}{(4)}&\multicolumn{1}{c}{(5)}&\multicolumn{1}{c}{(6)}&\multicolumn{1}{c}{(7)}\\
            &\multicolumn{1}{c}{Prob. mort}&\multicolumn{1}{c}{Prob. low cred lim}&\multicolumn{1}{c}{rev\_bal\_open}&\multicolumn{1}{c}{harsh def}&\multicolumn{1}{c}{home\_own}&\multicolumn{1}{c}{tot cred lim}&\multicolumn{1}{c}{m\_bal\_open}\\
\hline
Soft def &     -0.00561\sym{***}&      0.0101\sym{***}&     -4736.9\sym{***}&      0.185\sym{***} &     -0.0214\sym{***}&    -53200.3\sym{***}&    -33079.9\sym{***} \\
            &   (0.0013)         &   (0.00270)         &     (342.7)         &   (0.00225)    &   (0.00338)         &    (2141.7)         &    (1679.3)       \\
\hline
\(N\)        &      1044486         &     1044486         &     1044486         &     1044486      &     1044486         &     1044486         &     1044486     \\
\hline\hline
\multicolumn{8}{l}{\footnotesize Standard errors in parentheses}\\
\multicolumn{8}{l}{\footnotesize \sym{*} \(p<0.05\), \sym{**} \(p<0.01\), \sym{***} \(p<0.001\)}\\
\end{tabular}
}
  \end{table}

  In general, we deduce that the results do not change substantively if 2020 is used rather than 2019. 
  

\section{Additional Evidence in the Long Term \label{sec:long_termB}}

 

   \setcounter{table}{0}
\renewcommand{\thetable}{H\arabic{table}}

\setcounter{figure}{0}
\renewcommand{\thefigure}{H\arabic{figure}}
    
\subsection{Mobility, Income and Credit Score}

   \begin{figure}[H]
   \includegraphics[height=4cm, width=7cm]{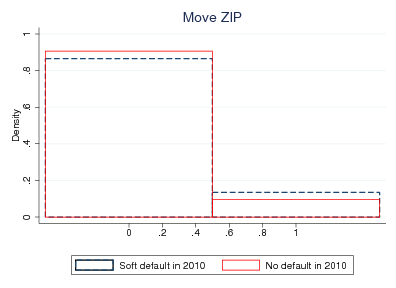}
      \includegraphics[height=4cm, width=7cm]{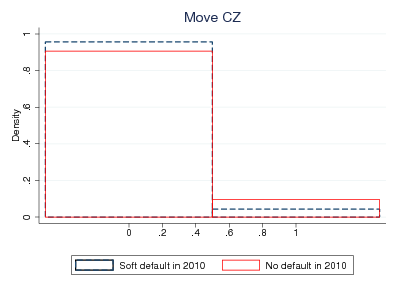} \\
      
        \includegraphics[height=4cm, width=7cm]{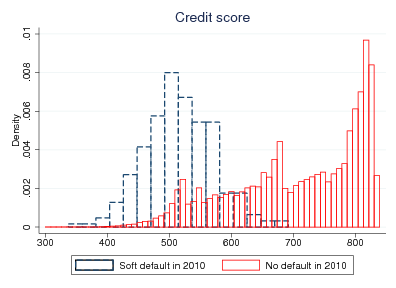}
      \includegraphics[height=4cm, width=7cm]{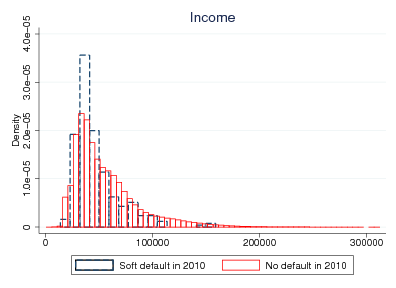} \\

        \caption{Comparison of the density (histogram) in 2020 of (i) probability of moving ZIP code (ii) probability of moving commuting zone (iii) Credit score (iv) income, for individuals who had a soft default in 2010 vs for those who hadn't.  Top 1\% of total credit amount, income, total revolving balance, revolving credit limit and mortgage balance open have been trimmed for readability of the graphs.}  \label{fig:distribution_soft1}
   
   \end{figure}

   \subsection{Credit}
   
    \begin{figure}[H]
   \includegraphics[height=4cm, width=7cm]{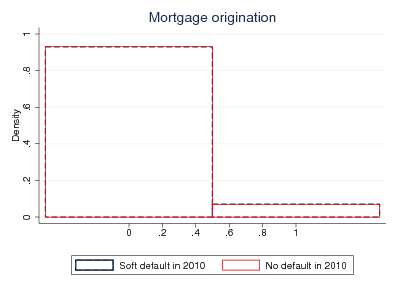}
      \includegraphics[height=4cm, width=7cm]{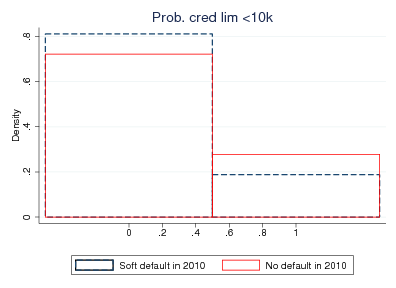} \\
      
        \includegraphics[height=4cm, width=7cm]{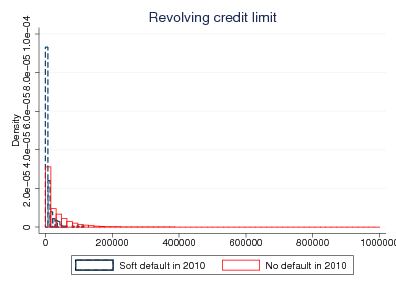}
      \includegraphics[height=4cm, width=7cm]{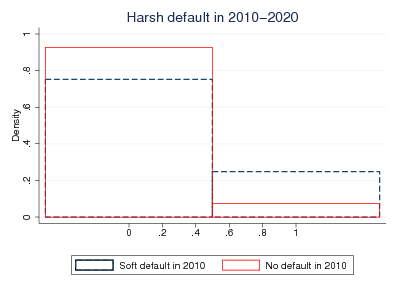} \\
      
              \includegraphics[height=4cm, width=7cm]{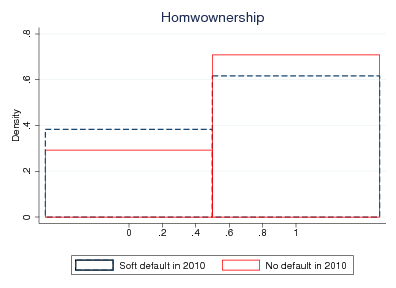}
      \includegraphics[height=4cm, width=7cm]{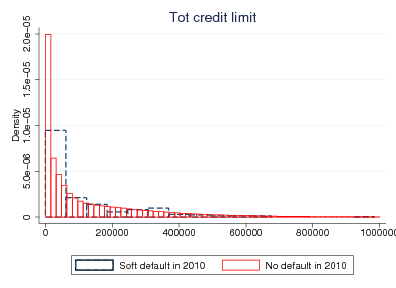} \\
      
       \includegraphics[height=4cm, width=7cm]{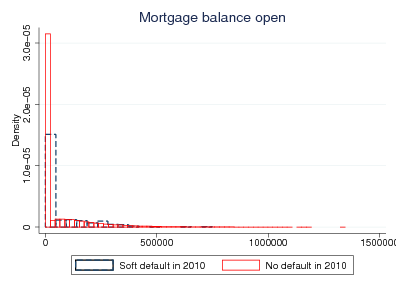}
   
        \caption{Comparison of the density (histogram) in 2020 of (i) mortgage origination, (ii) probability that the total credit limit is lower than 10,000USD, (iii) revolving credit limit, (iv) harsh default, (v) homeownership probability (comprehensive definition), (vi) total credit limit, (vii) mortgage balance open for individuals who had a soft default in 2010 vs for those who hadn't.  Top 1\% of total credit amount, income, total revolving balance, revolving credit limit and mortgage balance open have been trimmed for readability of the graphs.}  \label{fig:distribution_soft}
   
   \end{figure}

          \subsection{Treated in 2016 used as control group (5\% random sample)}    
          \label{sec:control2016}

          \begin{figure}[H] 
         
             	Panel (i) - Move Zip \hspace{4cm}Panel (ii) - Move CZ\\
                \includegraphics[height=4.5cm, width=7cm]{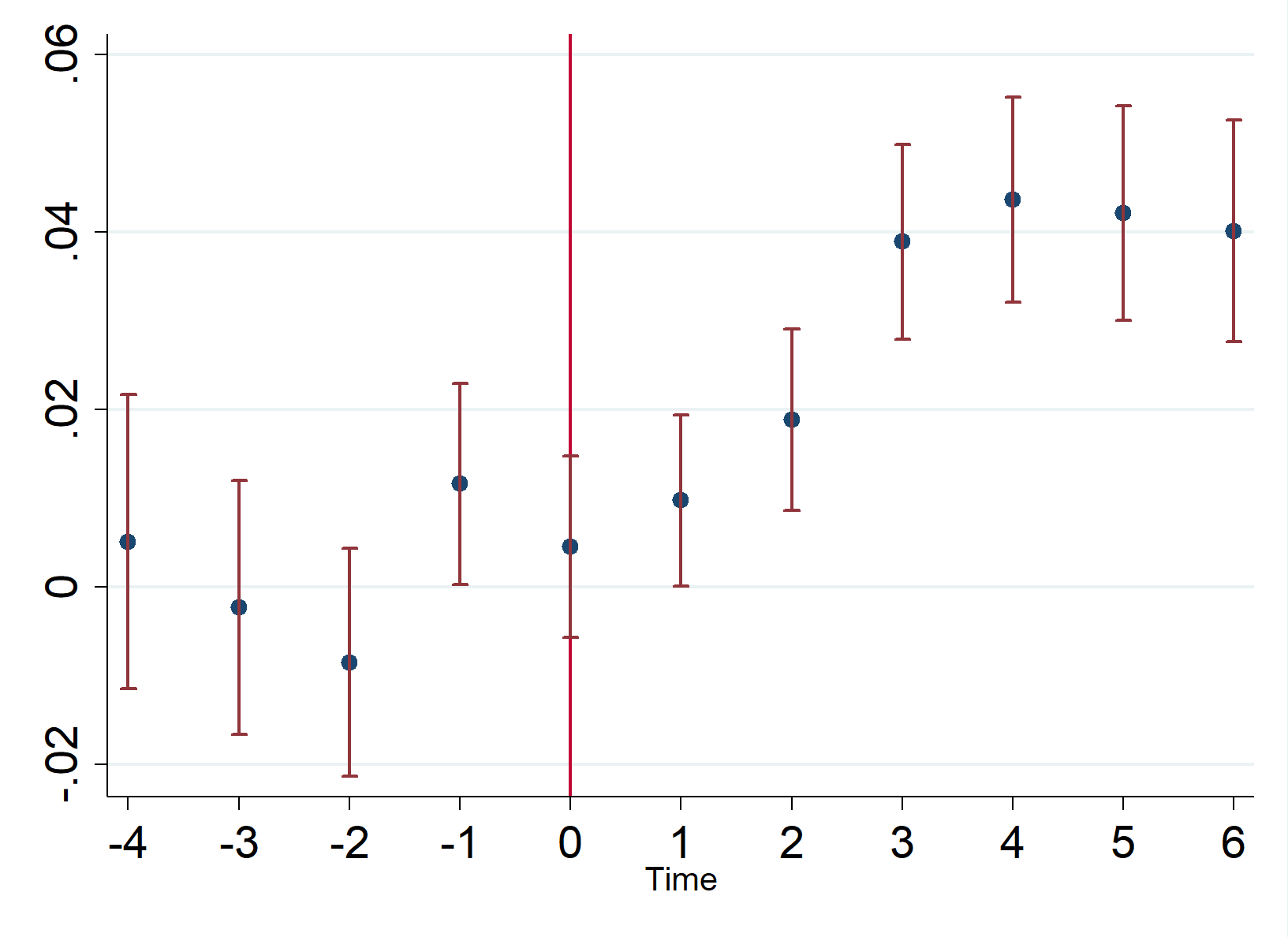}
             \includegraphics[height=4.5cm, width=7cm]{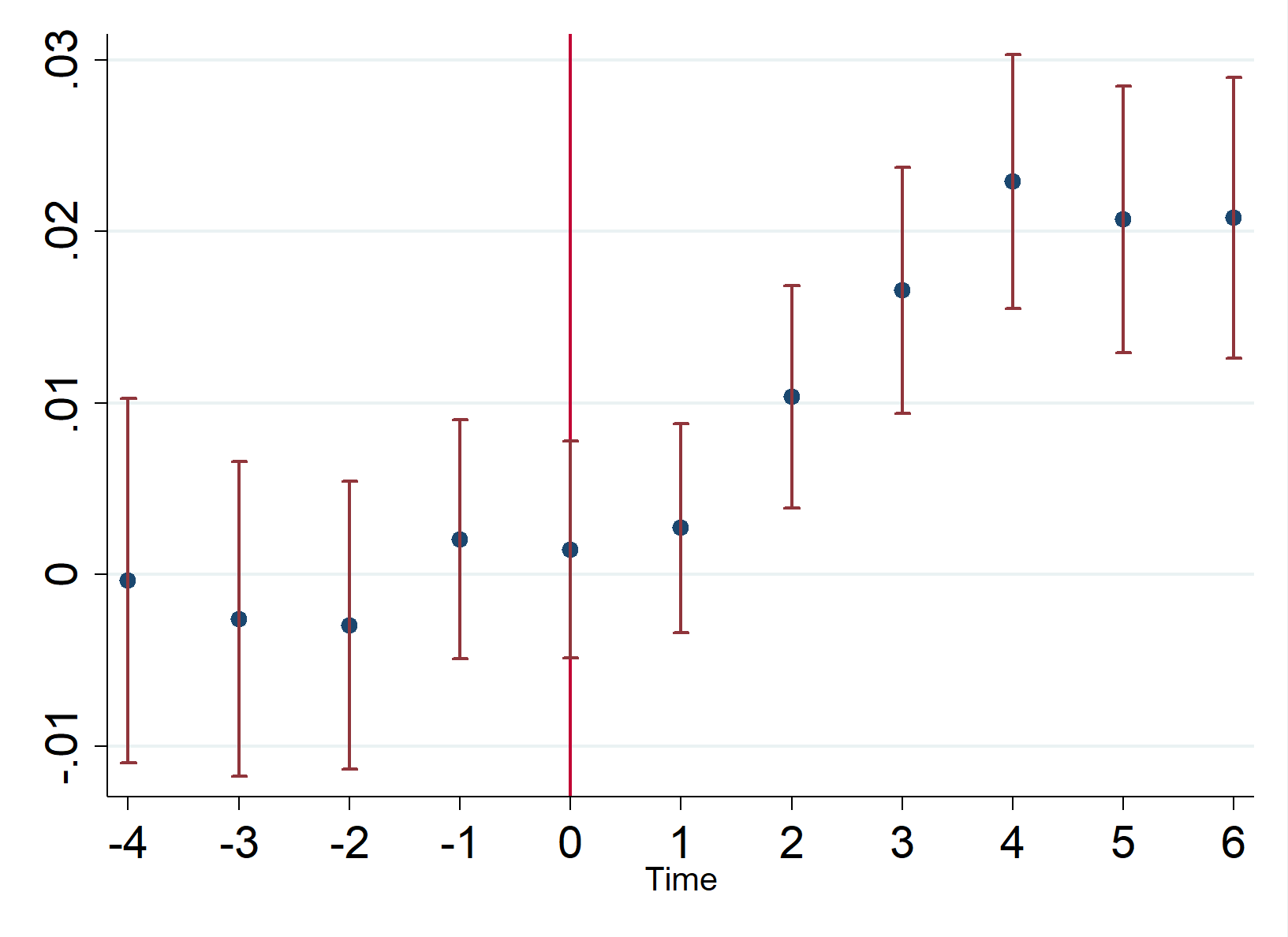}
             
                  	Panel (iii) - Median House Value \hspace{2cm} Panel (vi) - Income\\
                 \includegraphics[height=4.5cm, width=7cm]{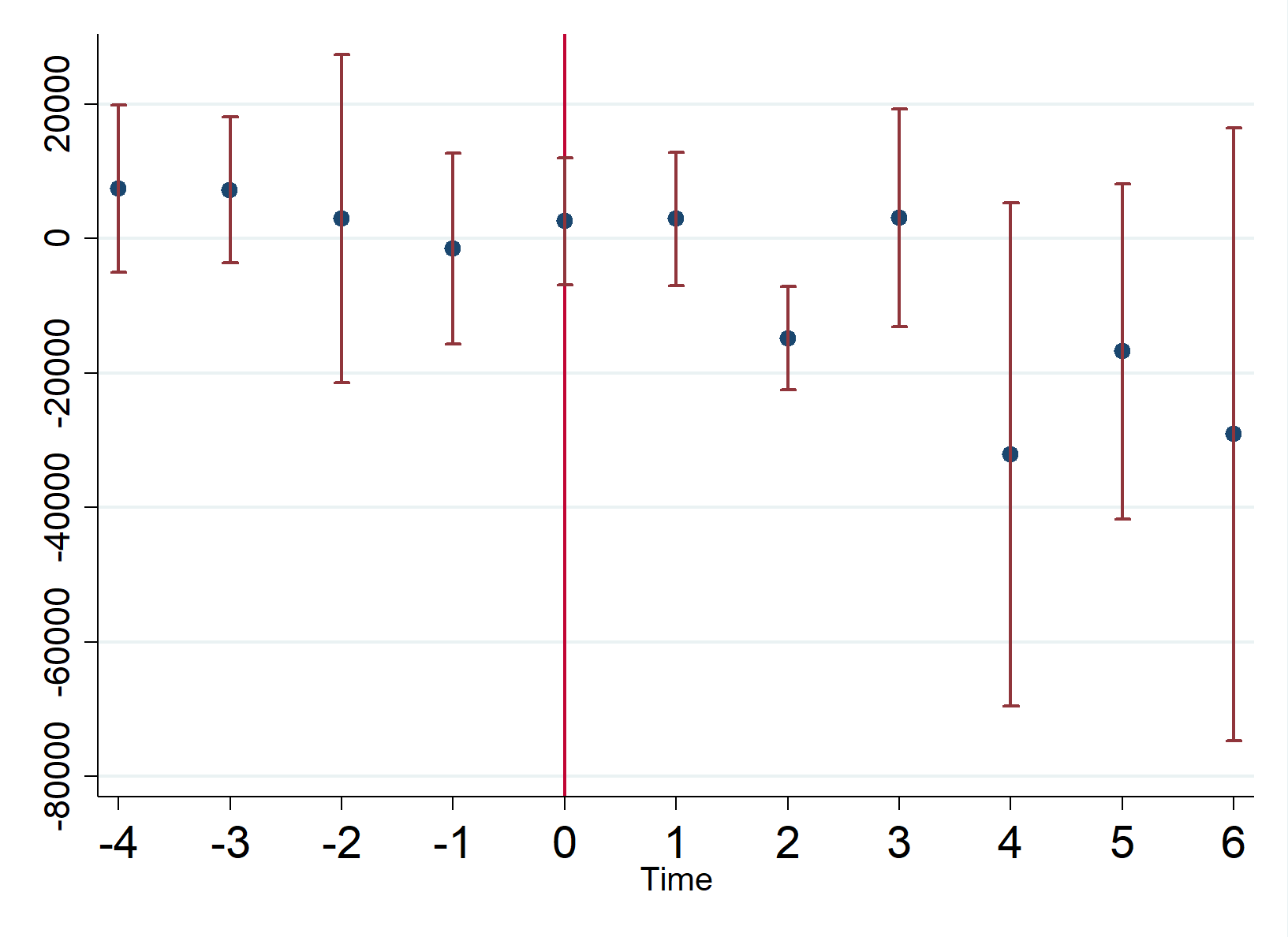}
             \includegraphics[height=4.5cm, width=7cm]{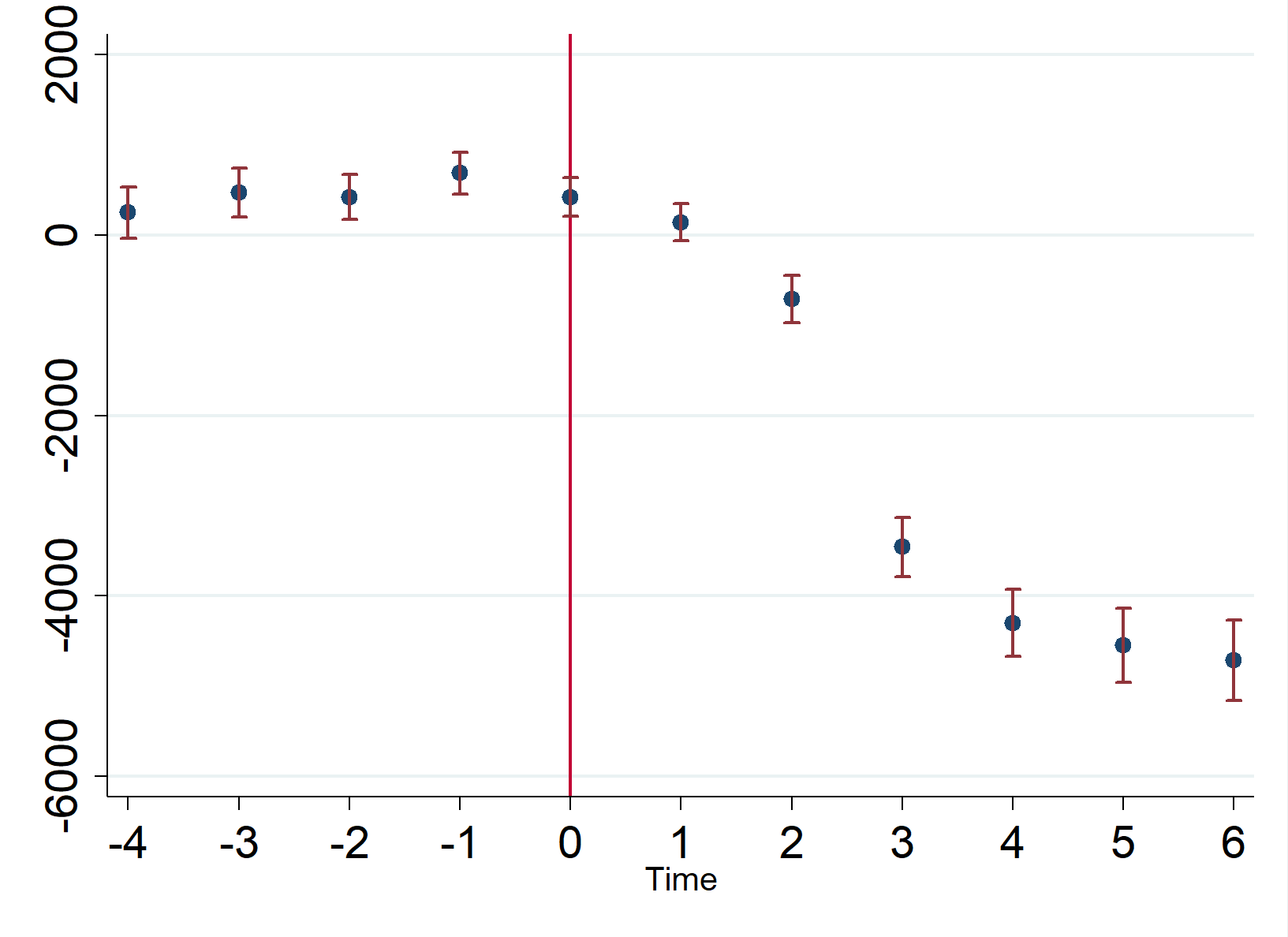}

        \caption{\small Event study: dependent variable is: (i) Probability of moving zip code. This variable takes value 1 if the individual is a different zip code in year $t$ than in year $t-1$, and zero otherwise, (ii) Probability of moving outside the commuting zone. This variable takes value 1 if the individual is in a different commuting zone in year $t$ than in year $t-1$, and zero otherwise, (iii) the Median House Value in the zip code of residence at year $t$ (iv) income imputed by Experian. The event considered is a soft default, i.e. a 90-day delinquency, but no Chapter 7, Chapter 13 or foreclosure taking place in the same year, neither before in the sample period. Other controls are age and age squared, credit score in 2004 and in 2005. 95\% confidence intervals around the point estimates. Estimation has been performed on a 5\% random sample of the full dataset}   \label{fig:event2b}   
     \end{figure}

  \section{Breakdown of the different types of harsh default}
  \label{sec:def}

  \setcounter{table}{0}
\renewcommand{\thetable}{I\arabic{table}}

\setcounter{figure}{0}
\renewcommand{\thefigure}{I\arabic{figure}}

Here we distinguish the impact of a soft default on each of the types of harsh default, i.e. Chapter 7, Chapter 13 and foreclosure\footnote{Chapter 7 bankruptcy implies the liquidation of assets: ...the bankruptcy trustee gathers and sells the debtor's non-exempt assets and uses the proceeds of such assets to pay holders of claims (creditors) in accordance with the provisions of the Bankruptcy Code... (see https://www.uscourts.gov/services-forms/bankruptcy/bankruptcy-basics/chapter-7-bankruptcy-basics). A chapter 13 bankruptcy is also called a wage earner's plan. It enables individuals with regular income stream to develop a plan to repay all or part of their debts. Under this chapter, debtors propose a repayment plan to make instalments to creditors over three to five years. If the debtor's current monthly income is less than the applicable state median, the plan will be for three years unless the court approves a longer period "for cause." (1) If the debtor's current monthly income is greater than the applicable state median, the plan generally must be for five years. In no case may a plan provide for payments over a period longer than five years. 11 U.S.C. 1322(d). During this time the law forbids creditors from starting or continuing collection efforts. This chapter discusses six aspects of a chapter 13 proceeding: the advantages of choosing chapter 13, the chapter 13 eligibility requirements, how a chapter 13 proceeding works, making the plan work, and the special chapter 13 discharge. https://www.uscourts.gov/services-forms/bankruptcy/bankruptcy-basics/chapter-13-bankruptcy-basics}. In Table \ref{tab:def} we report descriptive statistics of the variable that we use for the definition of a harsh default in year 2010. Note that the sum of their means is not exactly equal to the mean of our harsh default variable in year 2010 reported in Table 1, both Chapter 7 and Chapter 13 declarations may take place at the same time of a foreclosure.

 \begin{table}[H]\centering \caption{Summary statistics of our harsh default variables, 2010, balanced panel. Individuals who experienced a harsh default before or in the same year as a soft default in the sample period (i.e. from 2004 onwards) have been dropped.}
\normalsize
\begin{tabular}{l c c c c c}\hline\hline
\multicolumn{1}{c}{\textbf{Variable}} & \textbf{Mean}
 & \textbf{Std. Dev.}& \textbf{Min.} &  \textbf{Max.} & \textbf{N}\\ \hline
New Forecl. & 0.0073 & 0.0853 & 0 & 1 & 2162467\\
New Ch. 7 &0.0077 & 0.0874 & 0 & 1 & 2162467\\
New Ch. 13 & 0.0026 & 0.0513 & 0 & 1 & 2162467\\
\hline\end{tabular} \label{tab:def}
\end{table}

\subsection{Year-by-year DML results}

\begin{figure}[H]
      	Panel (i) - Chapter 7 \hspace{2cm}Panel (ii) - Chapter 13\\
           
                \includegraphics[height=4cm, width=7cm]{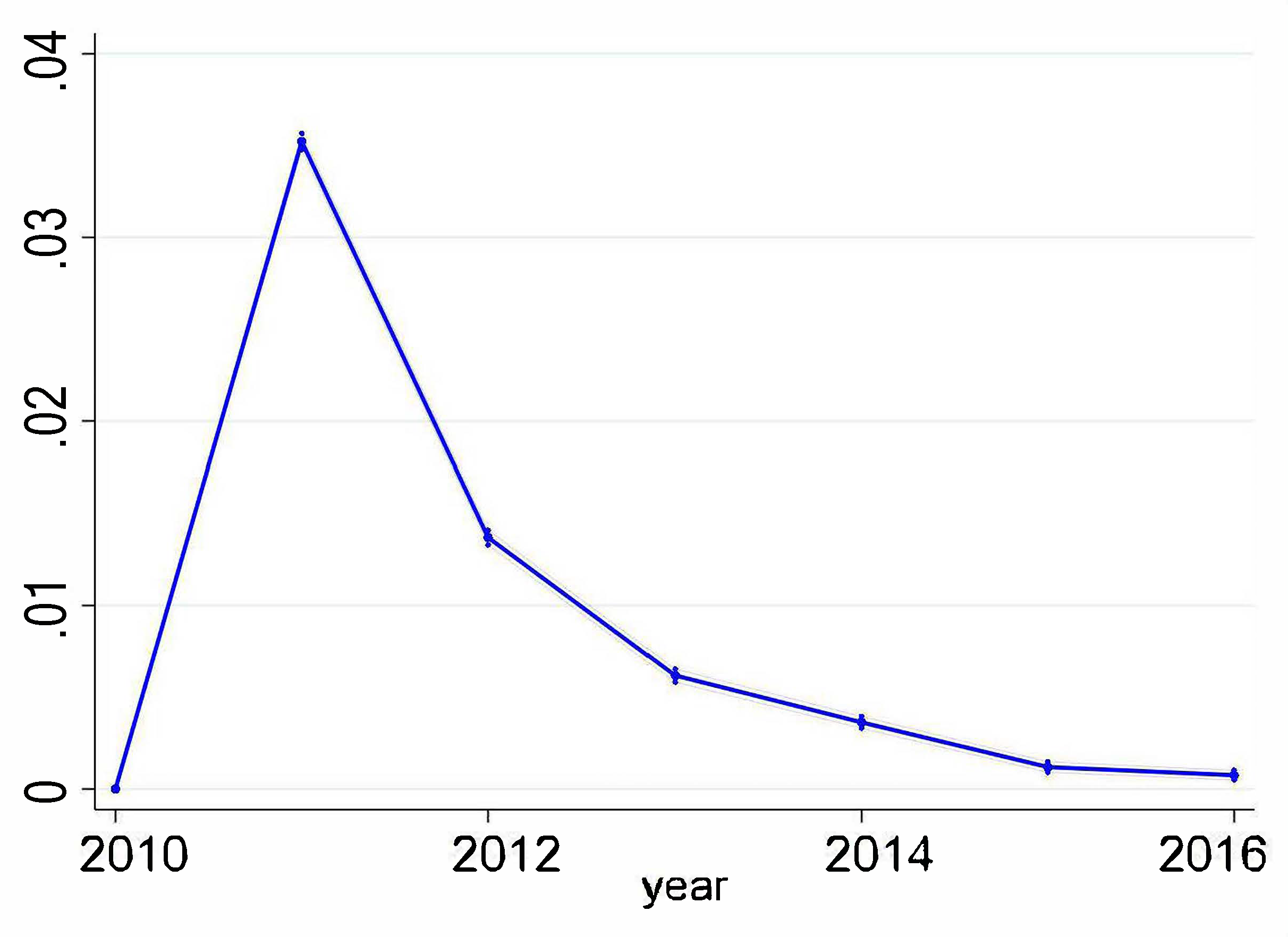} 
             \includegraphics[height=4cm, width=7cm]{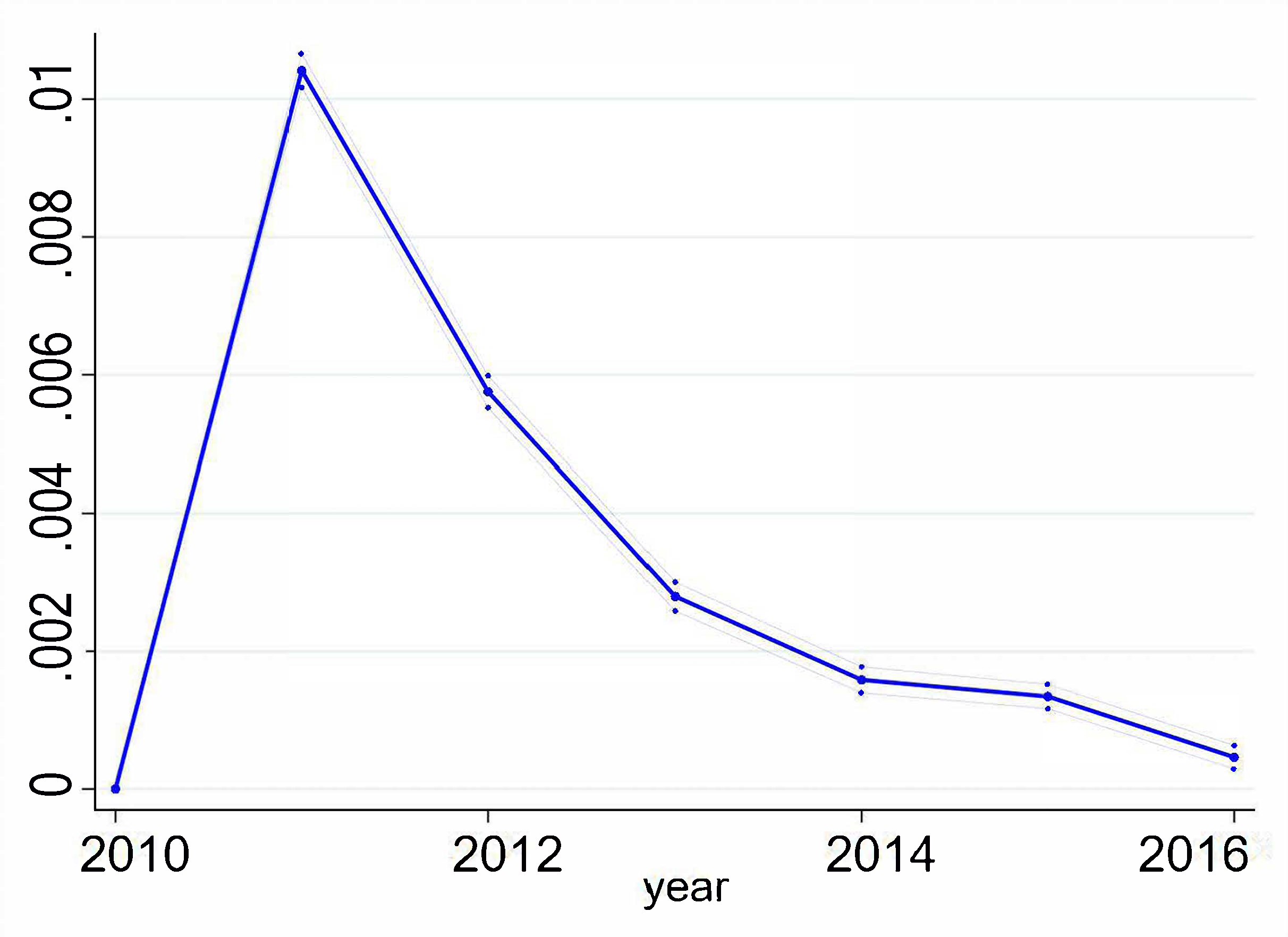} \\
          
          	Panel (iii) - Foreclosure \\
          	\centering
             \includegraphics[height=4cm, width=7cm]{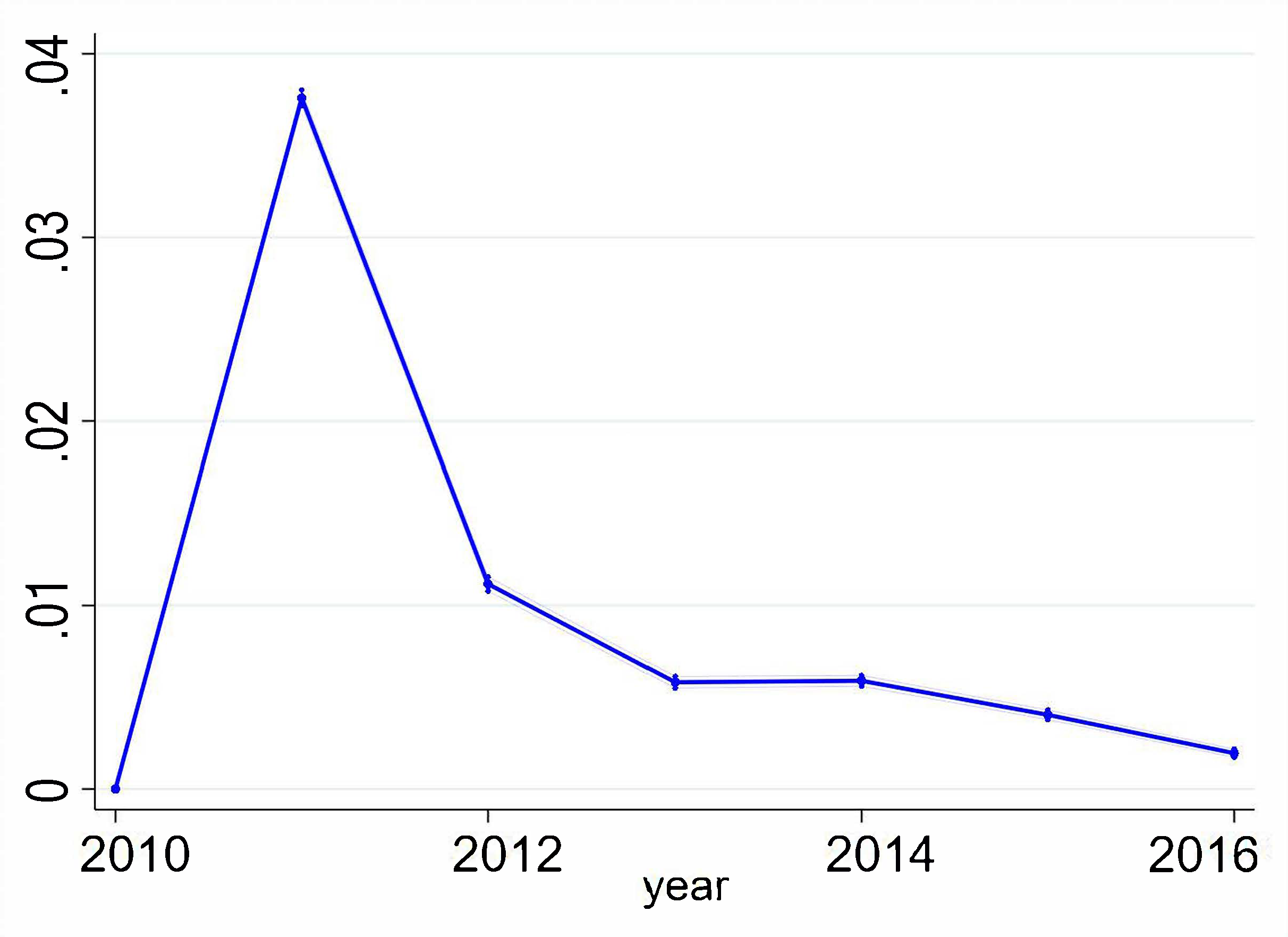}

        \caption{ DML year-by-year impacts of soft default on: (i) Chapter 7 declaration (new) (ii) Chapter 13 declaration (new) (iii) New foreclosure. Controls: age, age squared, commuting zones fixed effects, credit score, the amount of open mortgages and car loan, county unemployment rate, number of bank closings in the county pre-event (2004-2009), maximum interest rate allowed by law. 95\% confidence intervals around the point estimates. Individuals recording a harsh default between 2004 and 2010 (extremes included) and those recording a soft default between 2004 and 2009 (extremes included) have been dropped from the estimation sample.}
       \end{figure}

\subsection{Long term DML results}

   \begin{table}[H]
  \caption{DML long-run impact of a soft default  2020. Controls: age, age squared,  commuting zones fixed effects, credit score, the amount of open mortgages and car loan, as well as county unemployment rate and number of bank closings in the county pre-event (years 2004-2009), plus max interest rate. Individuals recording a harsh default between 2004 and 2010 (extremes included) and those recording a soft default between 2004 and 2009 (extremes included) have been dropped from the estimation sample.}
  \footnotesize
  {
\def\sym#1{\ifmmode^{#1}\else\(^{#1}\)\fi}
\begin{tabular}{c c c c }
\hline\hline
            &\multicolumn{1}{c}{(1)}&\multicolumn{1}{c}{(2)}&\multicolumn{1}{c}{(3)}\\
            &\multicolumn{1}{c}{Chapter 7}&\multicolumn{1}{c}{Chapter 13}&\multicolumn{1}{c}{Foreclosure}\\
\hline
Soft def &        0.0714\sym{***}&      0.0231\sym{***}&      0.0642\sym{***}\\
            &   (0.00142)         &  (0.000735)         &   (0.00114)         \\
\hline
\(N\)       &     1044486         &     1044486         &      1044486           \\
\hline\hline
\multicolumn{4}{l}{\footnotesize Standard errors in parentheses}\\
\multicolumn{4}{l}{\footnotesize \sym{*} \(p<0.05\), \sym{**} \(p<0.01\), \sym{***} \(p<0.001\)}\\
\end{tabular}
}
  \end{table}
  
  We know from previous results that a soft default increases by about 10pp the probability of experiencing a harsh default in the following years. From this Table we deduce that, more specifically, a soft default is associated with an increase by about 7pp in the probability of experiencing at least one new Chapter 7 declaration in the 10 years following the soft default. The increase is only equal to about +2pp for the probability of declaring Chapter 13 bankruptcy. Finally, a soft default is associated with a rise by about 6pp in the probability of experiencing at least one foreclosure in the next 10 years following a soft default.

  \subsection{Event study results results}

\begin{figure}[H]
      	Panel (i) - Chapter 7 \hspace{2cm}Panel (ii) - Chapter 13\\
           
                \includegraphics[height=4cm, width=7cm]{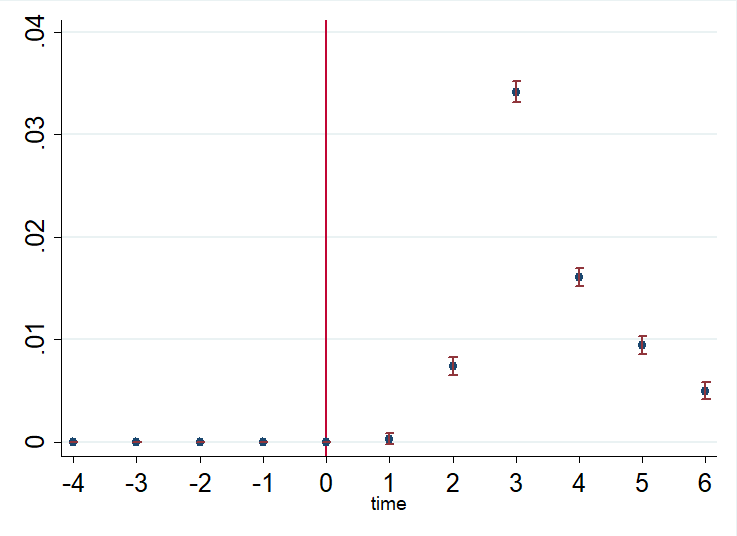} 
             \includegraphics[height=4cm, width=7cm]{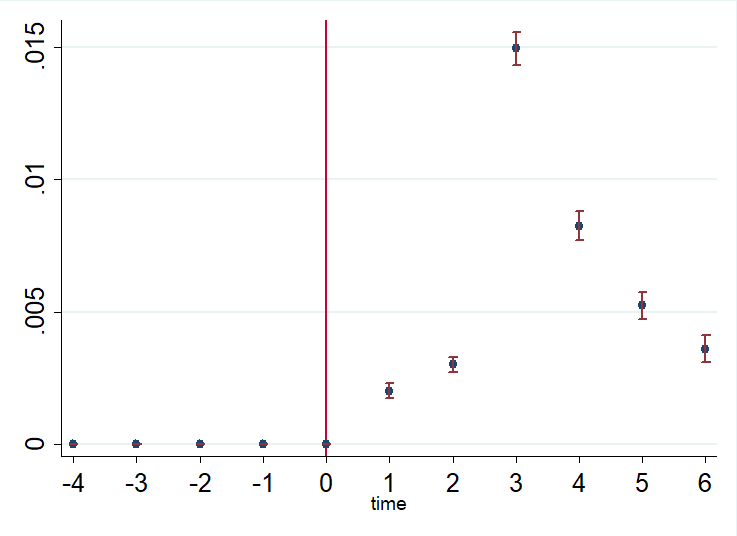} \\
          
          	Panel (iii) - Foreclosure \\
          	\centering
             \includegraphics[height=4cm, width=7cm]{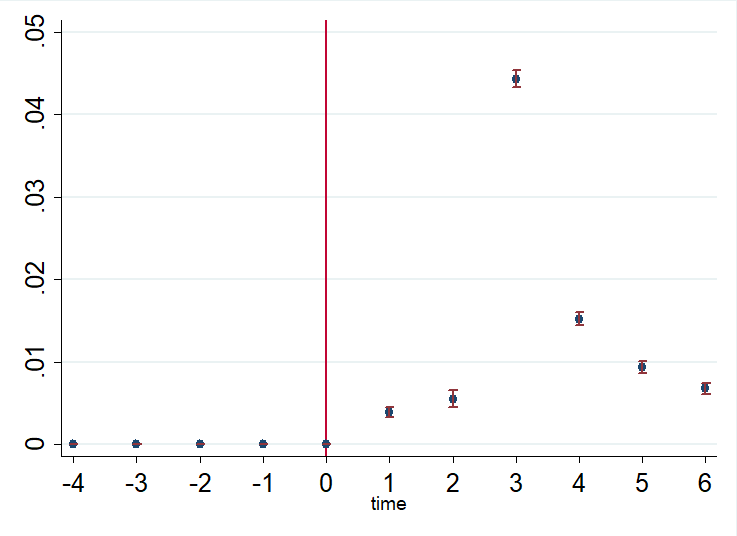}

        \caption{ Event study: dependent variable is: (i) Chapter 7 declaration (new) (ii) Chapter 13 declaration (new) (iii) New foreclosure. The event considered is a soft default, i.e. a 90-day delinquency, but no
Chapter 7, Chapter 13 or foreclosure taking place in the same year, neither before in the sample period. Other controls are
age and age squared, credit score in 2004 and in 2005. 95\% confidence intervals around
the point estimates.} \label{breakdown}
       \end{figure}

 The Callaway Sant'Anna event studies reported in Figure \ref{breakdown} confirm that a soft default is associated to a rise by about 4pp in both Chapter 7 declarations and foreclosures (peaking 3 years after the event, as expected from the baseline results), but to an increase by about 1.5pp only in Chapter 13 declarations.

  \section{Additional evidence on the impact of a soft default on zip code quality}
\label{sec:quality} 

\setcounter{table}{0}
\renewcommand{\thetable}{J\arabic{table}}

\setcounter{figure}{0}
\renewcommand{\thefigure}{J\arabic{figure}}

  In this Section we go deeper into the question of which is the effect of a soft default on the zip code of residence of the individual. In order to answer this question we present in the following event studies that show the impact of a soft default on: (i) the log annual payoll paid by firms in the zip code, (ii) the average zip code wage, (iii) the number of employees per zip code and (iv) the number of firms per zip code.
  
 All these variables are recorded at the level of 5-digit zip codes. Data comes from the ZIP Codes Business Patterns dataset\footnote{https://www.census.gov/data/developers/data-sets/cbp-nonemp-zbp/zbp-api.html} from US Census Bureau. Wages is a constructed variable defined as the total annual payroll per zip code divided by the number of employees per zip code. Descriptive statistics of these variables are reported in Table \ref{tab:zip}.
 
 \begin{table}[H]\centering \caption{Summary statistics on zip code characteristics in 2010, balanced panel. Individuals who experienced a harsh default before or in the same year as a soft default in the sample period (i.e. from 2004 onwards) have been dropped. }
\normalsize
\begin{tabular}{l c c c c }\hline\hline
\multicolumn{1}{c}{\textbf{Variable}} & \textbf{Mean}
 & \textbf{Std. Dev.}& \textbf{Min.} &  \textbf{Max.} \\ \hline
Number of employees & 9315 & 9854 & 0 & 142950   \\
Annual payroll (in 1,000 USD)& 399058 & 669285 & 0 & 18817294\\
Number of establishments &  640 & 535 & 1 & 7241\\
Wage (in 1,000 USD) & 36.4848 & 12.5249 & 2.3492 & 356.7019 \\
\hline \hline
\multicolumn{1}{c}{N} & \multicolumn{4}{c}{2161299}\\ \hline
\end{tabular} \label{tab:zip}
\end{table}

 \begin{figure}[H]
      	Panel (i) - Annual payroll (in mio USD) \hspace{1cm}Panel (ii) - Number of employees\\
           
                \includegraphics[height=4.5cm, width=7cm]{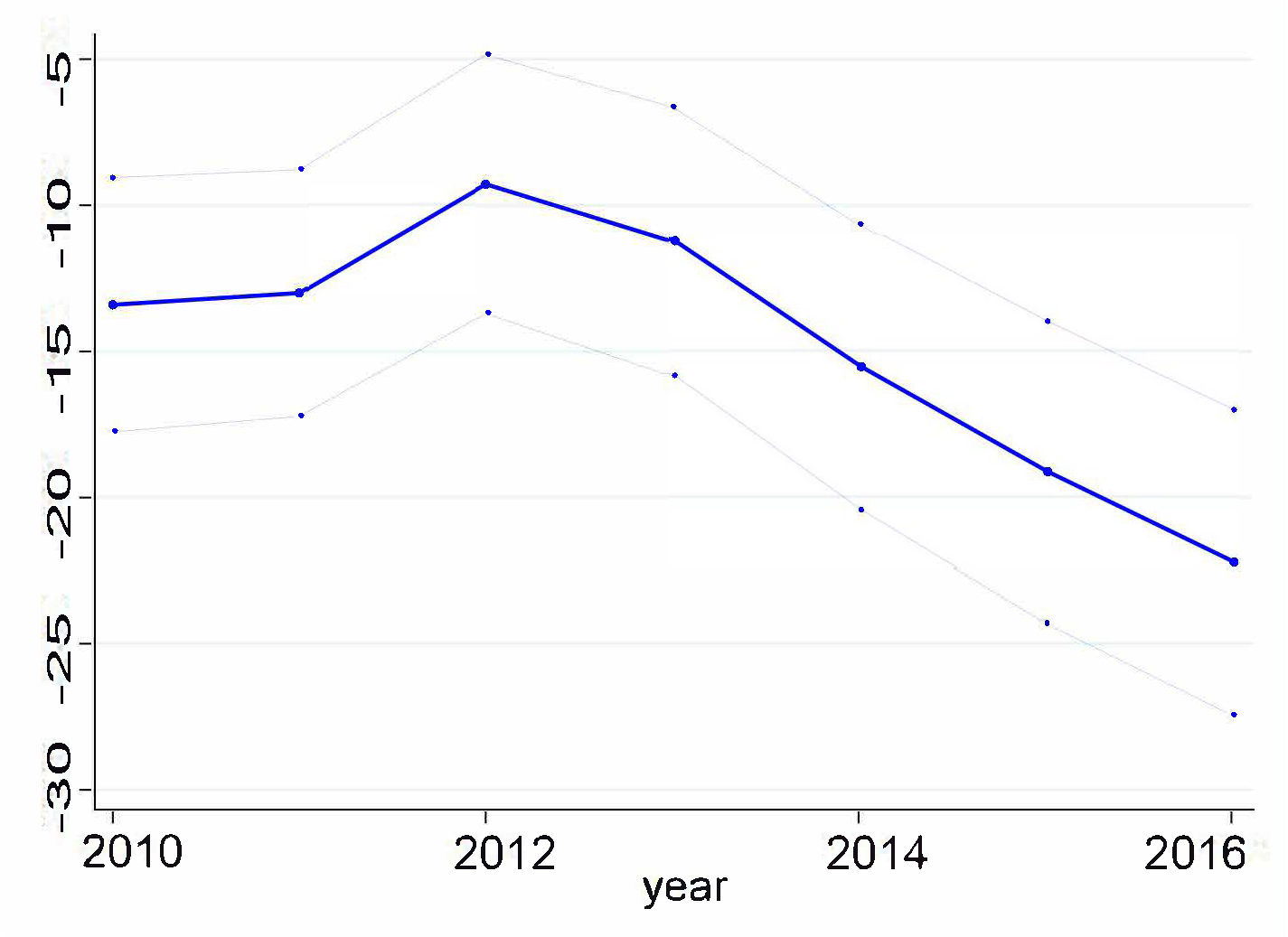} 
             \includegraphics[height=4.5cm, width=7cm]{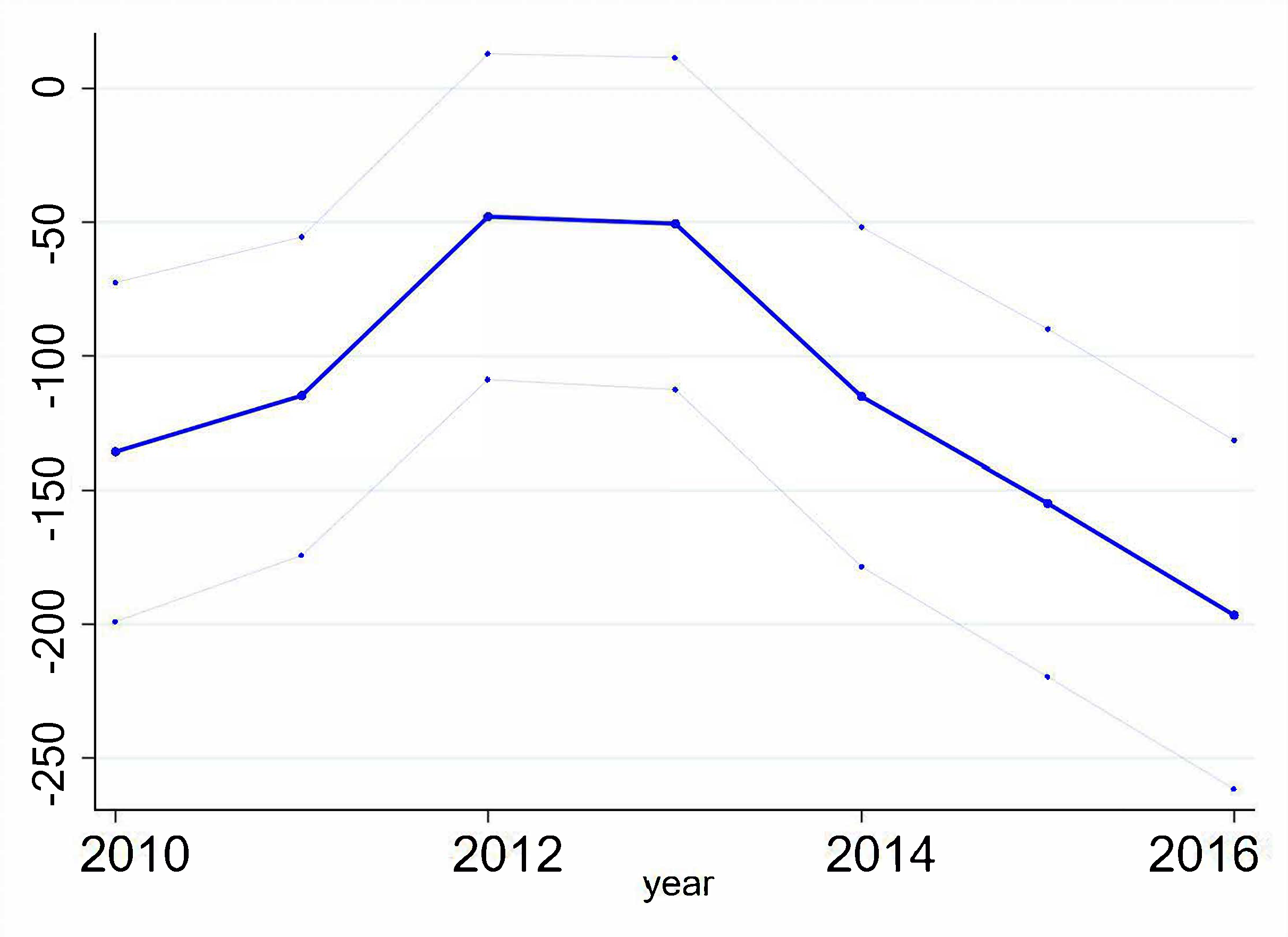} \\
          
          	Panel (iii) - Number of establishments \hspace{2cm}Panel (iv) Average wage (in 1'000)  \\
          	\centering
             \includegraphics[height=4.5cm, width=7cm]{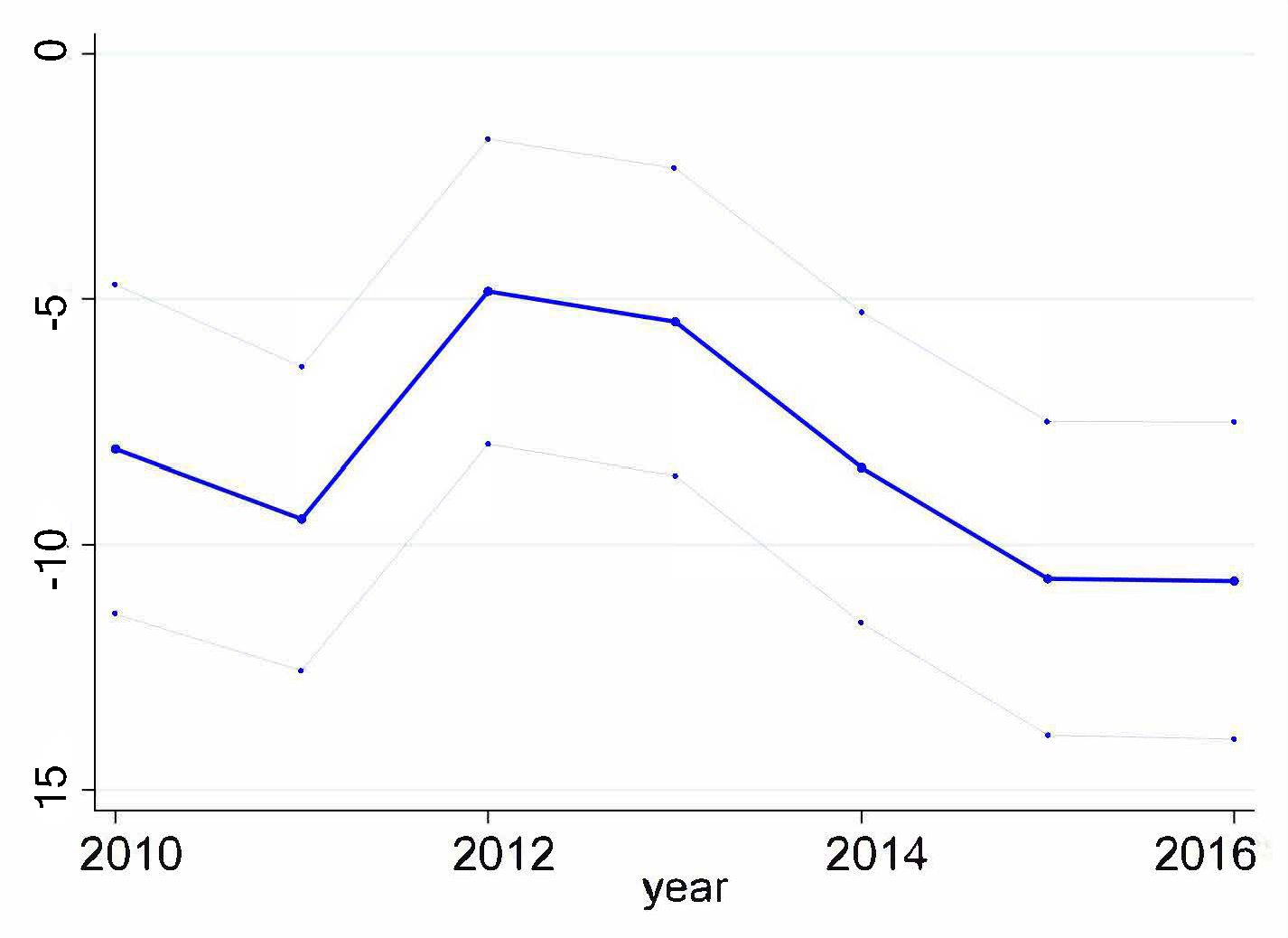} 
                 \includegraphics[height=4.5cm, width=7cm]{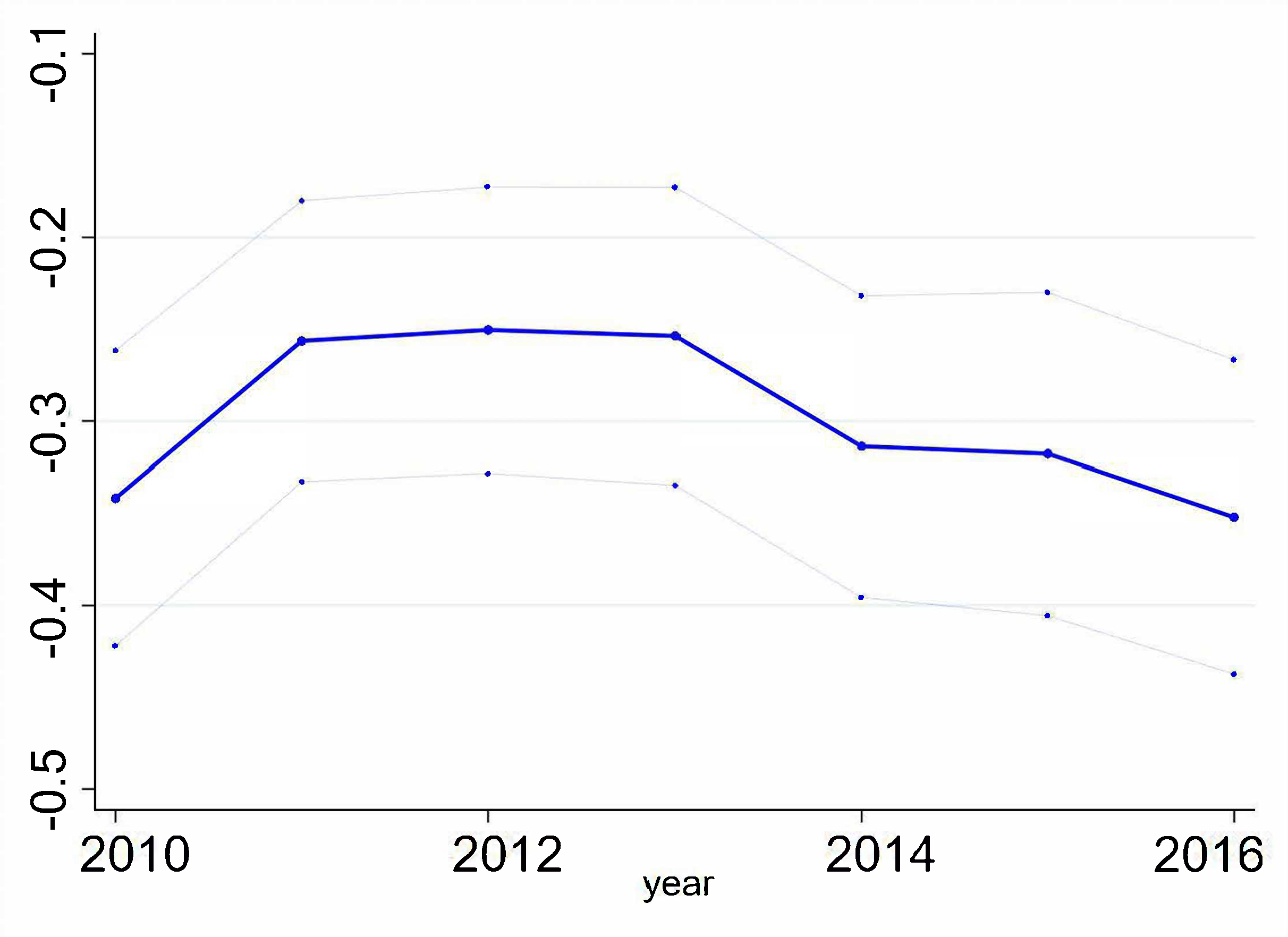}

        \caption{  DML year-by-year impacts of soft default on: (i) zip code annual payroll (ii) number of employees in the zip code (iii) number of establishments in the zip code (iv) zip code average wage (i.e. annual payroll divided by the number of employees).
   Controls: age, age squared, commuting zone's fixed effects, credit score, the amount of open mortgages and car loan, county unemployment rate, number of bank closings in the county pre-event (2004-2009), maximum interest rate allowed by law. 95\% confidence intervals around the point estimates. Individuals recording a harsh default between 2004 and 2010 (extremes included) and those recording a soft default between 2004 and 2009 (extremes included) have been dropped from the estimation sample.}\label{fig:zipcode_quality}
       \end{figure}
 
 Those who experience a soft default end up in zip codes where the annual payroll is about 15-20 mio USD lower than what they used to be in their origin zip code. Also, on average, the number of employees in the zip code declines by 150 and the number of establishments by about 10. Average zip code annual wage is lower by about  300-400 USD.
 
 \begin{figure}[H]
      	Panel (i) - Share of non-White residents\\
           
                \includegraphics[height=5.4cm]{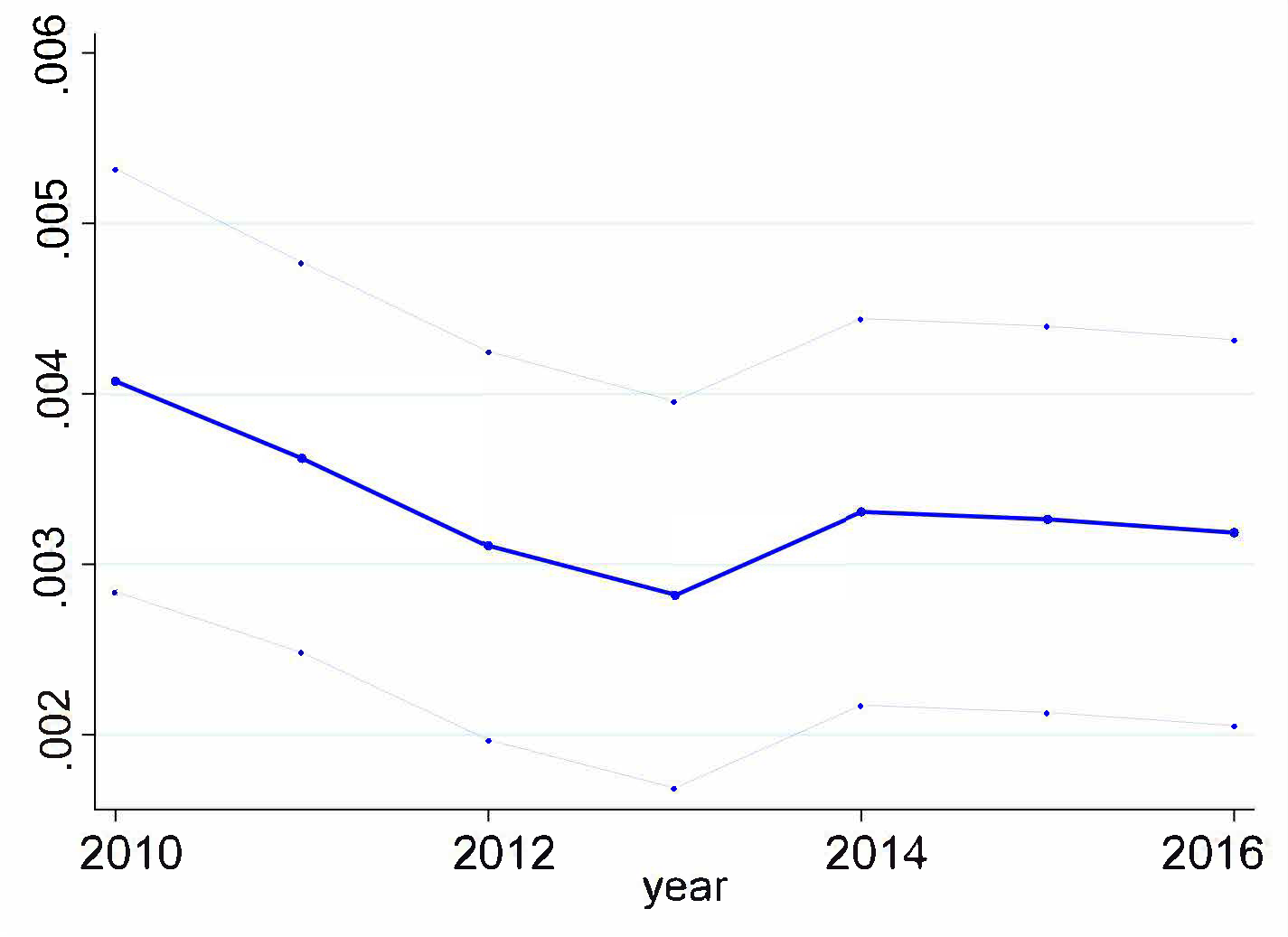}

        \caption{  DML year-by-year impacts of soft default on the share of non-White residents.
   Controls: age, age squared, commuting zone's fixed effects, credit score, the amount of open mortgages and car loan, county unemployment rate, number of bank closings in the county pre-event (2004-2009), maximum interest rate allowed by law. 95\% confidence intervals around the point estimates. Individuals recording a harsh default between 2004 and 2010 (extremes included) and those recording a soft default between 2004 and 2009 (extremes included) have been dropped from the estimation sample.} \label{fig:zip code_minority}
       \end{figure}
       
Finally, from Figure \ref{fig:zip code_minority}, we notice that a soft default is associated with an increase by 0.3-0.4pp in the share of non-White zip code residents (as measured by the 2010 Census), we take that as a proxy for more disadvantaged zip codes.

  \begin{table}[H]
  \caption{DML long-run impact of a soft default  2020. Controls: age, age squared,  commuting zones fixed effects, credit score, the amount of open mortgages and car loan, as well as county unemployment rate and number of bank closings in the county pre-event (years 2004-2009), plus max interest rate. Individuals recording a harsh default between 2004 and 2010 (extremes included) and those recording a soft default between 2004 and 2009 (extremes included) have been dropped from the estimation sample.}
  \footnotesize
  {
\def\sym#1{\ifmmode^{#1}\else\(^{#1}\)\fi}
\begin{tabular}{c c c c c c }
\hline\hline
          &\multicolumn{1}{c}{Annual payroll (1,000)}&\multicolumn{1}{c}{N. of establishments}&\multicolumn{1}{c}{N. of employees}&\multicolumn{1}{c}{Wage (1,000)}&\multicolumn{1}{c}{Share of non-Whites} \\
\hline
Soft def &   -32336.8\sym{**} &      -10.55\sym{*}  &      -194.2         &      -0.519\sym{**} &     0.00424\sym{*}  \\
            &   (10993.1)         &     (5.153)         &     (108.8)         &     (0.166)         &   (0.00165)         \\    
\hline
\(N\)       &     1040916         &     1040916         &     1040916         &     1040910         &     1043406         \\
\hline\hline
\multicolumn{6}{l}{\footnotesize Standard errors in parentheses}\\
\multicolumn{6}{l}{\footnotesize \sym{*} \(p<0.05\), \sym{**} \(p<0.01\), \sym{***} \(p<0.001\)}\\
\end{tabular}
} \label{lassoquality}
  \end{table}

The long term lasso results (10 years after the event) reported in Table \ref{lassoquality} confirm our previous findings. Ten years after a soft default individuals on average live in zip codes were the total annual payroll is about 32 million smaller, there are about 11 establishments less and almost 200 employees less. Further, annual average wage is about 500USD lower and the share of non-White residents is higher by about 0.4pp. Most of these results keep their statistical significance in the long run.

 \section{Robustness check: Matching on more observables in the event studies}
 \label{sec:robustness}

\setcounter{table}{0}
\renewcommand{\thetable}{K\arabic{table}}

\setcounter{figure}{0}
\renewcommand{\thefigure}{K\arabic{figure}}

 In this Section we repeat the exercise of Section \ref{sec:exploratory}, i.e. we perform event studies à la Callaway and Sant'Anna, by using more matching variables. The estimation results presented in Section \ref{sec:exploratory} are based on matching only done on age, age squared, credit score in 2004 and in 2005 (i.e. the first two years available in our sample). In contrast, in this Section we perform the matching on the basis of age, age squared, credit score in the year before the soft default and credit score two years before the soft default, as well as state dummies. Further, we consider here the group of the last treated in our dataset (i.e. those who recorded a soft default in 2016) as the control group, whereas we drop the never treated\footnote{This is done in order to be able to consistently define for all individuals the variables "credit score in the year before the event" and "credit score two years before the event".}. 
 The results we obtain are consistent with those presented in Section \ref{sec:exploratory}. Hence, since this second type of matching on more variables does not allow us to identify the effects for all of our outcome variables of interest\footnote{The model did not converge for the variables Median house value of the zip code of residence and probability of moving out of the commuting zone.}, we keep results based on matching on credit score in 2004 and 2005 and age only as our baseline results.
 
 
   \begin{figure}[H] 
         
             	Panel (i) - Move Zip \hspace{4cm}Panel (ii) - Income\\
                \includegraphics[height=4.5cm, width=7cm]{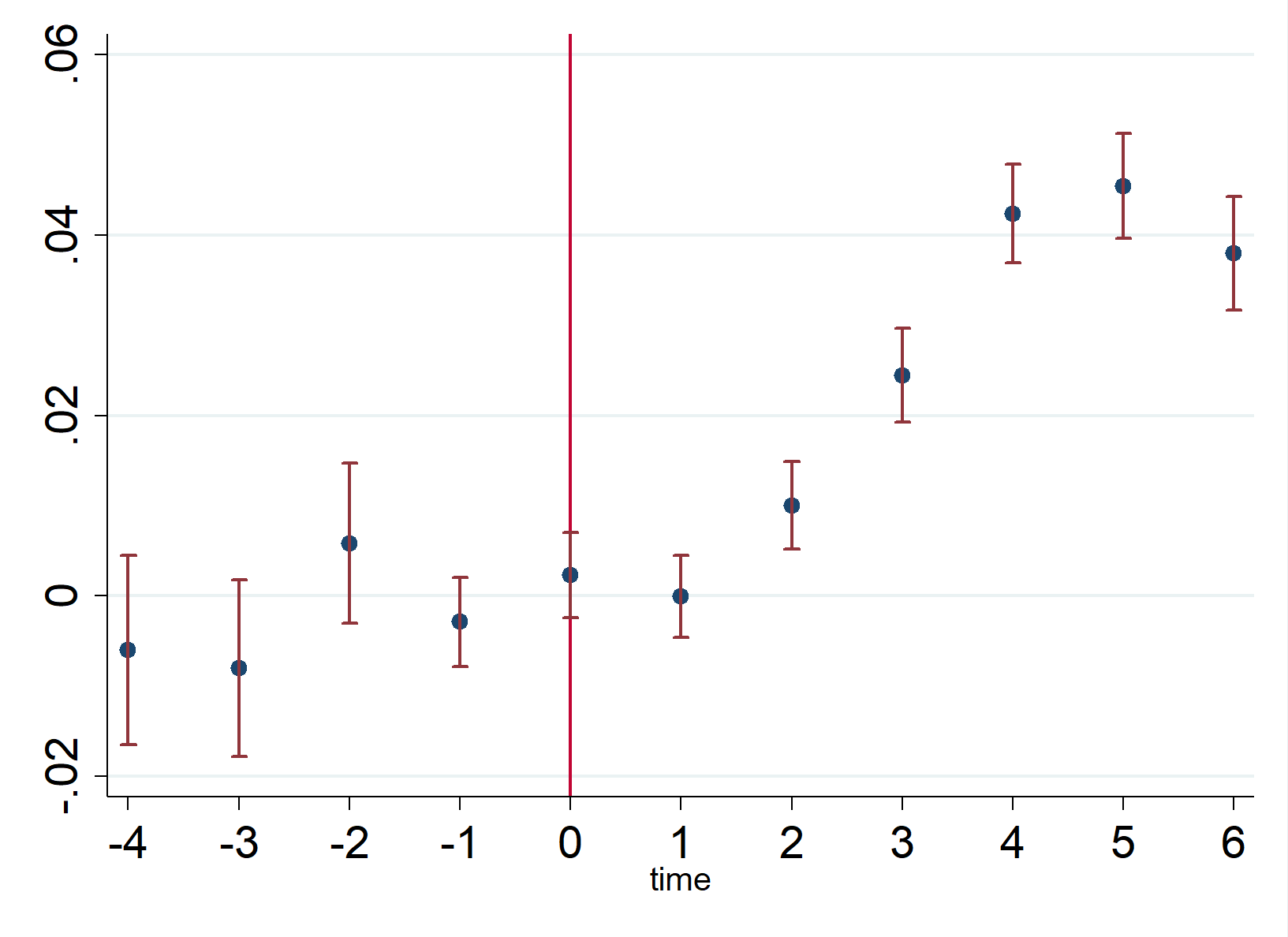}
             \includegraphics[height=4.5cm, width=7cm]{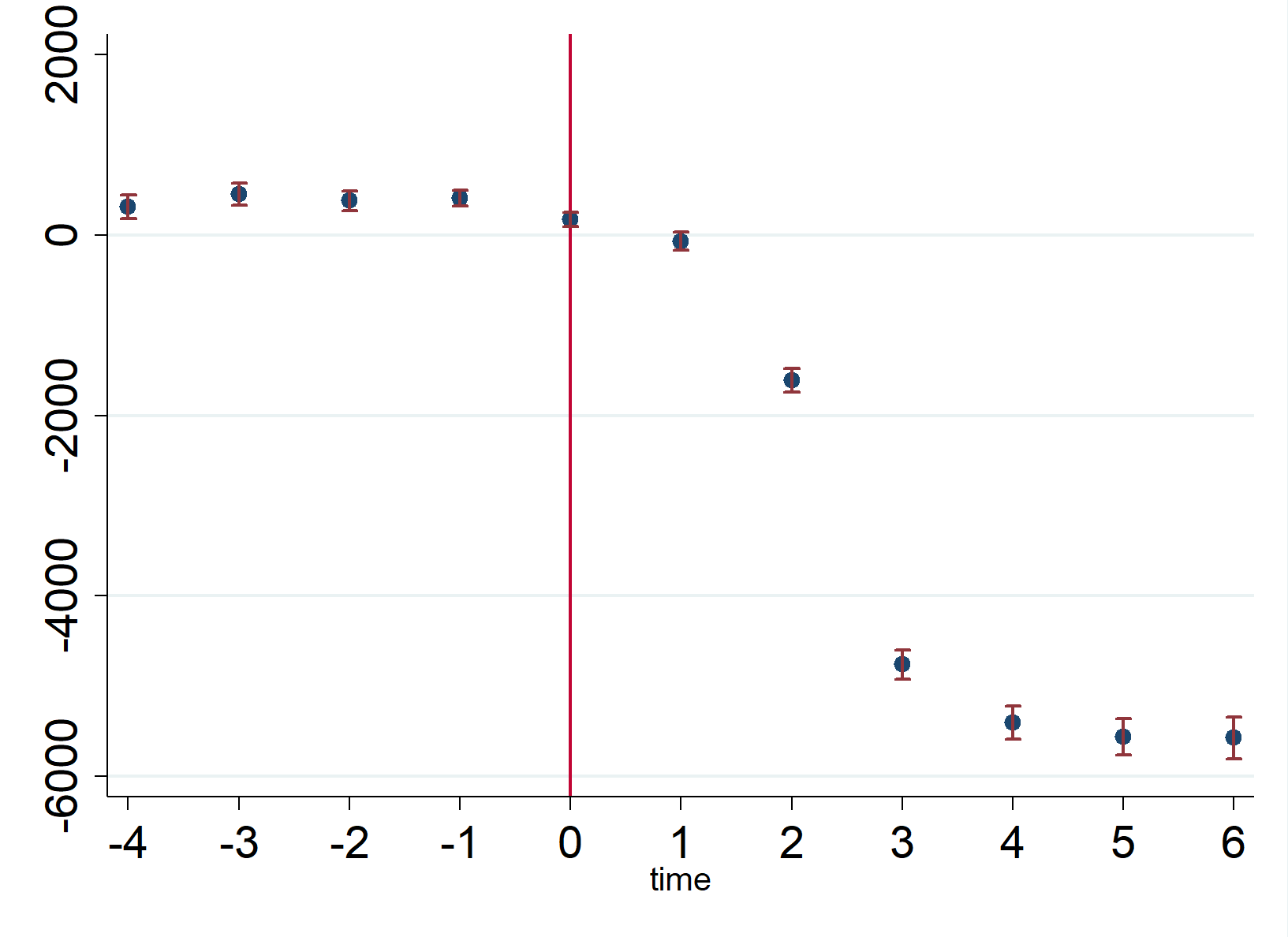}

        \caption{Event study: dependent variable is: 
        (i) Probability of moving zip code. This variable takes value 1 if the individual is a different zip code in year $t$ than in year $t-1$, and zero otherwise, (ii) income imputed by Experian. The event considered is a soft default, i.e. a 90-day delinquency, but no Chapter 7, Chapter 13 or foreclosure taking place in the same year, neither before in the sample period. Other controls are age and age squared, value of the credit score in the two years before the soft default, state dummies. Treated in 2016 (last year in the sample) used as control group. 95\% confidence intervals around the point estimates.}    \label{fig:fig5} 
     \end{figure}

From Figure \ref{fig:fig5}  we deduce that, similarly to the results presented in Figure \ref{fig:event2}, a soft default is associated with an increase, by about 4pp, in the probability of moving zip code\footnote{As mentioned above, the event studies for the outcome variables representing the probability of moving commuting zone and the Median House Value of the zip code of residence did not converge.}, and most importantly to a 5,000 USD drop in income. 

\begin{figure}[H] 
 
             	 Credit Score\\
                \includegraphics[ scale=.175]{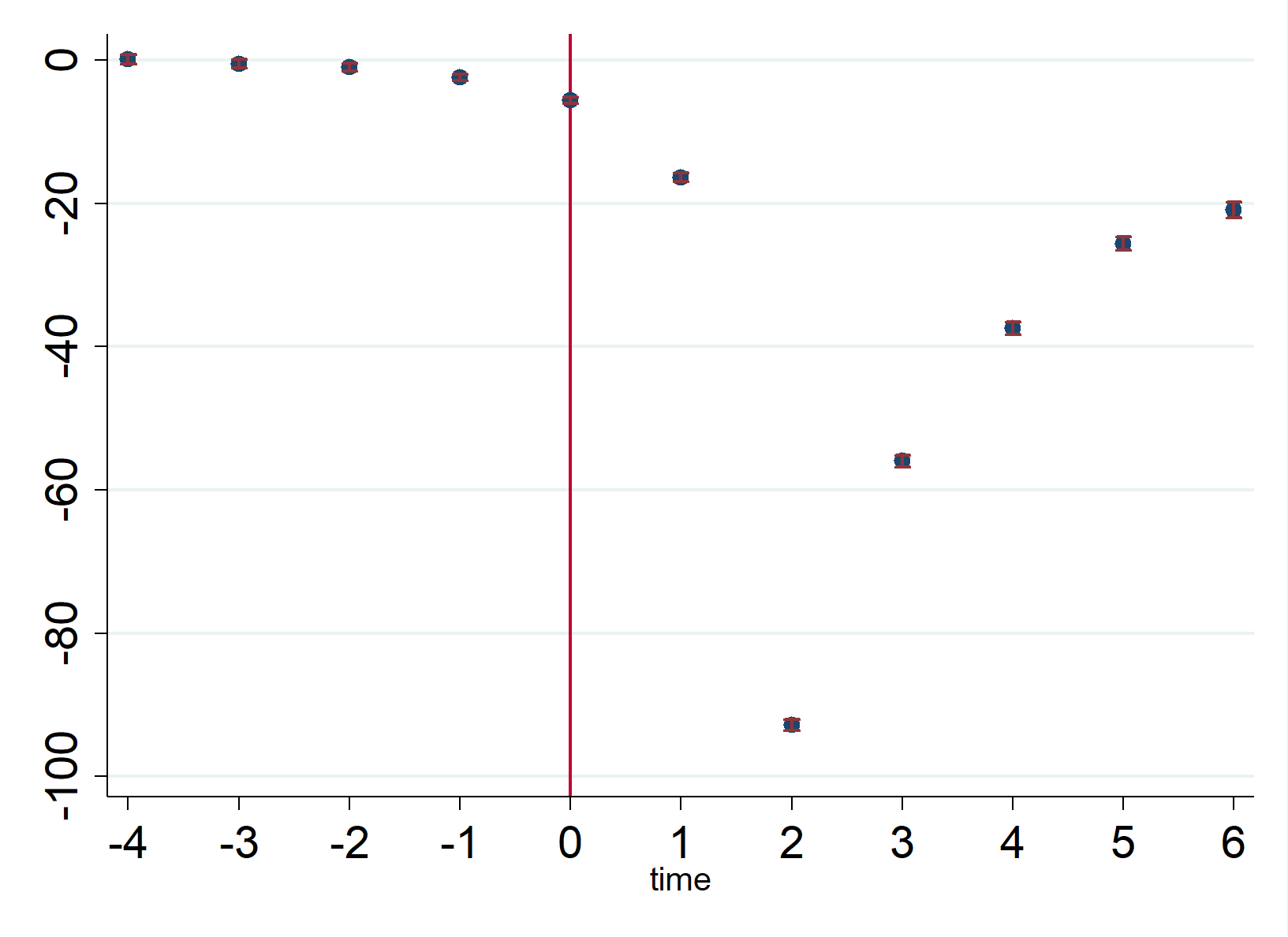}
    \caption{Event study: dependent variable is Credit Score. The event considered is a soft default, i.e. a 90-day delinquency, but no Chapter 7, Chapter 13 or foreclosure taking place in the same year, neither before in the sample period. Other controls are age and age squared, value of the credit score in the two years before the soft default, state dummies. Treated in 2016 (last year in the sample) used as control group. 95\% confidence intervals around the point estimates.}   \label{fig:fig6}
     \end{figure}

  As far as credit score is concerned, from Figure \ref{fig:fig6} we notice a negative impact of the soft default equal to about minus 100 points, bouncing back to about -20 over time. This is essentially the same results, in sign and approximate size, that we found when performing the matching on the smaller set of variables.

\begin{figure}[H] 
        
             	Panel (i) - Prob. new mortgage \hspace{1cm}Panel (ii) - Prob. Credit Limit $<$10k\\
                          
                            \includegraphics[height=4cm, width=7cm]{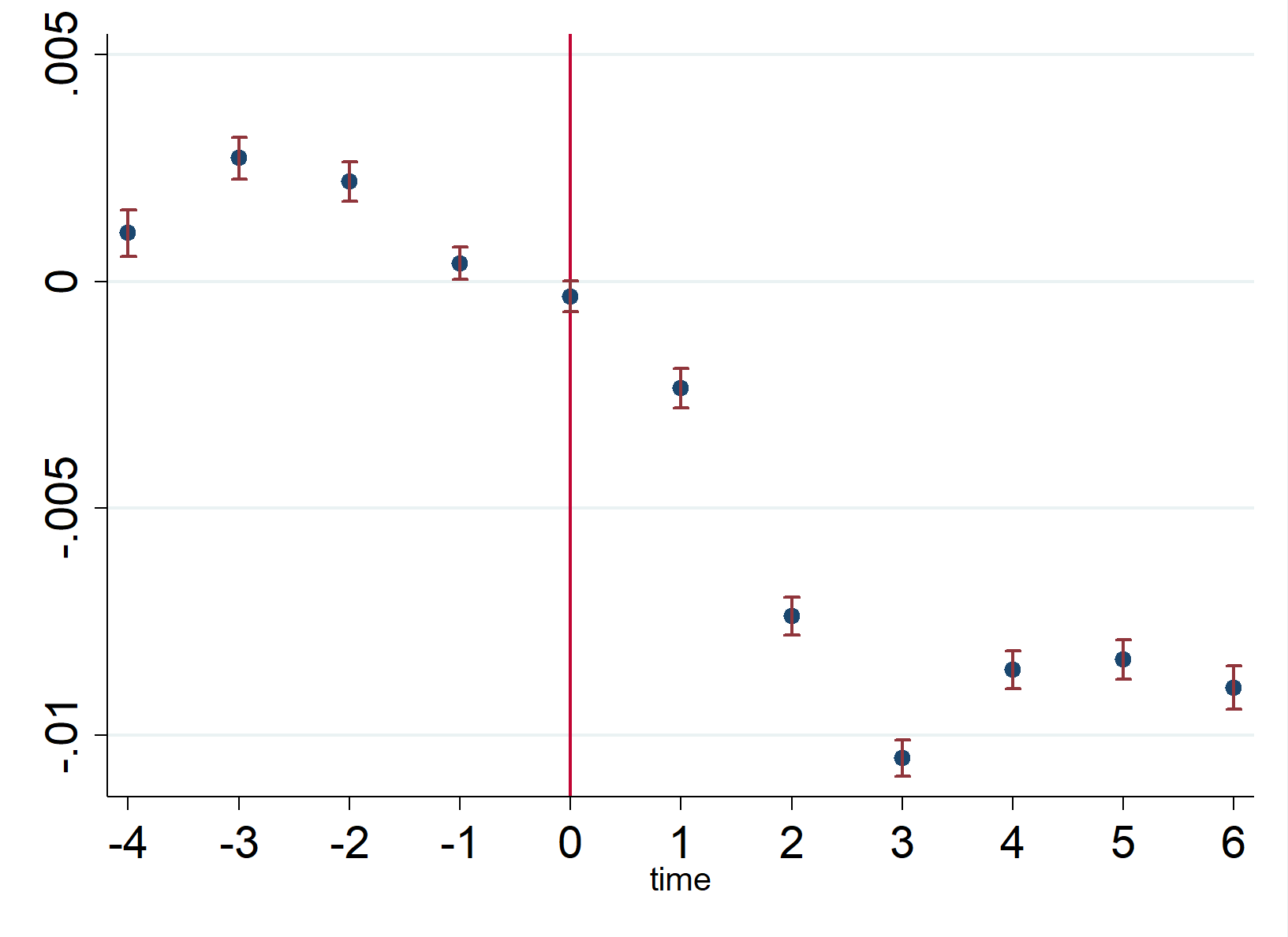}   
              \includegraphics[height=4cm, width=7cm]{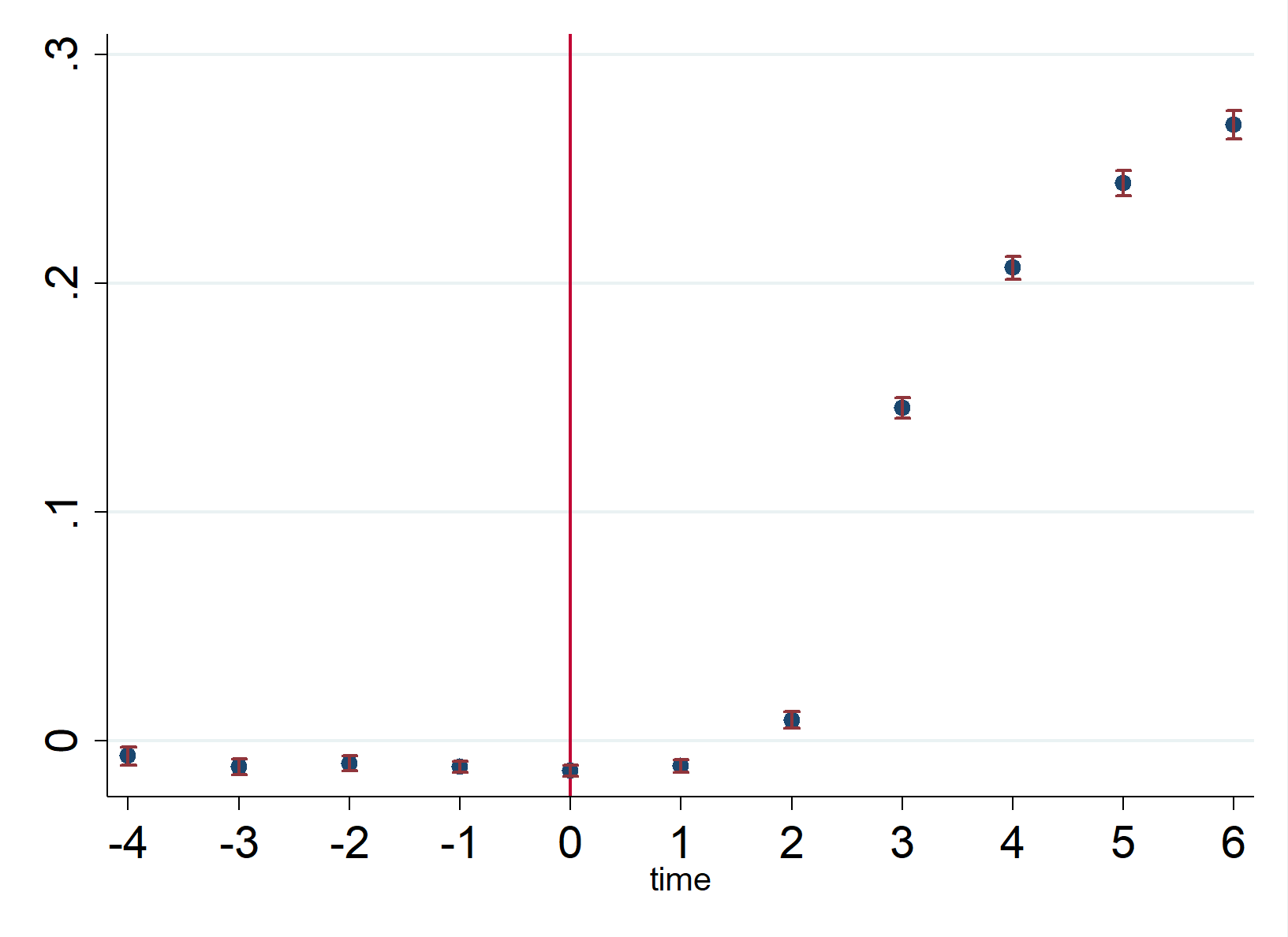}
              
                  	Panel (iii) - Revolving Balance \hspace{1cm}Panel (iv) - Harsh default\\
                     
             \includegraphics[height=4cm, width=7cm]{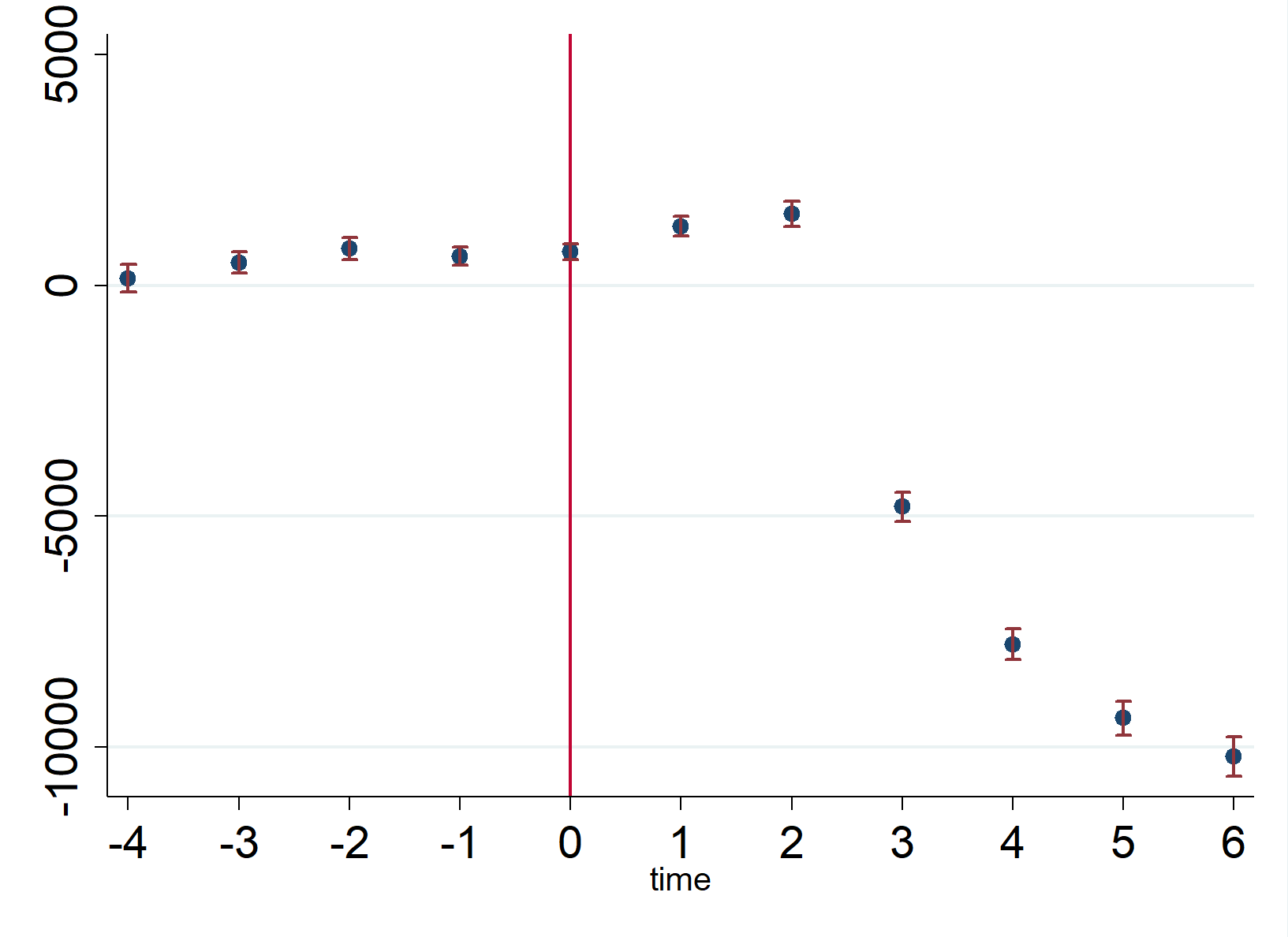}          	            	                \includegraphics[height=4cm, width=7cm]{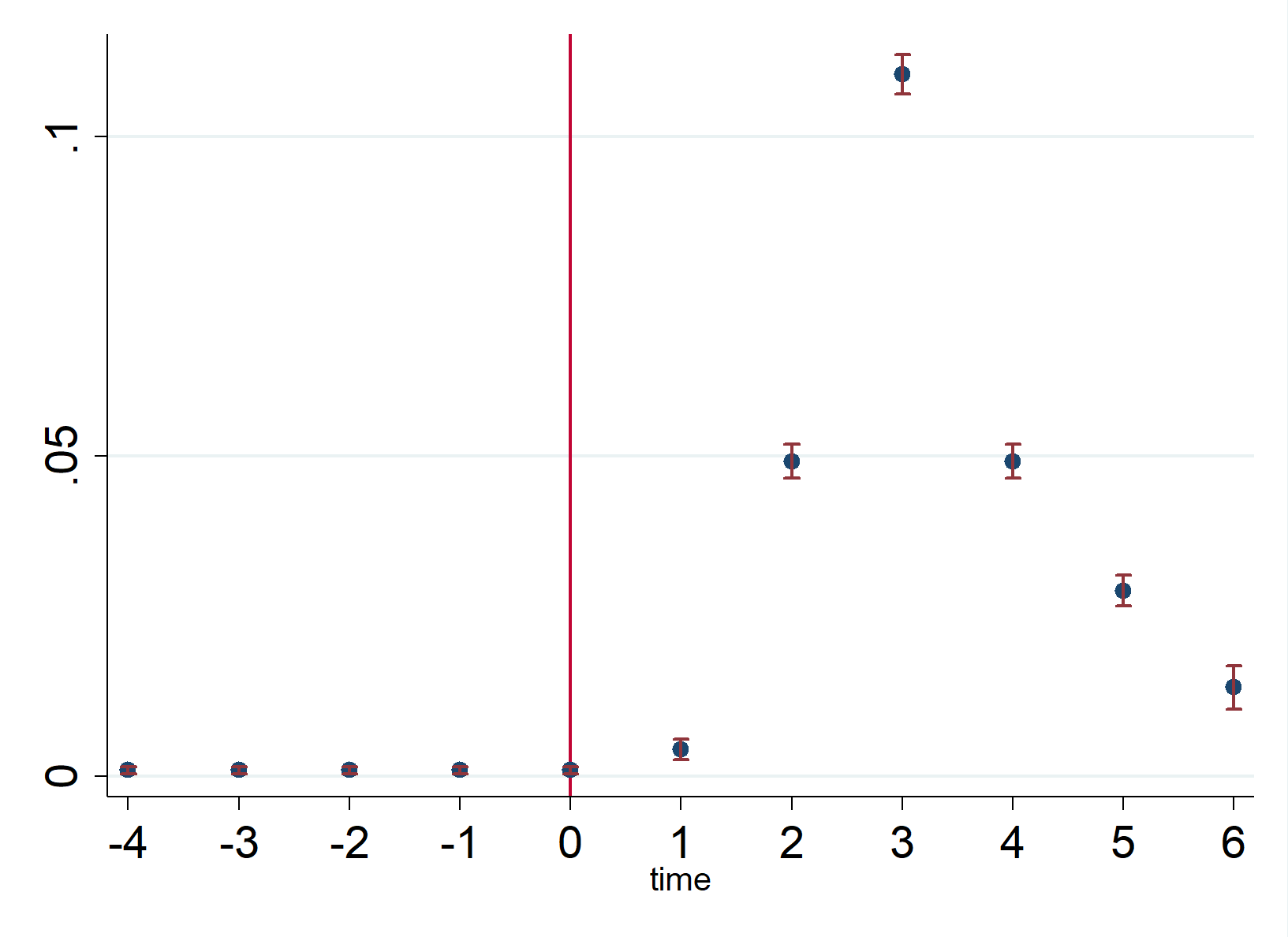}
                
                	Panel (v) - Home own \hspace{1cm}Panel (vi) - Total credit limit\\
                     
             \includegraphics[height=4cm, width=7cm]{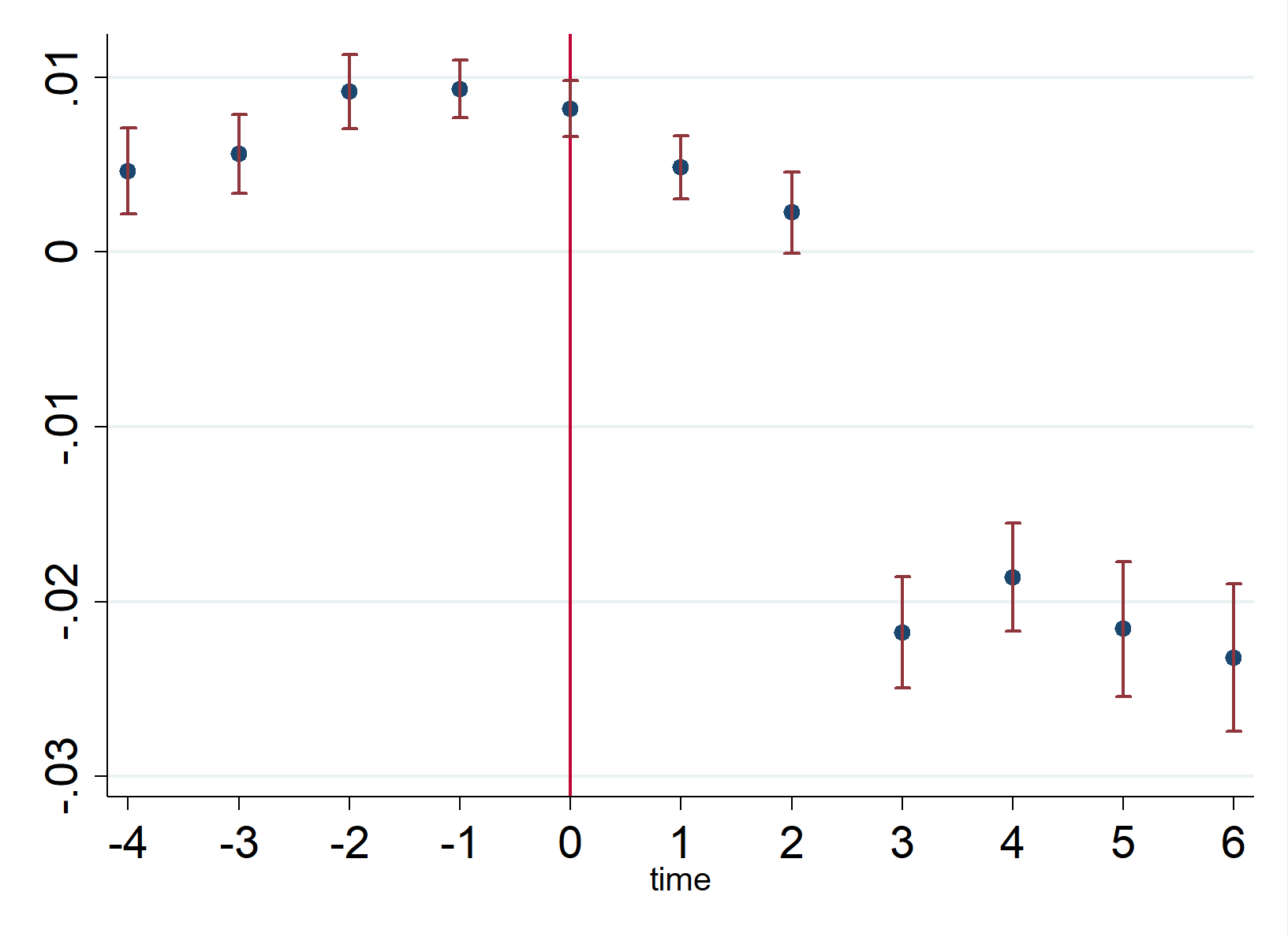}          	            	                \includegraphics[height=4cm, width=7cm]{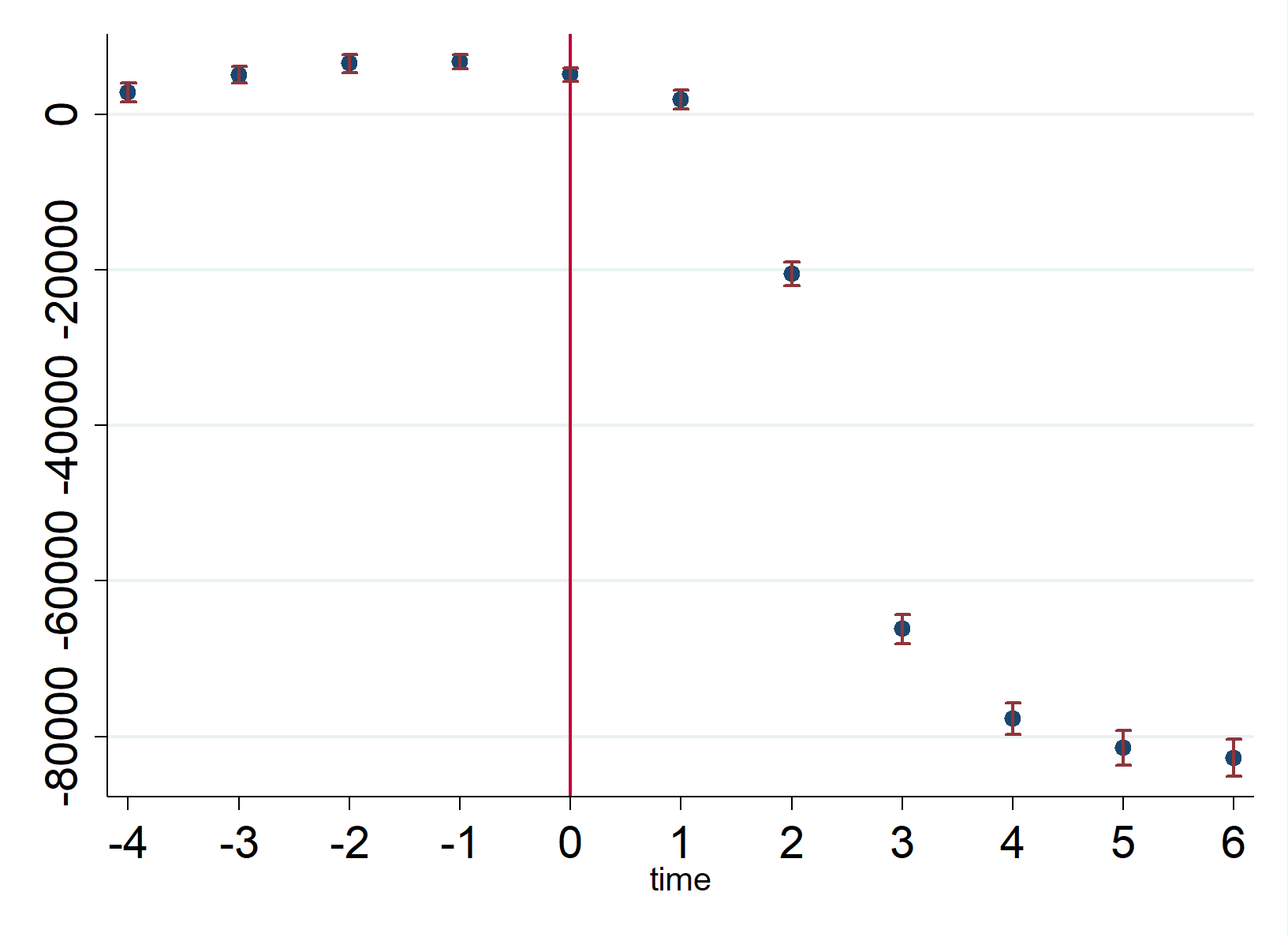}
                
                Panel (vii) - Mortgage balance open\\
                     
            \includegraphics[height=4cm, width=7cm]{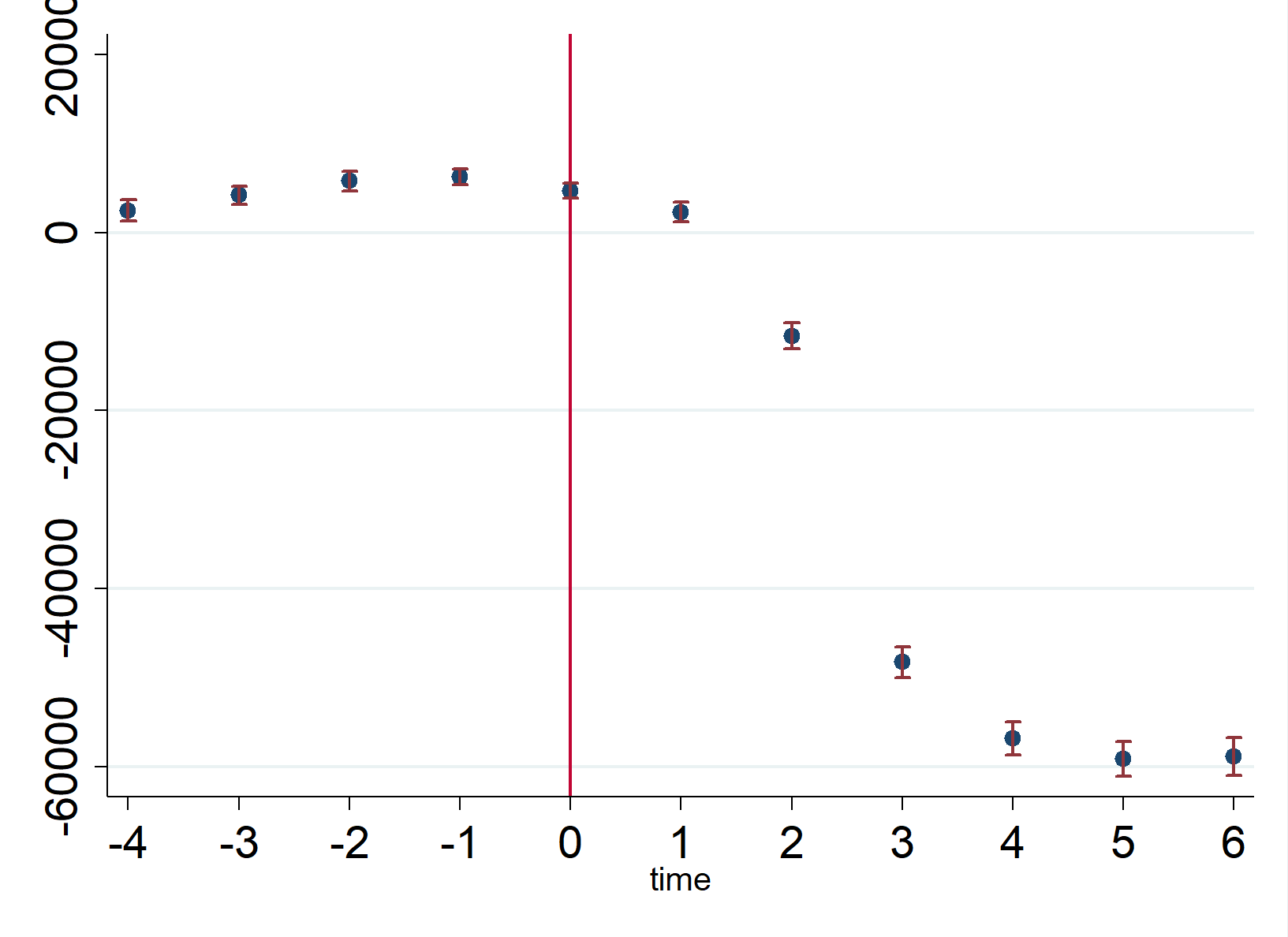}

        \caption{\small Event study: dependent variable is: 
        (i) Mortgage origination: this variable takes value 1 if the individual has a higher number of mortgage trades in year $t$ than in year $t-1$ or if the number of months since the most recent mortgage has been opened is less than 12, and zero otherwise, (ii) probability that total credit limit is lower than 10,000USD, (iii) total amount open on all revolving credit trades, (iv) probability of experiencing a harsh default (Chapter 7, Chapter 13 or foreclosure), (v) probability of being homeowner, i.e. either being recorded as a homeowner by Experian or having ever had a mortgage open (vi) total credit limit on all trades, (vii) open amount of mortgage balance. 
        The event considered is a soft default, i.e. a 90-day delinquency, but no Chapter 7, Chapter 13 or foreclosure taking place in the same year, neither before in the sample period. Other controls are age and age squared, value of the credit score in the two years before the soft default, state dummies. Treated in 2016 (last year in the sample) used as control group. 95\% confidence intervals around the point estimates.}     \label{fig:fig7}
     \end{figure}

      Finally, from the comparison of Figure \ref{fig:event3} with Figure \ref{fig:fig7}, we notice that the results relative to the other credit variables are also consistent across the two model specifications. In particular, we still find that the probability of opening a new mortgage significantly decreases by about 1pp after a soft default, whereas the probability of having a low credit limit still increases by about 20pp. The negative impacts on the revolving credit balance and on the total credit limit are, respectively, minus about 7,000USD and minus about 75,000USD, i.e. notably close to the baseline results presented in the previous Section. 
     Further, we still find that a soft default is associated with an increase by about 10pp in the probability of experiencing a harsh default, as well as with a relevant (about minus 2pp) decrease in the probability of being homeowners.

\begin{figure}
    	Panel (i) - Credit card consumption \hspace{1cm}Panel (ii) - Amount 90-180 days delinquent\\
             
            \includegraphics[height=4.5cm, width=7cm]{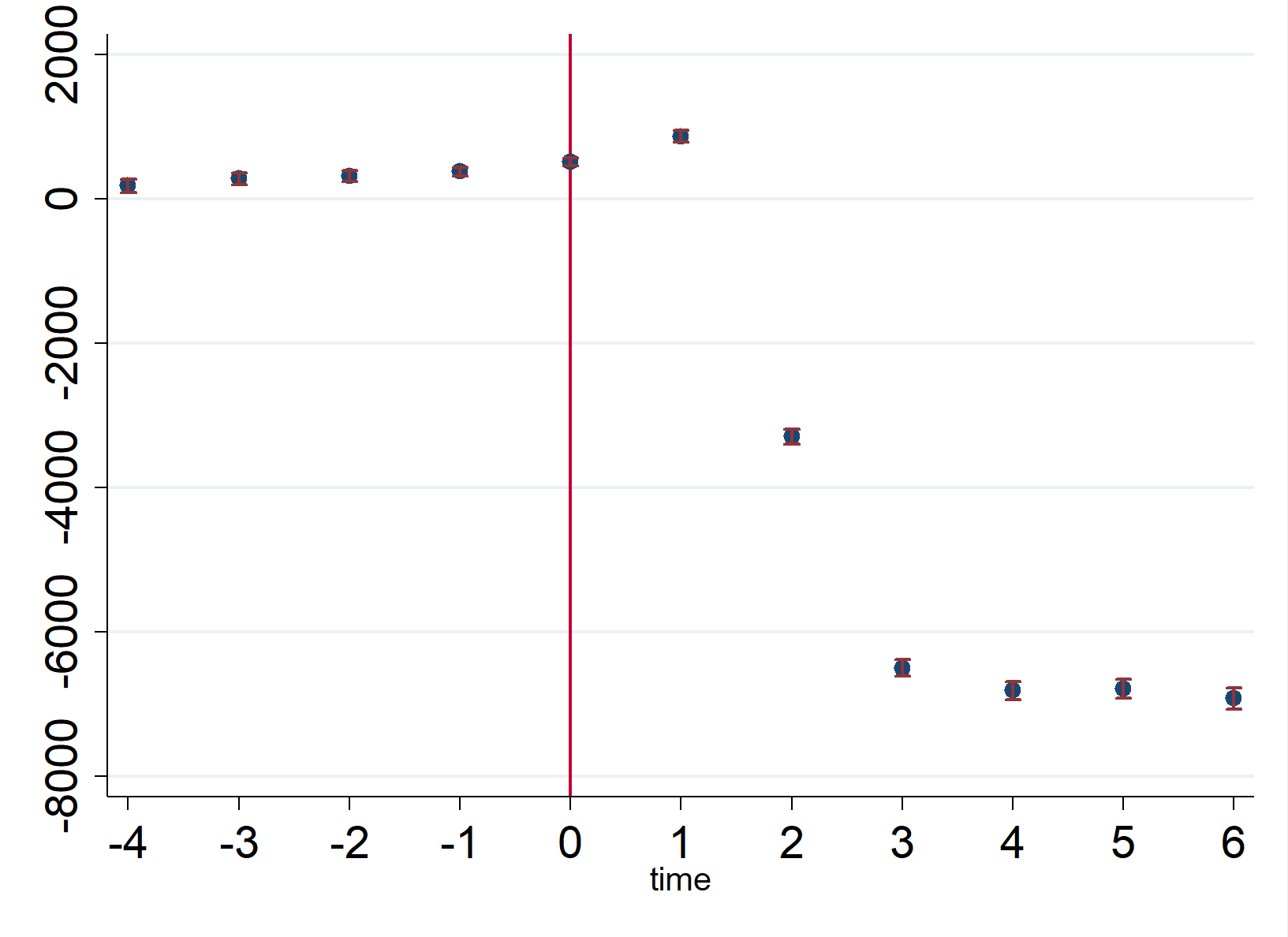}
                \includegraphics[height=4.5cm, width=7cm]{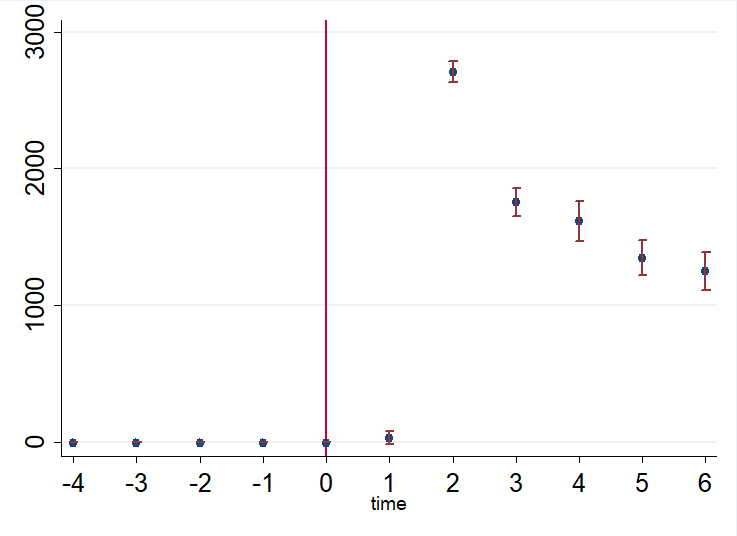}
             
                  	Panel (iii) - Number of collections\\
                              \includegraphics[height=4.5cm, width=7cm]{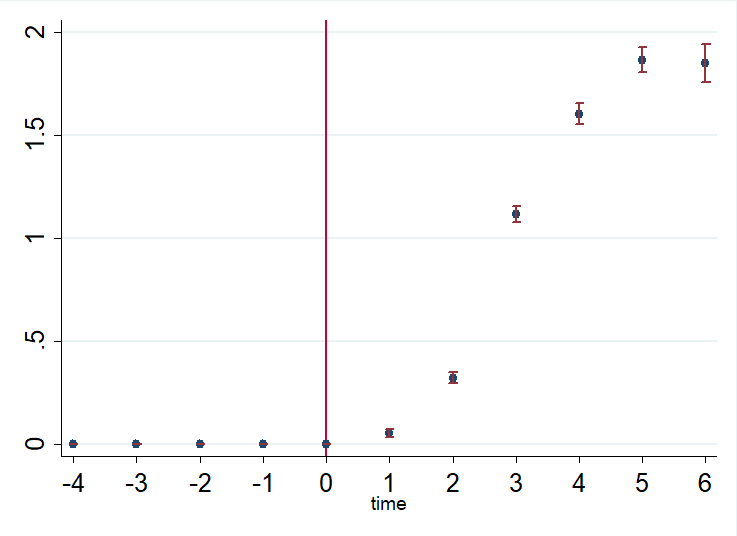}

        \caption{Event study: dependent variable is: 
        (i) credit card consumption: total balance on all open credit card trades reported in the last 6 months, (ii) amount 90-180 days delinquent, (iii) Number of collections. 
         The event considered is a soft default, i.e. a 90-day delinquency, but no Chapter 7, Chapter 13 or foreclosure taking place in the same year, neither before in the sample period. Other controls are age and age squared, value of the credit score in the two years before the soft default, state dummies. Treated in 2016 (last year in the sample) used as control group. 95\% confidence intervals around the point estimates.}      \label{fig:fig8}
     \end{figure}

Finally, the effect on the amount of mortgage balance, about minus 60,000USD, is also fully in line with the baseline results.
     Finally, from Figure \ref{fig:fig8}, we find evidence that credit card consumption declines by about 6,000USD following a soft default, whereas the amount delinquent rises by about 2,500USD and the number of collections grows by about 1-2. All these results are broadly consistent with our baseline results presented in Section \ref{sec:exploratory}.
     
\end{document}